\documentclass[a4paper,12pt,openany]{book}

\usepackage{blindtext} 
\usepackage{amsmath} 
\usepackage{amssymb} 
\usepackage{braket}
\usepackage[
	colorlinks=true
	,breaklinks
	]{hyperref} 
\usepackage[hyphenbreaks]{breakurl}  
\usepackage{xcolor}
\definecolor{c1}{rgb}{0,0,1} 
\definecolor{c2}{rgb}{0,0.3,0.9} 
\definecolor{c3}{rgb}{0.3,0,0.9} 
\definecolor{brown(traditional)}{rgb}{0.59, 0.29, 0.0}
\hypersetup{
    linkcolor={c1}, 
    citecolor={c2}, 
    urlcolor={c3} 
}
\usepackage[round,authoryear]{natbib} 
\usepackage[nottoc]{tocbibind} 
\usepackage{graphicx} 
\usepackage{longtable} 
\usepackage{bigstrut} 
\usepackage{enumerate} 
\usepackage{todonotes} 
\usepackage{makeidx} 
\usepackage{float}
\usepackage{subcaption}
\usepackage{enumitem}
\usepackage{lscape}
\newtheorem{theorem}{Theorem}
\usepackage{arydshln}
\usepackage[british]{babel}
\usepackage[autostyle]{csquotes}
\setlist{nolistsep}
\newsavebox{\bigimage}
\MakeOuterQuote{"}

\makeindex

\usepackage[top=2cm, bottom=1.8cm,left=2.5cm,right=2.5cm]{geometry} 
\usepackage{fancyhdr} 

\begin{document}

\begin{titlepage}

\newcommand{\HRule}{\rule{\linewidth}{0.5mm}} 

\center 
 

\textsc{\LARGE University of Cape Town}\\[1.5cm] 


\HRule \\[0.4cm]
{ \huge \bfseries Chaotic behaviour of disordered\\[0.5cm]
  nonlinear lattices}\\[0.4cm] 
\HRule \\[1.5cm]
 
{\huge
Bob \textsc{Senyange}} \\[2cm]

{\Large
A thesis presented for the degree of \\[0.3cm]
Doctor of Philosophy}\\[1cm]

\includegraphics[width=0.4\textwidth]{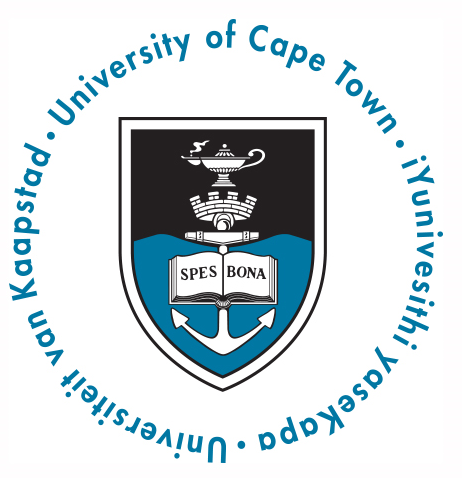}\\[1cm] 
 
\Large Department of Mathematics and Applied Mathematics\\[1.5cm]


{\large April 2021}\\[1.5cm] 

 \Large
Supervisor: Associate Professor Haris Skokos

\vfill 

\end{titlepage}

\pagestyle{plain}
\pagenumbering{roman}
{\LARGE The copyright of this thesis vests in the author. No quotation from it or information derived from it is to be published without full acknowledgement of the source. The thesis is to be used for private study or non-commercial research purposes only.
\newline\newline
Published by the University of Cape Town (UCT) in terms of the non-exclusive license granted to UCT by the author.}
\newpage
\section*{Abstract}
In this work we systematically investigate the chaotic energy spreading 
in prototypical models of disordered nonlinear lattices, 
the so-called disordered Klein-Gordon (DKG) system, in one 
(1D) and two (2D) spatial dimensions. The normal modes’ exponential 
localization in 1D and 2D heterogeneous linear media explains 
the phenomenon of Anderson Localization. Using a modified version 
of the 1D DKG model, we study the changes in the properties of the 
system’s normal modes as we move from an ordered version 
to the disordered one. We show that for the ordered case, the 
probability density distribution of the normal modes’ frequencies 
has a \textquoteleft{U}\textquoteright-shaped profile that gradually turns into a plateau for 
a more disordered system, and determine the dependence of two 
estimators of the modes’ spatial extent (the localization volume 
and the participation number) on the width of the interval from 
which the strengths of the on-site potentials are randomly selected. 
Furthermore, we investigate the numerical performance of several 
integrators (mainly based on the two part splitting approach) for 
the 1D and 2D DKG systems, by performing extensive numerical 
simulations of wave packet evolutions in the various dynamical 
regimes exhibited by these models. In particular, we compare 
the computational efficiency of the integrators considered by 
checking their ability to correctly reproduce the time evolution 
of the systems’ finite time maximum Lyapunov exponent estimator $\Lambda$ 
and of various features of the propagating wave packets, and determine 
the best-performing ones. Finally we perform a numerical investigation 
of the 
characteristics of chaos evolution for a spreading wave packet in 
the 1D and 2D nonlinear DKG lattices. We confirm the slowing down of the 
chaotic dynamics 
for the 
so-called weak, strong and selftrapping chaos 
dynamical regimes encountered in these systems, without showing any 
signs of a crossover to regular behaviour. We further substantiate 
the dynamical dissimilarities between the weak and strong chaos 
regimes by establishing different, but rather general, values for 
the time decay exponents of $\Lambda$. In addition, the spatio-temporal 
evolution of the deviation vector associated with $\Lambda$ reveals the 
meandering of chaotic seeds inside the wave packets, supporting 
the assumptions for chaotic spreading theories of energy.

\newpage
\section*{Declaration}
I know the meaning of plagiarism and declare that all the work on the 
\textit{Chaotic behaviour of disordered nonlinear lattices} contained in this 
thesis except that which is properly acknowledged as references is my own work 
and  
that it has not been submitted before for any degree or examination in any 
University.

 \vspace{0.6in}


\vspace{0.3in}

Bob Senyange \qquad \date{today}
\newpage
\section*{Acknowledgements}

I would like to firstly express my deepest gratitude to my thesis 
advisor Assoc.~Prof.~Haris Skokos for having granted me the chance to 
work with him, his collaborators and team of students. Most notably 
his guidance and mentorship, facilitation, availability for 
discussion at all times and 
patience whenever things did not go as planned.
\vspace{0.15in}

I am also grateful to all colleagues and visitors of the research group, 
led by Assoc. Prof. Skokos,
most especially: A.~Schwellnus, 
B.~K.~Mfumadi, 
A.~Ngapasare, M.~Hillebrand, B.~M.~Many, A.~J.~Chinenye, 
H.~Moges and J-J.~Du~Plessis 
for their academic and social contribution during the course of my studies 
and in particular for the insightful scientific discussions. Fellow 
colleagues and 
friends, Dr.~H.~J.~B.~Njagarah, Dr.~B.~Aineamani, J.~Nagawa, M.~Isiagi 
and L.~Mthombeni, 
are much appreciated 
for 
welcoming me to Cape Town and ensuring my well-being at all times.
\vspace{0.15in}

I additionally express my thanks to the members of the Center for High 
Performance Computing (CHPC, South Africa), whose facility 
I used for my computations, especially  
Mr. Kevin Colville who 
attended to me from the time I started learning scientific computing. I express 
my gratitude to the 
University of Cape Town’s (UCT) ICTS High Performance Computing team
whose services offered HPC training courses and a working space on the 
University 
facilities for my computations. I am also grateful to 
the staff of 
the Department of 
Mathematics and Applied Mathematics (MAM) for providing a pleasant working 
environment. Great thanks to Mr. Nimrod Matotong of MAM for ensuring my working 
space and data stored on ‘Zeus’ MAM computing facility are as required and for 
attending to my immediate computer needs especially software installations and 
repairs.
\vspace{0.15in}

I am so grateful to my family, especially my mother 
for her encouraging words, my dear wife Lilian for her love, patience,
and great support during the time I was away for studies and the period 
of 
writing this thesis, my daughters Josephine Gemma and Mariah Antonia, my son 
Leroy and all my siblings for the prayers and having to bear with my absence. 
‘Eri bazadde bange, Taata (Omukama akuwummuze mirembe) ne Maama Nyonjo, 
mweebale kunteekerateekera kkubo lino eryatandikira ddala okuva mu buto 
bwange’. 
I am grateful to God for He heard our prayers and made a way for me even when 
it seemed impossible.
\vspace{0.15in}

Last but not the least, I also thank Muni University (Arua, Uganda) 
for awarding me the African Development Bank (ADB V-HEST) scholarship 
($2016$-$2019$) towards my studies, the UCT 
for awarding me the International \& Refugee students’ scholarship 
$2018$ 
and the various institutions or 
persons that facilitated my travels to 
the conferences I attended.
\newpage
\section*{Publications and Talks related to this work}
Parts of the work in this thesis have been presented in the 
international peer reviewed journal papers listed below.
\vspace{0.2in}
\begin{itemize}
	\item [P-1] {\bf B. Senyange} and Ch. Skokos, \textit{Computational efficiency of symplectic integration schemes: Application to multidimensional disordered Klein–Gordon lattices}, Eur.~Phys.~J.~Spec.~Top. {\bf 227}, 625 (2018).
	\vspace{0.1in}
	\item [P-2] {\bf B. Senyange}, B. Many Manda, and Ch. Skokos, \textit{Characteristics of chaos evolution in one-dimensional disordered nonlinear lattices}, Phy.~Rev.~E. {\bf 98}, 052229 (2018).
	\vspace{0.1in}
	\item [P-3] B. Many Manda, {\bf B. Senyange}, and Ch. Skokos, \textit{Chaotic wave-packet spreading in two-dimensional disordered nonlinear lattices},  Phy.~Rev.~E. {\bf 101}, 032206 (2020).
	\vspace{0.1in}
	\item [P-4] {\bf B. Senyange}, J.-J. du Plessis, B. Many Manda, and Ch. Skokos, \textit{Properties of normal modes in a modified disordered Klein–Gordon lattice: From disorder to order}, Nonlinear Phenomena in Complex Systems, {\bf 23}, $165{-}171$ (2020).
\end{itemize}

\vspace{0.3in}

Some results from this work have also been included in the following oral 
and poster conference presentations delivered by B. Senyange.
\vspace{0.15in}
\begin{itemize}
	\item [C-1] \textquotedblleft On the symplectic integration of the Klein-Gordon lattice model\textquotedblright, \textit{The $3^{rd}$ EAUMP conference}, Makerere University, Kampala (Uganda), October $26{-}28$, 2016 (Oral presentation).
	\vspace{0.1in}
	\item [C-2] \textquotedblleft Chaotic dynamics of the disordered Klein-Gordon lattice\textquotedblright, \textit{The International Scientific Workshop: Recent Advances in Hamiltonian and Nonholonomic Dynamics}, Moscow (Russia), June $15{-}18$, 2017 (Oral presentation).
	\vspace{0.1in}
	\item [C-3] \textquotedblleft Chaotic dynamics of one-dimensional disordered nonlinear lattices\textquotedblright, \textit{Nonlinear Localization in Lattices}, Spetses (Greece), June $18{-}22$, 2018 (Poster presentation).
	\vspace{0.1in}
	\item [C-4] \textquotedblleft On characterising the dynamics of 1-D disordered nonlinear Hamiltonian lattices\textquotedblright, \textit{The annual Southern Africa Mathematical Sciences Association (SAMSA) conference}, Palapye (Botswana), November $19{-}22$, 2018 (Oral presentation).
	\vspace{0.1in}
	\item [C-5] \textquotedblleft Chaotic dynamics in disordered nonlinear lattices\textquotedblright, \textit{The International Congress on Industrial and Applied Mathematics (ICIAM)}, Valencia (Spain), July $15{-}19$, 2019 (Oral presentation). 
	\vspace{0.1in}
	\item [C-6] \textquotedblleft Chaotic dynamics in disordered nonlinear lattices: Symplectic integrators\textquotedblright, \textit{The International Congress on industrial and Applied Mathematics (ICIAM)}, Valencia (Spain), July $15{-}19$, 2019 (Poster presentation). 
	\vspace{0.1in}
	\item [C-7] \textquotedblleft Chaotic dynamics in disordered nonlinear lattices: Symplectic integrators\textquotedblright, \textit{The 62nd Annual Congress of South African Mathematical Society (SAMS)}, Cape Town (South Africa), December $2{-}4$, 2019 (Oral presentation). 
	\vspace{0.1in}
	\item [C-8] \textquotedblleft Symplectic integration techniques for disorderd nonlinear Hamiltonian lattices\textquotedblright, \textit{The annual Southern Africa Mathematical Sciences Association (SAMSA) conference}, Virtual, November $23{-}25$, 2020 (Oral presentation).
\end{itemize}
\newpage
\section*{List of abbreviations}
\begin{description}
	\item[1D -] one-dimensional
	\item[2D -] two-dimensional
	\item[AL -] Anderson Localization
	\item[BCH -] Baker Campbell Hausdorff
	\item[BEC -] Bose–Einstein condensate
	\item[DDLSE -] Disordered discrete linear Schr$\ddot{o}$dinger equation
	\item[DDNLS -] Disordered discrete
	nonlinear Schr\"{o}dinger
	\item[DFPUT -] Disordered Fermi Pasta Ulam Tsingou
	\item[DKG -] Disordered Klein-Gordon
	\item[DNA -] Deoxyribo-Nucleic Acid
	\item[DVD -] Deviation vector distribution
	\item[FPUT -] Fermi Pasta Ulam Tsingou
	\item[IC -] Initial condition
	\item[KG -] Klein-Gordon
	\item[LCE -] Lyapunov characteristic exponent
	\item[mLCE -] maximum Lyapunov characteristic exponent
	\item[NM -] Normal mode
	\item[PBD -] Peyrard–Bishop Dauxois
	\item[SI -] Symplectic integrator
	\item[TDH -] Tangent dynamics Hamiltonian
	\item[TM -] Tangent map
\end{description}
\input{ch_01_acknowledgements}
\tableofcontents
\pagestyle{plain}


\pagenumbering{arabic}
\chapter{Introduction}\label{chap:intro}

\pagestyle{fancy}
\fancyhf{}
\fancyhead[OC]{\leftmark}
\fancyhead[EC]{\rightmark}
\cfoot{\thepage}
Disordered models are extended systems in space with many degrees of 
freedom which try to mimic heterogeneous arrangements observed in natural 
phenomena. 
Heterogeneity is introduced in the model by attributing a unique 
random value (or a set of unique random values) for each degree of 
freedom 
in relation to one (or many)
of the system’s parameter(s). 
Anderson Localization (AL), a phenomenon where energy excitations remain localized in linear systems with sufficient disorder, 
was discovered about six decades ago by physicist 
Phillip Warren Anderson [\cite{Anderson1958}] in a 
work investigating the underlying mechanisms of electron transport 
in crystals, which won him the Nobel Prize in 1977. 
Anderson considered the tight-binding approximation for 
an electron on a lattice with randomly distributed on-site energies 
and nearest-neighbour tunnelling. 
Electron inability to diffuse away from an initial position 
was related to the wave function amplitude falling off exponentially 
with growing distance from its original location. 
This wave localization behaviour is a phenomenon of phase-coherence. Therefore, 
the electronic state keeps 
phase coherence, while loss of coherence leads to delocalization 
[\cite{Rayanov2013}].
AL manifests in many types of wave propagation in disordered systems, 
for example in the dynamics of 
Bose–Einstein condensates, the conductivity process of materials, etc.
AL has been extensively investigated in theoretical and numerical 
[\cite{Kopidakis2008,Pikovsky2008,Flach2009b,Laptyeva2014,Prat2019}], 
and experimental
[\cite{Hu2008,Lahini2008,Cobus2016}] studies 
in the context of metal-insulator transitions, dynamics of ultracold 
atoms in optical arrays, light propagation in photonic crystals and 
the quantum Hall effect [\cite{Kramer1993,Evers2008,Sanchez-Palencia2010}].
Heterogeneity in disordered systems can be either correlated or uncorrelated 
but still leads to localization. In his study, Anderson [\cite{Anderson1958}] used uncorrelated 
disorder.
In general, the normal modes (NMs) of linear 
systems with sufficiently strong 
disorder are localized and therefore any initially localized wave packet 
does not exhibit a spreading behaviour for all time. The introduction
of nonlinearities to such systems leads to NM interaction
and hence the emergence of chaos. 

The nonlinearity effect in disordered systems has attracted 
extensive attention in 
experiments 
[\cite{Dalichaouch1991,Schwartz2007,Billy2008,Lahini2008,Roati2008}], 
as well as in 
theory and simulations [\cite{Kopidakis2008,Pikovsky2008,Flach2009b,Skokos2009,Garcia-Mata2009,Veksler2009,Laptyeva2010,Basko2011,Mulansky2012,Mulansky2013}]
and it has been primarily investigated 
using 
the one-dimensional (1D)
disordered discrete
nonlinear Schr\"{o}dinger equation (DDNLS) and 
the disordered nonlinear Klein-Gordon (DKG) lattice models 
[\cite{Pikovsky2008,Flach2009b,Skokos2009,Flach2010,Laptyeva2010,Skokos2010a,Bodyfelt2011,Skokos2013,Laptyeva2014}].  
It was revealed in these research papers that AL is eventually 
destroyed by nonlinearity 
and the characteristics of different spreading behaviours, namely the 
so-called ‘weak’, ‘strong’ and  ‘selftrapping’ 
regimes of chaos, were identified and their 
appearance was theoretically explained. 
From the models' structural point of view, the appearance of these 
dynamical regimes can be explained using the systems' NMs properties, 
that is to say, 
the NMs' localization volume, the width of their frequency band, 
the average eigenvalue spacing of the modes which strongly 
interact with a particular 
NM, and the magnitude of these quantities compared to the nonlinearity induced 
frequency shift 
[\cite{Flach2009b,Laptyeva2010,Flach2010,Laptyeva2014}].

The studies performed on the 1D DKG model reported in \cite{Skokos2013} 
confirmed that nonequilibrium chaos and phase decoherence persist in a 
disordered nonlinear system thus fuelling the prediction of a complete 
delocalization.
However, the following are a number of interesting related open questions 
that have been raised in the past few years:
What is the connection between chaoticity and energy transport?
How does the initial excitation affect the chaoticity and the nature 
of it's evolution in time? 
How does the appearance of chaos depend on the basic 
	dynamical parameters of the system, i.e. the disorder and 
	nonlinearity strengths? 
Are wave packets homogeneously chaotic, or there exist 
	‘chaotic hot spots’, with some degrees of freedom behaving
	‘more’ chaotically than others? 
Will wave packets eventually exhibit a less chaotic behaviour, 
leading to the halt of spreading [\cite{Johansson2010,Aubry2011}], or will they 
spread indefinitely 
as recent numerical simulations have indicated 
[\cite{Kopidakis2008,Pikovsky2008,Flach2009b,Skokos2009,Garcia-Mata2009,Laptyeva2010,Skokos2010a,Bodyfelt2011,Skokos2013}]?
What are the basic principles and the intrinsic dynamical 
	components that define the observed chaotic behaviours? 
	How the observed dynamical behaviours depend on the dimensionality of 
	our lattice and more explicitly how all these behaviours alter when 
	we consider 2D disordered systems?

The scope of this thesis is to set the basis for a systematic study of 
these questions providing some concrete answers to them.
For this purpose we use the maximum Lyapunov characteristic exponent 
(mLCE) to characterize chaos 
along with the deviation vector distributions (DVDs) [\cite{Skokos2013}], which 
is used to identify possible ‘chaotic hot spots’. Extensive 
numerical simulations of wave packets evolution for different initial 
excitations and for several values of disorder and nonlinearity strengths are 
performed and their dynamics is analysed by the above mentioned techniques in 
an attempt to uncover the mechanisms that introduce chaos in disordered 
systems. 
We note that although we will mainly 
present results for a 1D disordered lattice 
namely the 1D DKG system, a 
significant part of our analysis will be devoted to the 
less studied case of 2D lattices in order to 
get a broader understanding of the generality of our results for 
the chaotic dynamics of disordered lattices in spatial dimensions greater than 
one.

The thesis is organised in six chapters as follows. 
In Chapter~\ref{chap:disor_sys} we present a review of 
some theoretical, experimental and numerical aspects of energy localization in 
linear disordered systems. We also discuss some theoretical results about 
the properties of energy 
spreading in typical 1D lattices followed by the structural properties of 
their NMs through a wide range of disorder strengths. 
In Chapter~\ref{chap:num_tech}, we present in detail 
the numerical techniques used to obtain our 
findings. 
In Chapter~\ref{chap_1d} we discuss results for the 1D DKG system while we extend our 
investigation to the computationally more challenging case of the 2D 
DKG model  
in Chapter~\ref{chap:2D}. 
Finally in Chapter~\ref{chap:summary} we summarize the results of our 
work. 
\chapter{Disordered lattices}\label{chap:disor_sys}
Disordered systems exist in nature, for example in cases where 
imperfections or impurities appear in materials, and disordered models 
try to capture their dynamics. We can therefore consider disordered 
systems to be models with many degrees of 
freedom, extended in space, exhibiting heterogeneity as observed in 
nature. 
For such a system, heterogeneity is modelled by allocating random values 
to one or more of the system's parameters in each degree of freedom.
In 1958, a phenomenon where in linear disordered systems, whose eigenstates 
are also localized, 
energy excitations do not propagate but remain localized was studied by 
Anderson [\cite{Anderson1958}].
This behaviour, a study which won him a Nobel Prize in 1977, is usually called 
AL. AL can be used to explain the evolution of many physical processes 
for example, the 
conductivity of materials and the dynamics of Bose–Einstein condensates 
(BECs).\\

In this chapter we present some examples of widely studied ordered models and 
also discuss briefly the theory of disordered lattices.
The 
chapter is presented in the following order: 
We briefly present some widely studied ordered models in 
Section \ref{sec:ordered} 
and discuss their dynamical 
properties and characteristics in relation to disordered 
systems. 
Section \ref{sec:lineardis} discusses the underlying dynamical 
properties of linear models 
using as an example the 
well known Anderson model [\cite{Anderson1958}] for two different 
forms of disorder. We then discuss the dynamics of nonlinear models in 
Section \ref{sec:nonlineardis} and give some 
examples emphasizing already published results for them. 
In particular, we focus on the two main models used in our research namely, the 
DKG chain of anharmonic oscillators and the 
DDNLS system. For small wave packet 
amplitudes, there exists a direct correspondence between the 
parameters of the DDNLS system and those of the DKG model. 
Because of this, we only 
theoretically discuss in detail 
the 
properties of disordered models using the DDNLS model. 
Section \ref{sec2:m2_P} describes some of the measures 
we use in quantifying the spatial extent of wave packets, and lastly, 
we present results on the properties of NMs for 
a modified Klein-Gordon model having a disorder set 
with a variable width in Section 
\ref{sec2:disord2ord}. Parts of the results 
discussed in Section \ref{sec2:disord2ord} were also presented 
in \cite{Senyange2020}.

\section{Ordered lattices}
\label{sec:ordered}
As a precursor to disordered systems, we present some examples of 
homogeneous Hamiltonian models which have been 
used to understand the phenomenon of energy 
propagation in physical processes.
\subsubsection{Fermi Pasta Ulam Tsingou (FPUT) models}
In 1955, a Los Alamos publication [\cite{Fermi1955}]
introduced and described the now famous FPUT (originally named as 
FPU) problem: a paradox in chaos theory where a number of 
physical systems exhibited almost exactly periodic behaviour. 
This discovery 
was a great contribution to the birth of nonlinear science, 
whose explorations led to the revelation of interesting phenomena like 
for example the existence of solitons (a wave packet propagating at a constant velocity 
while maintaining its shape) [\cite{Sirovich2005}] 
and to the deeper understanding of chaotic dynamics.
The main communication in \cite{Fermi1955} was
the observation of a practical 
recurrence of the initial state
as opposed to the expected state of energy equipartition amongst the 
modes of the FPUT 1D system.
Over the years, studies of the FPUT problem have led to 
deeper understanding of the dynamical behaviour of ordered 
discrete nonlinear lattices 
with many degrees of freedom. 
The FPUT model has been pivotal in the theoretical studies of solitons, 
inverse scattering, discrete breathers, fully integrable systems, 
equipartition of energy 
and the development of Chirikov’s criterion for 
the onset of chaotic motion [\cite{Ford1992,Campbell2005}] in lattice models.
Some of the most recent studies have led to the observation of 
the FPUT recurrence with dependence on space and time [\cite{Goossens2019}], 
regular motion in multidimensional models [\cite{Moges2020}] and dynamics of the model 
in thermal equilibrium [\cite{Amati2020}].

\subsubsection{Klein Gordon models}
The Klein-Gordon (KG) system 
without disorder has been investigated as far back as 1980. 
In \cite{Butera1980}, it was confirmed that for the KG model 
of a lattice field theory, the lower frequency oscillators have 
smaller critical energies (describing the stochasticity border 
separating the regions of chaos from those of order) 
than the higher frequency oscillators.
Two years later, the model was used to probe 
oscillator behaviour as the system approaches thermodynamic equilibrium
[\cite{Fucito1982}]. Further works 
regarding thermalization in the KG chain have been carried out 
including \cite{Giorgilli2015}, \cite{Paleari2016}, \cite{Pistone2018}, and 
\cite{Danieli2019}
among others.

\section{Linear disordered models and Anderson localization}
\label{sec:lineardis}
Linearity/Nonlinearity is a key ingredient to consider when dealing with transport 
properties of energy in disordered lattices. It is well known that 
linearity 
combined with a presence of sufficiently strong disorder
leads to  AL [\cite{Anderson1958}]. 
Wave delocalization, however, is due to loss of phase coherence and so for AL to occur, 
the state of the electrons has to preserve the coherence [\cite{Rayanov2013}]. 
However, in the general sense, AL occurs in many different scenarios of 
wave packet propagation involving a variety of disordered media.
Experimental observation of AL for matter, electromagnetic and acoustic waves 
in disordered systems has been reported. 
For example in \cite{John1983}, phonon localization in heterogeneous 
elastic media 
was predicted and four years later, in \cite{John1987} the critical behaviour of 
electromagnetic waves in heterogeneous dielectric systems was 
described, while in \cite{Wiersma1997} an 
optical experiment showing 
AL of light 
was reported. We now briefly describe the Anderson model in one and 
two spatial dimensions and mention some 
studies where AL has been observed.

\subsubsection{Anderson model}\label{sec:anderson}
In 1D, the $N$ oscillator Anderson model 
(a tight-binding approximation 
model 
for electron motion) is described using the 
Hamilton function [\cite{Anderson1958}] 
\begin{equation}\label{eq:anderson}
H_{1A} = \sum_{l=1}^N\left[\epsilon_l|\psi_l|^2 - \left(\psi_{l+1}\psi_l^* + \psi_{l+1}^*\psi_l\right)\right],
\end{equation}
with $\psi_l$ being a complex variable denoting the $l^{th}$ site 
wave 
function and $\psi_l^*$ denotes the corresponding complex conjugate.
The random on-site energies $\epsilon_l$, which introduce the disorder, 
are chosen from the distribution 
\[ \mathcal{P}(\epsilon_l)=\begin{cases} \frac{1}{W}, & {\textnormal{for}} \,\,\,|\epsilon_l| < \frac{W}{2};\\
				0, & {\textnormal{otherwise.}} \end{cases}\]
In this case, parameter $W \ne 0$ dictates the magnitude of the onsite 
energies.

\begin{figure}[H]
	\centering
	\includegraphics[width=0.47\textwidth,keepaspectratio]{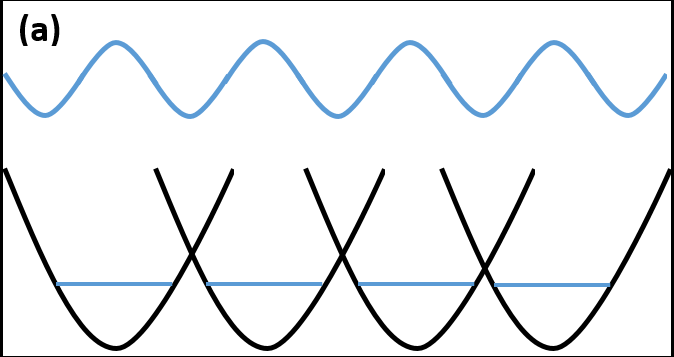}
	\includegraphics[width=0.442\textwidth,keepaspectratio]{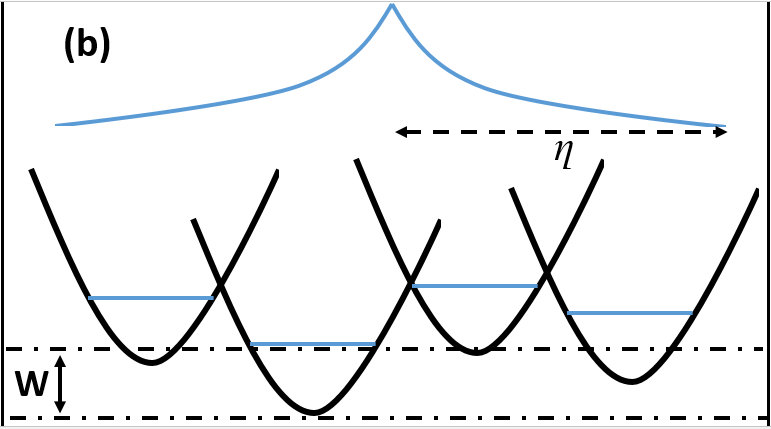}
	\caption{{\bf Bottom:} Representation of the Anderson electron 
		model \eqref{eq:anderson} for a lattice {\bf (a)} without disorder  
		and {\bf (b)} with disorder. {\bf Top:} The electronic wave 
		function is {\bf (a)} extended in the homogeneous case 
		and, {\bf (b)} localized with exponentially decaying tails and 
		localization length {$\eta$} in the disordered case.}
	\label{fig:2_and}
\end{figure}

The schematic representation of 
model \eqref{eq:anderson} where the electron 
tunnels with a unit constant rate between
immediate neighbouring sites is presented in \autoref{fig:2_and}. In 
\autoref{fig:2_and} {\bf (a)}, 
there is no disorder in the system, and the wave function (represented by the 
blue 
sinusoidal curve) is extended. The 
introduction of disorder in the system [\autoref{fig:2_and} {\bf (b)}], 
leads to a spatial wave function (represented by the 
upper blue 
curve) localization with exponentially decaying tails.
The equations of motion of system \eqref{eq:anderson} are generated by 
$\dot{\psi}_l=\frac{\partial H_{1A}}{\partial (i\psi_l^*)}$ and yield the 
1D 
disordered Discrete Linear 
Schr$\ddot{o}$dinger equation (DDLSE):

\begin{equation}\label{eq:andersonmotion}
i\dot{\psi_l} = - \psi_{l+1} + \epsilon_l\psi_l - \psi_{l-1}.
\end{equation}
Taking $\psi_l(t)=A_le^{-i\lambda t}$ and substituting in 
equation (\ref{eq:andersonmotion}) we get the linear eigenvalue problem 
\begin{equation}\label{eq:eigval_DDLSE}
\lambda A_l = \epsilon_l A_l - A_{l+1} - A_{l-1}.
\end{equation}
The system's NMs 
are the normalized eigenvectors $A_{\nu,l}$ (where $\sum_lA_{\nu,l}^2=1$) 
and the frequencies of these 
modes are the corresponding eigenvalues 
$\lambda_{\nu}$. Finding the solution of eigenvalue 
problem (\ref{eq:eigval_DDLSE}) is equivalent to diagonalizing
the tridiagonal matrix ${\bf A}$ with elements
\begin{equation}\label{eq:A_epsilon}
	a_{k,l}=\begin{cases} \epsilon_k, & {\textnormal{for}}\,\,\, k=l;\\
		-1, & {\textnormal{for}}\,\,\, k=l\pm1;\\
	0, & {\textnormal{otherwise.}} \end{cases}
\end{equation}
\begin{theorem}[Gershgorin circle
	theorem]\label{thm:gersh}
	Let ${\bf A}$ be a matrix whose entries are defined in equation 
	\eqref{eq:A_epsilon}. Define $R_k$ as $R_k=\sum_{l}\left| a_{k,l} \right|$. 
	Then, the
	eigenvalues $\lambda_{\nu}$ of
	matrix ${\bf A}$ are bounded as 
	$\left| \lambda_{\nu}-a_{k,k} \right| \leq R_k-\left|a_{k,k} \right|$.
\end{theorem}
Using \autoref{thm:gersh} [\cite{Gerschgorin1931,Wolkowicz1980}], 
we find the upper and lower 
bounds of the eigenvalues 
$\lambda_{\nu}$ to be respectively  
$2+\epsilon_l$ and $-2+\epsilon_l$. Since
$\epsilon_l\in [-W/2,W/2]$, the maximum 
and
minimum possible values of the frequencies are $2+\frac{W}{2}$
and $-2-\frac{W}{2}$ respectively. The 
eigenfrequency spectrum width $\Delta$ therefore is 
\begin{equation*}
\Delta = W+4.
\end{equation*}
Due to the localization nature of the system's eigenstates, the
exponential decay
\begin{equation*}
|A_{\nu,l}| \propto \exp\left({\frac{-\left|l-l_0\right|}{\eta\left(\lambda_{\nu}\right)}}\right)
\end{equation*} describes their asymptotic behaviour,
with $\eta(\lambda_{\nu})$ denoting 
the localization length and $l_0$ the position of 
the spatial centre of the NM. 
The localization length for 
\eqref{eq:anderson} with weak uncorrelated disorder $W\le4$ and for 
eigenfrequencies close to the centre of the spectrum can be estimated as
[\cite{Kramer1993}] 
\begin{equation*}
\eta(\lambda_{\nu}) \approx 24\frac{4-\lambda_{\nu}^2}{W^2}.
\end{equation*}
An alternative computation approach of $\eta(\lambda_{\nu})$  is by 
using the 
transfer matrix method, basing on the 
random matrix theory and various perturbative techniques 
[\cite{Ishii1973,Crisanti1993,Kramer1993}]. Localized states 
of disordered systems can also be quantified 
using Lyapunov characteristic exponents (LCEs), 
intensity-intensity correlation 
functions and the participation number [\cite{Crisanti1993,Kramer1993,Laptyeva2014}].
As a disorder dependent quantity, the localization length is maximized in 
the centre, $\lambda_{\nu} = 0$, of the frequency band where 
$\eta(0)\approx 96/W^2$.
Using the approach of perturbation theory, $\eta(0)\approx [\ln(W/(2e))]^{-1}$ 
for strong disorder $W\ge4$ [\cite{Krimer2010,Laptyeva2014}].

The level of localization of the modes can also be 
characterised by two commonly used quantities, namely, 
the second moment $m_2^{(\nu)}$ which is computed as 
\begin{equation*}
	m_2^{(\nu)} = \sum_l(l_0-l)^2|A_{\nu,l}|^2
\end{equation*}
with 
$l_0 = \sum_ll|A_{\nu,l}|^2$ denoting the centre position of eigenstate 
$\nu$. $m_2^{(\nu)}$ quantifies 
the degree of spreading of the 
$\nu^{th}$ eigenstate. 
The second quantity is the participation number $P_{\nu}$ 
which is computed as
\begin{equation}\label{eq:P_mode}
P_{\nu} = \frac{1}{\sum_l|A_{\nu,l}|^4}
\end{equation} and
estimates the number of sites in the mode with high energy density. 
Additionally, the effective distance $V_{\nu}$ between the exponential 
tails of the eigenstate $\nu$ can be computed as 
[\cite{Krimer2010,Laptyeva2014}]
\begin{equation}\label{eq:V_mode}
V_{\nu}=\sqrt{12m_2^{(\nu)}} + 1.
\end{equation}
Quantities $P_{\nu}$ and $V_{\nu}$ give the exact 
width of a flat, compactly distributed norm density distribution 
and have 
been used to approximate the spatial extent (localization volume) of 
eigenstate $\nu$. However, for weak disorder and with the inclusion 
of fluctuations, the calculation of $P_{\nu}$ 
gives a reduced value 
compared to the correct volume, while the 
$V_{\nu}$ does not. Therefore, we use 
$V_{\nu}$ to
measure the localization volume of eigenstate $\nu$.
The averages 
\begin{equation}\label{eq:Vmod}
V=\overline{V}_{\nu}	
\end{equation} and 
\begin{equation}\label{eq:Pmod}
	P=\overline{P}_{\nu}
\end{equation} of 
these quantities over disorder realizations (sets of disorder numbers 
 $\epsilon_l$ of the model \eqref{eq:anderson})
and modes can be used to 
approximate the
$W$-dependent localization volume and participation number respectively, of 
the NMs.
For weak disorder, the average $V$ 
scales as $3.3\eta(0)$ 
and it tends to $1$ for sufficiently strong 
disorder [\cite{Krimer2010}]. 

This therefore implies that the average spacing 
\begin{equation}\label{eq:dmod}
	d \approx \frac{\Delta}{V}
\end{equation} of NM eigenfrequencies 
in the range
of the NM localization volume gives 
\begin{equation*}
	d \approx \frac{\Delta W^2}{300}
\end{equation*}
for $W\le4$. The scales 
$d$ and $\Delta$ are vital 
in determining the characteristics of wave packet evolution 
in nonlinear systems [\cite{Flach2010,Laptyeva2014}].

Since NMs of disordered systems are spatially exponentially localized, an 
initially localized wave packet which covers a finite 
number $L$ of sites will 
not spread beyond the extent of the NMs containing the excited sites.
For an excitation well within the localization volume (i.e. $L \ll V$), 
the wave packet spreads to fill up the localization volume 
in an approximate integration time of $2\pi/d$ units [\cite{Flach2010}]. 
However, there is no significant expansion of 
the wave packet observed when $L\geq V$. 
Wave 
packets which are under the influence of AL correspond 
to trajectories that evolve on tori in the phase space, following what is usually called quasi-periodic motion.

In two spatial dimensions, the Hamilton function for the 
Anderson model is [\cite{Laptyeva2012,Laptyeva2014}],
\begin{equation*}
H_{2A} = \sum_{k,l}\left[\epsilon_{k,l}|\psi_{k,l}|^2 - 
\left(\psi_{k,l+1}\psi_{k,l}^* +\psi_{k+1,l}\psi_{k,l}^* +
\psi_{k,l+1}^*\psi_{k,l} + \psi_{k+1,l}^*\psi_{k,l}\right)\right],
\end{equation*}
with $\psi_{k,l}$ being a complex variable denoting the wave 
function 
at lattice site $(k,l)$ and $\psi_{k,l}^*$ the corresponding complex conjugate.
The random uncorrelated disorder $\epsilon_{k,l}$  
is chosen uniformly from interval $[-W/2,W/2]$, where $W$ 
is the strength of disorder.
The corresponding equations of motion computed as 
$\dot{\psi}_{k,l}=\frac{\partial H_{2A}}{\partial (i\psi_{k,l}^*)}$ 
are 
\begin{equation}\label{eq:2Dandersonmotion}
i\dot{\psi}_{k,l} = \epsilon_{k,l}\psi_{k,l} - 
\left(\psi_{k-1,l} + \psi_{k+1,l} + \psi_{k,l-1} + \psi_{k,l+1}\right).
\end{equation}
Similar to the 1D case, taking 
$\psi_{k,l}(t)=A_{k,l}e^{-i\lambda t}$ and substituting it in 
equation (\ref{eq:2Dandersonmotion}) we get the linear eigenvalue problem 
\begin{equation}\label{eq:2Deigval_DDLSE}
\lambda A_{k,l} = \epsilon_{k,l} A_{k,l} - A_{k-1,l} - 
A_{k+1,l} - A_{k,l-1} - A_{k,l+1}.
\end{equation}
For a large lattice, based on \autoref{thm:gersh} 
[\cite{Gerschgorin1931}] we see that the bounds of the eigenvalues 
$\lambda_{\nu}$ are 
$-4+\epsilon_{k,l} \leq \lambda_{\nu} \leq 4+\epsilon_{k,l}$, and 
therefore the frequency band is defined by the interval 
$[-4-\frac{W}{2},4+\frac{W}{2}]$.
\newline\newline

We now highlight some literature on AL and the role of disorder in AL.

 Experimental studies for the observation 
of AL in 1D and 2D models have been 
reported in light waves [\cite{Wiersma1997,Scheffold1999,Storzer2006,Schwartz2007,Lahini2008}],
 microwaves [\cite{Dalichaouch1991,Chabanov2000}], sound waves
  [\cite{Weaver1990}], electron gases and 
  matter-waves [\cite{Akkermans2007}].
\cite{Billy2008}, reported the experimental study and observation 
of AL using 
dilute BECs of $^{87}Rb$ 
prepared in an isotropic opto-magnetic hybrid trap. 
The experiment was performed as follows: 
a BEC was loaded onto an optical lattice and its expansion was confined by 
a magnetic trap [\cite{Dalfovo1999,Morsch2006}]. 
A disordered potential 
was created by an optical speckle field when a laser beam was passed through 
a diffusive plate [\cite{Freund2007}].
The BEC was then allowed to diffuse across the optically 
generated disordered potential by turning off the magnetic trap.
Feshbach resonances [\cite{Timmermans1999,Chin2010}] are responsible for reducing the influence of 
interactions between atoms in the condensate to a negligible level. 
On the other hand, 
deep optical potentials are responsible for compensating the influence of 
these interactions. 
The bosonian gas is non-interacting under these conditions and 
so a single wave function can be used to 
characterize it. Localization of the atomic 
wave function 
can be studied in situ, using fluorescence or absorption 
approaches in imaging the 
atomic density. Through this method, findings 
showing the presence of AL for 
BEC in one [\cite{Clement2005,Fort2005,Billy2008}], and three 
[\cite{Jendrzejewski2012,Kondov2011}] spatial dimensions random potentials 
have been reported.

\begin{figure}[H]
	\centering
	\includegraphics[width=0.33\textwidth,keepaspectratio]{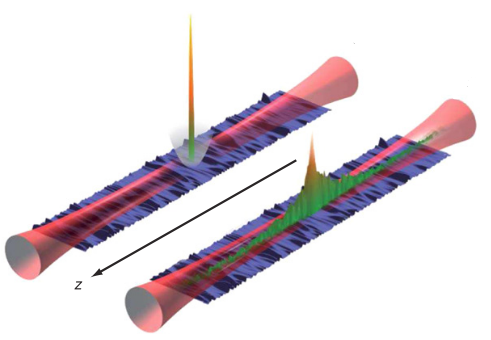}
	\vline
	\includegraphics[width=0.33\textwidth,keepaspectratio]{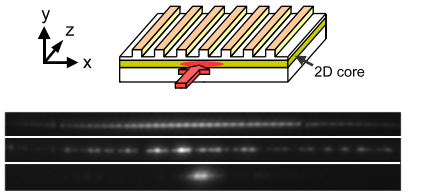}
	\vline
	\includegraphics[width=0.24\textwidth,keepaspectratio]{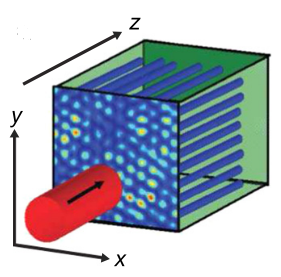}
	\put(-105,100){\bf (c)}
	\put(-260,100){\bf (b)}
	\put(-410,100){\bf (a)}
	\put(-422,14){\bf i}
	\put(-390,1){\bf ii}
	\put(-122,21){\bf i}
	\put(-122,12){\bf\small ii}
	\put(-122,2){\bf\small iii}
	\caption{Experimental observation of AL. 
		{\bf (a)} from \cite{Billy2008}: ({\bf i}) the BEC is initially prepared in a 
		magnetic trap and confined transversely to the $z$-axis in a 1D 
		optical waveguide. ({\bf ii}) When the trap is switched off, the 
		condensate is allowed to expand along the waveguide superimposed 
		with a speckle potential. The BEC stops its expansion 
		after about $0.5$s.
		{\bf (b)} from \cite{Lahini2008}: propagation of light in a 1D disordered waveguide lattice.
		The red arrow indicates the input beam, that generally covers a few 
		lattice sites. Lower panels {\bf i}–{\bf iii} display the output 
		light distribution in the case of ({\bf i}) a periodic lattice, and 
		({\bf ii}), ({\bf iii}) a disordered lattice. 
		{\bf (c)} from \cite{Schwartz2007}: a probe beam entering a disordered lattice, which is periodic 
		in the two transverse dimensions ($x$ and $y$) but invariant in the 
		propagation direction ($z$). In this experiment 
		a triangular (hexagonal) photonic lattice was used. The lattice is induced 
		optically, by transforming the interference pattern among three plane 
		waves into a local change in the refractive index, inside a 
		photorefractive crystal. The input probe beam is always launched at 
		the same location, while the disorder is varied in each realization 
		of the performed experiments.}
	\label{fig:bose1}
\end{figure}
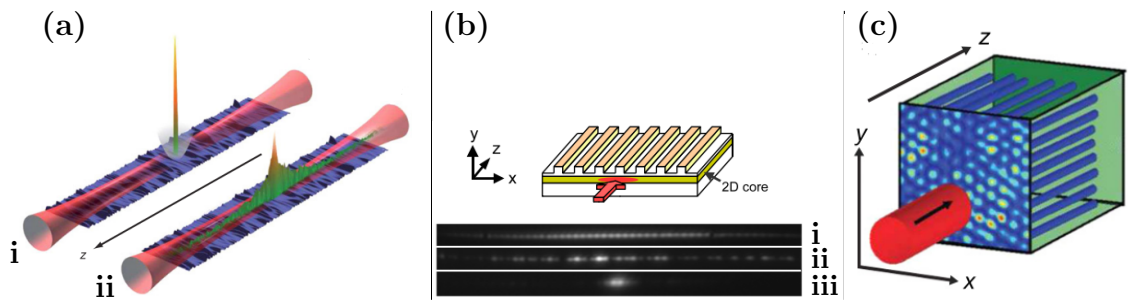

\autoref{fig:bose1} 
illustrates the setup for the observation of AL in {\bf (a)} the case of 
ultra-cold atoms in random 
potentials, {\bf (b)} 1D disordered photonic 
lattices and 
{\bf (c)} 2D disordered photonic lattices. 
In \autoref{fig:bose1}{\bf (a)} {\bf i} [from \cite{Billy2008}],
a small BEC, of approximately $2 \times 10^4$ atoms, was formed in a 
hybrid trap combination of a loose
magnetic longitudinal trap, and a horizontal optical 
waveguide ensuring a strong transverse confinement.
A weak disordered optical potential,
transversely invariant over the atomic cloud, was then superimposed. 
\autoref{fig:bose1}{\bf (a)} {\bf ii} shows the BEC 
expansion on switching off the longitudinal 
trap, which, eventually localizes.
Another experiment was performed in \cite{Lahini2008}, 
by introducing disorder using a randomly varying waveguide 
width. The authors 
demonstrated the wave packet transitions from 
a ballistic expansion to exponential localization in coupled optical
waveguide lattices in one dimension. 
The experimental setup of this investigation is shown 
in \autoref{fig:bose1}{\bf (b)} 
with the output observations in 
\autoref{fig:bose1}{\bf (b)} {\bf i}-{\bf iii}.
In this demonstration, 
photonic lattices were used to observe AL in light. 
The distribution of the light intensity 
was visualized and measured 
at the lattice output. 
A year before the two studies shown in \autoref{fig:bose1} {\bf (a)} and 
{\bf (b)} were performed, 
experiments 
on the transverse 
localization of light in the presence of disorder potentials 
in 2D photonic lattices were performed
[\cite{Schwartz2007}]. 
In the set up, shown in 
\autoref{fig:bose1} {\bf (c)},
the random potential was controlled and varied by 
introducing disorder through a speckle beam. 
For weak disorder, ballistic transport was reported to 
turn diffusive and AL was observed for stronger disorder.

Our last consideration is \cite{Dalichaouch1991} where the authors 
reported 
microwave localization in two dimensional random media 
of dielectric cylinders and placed between two parallel 
aluminium plates. Due to the presence of localized modes, 
sharp peaks were observed in the transmission 
spectrum. 
\newline\newline
Disorder in a system can be generated in ways other than 
using a random uncorrelated distribution for example as discussed in our next model. 
We consider 
the Aubry-Andr$\acute{e}$ model [\cite{Aubry1980}], 
a model with correlated disorder, whose Hamiltonian is:
\begin{equation}\label{eq:Aubryham}
	H_{AA} = \sum_l\left[\theta\cos(2\pi\alpha l)|\psi_l|^2 - \left(\psi_{l+1}\psi_l^* + \psi_{l+1}^*\psi_l\right)\right],
\end{equation}
with $\psi_l$ being a complex variable denoting the wave 
function 
for lattice site $l$ and $\psi_l^*$ the corresponding complex conjugate. $\alpha = \left(\sqrt{5}-1\right)/2$, and 
the parameter $\theta$ controls the 
magnitude of the on-site potentials, similar to the 
role of $W$ in equation (\ref{eq:anderson}). 
With this model, we can observe NMs due to quasi-periodic 
potentials displaying a transition between localized and 
extended modes 
for specific values of $\theta$.
The corresponding equations of motion of 
Hamiltonian \eqref{eq:Aubryham} are generated by 
$\dot{\psi}_l=\frac{\partial H_{1A}}{\partial (i\psi_l^*)}$ and yield the 
quasi-periodic version of the 1D 
DDLSE:

\begin{equation}\label{eq:aubrymotion}
i\dot{\psi_l} = \theta\cos(2\pi\alpha l)\psi_l - \psi_{l+1} - \psi_{l-1}.
\end{equation}
Setting $\psi_l(t)=A_le^{-i\lambda t}$ in 
equation (\ref{eq:aubrymotion}) we get the linear eigenvalue problem 
\begin{equation}\label{eq:eigval_aubry}
\lambda A_l = \theta\cos(2\pi\alpha l) A_l - A_{l+1} - A_{l-1}.
\end{equation}
In solving the eigenvalue problem (\ref{eq:eigval_aubry}) we use 
the matrix ${\bf A}$ whose entries $a_{k,l}$ are such that
\begin{equation*}
	a_{k,l}=\begin{cases} \theta\cos(2\pi\alpha k), & {\textnormal{for}}\,\,\, k=l;\\
		-1, & {\textnormal{for}}\,\,\, k=l\pm1;\\
		0, & {\textnormal{otherwise.}} \end{cases}
\end{equation*}
In this case, 
the eigenvalues 
$\lambda_{\nu}$ of ${\bf A}$ are bounded as 
$-2+\theta\cos(2\pi\alpha l) \leq \lambda_{\nu} \leq 2+\theta\cos(2\pi\alpha l)$. The minimum 
and
maximum possible values of the frequencies are therefore $-2-\theta$
and $2+\theta$ respectively.
This model has been used to study the 
so-called Aubry-Andr$\acute{e}$ localization [\cite{Roati2008,Edwards2008}] 
in ultra-cold atomic physics. 
In an experimental manifestation of the phenomenon, 
a primary periodic lattice was created via interference patterns of 
two counter-propagating beams and quasi-periodicity was introduced 
by superimposing a weak secondary lattice with incommensurate wavelength 
[\cite{Guidoni1997}].
For the case when $\theta=2$, Hamiltonian (\ref{eq:Aubryham}) is 
identical to the so-called Harper system [\cite{Harper1955}]. 
This is a critical point whereby for 
 $\theta<2$ the states are
localized and extended respectively in momentum and real space. 
For $\theta>2$ the states have an exponential localization 
in real space with length 
$\eta = [ln(\theta/2)]^{-1}$ while they are 
extended in the momentum space [\cite{Aubry1980,Sokoloff1981}].

\section{Nonlinear disordered models}
\label{sec:nonlineardis}
Having discussed energy localization for linear disordered systems 
in the previous subsection,
a fundamental question in this context is the effect 
of nonlinearities on the localization. This discussion has
attracted extensive 
attention and studies in experiments [\cite{Pertsch2004,Schwartz2007,Roati2008,Lahini2008,
Lucioni2011}], as  
well as in theory and simulations [\cite{Kopidakis2008,Pikovsky2008,
Flach2009,Skokos2009,Garcia-Mata2009,Veksler2009,Mulansky2009,Mulansky2010,
Laptyeva2010,Skokos2010a,Krimer2010,Flach2010,Johansson2010,
Basko2011,Bodyfelt2011,Bodyfelt2011a,Aubry2011,Ivanchenko2011,
Molina2012,Vermersch2012,Michaely2012,Basko2012,Mulansky2012,
Laptyeva2012,Milovanov2012,Lucioni2013,Vermersch2013,Mulansky2013,
Skokos2013,Laptyeva2013,Tieleman2014c,Ivanchenko2014,
Antonopoulos2014,Ermann2014,
Laptyeva2014,Basko2014,Besse2015,Martinez2016,Achilleos2016,Antonopoulos2017,
Cherroret2016,Iomin2017,Achilleos2018}].
As mentioned before, two typical 1D Hamiltonian lattice 
models, namely, the DDNLS equation and the DKG 
anharmonic oscillator chain, have been used to  
numerically investigate the nonlinearity effect on the 
evolution of initially localized wave packets in heterogeneous media
[\cite{Flach2009b,Skokos2009,Laptyeva2010,Skokos2010a,Flach2010,Bodyfelt2011,Bodyfelt2011a,Skokos2013}]. The 1D and 2D versions of the 
DDNLS and DKG models are described in the next sections of this chapter with 
the DKG systems also further studied in the next chapters of this thesis.

\subsection{The discrete nonlinear Schr\"{o}dinger (DDNLS) equations}
We start with the DDNLS system, whose Hamilton function 
is
\begin{equation}\label{eq:dnls1ham}
H_{1D}=\sum_l\left(\epsilon_l|\psi_l|^2-
\left(\psi_{l+1}\psi_l^*+\psi_{l+1}^*\psi_l\right)+\frac{\beta}{2}|\psi_l|^4\right),
\end{equation} where $\psi_l$ are complex variables, with 
$\psi_l^*$ being their complex conjugate values 
at lattice site $l$ and $\beta>0$ is quantifying 
the strength of the nonlinearity.
The random disorder numbers $\epsilon_l$,
are chosen from a uniform distribution over the interval 
$\left[-\frac{W}{2},\frac{W}{2}\right]$, where 
$W$ denotes the disorder strength.
Setting $\beta=0$ in equation (\ref{eq:dnls1ham}) gives the Anderson model 
of equation (\ref{eq:anderson}).
The model of equation (\ref{eq:dnls1ham}) describes a variety of nonlinear processes 
including localization of energy in chains of homogeneous anharmonic 
oscillators [\cite{Eilbeck1985,Kevrekidis2001,Eilbeck2003}] 
where $\epsilon_l=0$, 
the light evolution 
through coupled optical waveguides in Kerr media 
[\cite{Kivshar2003}] and two-body interactions in 
ultracold atomic gases on an optical 
lattice within a mean-field approximation
[\cite{Morsch2006}] among others.
The equations of motion are 
\begin{equation}\label{eq:dnls1motion}
i\dot{\psi_l}=\epsilon_l\psi_l-\psi_{l+1}-\psi_{l-1}+\beta|\psi_l|^2\psi_l.
\end{equation}

Besides the total energy $H_{1D}$ (\ref{eq:dnls1ham}), the equations 
(\ref{eq:dnls1motion}) conserve 
the total norm 
\begin{equation*}
S = \sum_l|\psi_l|^2.
\end{equation*} 
Since varying the total norm
is equivalent to changing the nonlinearity strength $\beta$, often 
the norm is fixed at $S=1$ and $\beta$ is left as the control parameter.
In a physical sense, the norm $S$ could be used to denote 
the intensity of light or the density of an atomic condensate.

Model (\ref{eq:dnls1ham}) can be naturally extended to two spatial 
dimensions over $N\times M$. 
The associated $2D$ DDNLS Hamiltonian is
\begin{equation}\label{eq:dnls2ham}
H_{2D}=\sum_{l,m}\left(\epsilon_{l,m}|\psi_{l,m}|^2-
\left[\psi_{l,m}^*\left(\psi_{l,m+1}+\psi_{l+1,m}\right)+
\psi_{l,m}\left(\psi_{l,m+1}^*+\psi_{l+1,m}^*\right)
\right]+\frac{\beta}{2}|\psi_{l,m}|^4\right),
\end{equation} with the disorder $\epsilon_{l,m}$ again being randomly 
chosen from $[-W/2,W/2]$ where $W$ and $\beta$ are respectively 
disorder and
nonlinearity strength.
The equations of motion of \eqref{eq:dnls2ham} are 
\begin{equation}\label{eq:dnls2motion}
i\dot{\psi}_{l,m}=\epsilon_{l,m}\psi_{l,m}-\left(\psi_{l-1,m}+
\psi_{l+1,m}+\psi_{l,m-1}+\psi_{l,m+1}\right) +
\beta|\psi_{l,m}|^2\psi_{l,m}.
\end{equation}
In general it is not possible to analytically 
integrate equations (\ref{eq:dnls1motion}) 
and (\ref{eq:dnls2motion}). 
However, for a coupled two oscillators lattice, system 
\eqref{eq:dnls2ham} becomes 1D and it is 
fully analytically integrable because of the DDNLS norm and energy conservation 
properties 
[\cite{Kenkre1986,Eilbeck2003}]. 
In \cite{Ablowitz1976} and 
\cite{Ablowitz1991} an integrable version of the DDNLS model was 
described, however, it has not yet been used to describe any physical 
processes.
We also have nonlinear quasi-periodic chains which are described by 
taking the disorder $\epsilon_l$ of the 1D 
Hamiltonian (\ref{eq:dnls1ham}) to be $\epsilon_l=\theta\cos(2\pi\alpha l)$, 
where the parameter $\theta$ and $\alpha$ 
are as described in equation (\ref{eq:Aubryham}). 
These quasi-periodic chains  
have been reported to describe the 
evolution of light pulses with high intensity 
in Kerr photonics [\cite{Lahini2009}] and the 
dynamical behaviour of interacting 
BECs in optic traps [\cite{Deissler2010,Lucioni2011}].

\subsection{The Klein-Gordon (DKG) lattice models}
\label{sec:kg}
The second model we focus on is the DKG 
lattice Hamiltonian model with a quartic nonlinearity in the potential and 
quadratic nearest-neighbour coupling. For physical 
applications, the DKG system is suitable for describing the dynamics  
of atomic arrays under the influence of external fields, e.g.~the conservative 
nonlinear  
optical lattice vibration dynamics in molecular 
crystals [\cite{Ovchinnikov2001}]. The 1D Hamiltonian 
of the DKG model 
is
\begin{equation}\label{eq:kg1dham}
H_{1K}=\sum_l\left[\frac{p_l^2}{2}+\frac{\tilde{\epsilon_l}q_l^2}{2}+
\frac{1}{2W}\left(q_{l+1}-q_l\right)^2+\frac{q_l^4}{4}\right],
\end{equation} where  $q_l$ 
is the generalized coordinate and 
$p_l$ 
the 
momentum of chain site $l$. 
$\tilde{\epsilon_l}\in[1/2, 3/2]$ are uncorrelated random values 
from a uniform distribution and 
$W$ is the disorder strength.
Therefore using Hamiltonian \eqref{eq:kg1dham}, the energy at 
site $l$ can be defined as 
\begin{equation}\label{eq:kg1en_per_site}
h_l=\frac{p_l^2}{2}+\frac{\tilde{\epsilon_l}q_l^2}{2}+
\frac{\left(q_l-q_{l-1}\right)^2+\left(q_{l+1}-q_l\right)^2}{4W}+
\frac{q_l^4}{4},
\end{equation} where the interaction component 
$\left(q_{l+1}-q_l\right)^2/2$ of the 
energy \eqref{eq:kg1dham} between sites $l+1$ and $l$ is shared equally 
to contribute $\left(q_{l+1}-q_l\right)^2/4$ to 
each of the two sites.

The corresponding equations of motion 
$\frac{d{\bf q}}{dt}=\frac{\partial H}{\partial {\bf p}}$; 
$\frac{d{\bf p}}{dt}= - \frac{\partial H}{\partial {\bf q}}$ of Hamiltonian 
\eqref{eq:kg1dham} lead to
\begin{equation}\label{eq:kg1dmotion}
\ddot{q}_l=-\tilde{\epsilon_l}q_l + 
\frac{1}{W}\left(q_{l+1}+q_{l-1}-2q_l\right)-q_l^3.
\end{equation}
The total energy $H_{1K}=\sum_lh_l$ of the system is conserved by the 
equations \eqref{eq:kg1dmotion}. 
We also use the parameter $H_{1K}$ to control the system's nonlinearity.
The 2D KG Hamiltonian takes the form
\begin{equation}\label{eq:kg2dham}
H_{2K}=\sum_{l,m}\left[\frac{p_{l,m}^2}{2}+\frac{\tilde{\epsilon}_{l,m}q_{l,m}^2}{2}+
\frac{1}{2W}\left(\left[q_{l,m+1}-q_{l,m}\right]^2+\left[q_{l+1,m}-q_{l,m}\right]^2\right)+\frac{1}{4}q_{l,m}^4\right],
\end{equation} where $q_{l,m}$ and $p_{l,m}$ are respectively the 
generalized coordinate and momentum 
of site
$(l,m)$, $\tilde{\epsilon}_{l,m}$ are random values chosen uniformly 
from $[\frac{1}{2},\frac{3}{2}]$ and $W$ is the 
disorder strength.
Using Hamiltonian \eqref{eq:kg2dham}, energy at site $(l,m)$ is defined as
\begin{equation}\label{eq:kg2dendens}
\begin{split}
h_{l,m}&=\frac{p_{l,m}^2}{2}+
\frac{\tilde{\epsilon}_{l,m}q_{l,m}^2}{2}+
\frac{1}{4W}\\
& \left(\left[q_{l,m+1}-
q_{l,m}\right]^2+\left[q_{l+1,m}-q_{l,m}\right]^2+\left[q_{l,m-1}-
q_{l,m}\right]^2+\left[q_{l-1,m}-q_{l,m}\right]^2\right)+\frac{1}{4}q_{l,m}^4,
\end{split}
\end{equation} and
the equations of motion corresponding to \eqref{eq:kg2dham} are
\begin{equation}\label{eq:kg2dmotion}
\ddot{q}_{l,m} = -\tilde{\epsilon}_{l,m}q_{l,m} +\frac{1}{W}\left(q_{l-1,m}+q_{l+1,m}+q_{l,m-1}+q_{l,m+1}-
4q_{l,m}\right)-q_{l,m}^3.
\end{equation}

Similar to the DDNLS Hamiltonian models in equations (\ref{eq:dnls1ham}) and 
(\ref{eq:dnls2ham}), the linear parts of Hamiltonians (\ref{eq:kg1dham}) and 
(\ref{eq:kg2dham}) can be reduced to the
Anderson model in one and two dimensions respectively.
This explains the exponential localization of all harmonic 
eigenstates of the DKG models. 
By neglecting the nonlinear terms $q_l^4$ in equation (\ref{eq:kg1dham}) 
and $q_{l,m}^4$ in equation (\ref{eq:kg2dham}) and taking the ansatz 
$q_l=A_le^{-i\omega t}$ and $q_{l,m}=A_{l,m}e^{-i\omega t}$ for 
the 1D and 2D models respectively, 
the DKG Hamiltonians are reduced to the same linear eigenvalue problems 
provided in  
equations (\ref{eq:eigval_DDLSE}) and (\ref{eq:2Deigval_DDLSE}), 
using the relations $\lambda =W\omega^2-W-2$, 
$\epsilon_l=W(\tilde{\epsilon}_l-1)$ and 
$\epsilon_{l,m}=W(\tilde{\epsilon}_{l,m}-1).$
The corresponding eigenvalues (square frequencies) $\omega_{\nu}^2$ 
lie in the interval $[1/2,4/W+3/2]$.
In the limit of small energies and amplitudes, a map
\begin{equation}\label{eq:dnls_kg_transform}
	\beta S\approx 3WH
\end{equation}
providing a relationship between parameters $W$ and $H$ of the DKG systems and 
$\beta$ and $S$ of the DDNLS model 
exists [\cite{Kivshar1992,Kivshar1993,Johansson2004,Johansson2006}]. 

In {\color{blue} Table} {\color{blue} $2.1$}, we present the 
exact correspondences (and the main characteristics) 
 between 
the 1D DDNLS and DKG models for small nonlinearities.  
Because of the existing mapping between these two models, we 
only give a more detailed discussion on 
theoretical estimates in Section \ref{sec:theories} 
for the DDNLS system and adapt the 
results for the DKG model. Existing simulations of these two models 
 show qualitatively similar 
dynamical behaviour for a variety of energy and disorder 
parameters [\cite{Flach2009,Skokos2009,Laptyeva2010,Bodyfelt2011}].
Compared to the DDNLS model, the DKG system allows up to a 
multiplicative factor 
of $10^2$ less numerical integration time for a fixed 
accuracy irrespective of the systems' final evolution time.

\begin{figure}[H]
	\centering
	\includegraphics[scale=0.6,keepaspectratio]{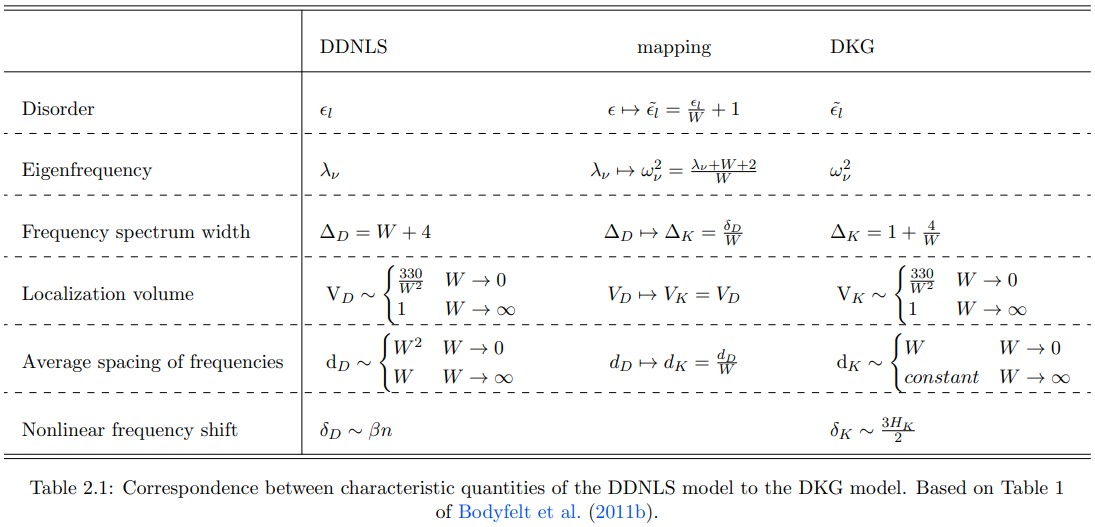}
\end{figure}

A quasi-periodic variation of the DKG chain [\cite{Laptyeva2014}] is governed 
by the Hamiltonian
\begin{equation*}
H_{q1K}=\sum_l\left[\frac{p_l^2}{2}+\frac{q_l^2(2+
	\cos(2\pi\alpha l))}{4}+\frac{1}{4\theta}(q_{l+1}-q_l)^2 +
\frac{q_l^4}{4}\right],
\end{equation*}
where the parameters $\alpha$ and $\theta$ have the same meaning as for the 
Aubry-Andr$\acute{e}$ model (\ref{eq:Aubryham}). The system's equations of motion are 
\begin{equation*}
\ddot{q}_l=-\frac{q_l(2+\cos(2\pi\alpha l))}{2} + \frac{1}{2\theta}(q_{l+1}+q_{l-1}-2q_l)-q_l^3.
\end{equation*}

We also note that the DDNLS and DKG 
systems discussed have quartic nonlinearity 
in their Hamiltonians. 
However, in some cases the 
index of nonlinearity may be 
parametrized to get a higher or fractional order 
for example in some optical materials, doped glasses, 
semiconductors and at the crossover 
between the molecule dimer BEC and Bardeen-Cooper-Schrieffer (BCS) 
pairs in ultra-cold Fermi 
gases [\cite{Yan2011}]. 
A fairly detailed coverage on parametrized nonlinearity index, 
which we recommend for further 
reading, has been included in 
\cite{Laptyeva2014}.

\subsection{Theoretical estimations}\label{sec:theories}
The localized nature of NMs in disordered media hinders the spreading 
of energy 
if modes do not interact with each other.
The presence of nonlinearity induces interactions between the NMs 
of the system, hence spreading of the energy is possible.
The theory of the spreading mechanism in such cases 
has been explained in \cite{Flach2009b}, 
\cite{Skokos2009}, \cite{Flach2010}, \cite{Laptyeva2010}, 
\cite{Bodyfelt2011}  and \cite{Besse2015}.
Here we highlight the key ideas discussed in these papers.

Extremely high energies lead to a strong nonperturbative effect of nonlinearity, and 
so spreading of the energy is not realized. On the other hand, very low energies 
create AL effects in the system in which case the energy never spreads 
or it takes a long time before spreading.
The fast spreading of the wave packet occurs at moderate energies of the system. 
In order to understand the energy spreading mechanism, we use the NM 
space in studying the system. 
The equations of motion 
for the DDNLS system in NM form can be written using the transformation 
$\psi_l=\sum_{\nu}A_{\nu,l}\phi_{\nu},$ where $\phi_{\nu}=\phi_{\nu}(t)$ 
is the amplitude of NM $\nu$ at time $t$. Using the equations 
of the eigenvalue problem \eqref{eq:eigval_DDLSE} and taking into account 
the orthogonality of the NMs, we revise the equations of motion 
(\ref{eq:dnls1motion}) to become
\begin{equation*}
	i\dot{\phi}_{\nu}=\lambda_{\nu}\phi_{\nu}+\beta\sum_{\nu_1,\nu_2,\nu_3}
	I_{\nu,\nu_1,\nu_2,\nu_3}\phi_{\nu_1}^*\phi_{\nu_2}\phi_{\nu_3},
\end{equation*} where $I_{\nu,\nu_1,\nu_2,\nu_3}$ is the overlap integral
\begin{equation*}
	I_{\nu,\nu_1,\nu_2,\nu_3}=\sum_lA_{\nu,l},A_{\nu_1,l},A_{\nu_2,l},A_{\nu_3,l}.
\end{equation*}
The nonlinear interactions of modes are of finite range, 
because of the exponential localized nature of the NMs, 
and thus each NM effectively couples with finitely 
many neighbouring modes. 
This finite number of neighbouring partner modes 
with effective coupling to a NM $\nu$ is on average 
the so-called localization volume $V_{\nu}$. 
The eigenvalues of the neighbouring eigenstates of $\nu$ will be different 
in general, but bounded within the frequency spectrum band which 
has length $\Delta$.
This therefore means that the eigenvalues of the NMs which are 
neighbours of $\nu$ 
within $V_{\nu}$ have an average frequency 
spacing $d_{\nu}\approx \Delta/V_{\nu}$.
Averaging over all NMs, gives an average spacing $d$ of the eigenvalues 
where $d\approx \Delta/V$, and $V=\overline{V}_{\nu}$. In this case 
$\overline{V}_{\nu}$ 
is used to denote the average of quantity ${V}_{\nu}$ over 
all disorder realisations. 

The strength of the NM interaction is proportional to the DDNLS norm density 
and to the energy density for the DKG model. 
In general, 
we can define two norm densities $n_l=|\psi_l|^2$ and $n_{\nu}=|\phi_{\nu}|^2$ 
in the real and NM spaces respectively, for the DDNLS 
system, if the wave packet has spread 
far enough. The averages of these two densities over a number of 
different disorder realizations give practically the same characteristic 
norm density $n$. Therefore the presence of 
nonlinearity induces a frequency shift 
$\delta_{1D}=\beta n$ for the DDNLS model 
and $\delta_{1K}=3H_{1K}/2$ for the DKG 
system 
on a single oscillator [\cite{Skokos2009,Laptyeva2010}]. 

Consider an excitation of $L$ central neighbouring 
oscillators with a constant total norm $n\neq0$, a random on-site 
phase for all sites and a vanishing norm for the remaining oscillators in the 
lattice. This, for the 1D DKG model, corresponds to 
giving the same energy
\begin{equation*}\label{eq:W_space}
h=\frac{H_{1K}}{L}
\end{equation*} to each of the $L$ oscillators. This can be easily achieved by 
setting the momentum of each oscillator to be $\sqrt{2h}$ 
with
randomly assigned signs while their initial displacements are set to 
zero.

For very high energies (or energy densities), i.e. when the size of the 
frequency shift $\delta_{1D}$ induced by the nonlinearity 
is larger than the frequency 
spectrum width $\Delta_D$, 
the sites where the excitation takes place will instantly 
be set out of resonance 
compared to the non excited neighbouring sites. This leads to the 
formation of structures similar to discrete breathers that persist for 
a very long time. That is to say, a large section of the 
wave packet will be self-trapped [\cite{Kopidakis2008,Skokos2009}]. Unlike AL, 
this phenomenon requires gaps to exist in the spectrum of the 
linear wave equations 
in order to occur [\cite{Flach1998,Flach2008}]. For a frequency shift 
$\delta_{1D}\geq 2$ in the 
DDNLS model, some wave packet sites are tuned out of resonance, 
therefore self-trapping behaviours can be observed.
However for relatively small frequency shifts $\delta_{1D} < \Delta_D < 2$, 
the self-trapping tendencies are avoided and propagation of the wave packet 
will be realised.
A NM in the wave packet resonantly interacts with it's neighbours 
when the size of the frequency shift $\delta_{1D}$ is greater 
than the average spacing 
$d_{1D}$ of the NMs. For this reason, this dynamical regime is 
called the ‘strong chaos’ regime. The NMs weakly interact 
with each other when $\delta_{1D}$ is smaller than $d_{1D}$ and so this 
is called the ‘weak chaos’ regime [\cite{Laptyeva2010}].
In general, the width of a propagating wave packet will increase 
(but with a 
fixed total norm) and so its norm density will decrease. This means that the 
nonlinear frequency shift will decrease and therefore the dynamics will 
crossover into the asymptotic (i.e.~limiting behaviour of the dynamics in time) 
weak chaos regime from the 
transient strong chaos regime 
at later times. 

\begin{figure}[H]
	\centering
	\includegraphics[width=0.47\textwidth,keepaspectratio]{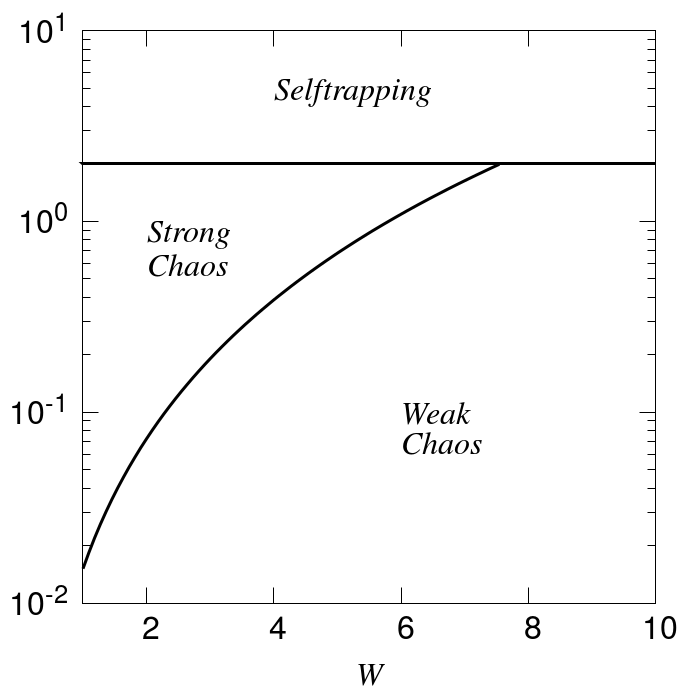}
	\includegraphics[width=0.47\textwidth,keepaspectratio]{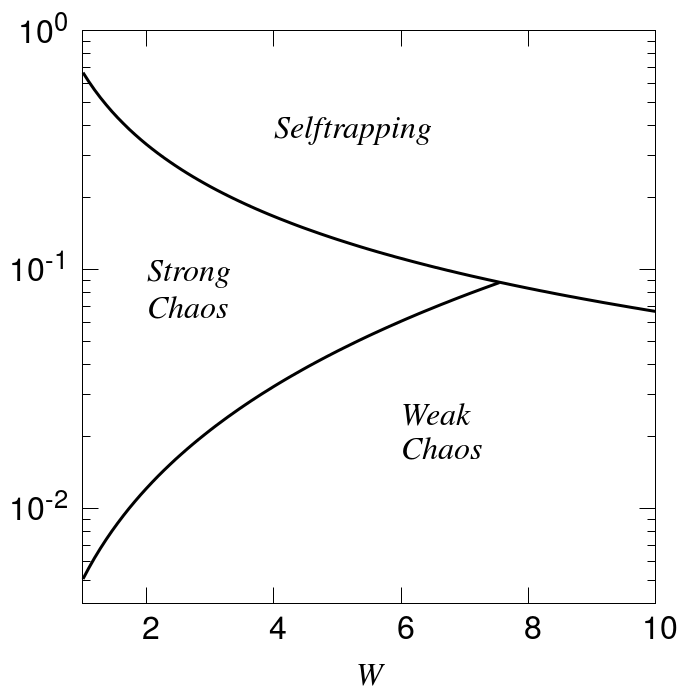}
	\put(-215,110){\bf $h$}
	\put(-440,110){\bf $\delta_{1D}$}
	\put(-360,210){DDNLS}
	\put(-120,210){DKG}
	\caption{The parameter space $(W,\delta_{1D})$ and $(W,h)$ respectively 
		showing the dynamical regimes of 
		the dynamics for the DDNLS model (\ref{eq:dnls1ham}) [left panel] and 
		the DKG system (\ref{eq:kg1dham}) [right panel]. Parameters $W$, 
		$\delta_{1D}$ and $h$ respectively denoting the disorder strength, the 
		DDNLS (model) nonlinear frequency shift and the uniformly distributed initial
		energy \eqref{eq:kg1en_per_site} at each site. The plots are based on Fig.~1 of \cite{Skokos2009}, Figure 1.2 of \cite{Gkolias2013} and 
		Figures 10 and 16 of \cite{Laptyeva2014}.}
	\label{fig:param_space}
\end{figure}

Only weak chaos and selftrapping regimes 
are present whenever only a single 
site is excited [\cite{Pikovsky2008,Skokos2009,Flach2009}]. 
For sites excitations of the order of
the localization 
volume, we can summarize 
the dynamical regimes as follows:
\begin{itemize}
	\item[i.] \textit{Selftrapping} regime: $\delta_{1D} > \Delta_D > 2$
	\item[ii.] \textit{Strong chaos} regime: $\Delta_D > \delta_{1D} > d_{1D}$
	\item[iii.] \textit{Weak chaos} regime: $\Delta_D > d_{1D} > \delta_{1D}$
\end{itemize}
In \autoref{fig:param_space} we illustrate these dynamical 
regimes in a parametric space for both 
the DDNLS and DKG models when the number of initially 
excited neighbouring sites is equal to the localization volume $V$. 
The DDNLS model's nonlinear 
frequency shift $\delta_{1D} \sim \beta n$ of the initial excitation 
is equivalent to the energy density $h$ in the DKG system.
The regime boundary curves for the DDNLS model 
(left panel of \autoref{fig:param_space}) are represented by 
the equations $\delta_{1D} =d_{1D}$ 
and $\delta_{1D}=2$. The lower regime boundary estimates for the weak chaos 
case are analytically found using 
$d_{1D}=\Delta_{1D}W^2/330$. To get the boundary curves for the DKG system, 
we use the equations involving 
$\delta_{1D}$ and the transformation (\ref{eq:dnls_kg_transform}) 
which practically maps the regime regions of the DDNLS model to the DKG system.

Chaotic dynamics takes place primarily due to the nonlinear 
interaction of 
modes [\cite{Skokos2009,Flach2009b}] in the 
lattice interior. As a consequence, the modes close 
to the wave packet boundaries get heated. 
The wave packet spreading mechanism depends on either 
a normal mode on the cold exterior (bordering the wave packet) being 
excited resonantly by some 
particular normal mode within the packet, or incoherently heated by the packet.
It was conjectured
[\cite{Laptyeva2010,Flach2010}] that an exterior mode $\mu$ is excited according 
to the equation
\begin{equation*}
	i\dot{\phi}_{\mu}\approx\lambda_{\mu}\phi_{\mu}+\beta n^{\frac{3}{2}}\mathcal{P}(\beta n)f(t)
\end{equation*}
with $\mathcal{P}(\beta n)\approx 1 - exp(\frac{-\beta n}{d})$ denoting 
the probability, 
for a mode excited to average packet norm density $n$, 
to be resonant with at least one triplet of other modes at a 
specific value of the interaction parameter $\beta$
[\cite{Flach2010,Krimer2010}] while 
$f(t)$ denotes a stochastic force. The evolution of the norm, averaged 
over all modes, is therefore 
given by 
\begin{equation*}
|\phi(t)|^2\propto\beta^2n^3[\mathcal{P}(\beta n)]^2t.
\end{equation*}
$|\phi(t)|^2$ will be equal to the wave packet norm $n$ at a time 
$T\propto\beta^{-2}n^{-2}[\mathcal{P}(\beta n)]^{-2}$, after which 
the mode forms part of the wave packet. We can characterise the 
rate of norm diffusion using the inverse of $T$ which we denote as $D$,
\begin{equation*}
	D(t)=T^{-1}\propto\beta^2n^2[\mathcal{P}(\beta n)]^2.
\end{equation*}
Thus, the resonance probability $\mathcal{P}(\beta n)$ 
largely dictates the nature of the chaotic 
dynamics whether it is characterised as strong or weak chaos. 
Since the second moment,
\begin{equation}\label{eq:m2_1D}
	m_2=\sum_l(l-\bar{l})^2n_l,
\end{equation} where $\bar{l}=\sum_lln_l$ is the 
centre of the norm distribution $\{n_l\}_l$ of the 
wave packet, is approximately equal to the inverse of the square packet norm, 
i.e. $m_2\approx1/n^2$, and we also have the diffusion equation $m_2\propto Dt$, then
\begin{equation}\label{eq:diffusion}
	\frac{1}{n^2} \propto\beta\left[1-\exp\left(-\frac{\beta n}{d}\right)\right]t^\frac{1}{2}.
\end{equation}
Equation (\ref{eq:diffusion}) also determines the subdiffusive 
spreading for the strong and weak chaos regimes 
as
\[  m_2(t)\propto\left\{
\begin{array}{lr}
\beta t^{\frac{1}{2}}, & {\textnormal {for~strong~chaos}},\\
d^{-\frac{2}{3}}\beta^{\frac{4}{3}}t^{\frac{1}{3}}, & {\textnormal {for~weak~chaos}}. \\
\end{array}\label{eq:m2_1d} 
\right. \]
Thus, for 1D systems, this yields propagation laws 
in terms of the second moment $m_2$ \eqref{eq:m2_1D} for the weak
and strong chaos cases of the  
wave packet. That is to say, the second moment increases 
as $m_2(t)\propto t^{\frac{1}{3}}$ in the weak chaos and 
$m_2(t)\propto t^{\frac{1}{2}}$ in the strong chaos regime.
\newline\newline
A similar theoretical analysis [\cite{Flach2010,Laptyeva2012}] 
for the 2D systems gives the following spreading laws
\[\label{eq:m2_2d} P(t),m_2(t)\propto\left\{
\begin{array}{ll}
	t^{\frac{1}{3}}, &  {\textnormal {for~strong~chaos}},\\
	t^{\frac{1}{5}}, &  {\textnormal {for~weak~chaos}}. \\
\end{array} 
\right. \] 
\subsection{Other disordered systems}
Besides the DDNLS and DKG models, various 1D coupled oscillatory 
lattice models have been intensively studied 
in a variety of physical disciplines for example 
biology, biophysics and material science 
[\cite{Ford1992,Braun2003,Peyrard2004}]. 
An enormous number of models and their applications have been 
reported, including but 
not limited to  
the Fermi–Pasta–Ulam-Tsingou (FPUT) chains 
[\cite{Fermi1955}], the Frenkel–Kontorova model 
[\cite{Kontorova1938,Braun2003}] and the Peyrard–Bishop Dauxois (PBD) 
Deoxyribo-Nucleic Acid (DNA) model [\cite{Dauxois1993}].

\subsubsection{The disordered Fermi Pasta Ulam Tsingou (DFPUT) model}
The disordered version of the FPUT model 
[\cite{Fermi1955}] was used in \cite{Li2001} and \cite{Dhar2008a} to 
investigate the heterogeneity effect 
on thermal conductivity in one dimensional mass-disordered harmonic and 
anharmonic lattices. 
In contrast to the conclusions of \cite{Li2001}, who reported that 
disorder induces a finite thermal conductivity at low temperatures, 
\cite{Dhar2008a} found no evidence of a finite-temperature transition 
in the conducting properties of the DFPUT model. They instead, revealed that 
disorder dominates transport properties of 
systems of small sizes at low temperatures. For systems of large size, the heat current $J$ was 
found to depend on the lattice size $N$ as $J\propto1/N^{2/3}$. 
A comparative study of chaos and AL in the DFPUT and 
Hertzian [\cite{Johnson1989}] models describing granular chains 
has been presented 
in \cite{Ngapasare2019}, where, 
the discontinuous nonlinearity of the Hertzian system has been found to 
trigger energy spreading at lower energies. The DFPUT model, however, 
was shown 
to exhibit an alternate behaviour between localized and 
delocalized chaos which is strongly dependent on the type of initial 
excitation.
\subsubsection{The Frenkel–Kontorova model}
Frenkel-Kontorova system is a simple model that has become one of 
the fundamental and universal tools of low-dimensional nonlinear physics. It
describes the 
dynamics of a chain of particles with nearest neighbour coupling 
and the influence of an external periodic potential [\cite{Oleg1998}]. 
This model was first mentioned by \cite{Prandtl1928} and \cite{Dehlinger1929}.
Then Frenkel and Kontorova [\cite{Kontorova1938}] worked independently 
on 
the same model. 
This 
model has been used in describing the dynamics of a crystal lattice near 
a soliton (a wave packet propagating at a constant velocity 
while maintaining its shape) [\cite{Sirovich2005}]. 
Frenkel and Kontorova derived an analytical solution 
for a moving single soliton which was generated for large enough 
displacement of the atoms [\cite{Filippov2010}].
A detailed discussion of the model including it's applications 
is given in \cite{Braun2003}.
\subsubsection{DNA models}
The dynamical behaviour of base pairs in the DNA molecule 
has been studied using various models.
A comprehensive review 
of some of the various models has been presented in \cite{Manghi2016}.
One of these models, namely the PBD model, was introduced 
in \cite{Dauxois1993}, and it has since been used to explain a number 
of observations from experiments performed in relation to 
base pair openings in DNA. 
For specific temperatures, numerical simulations showed good 
qualitative description of the amplitude fluctuation openings 
and emergence of denaturation bubbles from thermal fluctuations. 
Numerical studies of the PBD model 
in \cite{Hillebrand2019} revealed that chaoticity
increases with energy and this behaviour is 
independent of the heterogeneity composition of base 
pairs.

\section{Measures of spreading}\label{sec2:m2_P}
In our study we concentrate on the energy spreading behaviour for the 
DKG models 
in one and two spatial dimensions. We present here 
the spreading measures we use in our analysis.
\subsection{1D DKG model} 
As we explained in Section \ref{sec:kg}, the energy $h_l$ 
of the oscillator at site $l$ of the 1D DKG model \eqref{eq:kg1dham} is given by 
\begin{equation}\label{eq:en_pa_site_1dkg}
	h_l=\frac{p_l^2}{2}+\frac{\tilde{\epsilon_l}q_l^2}{2}+
	\frac{\left(q_{l-1}-q_l\right)^2+\left(q_{l+1}-q_l\right)^2}{4W}+
	\frac{q_l^4}{4},
\end{equation} and thus the energy density distribution is defined 
by 
\begin{equation}\label{eq:norm_en1d}
\xi_l=\frac{h_l}{H_{1K}}.
\end{equation}
The extent of the wave packet can be measured by the second moment $m_2$ 
and the participation number $P$. 
The second moment $m_2$ estimates the wave packet extent 
of spreading and is given by 
\begin{equation}\label{eq:m2_1dkg}
	m_2=\sum_l(l-\bar{l} )^2\xi_l
\end{equation} where $\bar{l}=\sum_ll\xi_l$ denotes the 
centre of the distribution $\{\xi_l\}_l$ \eqref{eq:norm_en1d}.

The participation number 
\begin{equation}\label{eq:P_1dkg}
	P=\left[\sum_l\xi_l^2\right]^{-1},
\end{equation}
on the other hand, estimates the number of sites with the 
highest energy. When the 
total energy $H_{1K}$ of the system is 
equally distributed amongst $L$ oscillators then $P=L$, with the 
particular case of a single site excitation giving $P=1$. In all other 
cases, $P$ lies between one and the lattice size. We note that $P$ is 
equivalent to the Renyi entropy $I_2$ where 
\begin{equation*}
	I_r=\frac{1}{1-r}\ln\sum_l\xi_l^r.
\end{equation*}

\subsection{2D DKG model}
For the 2D DKG model \eqref{eq:kg2dham}, the energy $h_{l,m}$ at site $(l,m)$ is given by 
\begin{equation}\label{eq:en_density2}
h_{l,m} =  \frac{p^2_{l,m}}{2} + \frac{\epsilon_{l,m}}{2} q^2_{l,m} +
\frac{q_{l,m} ^{4}}{4} + \frac{1}{4W}\big[(q_{l+1,m} - q_{l,m})^2 +
(q_{l,m+1} - q_{l,m})^2\big],
\end{equation} and thus the energy density distribution is defined 
by 
\begin{equation}
\label{eq:norm_en2d}
\xi_{l,m}=\frac{h_{l,m}}{H_{2K}}.
\end{equation}

The second moment $m_2$ is given by 
\begin{equation}\label{eq:m2_2dkg}
m_2=\sum_{l,m}||(l,m)-\overline{(l,m)}||^2\xi_{l,m}
\end{equation} with 
$\overline{(l,m)}=\sum_{l,m}(l\xi_{l,m},m\xi_{l,m})$ 
denoting the position of the
energy distribution $\{\xi_{l,m}\}_{l,m}$ centre  and $||\cdot||$ 
denoting the Euclidean norm.

The participation number $P$ is given by 
\begin{equation}\label{eq:P_2dkg}
P=\frac{1}{\sum_{l,m}\xi_{l,m}^2},
\end{equation}
such that $P=N{\cdot}M$ whenever the 
total energy of the system is equally distributed amongst an 
$N\times M$  rectangular array 
of $N{\cdot}M$ oscillators 
(where $N\times M$ is a sub-array of the entire lattice). 
For a general distribution of the energy amongst the oscillators, 
$P$ always lies between one and the lattice size.
\section{Transition from order to disorder}\label{sec2:disord2ord}
In Section \ref{sec:theories}, we used the theory of 
NMs to describe the temporal behaviour of the 
second moment $m_2$ (\ref{eq:m2_1dkg} and \ref{eq:m2_2dkg}) of the 
wave packet 
as a quantity for 
classifying the dynamics of 1D and 2D 
disordered Hamiltonian systems in categories of 
chaotic regimes. 
This shows the power of NMs, as inherent structures of a system, 
in revealing the characteristics a system's dynamics.
Here we probe the NM properties for 
Hamiltonian lattices over a transition in the heterogeneity spectrum 
from order at one end 
to disorder at the other end.
In order to do this, we consider 
a 1D 
linear version of the DKG model (\ref{eq:kg1dham}) 
to which we introduce parameters, 
$D$ (related to the site disorder) and $\mathcal{W}$ 
(regulates the size of neighbour interactions). $D$ together with 
$\mathcal{W}$ 
determine 
the strength of disorder in our model. 
Starting from an ordered version of this new model, we study the 
properties of the NMs as we move towards a disordered system by changing $D$ 
and keeping $\mathcal{W}$ constant. In the process, we monitor 
changes in the qualitative structure, eigenfrequencies, average 
participation number $P$ \eqref{eq:Pmod} 
and average localization volume $V$ \eqref{eq:Vmod} of the NMs, 
studying the changes in the width $\Delta$ of the frequency 
spectrum, and the average 
spacing $d$ \eqref{eq:dmod} of the frequencies of the modes.
%
%

\subsection{The modified Klein-Gordon model}
\label{sec:models}
We start by performing an analysis for the linear version LDKG of the 1D DKG 
model (\ref{eq:kg1dham}) whose Hamiltonian function is
\begin{equation}\label{eq:LDKG}
H_{K} = \sum _{l} \left[ \frac{p_l^2}{2} + 
\frac{\tilde{\epsilon}_lq_l^2}{2} + \frac{1}{2\mathcal{W}}\left(q_{l+1} - 
q_l\right)^2 \right],
\end{equation}
where again $q_l$ and $p_l$ are respectively representing the generalized
position and momentum of oscillator $l$.
The disorder coefficients $\tilde{\epsilon}_l$ take uncorrelated 
random values
chosen from the interval 
$\left[ 1-D, 1+D\right]$ following a probability distribution with function
$\mathcal{P}(\tilde{\epsilon}_l)=1/(2D)$. The 
parameter $D$ defines the interval width from which $\tilde{\epsilon}_l$ 
is chosen, while 
$\mathcal{W} > 0$ determines the strength of nearest neighbor interaction. 
In our investigation, we 
use $0 \leq D \leq 1/2$ 
with $D=0$ (i.e. setting $\epsilon_l=1$ for all sites) 
corresponding to the ordered version of Hamiltonian 
\eqref{eq:LDKG}. For a fixed $\mathcal{W}$, when $D$ increases from $0$, 
the system shifts from 
ordered to disordered while keeping the strength of 
nearest neighbor interactions constant.

The equations of
motion of Hamiltonian (\ref{eq:LDKG}) are
\begin{equation}\label{eq:eqmot_LDKG}
\dot{q_l}=p_l \,;\qquad\qquad\dot{p_l}= -\left[\tilde{\epsilon}_lq_l+
\frac{1}{\mathcal{W}}\left(2q_l-q_{l-1}-q_{l+1}\right)\right],
\end{equation}
where as usual the dot denotes the time $t$ derivative. 
Using the ansatz $q_l=A_le^{i\omega t}$,
where $A_l$ is the amplitude of oscillator $l$, 
equation~\eqref{eq:eqmot_LDKG} leads to the eigenvalue problem
\begin{equation}\label{eq:eigval_LDKG}
\omega^2A_l=\frac{1}{\mathcal{W}}\left[(\mathcal{W}\tilde{\epsilon}_l+2)A_l-A_{l-1}-A_{l+1}\right].
\end{equation}
The system's NMs are the normalized eigenvectors 
$A_{\nu,l}$, $\nu =1,2, \ldots,N$, with $\sum_lA_{\nu,l}^2 = 1$,
and the eigenvalues $\omega^2_{\nu}$ are the corresponding
squared frequencies of these modes.
System \eqref{eq:LDKG} corresponds to the Anderson model \eqref{eq:anderson} 
by considering the 
relations
\begin{eqnarray}
\label{eq:change_lambda}
\lambda &=& \omega ^2 \mathcal{W} - \mathcal{W} - 2, \\
\label{eq:change_epsilon}
\epsilon_l &=& \mathcal{W}\left(\tilde{\epsilon}_l - 1\right),\\
\label{eq:change_W}
W &=& 2D\mathcal{W},
\end{eqnarray} where $\lambda$, $\epsilon_l$ and $W$ 
are as defined for the model \eqref{eq:anderson} with 
$0 < W\leq\mathcal{W}$. In this way the 
eigenvalue problems \eqref{eq:eigval_DDLSE} and 
\eqref{eq:eigval_LDKG} and consequently the Hamiltonian 
systems \eqref{eq:anderson} and \eqref{eq:LDKG}, become identical.
Finding the solution of the eigenvalue problem 
\eqref{eq:eigval_LDKG} involves the diagonalization of 
the $N \times N$ tridiagonal matrix ${\bf B}$ with elements
\begin{equation}
\label{eq:B_epsilon}
b_{l,l-1}=b_{l,l+1}=-\frac{1}{\mathcal{W}},\;\; b_{l,l}=\tilde{\epsilon}_l+\frac{2}{\mathcal{W}},
\end{equation}
and $b_{k,l} = 0$ otherwise, for $l, k=1,2, \ldots,N$. Once 
again, the direct application of 
\autoref{thm:gersh} [\cite{Gerschgorin1931,Wolkowicz1980}]
to matrix ${\bf B}$ \eqref{eq:B_epsilon} gives 
$\tilde{\epsilon}_l \leq \omega^2(D) \leq \tilde{\epsilon}_l+4/\mathcal{W}$. 
The minimum $\omega_{-}^2(D)$, and
maximum $\omega_{+}^2(D)$, possible values of the squared frequencies 
are respectively $\omega_{-}^2(D)=1-D$
and $\omega_{+}^2(D)=1+D+4/\mathcal{W}$ since
$\tilde{\epsilon}_l\in [1-D,1+D]$. The width of the squared 
frequency ($\omega_{\nu}^2$) spectrum therefore is
\begin{equation}
\label{eq:Delta_K}
\Delta _K(D) = 2D + \frac{4}{\mathcal{W}}.
\end{equation} 
Setting $\tilde{\epsilon}_l=1$ $\forall~l=1,2,\ldots, N$ in 
\eqref{eq:LDKG}, i.e.~$D=0$, gives an ordered linear system 
where the analytic solution of the corresponding eigenvalue problem 
[\cite{Losonczi1992,Yueh2005,Borowska2015,DaFonseca2019}]  is
\begin{equation*}
\omega_{\nu}^2(D=0)=1+\frac{2}{\mathcal{W}}\left[1-\cos\left(\frac{\nu \pi}{N+1}\right)\right], \,\,\, \nu = 1, 2, \ldots, N.
\end{equation*}
For each mode $\nu$, $\omega_{\nu}^2(0)$ is 
bounded below by $1$ and above by 
$1+4/\mathcal{W}$, thereby giving a maximum spectrum width 
$\Delta _K = \frac{4}{\mathcal{W}}$ in accordance with \eqref{eq:Delta_K}.

The presence of the disorder range width $D$ and the parameter 
$\mathcal{W}$ in the 
LDKG model \eqref{eq:LDKG} allows for the alteration of the
system's disorder strength by modifying $D$ or $\mathcal{W}$ or 
both parameters, while 
for the equivalent DDLSE system \eqref{eq:anderson} we can only 
change the disorder strength by varying $W$. 
In the studies of the DKG and LDKG models 
[\cite{Flach2009,Skokos2009,Laptyeva2010,Skokos2013}] performed so far, 
the value of $D$ was fixed at $D=1/2$ and various values of 
$W\geq 1$ (implying $\mathcal{W}\geq 1$ 
through \eqref{eq:change_W}) used. 
As $W\rightarrow 0$, the nearest neighbor interaction term 
becomes more significant compared to 
the on-site potential $\tilde{\epsilon}_l q_l^2/2$ in Hamiltonian 
\eqref{eq:LDKG} and the 
system tends to a more ordered one.
In [\cite{Krimer2010}], the focus was on strong disorder 
and so the properties of NMs for the DDLSE \eqref{eq:anderson} 
were discussed for disorder parameter values $W \geq 1$. 
For the modified model \eqref{eq:LDKG} [and equivalently for 
the DDLSE \eqref{eq:anderson}], by changing the disorder 
through increasing $D$ starting from $0$ while keeping 
$\mathcal{W}$ fixed, we obtain a transition from order to disorder.
By considering equations \eqref{eq:change_lambda}--\eqref{eq:change_W}, 
we can use analogous results obtained for the DDLSE \eqref{eq:anderson} 
[\cite{Kramer1993,Krimer2010}] to deduce the results for the LDKG 
model \eqref{eq:LDKG}.
A NM $\nu$ of the system \eqref{eq:LDKG} has an asymptotic exponential 
spatial decay described by 
\begin{equation*}
A_{\nu,l}\sim \exp\left(-\frac{|l-l_0|}{\eta_{\nu}}\right),
\end{equation*} with
$l_0$=$\sum_l lA_{\nu,l}^2$ denoting the mean spatial position of the NM and
$\eta_{\nu}$ is the localization length
[\cite{Anderson1958,Kramer1993,Krimer2010}], whose dependence on  
$D$ and $\omega_{\nu}^2$ 
is given by
\begin{equation}\label{eq:loc_length_LDKG}
\eta_{\nu}(D,\omega_{\nu}^2)  = \frac{24 \left[4 - (\omega_{\nu}^2 \mathcal{W} - \mathcal{W} - 2)^2\right]}{4D^2\mathcal{W}^2},
\end{equation} where $D\mathcal{W}\leq2$.
We note that $D$ and $\mathcal{W}$ affect $\Delta_K$ \eqref{eq:Delta_K} 
and $\eta_{\nu}$ \eqref{eq:loc_length_LDKG} in different ways, 
since they do not appear in each one of these two expression exclusively as 
a product $D\mathcal{W}$ or a quotient $D/\mathcal{W}$ (or $\mathcal{W}/D$).
By solving the equation $\frac{d}{d\omega_{\nu}^2}\eta_{\nu}=0$, we find that the 
most extended NMs (i.e. NMs with the largest localization 
length $\eta _0$ at a particular value of $D$) appear whenever 
$\omega_{\nu}^2 = 1 + \frac{2}{\mathcal{W}}$. In this case
\begin{equation}
\eta_{\nu} \left(D,\omega_{\nu}^2 = 1 + \frac{2}{\mathcal{W}}\right) = \eta _0 = \frac{24}{D^2\mathcal{W}^2},
\label{eq:loc_length_LDKG_max}
\end{equation} for $D\mathcal{W}\leq2$. This is a generalization over 
$D$, of the 
findings of \cite{Kramer1993} and \cite{Krimer2010} where the 
localization length is reported to be maximized for 
frequencies at the centre of the band width for $D=0.5$.
 
\subsection{Properties of NMs}
\label{sec:num_res}
In this investigation, 
we fix $\mathcal{W}=4$ and systematically change $D$ from 
values very close to $D=0$ (system's state of order) to 
$D=1/2$ (system's state of disorder).
This set up leads to $W=8D$ \eqref{eq:change_W} with the 
square frequency band width $\Delta_K=2D+1$ \eqref{eq:Delta_K}. 
Computation of NMs for a lattice of size $N$ implies working with 
an $N\times{N}$ matrix. Therefore very large sizes of $N$ are impossible 
to use in numerical computations. For this reason, 
we use $D-$dependent lengths of up to 
a maximum length $N=50{,}000$.
We note that the pair of parameters $\mathcal{W}=4$ and 
$D=1/2$ is a typical combination 
of values 
that have been used in several studies of disordered systems 
including \cite{Flach2009}, \cite{Skokos2009}, 
\cite{Laptyeva2010} and \cite{Skokos2013}.
In order to perform a statistical analysis of the NM properties, we perform
simulations over $n_d=100$ disorder realizations for a fixed  
value of $D\in(0,0.5]$.
We order the NMs either by increasing 
value of their squared frequency $\omega^2$, or of their mean spatial position
$l_0$ for each value of $D$. 
Since the frequency band and it's width $\Delta_K$ depend on $D$, 
we use the normalized square frequency 
\begin{equation} \label{eq:nomr_freq}
\omega_{\nu, n}^2 = \frac{\omega_{\nu}^2 - \omega_{\nu, -}^2}{\Delta _K} = \frac{\omega_{\nu}^2 - \omega_{\nu, +}^2}{\Delta _K}+1 = \frac{\omega^2_{\nu}+D-1}{2D+1},
\end{equation} of the NM to allow for a 
direct comparison of cases with different 
values of $D$.

\begin{figure}[H]
	\centering
	\includegraphics[width=0.8\textwidth,keepaspectratio]{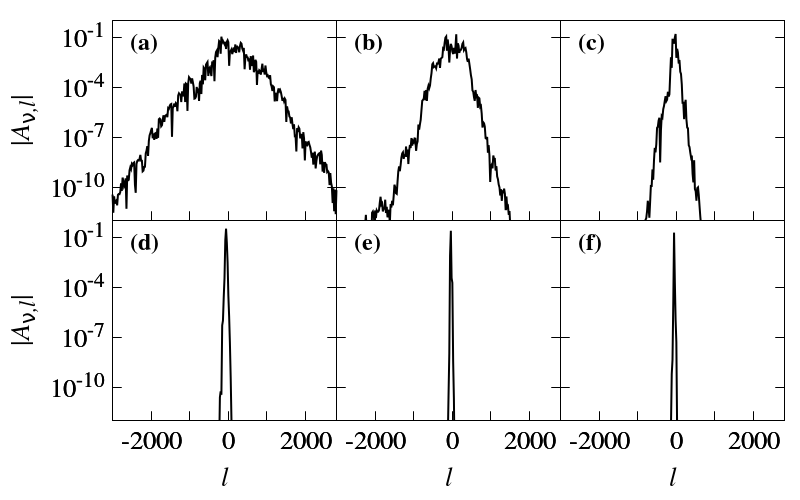}
	\caption{\label{fig:NMs_profiles} The profiles of representative 
		NMs of the LDKG model \eqref{eq:LDKG}, 
		whose mean spatial position $l_0$ is approximately
		positioned at the lattice centre and the corresponding normalized 
		square frequencies \eqref{eq:nomr_freq} $\omega_{\nu, n}^2 \approx 0.5$ 
		for {\bf(a)} $D=0.1$, {\bf(b)}
		$D=0.15$, {\bf(c)}  $D=0.2$, {\bf(d)}	$D=0.3$, 
		{\bf(e)}  $D=0.4$  and {\bf(f)}	$D=0.5$. The plots are in linear-log scale.}
\end{figure}
In \autoref{fig:NMs_profiles} we present the profiles 
(absolute values of amplitudes $A_{\nu,l}$) of the NMs 
against the lattice site $l$, for some centrally positioned 
(with mean spatial position, $l_0$, close to the lattice centre) 
representative modes in 
linear-log scales for six different values of $D$, namely, 
$D=0.1,~0.15,~0.2,~0.3,~0.4,~0.5$. The 
disorder realizations $\tilde{\epsilon}_l\in[1-D, 1+D]$ for 
the different values 
of $D$ 
are the same up to an appropriate 
scaling. In each panel of 
\autoref{fig:NMs_profiles} we plot a NM 
so that $\omega_{\nu, n}^2 \approx 0.5$, 
a frequency which is approximately 
centrally placed in the frequency band with a general correspondence 
to the most extended NMs [\cite{Anderson1958,Kramer1993,Krimer2010}]. 
All NMs are characterized by clearly defined exponential tails 
(i.e.~NMs are exponentially localized) with a spatial extent which 
decreases, as $D$ increases, from an order of thousand of sites for $D=0.1$ 
[\autoref{fig:NMs_profiles}{\bf(a)}] to a 
couple of dozen sites 
for $D=0.5$ [\autoref{fig:NMs_profiles}{\bf(f)}]. 
The lattice centre in \autoref{fig:NMs_profiles} 
has been shifted to $l=0$. We note that 
NMs with square frequencies approximately 
centrally placed in the frequency band, 
i.e.~$\omega_{\nu, n}^2 \approx 0.5$, 
are not uniquely associated to having their centre $l_0$ in the 
middle of the lattice as one may deduce from \autoref{fig:NMs_profiles}. 

\begin{figure}[H]
	\centering
	\includegraphics[width=0.34\textwidth,keepaspectratio]{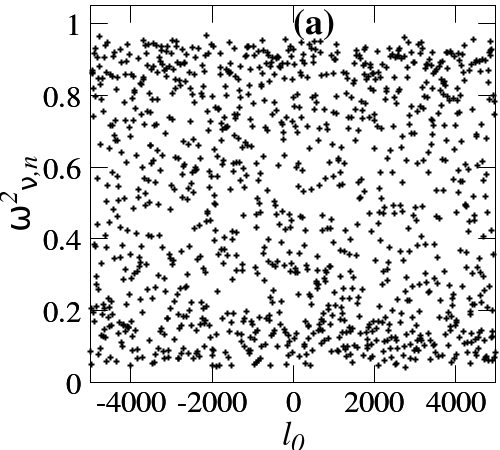}
	\includegraphics[width=0.65\textwidth,keepaspectratio]{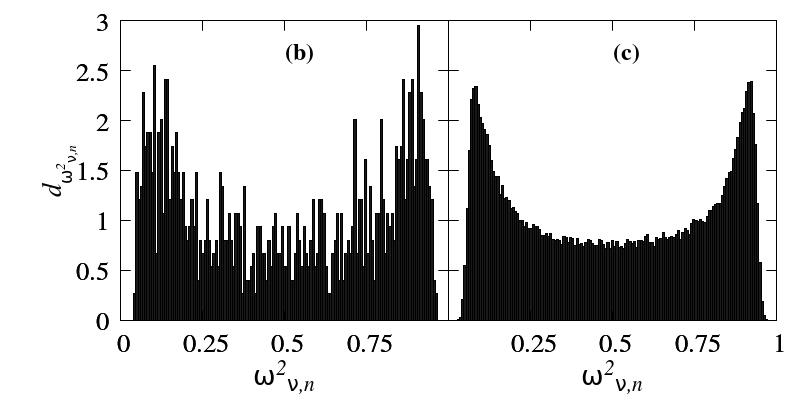}
	\caption{\label{fig:freq_1} Results for the distribution of the 
		square frequencies $\omega^2_{\nu}$ of the NMs of system \eqref{eq:LDKG} 
		with $N=10{,}000$ and $D=0.1$: 
		{\bf(a)} 
		the normalized square 
		frequencies $\omega_{\nu, n}^2$ \eqref{eq:nomr_freq} of 
		the NMs for a representative disorder realization 
		as a function of the 
		mean spatial position $l_0$ of the NMs,  
		{\bf(b)} the probability density distribution 
		$d_{\omega _{\nu, n}^2}$ of $\omega_{\nu, n}^2$ for the 
		realization shown in {\bf(a)} and, {\bf(c)} 
		similar to {\bf(b)} but the probability density distribution 
		is computed for
		for $n_d=100$ disorder realizations.}
\end{figure}

In \autoref{fig:freq_1}{\bf(a)} 
we show the normalized NM square 
frequencies $\omega_{\nu, n}^2$ \eqref{eq:nomr_freq} 
for a representative disorder realization of model \eqref{eq:LDKG} with $N=10{,}000$ 
and parameter $D=0.1$, as a function of the mean spatial position 
$l_0$ of the NMs. 
The distribution of square frequencies $\omega_{\nu}^2$ 
throughout the lattice is in such a way that they are 
concentrated at the borders of the spectrum as more data points 
are located in the regions around 
$\omega_{\nu, n}^2\approx 0.1$ and $\omega_{\nu, n}^2\approx 0.9$. 

\begin{figure}[H]
	\centering
	\includegraphics[width=0.4\textwidth,keepaspectratio]{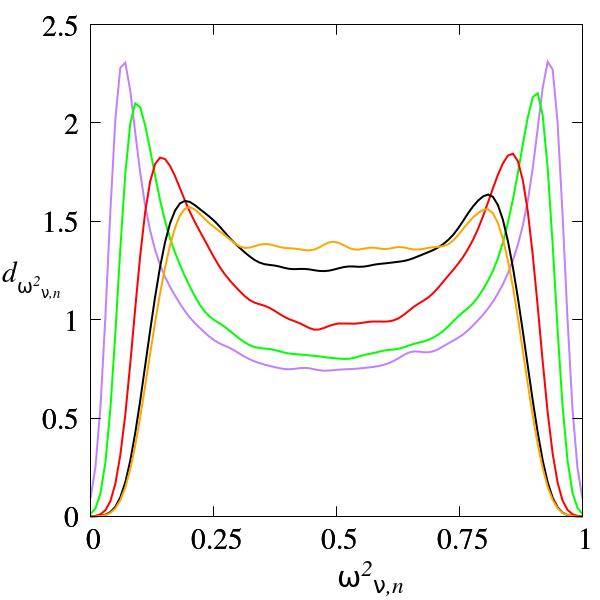}
	\caption{\label{fig:distributions_many}
		Results for the probability density distributions 
		$d_{\omega _{\nu, n}^2}$ of the normalized squared frequencies $\omega_{\nu, n}^2$ \eqref{eq:nomr_freq} of system 
		\eqref{eq:LDKG} for $D=0.06$ (purple curve), $D=0.1$ (green curve), $D=0.2$ (red curve), $D=0.4$ (black 
		curve) and
		$D=0.5$ (orange curve). Every curve is a result of 
		analysing $n_d=100$ disorder realizations. The plots are in linear-linear scale.}
\end{figure}

\begin{figure}[H]
	\centering
	\includegraphics[width=0.63\textwidth,keepaspectratio]{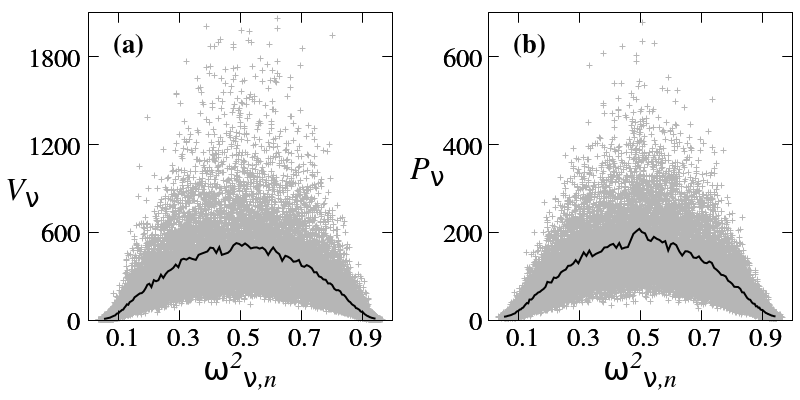}
	\includegraphics[width=0.34\textwidth,keepaspectratio]{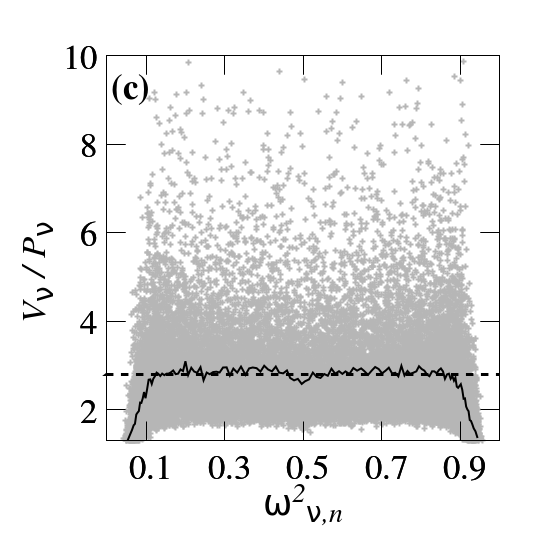}
	\caption{\label{fig:PV_one}Results for {\bf(a)} the localization volume $V_{\nu}$ \eqref{eq:V_mode}, {\bf(b)} the participation number 
		$P_{\nu}$ \eqref{eq:P_mode},
		and {\bf(c)} the scaling 
		$V_{\nu}/P_{\nu}$ of the 
		NMs of Hamiltonian model \eqref{eq:LDKG} for $D=0.1$ 
		and $n_d=100$ 
		disorder realizations, as a function of the NMs' normalized
		square frequency $\omega^2_{\nu,n}$ \eqref{eq:nomr_freq}.
		The black curves show running averages of 
		the plotted quantities, i.e.~{\bf(a)} $\langle V\rangle$, 
		{\bf(b)} $\langle P\rangle$ 
		and {\bf(c)} $\langle V/P\rangle$. The straight dashed line in 
		{\bf(c)} 
		shows the value $V_{\nu}/P_{\nu}=2.8$. The plots are in linear-linear scale.
	}
\end{figure}

This frequency distribution is substantiated in 
\autoref{fig:freq_1}{\bf(b)} where we present a histogram showing 
the probability density distribution $d_{\omega _{\nu, n}^2}$ of 
the square frequencies $\omega_{\nu, n}^2$ of \autoref{fig:freq_1}{\bf(a)}. 
The maximum values of $d_{\omega _{\nu, n}^2}$ appearing 
at the edges of the distribution are evidence of the 
concentration of the square frequencies at the borders of the spectrum.  
By considering similar findings to the ones of \autoref{fig:freq_1}{\bf(b)} 
over $n_d=100$ disorder realizations of 
parameter $D=0.1$, we obtain the squared 
frequency distribution shown in \autoref{fig:freq_1}{\bf(c)} where we 
have a smoother profile. Here the distribution is seen to have 
a `U' shape with peaks of the same distribution value 
at the two edges. As seen in 
\autoref{fig:freq_1}{\bf(a)}, where only one disorder 
realisation is considered, the frequencies avoid the extreme 
ends $\omega_{\nu, n}^2 \approx 0$ and $\approx 1$ of the band, 
something which is also visible in the more general case of  
\autoref{fig:freq_1}{\bf(c)}.
The observation of a `U' shaped distribution 
($d_{\omega _{\nu, n}^2}$) of the square 
frequencies $\omega^2_{\nu}$ 
for $D=0.1$, as seen in \autoref{fig:freq_1}{\bf(c)},
motivates us to investigate the general dependence 
of the probability density distribution $d_{\omega _{\nu, n}^2}$ on $D$.
Therefore in \autoref{fig:distributions_many} we show this dependence. 
We present results for $D=0.06$ (purple curve), $D=0.1$ 
(green curve), $D=0.2$ (red curve), $D=0.4$ (black curve) and $D=0.5$ 
(orange curve). 
For lower values of $D$, the distributions have a `bowl' 
shape feature at its centre with peaks at the 
edges of the 
distribution.
The distribution develops a 
chapeau-like shape with higher values obtained 
at plateau edges corresponding to $\omega_{\nu, n}^2 
\approx 0.2$ and 
$\omega_{\nu, n}^2\approx 0.8$ for higher values of $D$ ($D=0.4$ and $D=0.5$).
We have seen in \autoref{fig:NMs_profiles} 
the spatial extent of NMs decrease 
as $D$ increases for a representative mode at each 
value of $D$ that was considered. We now give a generalised estimate of the NM 
extent using a large 
number of disorder realisations for various $D$ values.
\newline
The extent of NMs can be numerically estimated using a number of 
different approaches [\cite{Kramer1993,Krimer2010}]. For this 
purpose, we use the localization volume $V_{\nu}$ \eqref{eq:V_mode} and 
participation number $P_{\nu}$ \eqref{eq:P_mode} 
of the NMs [\cite{Krimer2010}].
$V_{\nu}$ and $P_{\nu}$, which are proportional to the 
average localization length $\eta_{\nu}$ 
\eqref{eq:loc_length_LDKG}, were reported to capture correctly the main 
features of the NM extent [\cite{Kramer1993,Krimer2010}].
We compute $V_{\nu}$ \eqref{eq:V_mode} and $P_{\nu}$ \eqref{eq:P_mode} for 
various values of $D$ and for many disorder realisations 
in order to get better statistics. 
In \autoref{fig:PV_one} we present the localization volume 
$V_{\nu}$ \eqref{eq:V_mode}, the participation number $P_{\nu}$ \eqref{eq:P_mode} 
and the ratio  $V_{\nu}/P_{\nu}$
of the NMs of 
Hamiltonian \eqref{eq:LDKG} for $D=0.1$,
with respect to the frequency 
$\omega^2_{\nu,n}$ \eqref{eq:nomr_freq}.
All results are obtained from the analysis of 
$n_d=100$ disorder realisations. 
In creating these plots, we only consider modes with a 
mean position $l_0$ located in the central 
one-third of the lattice in order to avoid any boundary effects.
The black curves represent 
running averages $\langle V\rangle$ [\autoref{fig:PV_one}{\bf(a)}], $\langle P\rangle$ 
[\autoref{fig:PV_one}{\bf(b)}] 
and $\langle V/P\rangle$ 
[\autoref{fig:PV_one}{\bf(c)}] 
of, respectively, quantities $V_{\nu}$, $P_{\nu}$ and $V_{\nu}/P_{\nu}$.
We see that the maxima of $\langle V\rangle$  and  $\langle P\rangle$ 
are obtained 
at the frequency band centre 
(i.e.~$\omega^2_{\nu,n}\approx 0.5$), as expected from equations 
\eqref{eq:loc_length_LDKG} and \eqref{eq:loc_length_LDKG_max}.
The black continuous curve [\autoref{fig:PV_one}{\bf(c)}] 
of the ratio $\langle V/P\rangle$, is  
seen to be independent of  $\omega^2_{\nu,n}$ especially in the middle 
of the frequency band where we find the most extended modes. 
This shows that, generally the ratio 
$V_{\nu}/P_{\nu}\approx 2.8$ for all NMs 
except those whose frequencies are at the band edges.
We extend the study whose results are shown in \autoref{fig:PV_one} 
to other values of $D$ in order to understand their dependence on $D$. 
  
\begin{figure}[H]
	\centering
	\includegraphics[width=0.7\textwidth,keepaspectratio]{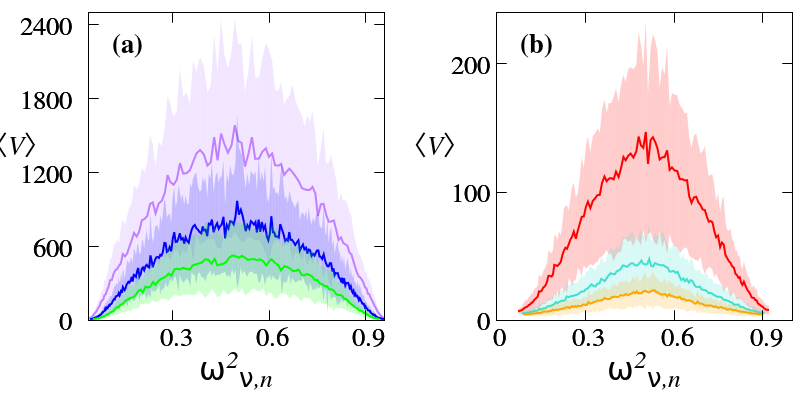}
	\caption{\label{fig:V_many} Results for the average localization 
		volume $\langle V\rangle$ of NMs located at the central 
		one-third of the lattice of the LDKG system \eqref{eq:LDKG} 
		as a function of  the normalized 
		square frequency $\omega^2_{\nu,n}$ \eqref{eq:nomr_freq}. 
		{\bf(a)} $D=0.06$ (purple curve), $D=0.08$ (blue curve) 
		and $D=0.1$ (green curve). {\bf(b)} $D=0.2$ 
		(red curve), $D=0.35$ (turquoise curve) and $D=0.5$ 
		(orange curve). The shaded area around 
		each curve indicates 1 standard deviation of $V_{\nu}$ \eqref{eq:V_mode}. 
		The plots are in linear-linear scale.
	}
\end{figure}

In \autoref{fig:V_many} we show how the dependence of $\langle V\rangle$ 
on $\omega^2_{\nu,n}$ changes with respect to $D$ for the cases 
{\bf(a)} $D=0.06$ 
(purple curve), $D=0.08$ (blue curve), $D=0.1$ 
(green curve) and {\bf(b)} $D=0.2$ 
(red curve), $D=0.35$ (turquoise curve), $D=0.5$ (orange curve).
The shaded area around the $\langle V\rangle$ curve in each case 
indicates one standard deviation.
As in \autoref{fig:PV_one}, 
we obtain the results of \autoref{fig:V_many} only for 
NMs whose 
mean positions are located in the middle one-third of the lattice. From the results 
of \autoref{fig:V_many} we observe that both the average value 
$\langle V\rangle$ and the corresponding standard deviation 
of the NM localization volume $V_{\nu}$ 
decrease as $D$ increases.

\begin{figure}[H]
	\centering
	\includegraphics[width=0.5\textwidth,keepaspectratio]{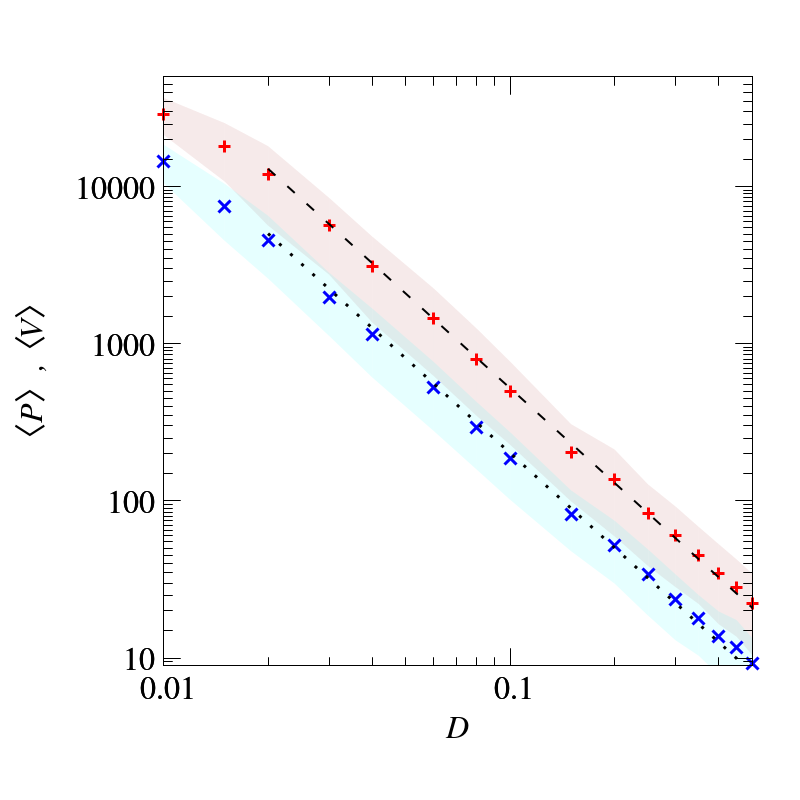}
	\caption{\label{fig:PV_D}Results for the relation between 
		the spatial extent of the NMs and parameter $D$: The average 
		participation number $\langle P\rangle$ (blue points) and 
		localization volume $\langle V\rangle$ (red points) of 
		the NMs positioned at the central one-third of the lattice 
		and with square frequencies located in the 
		middle one-third of the frequency band, 
		as a function of $D$, over $n_d=100$ disorder realizations. 
		The shaded area  indicates 1 standard deviation of $\langle V\rangle$ 
		(red shading) and $\langle P\rangle$ 
		(blue shading). The straight 
		lines correspond to the functions $\langle P\rangle= a_p D^{-2}$ (dotted lower line) and $\langle V\rangle= a_v D^{-2}$ (dashed upper line) with $a_p=2.01$ and $a_v=5.21$. 
		The plot has logarithmic axes.
	}
\end{figure}

Next we investigate further the dependence of the 
average spatial extent of the NMs on 
the disorder parameter $D$ by restricting our analysis to 
modes whose square frequencies correspond to 
the middle one-third of the frequency band (i.e.~the 
more extended modes) and in addition are also positioned at the lattice's middle 
one-third.
We compute the average participation number $\langle P\rangle$ 
and localization volume $\langle V\rangle$ for these 
modes, along with an 
estimation of the computation errors quantified 
by one standard deviation, for $n_d=100$ disorder realizations 
and present the results in \autoref{fig:PV_D}. 
We see a decrease of the spatial extent 
quantities $\langle P\rangle$ and $\langle V\rangle$ when 
$D$ increases (i.e.~as system \eqref{eq:LDKG} moves from an ordered towards 
a more disordered lattice).
Since the localization length $\eta_{\nu}$ scales as $\propto D^{-2}$ in 
equations \eqref{eq:loc_length_LDKG} and \eqref{eq:loc_length_LDKG_max}, 
we expect $\langle V\rangle$ and $\langle P\rangle$ to have a similar behaviour.
We see that this scaling is true from the results of 
\autoref{fig:PV_D} since the data 
$\langle V\rangle$ and $\langle P\rangle$ are respectively well fitted 
by the functions $a_vD^{-2}$ 
with $a_v=5.21 \pm 0.09$ shown by a dashed line
and $a_pD^{-2}$ with
$a_p=2.01 \pm 0.05$ shown by a dotted line. In 
particular, for $D=0.5$ 
we have $\langle V\rangle\approx a_vD^{-2}\approx21$ and 
$\langle P\rangle\approx a_pD^{-2}\approx8$, which is in agreement with the 
findings of \cite{Krimer2010}.	
The fitting of data shown in \autoref{fig:PV_D} gives a ratio 
$\langle V\rangle/\langle P\rangle \approx 2.6$, a value close to 
$2.8$ obtained in \autoref{fig:PV_one} for $D=0.1$. 
In \autoref{fig:PV_D} we show results for $0.01 \leq D \leq 0.5$, but 
for obtaining fittings $a_vD^{-2}$ and $a_pD^{-2}$ mentioned above, we only use 
results for $0.03 \leq D \leq 0.5$. 
This is because for $D<0.03$, the NMs are characterized by 
very large spatial extent 
which requires a 
lattice size greater than the maximum size $N=50{,}000$ 
which we used for these computations. 
Using the results of \autoref{fig:PV_D} and the frequency scale 
$\Delta_K$ \eqref{eq:Delta_K} of the LDKG system 
we can now define the average NM frequency spacing $d$ as  
\begin{equation}\label{eq:d_W4}
	d=\frac{\Delta_K}{\langle V\rangle} \approx \frac{D^2 (2D+1)}{5.21},
\end{equation} for $\mathcal{W}=4$,
where $\langle V\rangle= 5.21D^{-2}$ is the fitting 
shown in \autoref{fig:PV_D}. 
The quantities $\Delta_K$ \eqref{eq:Delta_K} and $d$ \eqref{eq:d_W4} are key in 
determining the wave packets' evolution in 
nonlinear systems.

\begin{figure}[H]
	\centering
	\includegraphics[width=0.49\textwidth,keepaspectratio]{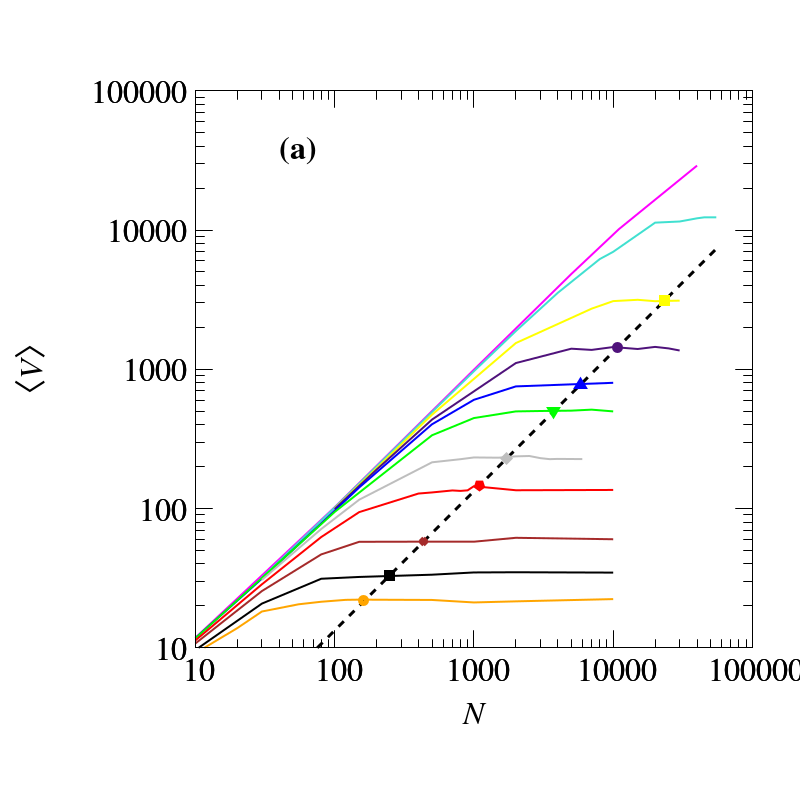}
	\includegraphics[width=0.49\textwidth,keepaspectratio]{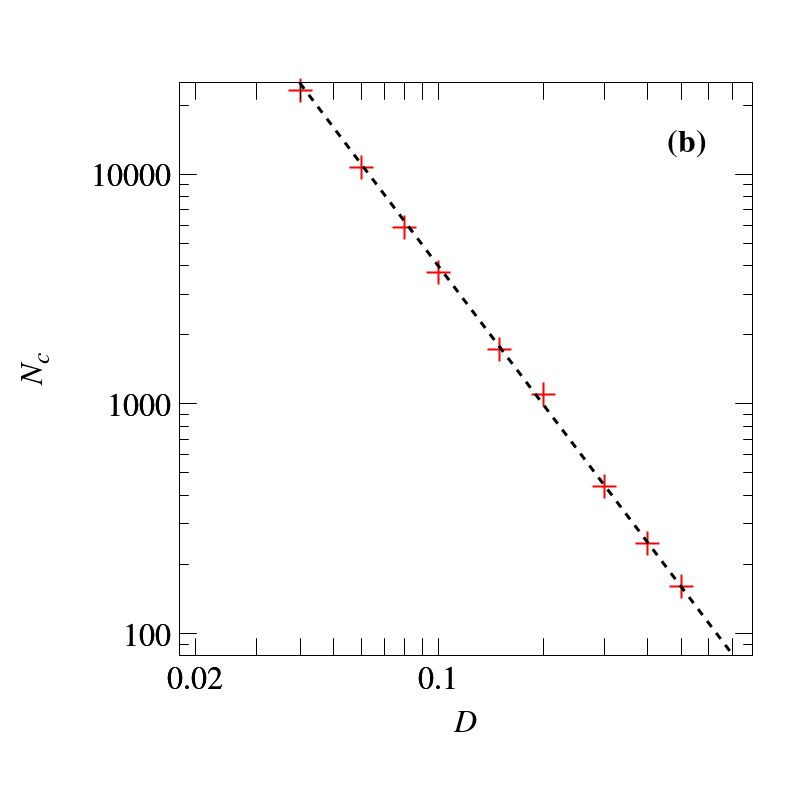}
	\caption{\label{fig:N_min} Results for lattice size $N$ 
		to use for computing NM extent. {\bf (a)} the dependence of 
		the average NM localization volume $\langle V\rangle$
		on lattice size $N$ for different values of $D=0.01$ (pink curve), 
		$0.02$ (turquoise 
		curve), $0.04$ (yellow 
		curve), $0.06$ (purple 
		curve), $0.08$ (blue curve), $0.1$ (green 
		curve), $0.15$ (gray curve), 
		$0.2$ (red curve), $0.3$ (brown curve), $0.4$ (black curve) and $D=0.5$ (orange curve).
		The straight dotted line shows the points corresponding to the critical 
		lattice size values $N_c$. {\bf (b)} the dependence 
		of $N_c$ on $D$ where the straight 
		line corresponds to the function $N_c= a_N D^{-2}$ 
		with fitting constant $a_N=40$. The plots have logarithmic axes.}
\end{figure}

As mentioned before for the results of \autoref{fig:PV_D}, lattice size 
is a restricting factor when numerically investigating NMs and their properties 
especially when $D$ is very small. 
We now present a brief discussion for an 
estimate of the lattice size $N$ which will produce 
reliable results for the study of NMs. 
We perform the study as follows:

For a particular value of $D$ (say $D=0.03$) and a fixed lattice size 
$N$ (starting with a minimum size $N=30$), 
we compute the average 
localization volume $\langle V\rangle$ of NMs using $n_d=100$ 
disorder realizations. We 
then repeat this computation procedure for an increasing value of $N$ 
(say $N=50$, $80$, $100$, etc.) until when the value of $\langle V\rangle$ 
becomes independent of $N$.
This will eventually happen when $N$ exceeds a 
critical value we denote as $N_c$. In other words, at the lattice size $N_c$ 
the system has it's maximum possible value of $\langle V\rangle$. 
We do this for other values of $D>0.03$. 
In \autoref{fig:N_min}{\bf (a)}, we show numerical findings for 
the dependence of $\langle V\rangle$ 
on $N$ when different values of $D=0.01, 0.02,\ldots,0.5$ are used. 
The dashed line shows points that correspond to 
suggested critical values $N_c$ of lattice size below which the computation 
results must be treated with utmost caution. That is to say, 
in order to 
eliminate the interference by lattice boundaries on the study of 
the NM properties and shapes, a lattice size $N>N_c$ should be used.
\autoref{fig:N_min}{\bf (b)} shows the exponential dependence 
of $N_c$ on $D$, where it clearly reveals that 
very large lattice sizes are required for 
computations involving small values of $D$. Small values of $D$ 
would therefore 
involve computing the eigenvectors and eigenvalues of a matrix 
${\bf B}$ \eqref{eq:B_epsilon}, with high order dimensions, 
and thus an extremely hard computational task.
More precisely, the critical lattice size,
$N_c$, depends on $D$ as 
\begin{equation*}
	N_c= a_N D^{-2},
\end{equation*} where $a_N=40$. This shows that for the ordered system 
(when $D$ is zero), one 
requires an infinite lattice while for 
$D=0.01$, a very big size of the order of $100{,}000$ 
is necessary to compute reliable properties of the NMs.

\section{Summary}\label{sec:res_qns}
In this chapter we gave a summarised review on properties and 
behaviour of disordered systems starting with 
the linear disordered models and discussing the localization 
of their NMs for both correlated and uncorrelated 
on-site potentials. We described some nonlinear 
disordered models with emphasis on two typical 
models, namely, the DKG and DDNLS systems. Based on the dynamics of 
the DDNLS model we 
highlighted the theoretical estimations about the energy 
spreading mechanisms induced by nonlinearities in disordered lattices.
We also mentioned some rather important findings on ordered systems like 
the FPUT and the Frenkel-Kontorova system as well as the DNA and DKG models. 
We saw that the NMs of linear disordered systems with sufficiently 
strong disorder only extend to a finite number of sites and any initially localized 
wave packet will not spread to an entire lattice of very large size. 
The introduction of nonlinearities to 
such models destroys AL as the 
NMs interact with each other and chaos emerges in 
the system.
The characteristics of the different chaotic 
dynamical behaviours, 
namely, the weak, strong and selftrapping 
regimes were identified. 
The weak chaos regime is characterised by the wave packet second moment 
growing subdiffusively following a power law $m_2(t)\propto t^{1/3}$. The strong chaos regime 
on the other hand has a faster subdiffusion as $m_2(t)\propto t^{1/2}$ 
which gradually slows down to the weak chaos spreading rate. 
The 
appearance of these various dynamical regimes is dependant on 
the properties of systems' NMs, namely, 
the frequency band width $\Delta$, the NM localization volume $V$, 
the average spacing $d\approx\Delta/V$ of the modes and the 
nonlinearity dependent frequency shift due to the energy.
In addition, we introduced and studied the qualitative and statistical NM 
properties for a modified 1D disordered DKG model \eqref{eq:LDKG}, 
whose disorder strength depends on two parameters, namely, $D$ which 
defines the interval from which the disorder values are chosen and 
$\mathcal{W}$ which regulates the size of nearest-neighbour interactions.
For a fixed  $\mathcal{W}=4$, we investigated the qualitative structural 
changes in the NMs as the model becomes disordered by letting 
$D$ approach $0.5$ from $0$. 
The NMs become more spatially localized in the lattice 
and the number of the NMs' squared frequencies 
become less clustered on the ends of the 
frequency band losing an initial `U' shape to 
a `plateau'-like distribution when $D$ increases.
Analysing further the model, we computed the NM localization volume $V_{\nu}$ \eqref{eq:V_mode} and 
participation number $P_{\nu}$ \eqref{eq:P_mode} 
for different values of $D$ in the interval 
$0<D\leq0.5$. The corresponding average values, 
over $100$ disorder realizations,
$\langle V\rangle$ and 
$\langle P\rangle$ are respectively governed by the 
laws $\langle P\rangle \propto  D^{-2}$ and 
$\langle V\rangle \propto  D^{-2}$. A correction scaling 
to cater for the 
fluctuations and thus underestimate of $\langle P\rangle$ is 
given by 
$\langle V\rangle \approx 2.6\langle P\rangle$.
The control of the on-site potentials by the two parameters 
$D$ and $\mathcal{W}$ provides more flexibility in ways 
of working with the system, as the impact of heterogeneity in the model 
can either be altered through adjusting the sample set of on-site potentials 
(by modifying 
the values of $D$) or the strength of the nearest-neighbour interactions 
(by changing the values of $\mathcal{W}$). The presence of these 
two parameters therefore 
enables the investigations on the effect of 
different physical processes on the 
system's dynamics theoretically, numerically and experimentally.
This analysis constitutes a first step 
towards an in-depth study of the dependence of the DKG system's 
dynamics on the disorder when $D$ is changed.
\chapter{Numerical techniques}\label{chap:num_tech}

\pagestyle{fancy}
\fancyhf{}
\fancyhead[OC]{\leftmark}
\fancyhead[EC]{\rightmark}
\cfoot{\thepage}

In this chapter we present the main numerical techniques we 
use to answer the research questions of our study as stated in 
Chapter \ref{chap:intro}. Parts of the findings in this chapter have been 
reported in \cite{Senyange2018}.
The Chapter is presented as 
followings: In Section \ref{sec3:dev_vec} we discuss the 
concepts of deviation vectors, 
the corresponding deviation vector distribution (DVD) and 
variational equations, which are then followed by 
a concise theoretical review on Lyapunov characteristic exponents 
(LCEs) in Section \ref{sec3:LCEs}. 
We then provide 
a detailed theoretical description of various numerical integration 
techniques
in 
Section \ref{sec:numerical_integration}, mainly
 focussing on the class known as symplectic integrators (SIs). There we 
 give a description of some already existing SIs 
 which have been previously used to integrate Hamiltonian systems 
 (e.g.~in solar system and particle accelerator dynamics) and we 
 also construct some new SIs using composition techniques. 
 We then implement in Section \ref{numerical_results}, the 
 methods discussed in Section \ref{sec:numerical_integration} to compare the 
 computational efficiency of many different integrators in 
 following the dynamical evolution of  
 Hamiltonian systems. Finally in Section \ref{sec3:conclusions} we 
 discuss and summarize our findings.


\section{Deviation vectors}\label{sec3:dev_vec}
A variety of techniques used to study dynamical systems 
involve the consideration of the relative position (or positions) 
of a pair (or pairs) of points in their respective time-trajectories 
in a phase space. An infinitesimal perturbation 
vector which describes the relative position of 
two phase space points is called a 
\textit{deviation vector}. For 
any
pair of points $\mathbf{y}=(y_1,y_2,y_3,\cdots,y_N,p_{y_1},p_{y_2},\cdots,p_{y_N})$
and $\mathbf{x}=(x_1,x_2,x_3,\cdots,x_N,p_{x_1},p_{x_2},\cdots,p_{x_N})$ 
in a $2N$-D phase space, 
the deviation of $\mathbf{y}$ from $\mathbf{x}$ is 
the vector $\mathbf{w}=\mathbf{y}-\mathbf{x}$ where 
\begin{equation}\label{eq:dev_vec}
\mathbf{w}=(w_{_\mathbf{r1}},w_{_\mathbf{r2}},\cdots,w_{_\mathbf{r2N}})=(\delta q_{_\mathbf{r1}},\delta q_{_\mathbf{r2}},\cdots,
\delta q_{_\mathbf{rN}},\delta p_{_\mathbf{r1}},\delta p_{_\mathbf{r2}},
\cdots,\delta p_{_\mathbf{rN}})
\end{equation} with $\mathbf{ri}$ denoting the $n$-tuple 
spatial-dimension index 
ranked in position \textbf{i}.
That is to say, the indexing for an $N$ oscillator 
1D Hamiltonian model is such that 
$\mathbf{r1}=1$ and $\mathbf{rN}=N$, while for a $N=KM$ oscillator 
2D Hamiltonian model (array of $K$ oscillators along one direction and $M$ 
along a perpendicular direction), $\mathbf{r1}=(1,1)$ and $\mathbf{rN}=(K,M)$. 
More particularly, $\delta q_\mathbf{ri}=y_i-x_i=w_\mathbf{ri}$ for 
$i\leq N$ and $\delta p_\mathbf{ri}=p_{y_i}-p_{x_i}=w_\mathbf{ri}$ for 
$i > N$.
We use Equation (\ref{eq:dev_vec}) of a deviation vector to define a 
normalized distribution, the DVD.

\subsection{Deviation vector distribution (DVD)}
In order to study extensively the 
chaotic behaviour of the DKG systems 
(\ref{eq:kg1dham}) and (\ref{eq:kg2dham}), 
we monitor the evolution of time dependent normalized
DVD
$\{\xi^D_{_\mathbf{r}}\}_{_{\mathbf{r}\in\mathbb{N}^{^n}}}$ of the 
system, for a lattice of $n$ spatial dimensions.
\begin{equation}\label{eq:dvd}
	\xi^D_{_\mathbf{r}} = \frac{\delta q_{_\mathbf{r}}^2+\delta p_{_\mathbf{r}}^2}
	{\sum_\mathbf{k}\delta q_{_\mathbf{k}}^2+\delta p_{_\mathbf{k}}^2},	
\end{equation} where $\mathbf{k}$ and $\mathbf{r}$ are $n$-tuples whose components are positive integers.

Through the DVD we monitor the position of the 
degrees of freedom where the chaotic dynamics is concentrated 
(chaotic hotspots) [\cite{Skokos2013}] as time evolves. We also use the deviation 
vector for the computation of the LCEs 
[\cite{Lyapunov1992,Oseledec1968,Benettin1976,Pesin1977,Skokos2010c}]. 
LCEs, which we describe in later in Section \ref{sec3:LCEs}, are 
a measure at relatively large times. Due to computational limitations, 
we are unable to compute the LCEs for very long times. We therefore seek 
deviation vectors that will give as much information about the exponents 
as possible in the shortest time possible.
For 
the DVD \eqref{eq:dvd}, we define the associated 
second moment $m_{_2}^D$ and 
participation number $P^D$ as follows:

The participation number $P^D$ of the DVD (\ref{eq:dvd}),
\begin{equation}\label{eq:P_Dvd}
P^D=\frac{1}{\sum_\mathbf{r}{\xi^D_{_\mathbf{r}}}^2},
\end{equation}
where $\mathbf{r}$ is a spatial dimension dependent index, 
estimates the number of lattice sites with the strongest deviation. 
The second moment $m_{_2}^D$ of the DVD (\ref{eq:dvd}) which quantifies 
the spatial extent of spreading of the distribution is given by 
\begin{equation}\label{eq:m2_dvd}
m_{_2}^D=\sum_{\bf r}||{\bf r}-\overline{\bf r}||^2\xi^D_{_{\bf r}}
\end{equation} where 
$\overline{\bf r}=\sum_{\bf r}{\bf r}\xi^D_{_{\bf r}}$ 
is the 
centre of the distribution $\{\xi^D_{_{\bf r}}\}_{_{{\bf r}}}$ and $||\cdot||$ 
denotes the Euclidean norm.
\subsection{Equations of motion and variational equations}\label{sec3:var_eq}
Let $H({\bf q},{\bf p})$ be a Hamiltonian function 
of a time independent model with 
$N$ degrees of freedom. For generalised position 
vector ${\bf q}$ and conjugate 
momenta vector ${\bf p}$, the system configuration 
in the $2N{-}D$ 
phase space is described by the point 
${\bf z}(t)=\left({\bf q}(t),{\bf p}(t)\right)$.
The resulting equations of motion
\begin{equation}\label{eq:motion}
\frac{d{\bf p}}{dt} = - \frac{\partial H}{\partial{\bf q}},\qquad\frac{d{\bf q}}{dt} = \frac{\partial H}{\partial{\bf p}}
\end{equation}
can be written using matrices as $ \frac{d{\bf z}}{dt}= J_{_{2N}}D$, 
where $D$ is the transpose of 
$\begin{pmatrix}
\frac{\partial H}{\partial{\bf q}} & \frac{\partial H}{\partial{\bf p}}
\end{pmatrix}$.
$J_{_{2N}}$ is a matrix of the form
\begin{equation*}
J_{_{2N}}=\begin{pmatrix}
{\bf 0}_{_N} & {\bf I}_{_N} \\
{\bf - I}_{_N} & {\bf 0}_{_N}
\end{pmatrix}
\end{equation*} with $N\times N$ identity matrix ${\bf I}_{_N}$ and 
$N\times N$ zero matrix ${\bf 0}_{_N}$.

A deviation vector stationed at the point ${\bf z}(t)$ evolves 
in the tangent space following the so called 
\textit{variational equations} [see e.g. 
\cite{Skokos2010,Gerlach2011,Gerlach2012}]
\begin{equation}\label{var_eq}
\frac{d {\bf w}(t)}{dt} =  A(t){\bf w}(t),
\end{equation} where ${\bf w}=(\delta q_{_{\bf r1}},\delta q_{_{\bf r2}},\cdots,
\delta q_{_{\bf rN}},\delta p_{_{\bf r1}},\delta p_{_{\bf r2}},
\cdots,\delta p_{_{\bf rN}})$ denotes the deviation vector 
(\ref{eq:dev_vec}) of small 
perturbations, $A(t)=J_{_{2N}}D^2({\bf z}(t))$ and 
$D^2({\bf z}(t))=\left[\frac{\partial^2H({\bf z})}{\partial q_{_{\bf j}}\partial q_{_{\bf k}}}\right]$ is 
the 
Hessian matrix of the Hamiltonian calculated at ${\bf z}(t)$.
The differential equations (\ref{var_eq}) are linear with respect to 
${\bf w}$ with coefficients which are elements of matrix $A(t)$.

Let us now compute the variational equations for an 
autonomous Hamiltonian $H({\bf z})$ which can be split in two parts as 
\begin{equation}\label{eq:hamp}
H({\bf z}) = T({\bf p}) + V({\bf q})
\end{equation} where $H({\bf q},{\bf p})$ is the energy, $V$ is the 
potential energy and $T=\sum_{\bf r}\frac{p_{_{\bf r}}^2}{2}$ is the kinetic energy. 
The corresponding equations of motion (\ref{eq:motion}) are 
\begin{equation}\label{eq:motionp}
\frac{d{\bf q}}{dt} = {\bf p},\qquad
\frac{d{\bf p}}{dt} = - \frac{\partial V({\bf q})}{\partial{\bf q}}
\end{equation} and the variational equations (\ref{var_eq}) become 
\begin{equation}\label{eq:variational}
\frac{d}{dt} {\bf w}(t) = \begin{pmatrix}
\frac{d}{dt}(\delta{\bf q}_{_{\bf r}}) \\
\\
\frac{d}{dt}(\delta{\bf p}_{_{\bf r}})
\end{pmatrix} = \begin{pmatrix}
{\bf 0}_{_N} & {\bf I}_{_N} \\
\\
-D^2_V({\bf q}(t)) & {\bf 0}_{_N}
\end{pmatrix}\begin{pmatrix}
\delta{\bf q}_{_{\bf r}} \\
\\
\delta{\bf p}_{_{\bf r}}
\end{pmatrix},
\end{equation}
where the element with indices ${\bf j}, {\bf k}$ of 
sub matrix $D^2_V({\bf q}(t))$ is
$D^2_V({\bf q}(t))_{_{\bf j,k}}=\frac{\partial^2V({\bf q})}{\partial q_{_{\bf j}}\partial q_{_{\bf k}}}.$
Equivalently, Equation (\ref{eq:variational}) can be written as 
\begin{equation*}
\frac{d}{dt}(\delta{\bf q}_{_{\bf r}})=\delta{\bf p}_{_{\bf r}};\qquad
\frac{d}{dt}(\delta{\bf p}_{_{\bf r}})=-D^2_V({\bf q}(t))\delta{\bf q}_{_{\bf r}}.
\end{equation*}
We use the \textit{tangent map method} (TM) technique 
[\cite{Skokos2010,Gerlach2011,Gerlach2012}] to integrate the 
variational equations (\ref{eq:variational}).
The dynamics on the tangent space for the Hamiltonian (\ref{eq:hamp}) is 
defined by the non autonomous Hamiltonian function 
\begin{equation}\label{eq:tdh}
H_{\mathcal{V}}(\delta{\bf q},\delta{\bf p},t)=
\frac{1}{2}\sum_{\bf i}\delta{p_{_{\bf i}}}^2+\frac{1}{2}\sum_{\bf j,k}D^2_V({\bf q}(t))_{_{\bf j,k}}\delta q_{_{\bf j}}\delta q_{_{\bf k}},
\end{equation} called the \textit{tangent dynamics Hamiltonian} (TDH). 
The equations of motion 
for the TDH (\ref{eq:tdh}) are the 
variational equations (\ref{eq:variational}).
We integrate equations (\ref{eq:variational}) along with the 
equations of motion 
(\ref{eq:motionp}) since 
the variational equations depend on the position ${\bf q}$.

\section{Lyapunov characteristic exponents (LCEs)}\label{sec3:LCEs}
LCEs are asymptotic measures which are used to 
characterise the average growth of small perturbations to the solutions of a 
dynamical system.
Lyapunov [\cite{Lyapunov1992}] introduced the concept of LCEs 
in $1892$ when analysing the stability of non-stationary 
solutions of ordinary differential 
equations. Since then, LCEs have been widely used by researchers 
in studying dynamical 
behaviour in various systems. A detailed survey of the 
theoretical and application aspects of this topic is 
presented and can be found in \cite{Skokos2010c}.

The theory of LCEs was applied to characterise chaotic orbits by \cite{Oseledec1968}, and the connection between LCEs and exponential 
divergence of nearby orbits was given in \cite{Benettin1976} and 
\cite{Pesin1977}. For a chaotic 
orbit, at least one LCE is positive, implying exponential divergence of nearby orbits, 
while in the case of regular orbits all LCEs are zero or negative. 
Therefore, since the presence of a positive exponent 
guarantees chaoticity of an orbit, the computation of 
the \textit{maximum Lyapunov Characteristic Exponent} (mLCE) $\lambda$  
suffices for purposes of 
determining the chaotic nature of the orbit.

The mLCE is computed as the limit for $t\rightarrow\infty$ of the quantity

\begin{equation}\label{eq:ftmLCE}
\Lambda(t)=\frac{1}{t}\ln\frac{\|{\bf w}(t)\|}{\|{\bf w}(0)\|},
\end{equation} often called the finite time mLCE, where ${\bf w}(0)$ and ${\bf w}(t)$ 
are deviation vectors from a given orbit at times $t=0$ and $t >0$,
respectively. As usual $\|.\|$ denotes the norm of a vector. So, we have 
\begin{equation}\label{eq:mLCE}
\lambda=\lim_{t\rightarrow\infty}\Lambda(t).
\end{equation}
The existence of this limit is guaranteed by the Multiplicative 
Ergodic Theorem [\cite{Oseledec1968,Benettin1980,Benettin1980b,Skokos2016}]. 
$\Lambda(t)$ tends to zero 
in the case of regular orbits following a power law [\cite{Benettin1976}],
\begin{equation*}
\Lambda(t)\propto t^{-1},
\end{equation*} while it tends to non-zero values in the case of chaotic orbits.

A Hamiltonian of $N$ degrees of freedom has at most $2N$ different 
LCEs, which are ordered as 
$\lambda=\lambda_1\geq\lambda_2\geq\ldots\geq\lambda_{2N}$.
In \cite{Benettin1978} a theorem was formulated, which led directly to the development 
of a numerical technique for the computation of all LCEs. This technique 
which is based on the time evolution of multiple deviation vectors
keeps them linearly 
independent by the Gram-Schmidt orthonormalization procedure.
The theoretical framework, as well as the corresponding numerical 
method for the computation of all LCEs (usually called the standard method), was 
given in \cite{Benettin1980,Benettin1980b}. According to this method all LCEs 
$\lambda_2, \lambda_3,$ etc. are computed respectively as the 
limit for $t\rightarrow\infty$ of some 
appropriate finite-time dependent quantities $\Lambda_2, \Lambda_3,$ etc 
[\cite{Benettin1980b,Skokos2010}].
It has been shown in \cite{Benettin1980} that for an autonomous 
Hamiltonian flow, the LCEs are such that, 
\begin{equation*}
\lambda_i=-\lambda_{k}, \,\,\,{\textnormal{for}}\,\,\, i=1,2,\cdots,N\,\,\, {\textnormal {and}}\,\,\, i+k=2N+1.
\end{equation*}
In addition, since the Hamiltonian function is an integral of motion, at least two LCEs vanish, 
i.e.,
\begin{equation*}
\lambda_N=-\lambda_{N+1}=0,
\end{equation*} while the presence of any additional independent integral of motion 
leads to the vanishing of another pair of the LCEs.

\section{Numerical Integration}\label{sec:numerical_integration}

Ordinary differential equations can be integrated using 
various numerical integration 
techniques. 
Unfortunately, some integration techniques 
fail to preserve important structural properties that define the 
systems of differential equations, for example Hamiltonian 
system conservation 
properties. 
However, SIs 
conserve the integrals of motion in Hamiltonian systems for long times
and so are a preference for the 
integration of these systems.
Starting with the work of \cite{Wisdom1991}, SIs 
have been widely used for long-term integrations of the solar system. 
In such studies, for example \cite{Mclachlan1995} and \cite{Morbidelli2002}, 
low order schemes 
were constructed and used for solving Hamiltonian systems.
For our work, we need schemes that are computationally fast 
and efficient thereby enabling us to perform a long term 
analysis of the dynamics of the studied systems.

\subsection{Symplectic integrators (SIs)}\label{sec3:SIs}
We now discuss in more detail the theory of SIs, mention some efficient 
schemes that have been previously used and recommended, 
describe composition techniques for the construction of new schemes 
and define some new SIs. 

Let $f({\bf q},{\bf p})$ and $g({\bf q},{\bf p})$ 
be real valued differentiable functions of ${\bf q}$ and ${\bf p}$ 
defined on $\mathbb{R}^{2m}$ for some positive integer $m$. 
Then, the 
\textit{Poisson bracket} $\{\cdot,\cdot\}$ of
$f({\bf q},{\bf p})$ and $g({\bf q},{\bf p})$ is defined as
\begin{equation*}
\{f,g\}=\sum_{i=1}^{m} \left( \frac{\partial f}{\partial q_i}
\frac{\partial g}{\partial p_i} - \frac{\partial f}{\partial p_i}
\frac{\partial g}{\partial q_i}\right).
\end{equation*} 
For a point ${\bf z} = ({\bf q},{\bf p})$ on the Hamiltonian 
surface defined by $H({\bf q},{\bf p})$,  we can write the 
Hamilton equations of motion (\ref{eq:motion}) in the form
\begin{equation} \label{eq:poisson}
\frac{d {\bf z}}{dt}=
  \{{\bf z},H({\bf z})\} =: L_H{\bf z},
\end{equation} where $L_H$ is a differential operator.
For an initial condition ${\bf z}(t)$, the formal solution of 
 Equation (\ref{eq:poisson}) is 
\begin{equation}\label{eq:f_sol}
{\bf z}(t+\tau)=\sum_{i\geq 0} \frac{\tau^i}{i!}  L_H^i{\bf z}(t)=e^{\tau L_H} {\bf z}(t).
\end{equation}
If the Hamiltonian function $H({\bf q},{\bf p})$ can be split into 
two integrable parts like for example $T=T({\bf p})$ and 
$V=V({\bf q})$, 
namely $H=T+V$ (for example in Equation (\ref{eq:hamp})), then the action of 
the operators $e^{\tau L_T}$ and 
$e^{\tau L_V}$ (obtained by splitting $H({\bf q},{\bf p})$) is 
explicitly known through finding 
the analytic solution of the differential equations (\ref{eq:motion}). 
An explicit SI of order $k$ 
(where $k\in \mathbb{N}$) 
approximates the action of operator $e^{\tau L_H}$ by a series of products of 
operators $e^{a_i\tau L_V}$ and $e^{b_i\tau L_T}$ for $i \in \{1,2,3 \cdots, j\}$ for 
some integer $j\ge1$, $a_i, b_i$ being coefficients that are determined 
so as to 
minimize the error in the approximation.
The Baker-Campbell-Hausdorff (BCH) relation enables us to find 
order conditions
which coefficients $a_i$, $b_i$ satisfy for high order 
schemes [\cite{Koseleff1993}, \cite{Koseleff1996}].

The energy components $T({\bf p})$ and $V({\bf q})$ can be 
considered as Hamiltonians and so the operators 
$e^{\tau L_V}$ and $e^{\tau L_T}$ are symplectic maps which are 
acting on the phase space point {\bf z}. This therefore means 
that the operator  $e^{\tau L_H}$ is approximated by 
a symplectic map which is 
the product composition of 
$e^{a_i\tau L_V}$ and $e^{b_i\tau L_T}.$ That is to say 
 \begin{equation*}
 e^{\tau L_H} = e^{\tau (L_V+L_T)} = \prod_{i=1}^{j}e^{a_i\tau L_V}e^{b_i\tau L_T} + \mathcal{O}(\tau^{n+1}),
 \end{equation*} where the positive integer $n$ 
 corresponds to the order of the integrator; 
 and the number ($\leq 2j$) of appearances 
 of $e^{a_i\tau L_V}$ and $e^{b_i\tau L_T}$ in the product 
 approximation of $e^{\tau L_H}$ is called the number of steps of the 
 SI.
 In the next Sections we study various SIs that have been developed and used by 
 different researchers, as well as some new SIs we construct through composition 
 techniques.
 
\subsubsection{SIs of order one}
\label{sec3:SI1}
If a Hamiltonian $H({\bf q},{\bf p})$ is analytically integrable (e.g for 
linear systems), then without having to split $H$ into it's potential and 
kinetic components, we can construct the 
simplest single step 
SI $S_1$ as $S_1=e^{\tau L_H}$ whose application in solving 
the Hamiltonian equations gives no error in the energy.
For non integrable Hamiltonians however, we think of SIs 
that make use of the part splitting 
techniques. A simple such first order SI is the 
two step symplectic Euler method 
[\cite{Hairer2002}] defined as
\begin{equation*}
 E(\tau) = e^{\tau L_V}e^{\tau L_T}.
\end{equation*} Unfortunately, SIs of order one have a limited 
application in solving 
general Hamiltonian systems because of 
their very low accuracy in preserving the energy 
$H({\bf q},{\bf p})$ of the system. 
That is to say, the approximation of the solution in 
Equation (\ref{eq:f_sol}) leaves 
a relatively large error term especially when a large time step 
$\tau$ is used. A small time step reduces the error in computation 
but greatly increases the time required to complete the integration. We therefore focus more on 
schemes of order greater than one.
\subsubsection{SIs of order two}
\label{sec3:SI2}
A splitting of the Hamiltonian to have the position and momentum 
operators $e^{\tau L_V}$ and $e^{\tau L_T}$ respectively contribute one 
and two integration steps gives a three step SI: 
\begin{equation}
\label{eq:Leap}
 LF(\tau) = e^{a\tau L_T}e^{b\tau L_V}e^{a\tau L_T}.
\end{equation}
This integrator, which has been 
referred to as the Leap frog or St\"{o}rmer/Verlet [\cite{Ruth1983}, 
\cite{Hairer2002}], has positive coefficients $a = \frac{1}{2}$ and $b = 1$. 
This is one of the simplest splitting we can have for the integration 
operators $e^{\tau L_V}$ and $e^{\tau L_T}$ as other possible 
splittings will either generate the same number of steps or more 
than three steps.
A second class of integrators with order $n \geq 2$ 
that has been frequently used to integrate Hamiltonian 
systems and symplectic maps is the $SABA$ and $SBAB$ schemes 
[\cite{Laskar2001}] 
in which only positive 
integration steps are used. These integrators are in general used 
to integrate 
Hamiltonian systems of the form $H=T + \epsilon V$ for $T$ and $V$ integrable 
and $\epsilon$ a small perturbation parameter. 
Second order SIs
$SABA_1$ and $SBAB_1$ of $3$ steps, are identical to the 
Leap frog integrator $LF$ (\ref{eq:Leap}). That is to say,
\begin{equation*}
SABA_1=e^{\frac{1}{2}\tau L_T}e^{\tau L_V}e^{\frac{1}{2}\tau L_T}\,\,\,
\textnormal{and}\,\,\,
SBAB_1=e^{\frac{1}{2}\tau L_V}e^{\tau L_T}e^{\frac{1}{2}\tau L_V}.
\end{equation*}
Other second order SIs are the $SABA_2$ and $SBAB_2$ 
(under SI classes $SABA$ and $SBAB$ respectively)
[\cite{Laskar2001}] and have the forms
\begin{equation}
\label{eq:SABA2}
SABA_2(\tau)=e^{a_1\tau L_T}e^{b\tau L_V}e^{a_2\tau L_T}e^{b\tau L_V}e^{a_1\tau L_T},
\end{equation}
with $a_1=\frac{1}{2} - \frac{1}{2\sqrt{3}}$, $a_2 = 1 - 2a_1$, $b = \frac{1}{2}$, and
\begin{equation}
\label{eq:SBAB2}
SBAB_2(\tau) = e^{b_1\tau L_V}e^{a\tau L_T}e^{b_2\tau L_V}e^{a\tau L_T}e^{b_1\tau L_V},
\end{equation}
for $a = \frac{1}{2}$, $b_1 = \frac{1}{6}$ and $b_2 = 1 - 2b_1$. Each one of these 
schemes has $5$ steps and an error of the order $\mathcal{O}(\tau^4+\tau^2)$ for an integration time step $\tau$. 

A second order integrator with $9$ steps, $ABA82$ was studied in
 \cite{Mclachlan1995} and \cite{Farres2013}, namely
\begin{equation}
\label{eq:ABA82}
ABA82(\tau) = e^{a_1\tau L_T}e^{b_1\tau L_V}e^{a_2\tau L_T}e^{b_2\tau
  L_V}e^{a_3\tau L_T}e^{b_2\tau L_V} e^{a_2\tau L_T}e^{b_1\tau
  L_V}e^{a_1\tau L_T},
\end{equation}
where the coefficients $a_k$, $b_k$, $k=1,2,3$ are specified in 
\cite{Farres2013}. We note that the SI 
$ABA82$ has been referred to as $SABA_4$ in \cite{Laskar2001}.

%
%
%
\subsubsection{SIs of order four}
\label{sec3:SI4}
The order two SIs $SABA_2$ (\ref{eq:SABA2}) and $SBAB_2$ (\ref{eq:SBAB2})
can be made more accurate by increasing their order and thereby reducing 
the error magnitude for their computations  
if the Poisson bracket $\{
V,\{ V,T \}\}$  leads to an 
integrable Hamiltonian [\cite{Laskar2001}]. 
The systems considered 
in this work are such that 
$T$ is quadratic in momenta ${\bf p}$ and independent of ${\bf q}$ 
while $V$ only depends 
on the positions ${\bf q}$. Therefore $\{V,\{ V,T \}\}$ is integrable 
since it only depends on ${\bf q}$. 
Thus, the accuracy of (\ref{eq:SABA2}) and 
(\ref{eq:SBAB2}) can be improved by the application of 
a corrector term 
\begin{equation}\label{eq:cor_term}
C(\tau)=e^{-\tau^3 \frac{c}{2}
	L_{\{ V,\{ V,T \}\}}}
\end{equation} before and after the application of the main
body of these integrators, where $c=\frac{(2-\sqrt{3})}{24}$
for $SABA_2$ and $c=\frac{1}{72}$ for $SBAB_2$.
The subsequent application of \eqref{eq:cor_term} gives two SIs, 
each with seven steps and error of order 
$\mathcal{O}(\tau^4)$, 
which we name $SABA_2C$ for $SABA_2$ and $SBAB_2C$ for $SBAB_2$. That is 
to say,
\begin{equation*}
SABA_2C=e^{-\tau^3\frac{c}{2}
	L_{\{ V,\{ V,T \}\}}}e^{a_1\tau L_T}e^{b\tau L_V}e^{a_2\tau L_T}e^{b\tau L_V}e^{a_1\tau L_T}e^{-\tau^3\frac{c}{2}
	L_{\{ V,\{ V,T \}\}}}
\end{equation*} and
\begin{equation*}
SBAB_2C=e^{-\tau^3\frac{c}{2}
	L_{\{ V,\{ V,T \}\}}}e^{b_1\tau L_V}e^{a\tau L_T}e^{b_2\tau L_V}e^{a\tau L_T}e^{b_1\tau L_V}e^{-\tau^3\frac{c}{2}
	L_{\{ V,\{ V,T \}\}}}
\end{equation*} with coefficients $a_i$, $b$ and $a$, $b_i$ respectively corresponding 
to the ones of Equations (\ref{eq:SABA2}) and (\ref{eq:SBAB2}).
We note that SIs in the classes $SABA$ and $SBAB$ of order greater than two,  
were constructed and tested in \cite{Laskar2001}. For small time steps, the 
SI $LF(\tau)$ gave the highest error in the energy compared 
to all the other SIs which all gave practically the 
same energy error. For each of the SIs investigated, the energy error 
was found to grow linearly with the increase in time step for moderate 
time steps. This linearity was found to be lost in almost 
all higher order SIs for larger time steps with $LF(\tau)$ giving 
the smallest error in energy for very large time steps. 
The SIs $SABA_2C$ and $SBAB_2C$, which 
require additional 
computations for the corrector terms $C(\tau)$ (\ref{eq:cor_term}), 
have been reported to 
be more efficient compared to other schemes of order less or equal to 
four except for cases where very high accuracy is desired and 
the time step used is small.

A second class of fourth order SIs we consider for our study 
was presented 
in \cite{Farres2013} and also in \cite{Blanes2013} as SIs of generalised order.
 We consider the relatively efficient SIs $ABA864$ of 15 steps 
and $ABAH864$ of 17 steps.
A detailed description of these integration methods and the values of 
the coefficients $a_i$, $b_i$ 
 can be found in \cite{Blanes2013}.

In \cite{Yoshida1990}, a method to construct SIs of 
higher order by successively applying lower order schemes is 
proposed. According to that approach, we can construct 
a SI $S_{2n+2}(\tau)$ of even order $2n+2$ by applying 
SI $S_{_{2n}}(\tau)$ of even order $2n$ in the following way:
\begin{equation}
\label{eq:CompYosh}
    S_{2n+2}(\tau)=S_{_{2n}}(d\tau)S_{_{2n}}(\tau-2d\tau)S_{_{2n}}(d\tau),
\end{equation}
with coefficient
\begin{equation*}
d=\frac{-2^{\frac{1}{2n+1}}}{2-2^{\frac{1}{2n+1}}}. 
\end{equation*}
This means that we have to successively apply 3 times the SI $S_{_{2n}}$ when 
generating the scheme $S_{2n+2}$. Hence,
 a successive application of $S_2$, $3^n$ times, with the 
 appropriate coefficients 
gives the SI $S_{2n+2}$.
In this way the number of steps of the scheme $S_{2n+2}$ grows rapidly when $n$
increases, even when the compositions can be optimised by grouping together 
similar 
adjacent elementary operators 
for example combining 
$e^{a_i\tau L_V}$ with $e^{b_i\tau L_V}$ in the two step operator 
$e^{a_i\tau L_V}e^{b_i\tau L_V}$ 
to get a one step operator $e^{(a_i+b_i)\tau L_V}$ (or $e^{a_i\tau L_T}$ with $e^{b_i\tau L_T}$ in $e^{a_i\tau L_T}e^{b_i\tau L_T}$ 
to get $e^{(a_i+b_i)\tau L_T}$).

Using the composition technique (\ref{eq:CompYosh}), we now demonstrate 
how to construct several 
SIs of order four by using order two schemes. In \cite{Forest1990} 
and \cite{Yoshida1990} a fourth 
order SI of 7 steps was constructed using the SI $LF$ (\ref{eq:Leap}).
This scheme, which we refer to as $FR_4$ is defined by
\begin{equation*}
    FR_4(\tau)=
    e^{a_1 \tau L_T}
    e^{b_1 \tau L_V}
    e^{a_2 \tau L_T}
    e^{b_2 \tau L_V}
    e^{a_2 \tau L_T}
    e^{b_1 \tau L_V}
    e^{a_1 \tau L_T},
\end{equation*}
with
\begin{equation*}
a_1=\frac{1}{2\left(2-2^{1/3}\right)},\,\,\,
a_2=\frac{1-2a_1}{2},\,\,\,
b_1=\frac{1}{2-2^{1/3}},\,\,\,
b_2= 1-2b_1.
\end{equation*}  
We note that $FR_4$ corresponds to $SABA_3$ or $SBAB_3$ under, 
respectively, the 
class of $SABA$ and $SBAB$ [\cite{Laskar2001}] SIs.

In a similar way we also construct fourth order schemes $SABA_2Y_4$,
$SBAB_2Y_4$ and $ABA82Y_4$ (of steps $13$, $13$ and $25$ respectively) 
by applying the 
composition method (\ref{eq:CompYosh}) to respectively 
the SIs
$SABA_2$ (\ref{eq:SABA2}), $SBAB_2$ (\ref{eq:SBAB2}) and $ABA82$
(\ref{eq:ABA82}) of order two.

%
\subsubsection{SIs of order six}
\label{sec3:SI6}

The composition technique (\ref{eq:CompYosh}) allows us to construct SIs of 
order six from integrators of order four.
In particular, we use $FR4$, $SABA_2Y_4$, $SBAB_2Y_4$, $ABA82Y_4$,
$SABA_2C$ and $ABA864$ to create respectively 
$FR4Y_6$, $SABA_2Y_4Y_6$, $SBAB_2Y_4Y_6$, $ABA82Y_4Y_6$, 
$SABA_2CY_6$ and $ABA864Y_6$.
The new schemes $FR4Y_6$, 
$SABA_2Y_4Y_6$, $SBAB_2Y_4Y_6$, $ABA82Y_4Y_6$,
$SABA_2CY_6$ and $ABA864Y_6$ respectively have
19, 37, 37, 73, 19 and 43 steps.

Besides the composition technique (\ref{eq:CompYosh}), another 
composition method 
for constructing an order six SI $S_6$ from an order two SI $S_2$ 
was presented in \cite{Yoshida1990}. This scheme requires fewer 
steps than a sixth order SI generated from an order two SI by 
equation (\ref{eq:CompYosh}) and  
is of the form
\begin{equation}
\label{eq:CY6}
 S_6(\tau)=S_2(w_3\tau)S_2(w_2\tau)S_2(w_1\tau)S_2(w_0\tau)
 S_2(w_1\tau)S_2(w_2\tau)S_2(w_3\tau),
\end{equation}
where the coefficients $w_i$ are specified in 
\cite{Yoshida1990}. According to \cite{Mclachlan1995} the set of coefficients in 
\cite{Yoshida1990} that corresponds to what is referred to as \textit{solution $A$} 
lead to the most numerically efficient composition schemes of the form 
(\ref{eq:CY6}). We note that in \cite{Kahan1997} the composition method with 
\textit{solution $A$} has been named $s7odr6$. We construct more SIs of order 
six by applying this composition method to some integrators of order two.
From $SABA_2$
(\ref{eq:SABA2}), $SBAB_2$ (\ref{eq:SBAB2}) and $ABA82$
(\ref{eq:ABA82}), we respectively obtain $SABA_2Y_6$ and
$SBAB_2Y_6$ with $29$ steps each and $ABA82Y_6$ with $57$ steps.

We also consider order six integration schemes $s11odr6$ of \cite{Sofroniou2005} and, $s9odr6b$ of
\cite{Kahan1997} which 
respectively are based on 
$11$ and $9$ successive applications of
$S_2(\tau)$. That is to say,
\begin{equation}
\label{eq:C6SS}
s11odr6(\tau)=S_2(\gamma_1\tau)S_2(\gamma_2\tau)\cdots
S_2(\gamma_5\tau)S_2(\gamma_6\tau)S_2(\gamma_5\tau) \cdots
S_2(\gamma_2\tau)S_2(\gamma_1\tau).
\end{equation}
\begin{equation}
\label{eq:C6KL}
s9odr6b(\tau)=S_2(\delta_1\tau)S_2(\delta_2\tau)S_2(\delta_3\tau)S_2(\delta_4\tau)
S_2(\delta_5\tau) S_2(\delta_4\tau) S_2(\delta_3\tau)
S_2(\delta_2\tau)S_2(\delta_1\tau),
\end{equation} 
where the values of
$\gamma_i$, $i=1,\ldots,6$ and $\delta_i$, $i=1,\ldots,5$ are respectively 
specified in \cite{Sofroniou2005} and \cite{Kahan1997}. 
When we use order two schemes $SABA_2$ (\ref{eq:SABA2}) and $ABA82$ 
(\ref{eq:ABA82}) in Equation (\ref{eq:C6KL}) we construct the order six SIs 
$s9SABA_26$ ($37$ steps) and $s9ABA82\_6$ ($73$ steps) 
respectively. Similarly we construct SIs 
of order six using the order two 
integrators $SABA_2$ (\ref{eq:SABA2}) and $ABA82$ 
(\ref{eq:ABA82}) in Equation (\ref{eq:C6SS}). The obtained integrators 
are namely $s11SABA_26$ ($45$ steps) and 
$s11ABA82\_6$ ($89$ steps) respectively.

%
%
\subsubsection{SIs of order eight}
\label{sec3:SI8}
We also consider in our study some order eight SIs using the composition 
schemes presented in 
\cite{Yoshida1990} with 15 applications of an order two integrator. For 
order two SIs 
$SABA_2$ (\ref{eq:SABA2}) and $ABA82$
(\ref{eq:ABA82}) we implement the schemes, of \cite {Yoshida1990}, 
whose coefficients are referred to 
as \textit{Solution $A$} and \textit{solution $D$}.
More specifically we apply \textit{solution $A$} to generate the integrators 
$SABA_2Y8A$ of $61$ steps and $ABA82Y8A$ of $121$ steps 
from respectively SIs $SABA_2$ and $ABA82$. 
Solution D has been reported by \cite{Mclachlan1995} and 
\cite{Sofroniou2005} to perform better than the other composition 
schemes generated by this method. 
We therefore  
use solution D to generate integrators $SABA_2Y8D$ of 61 steps and $ABA82Y8D$ 
of 121 steps 
from SIs $SABA_2$ and $ABA82$ respectively.

In \cite{Kahan1997}, various schemes of order 8 with different 
number of steps were constructed. From these, we consider the composition scheme $s15odr8$. Using
$SABA_2$ (\ref{eq:SABA2}) and $ABA82$ (\ref{eq:ABA82}) in the place of $S_2$ 
we get the order eight SIs $s15SABA_28$ of 61 steps and $s15ABA82\_8$ of 
121 steps respectively.
In Section 4.3 of \cite{Sofroniou2005}, two schemes of 
order eight which require 
$19$ and $21$ applications of $S_2$ were given. We include in our analysis 
the scheme with $19$ applications 
of $S_2$ because it requires fewer steps for a particular order two SI compared 
to the method where $S_2$ is applied $21$ times. 
We call this scheme $s19odr8$. We now use $s19odr8$ to 
construct the order eight 
schemes $s19SABA_28$ of 
$77$ steps and $s19ABA82\_8$ of $153$ steps when we replace $S_2$ with 
$SABA_2$ (\ref{eq:SABA2}) and $ABA82$ (\ref{eq:ABA82})
respectively.
\subsubsection{Symplectic integration schemes of order higher than eight}
\label{sec3:SI9}
Numerical experiments performed [\cite{Sofroniou2005}] have shown that composition 
of schemes of higher order would require double precision arithmetic in 
order to be efficient, something which 
would slow down the numerical computation. Secondly, 
unlike schemes of order smaller or equal to eight, there were inconsistencies in 
the solutions for higher order schemes depending on the processor used.
We therefore do not include 
composition schemes of order greater than eight in our work.
%
\subsection{Solving the equations of motion and the tangent map (TM) method}\label{sec3:tangent_map}
The equations of motion (\ref{eq:motionp}) can be written in 
terms of the differential operator \eqref{eq:poisson} as 
\begin{equation*}
\left.
\begin{array}{ll}
\frac{d{\bf q}}{dt} = {\bf p}\\
\\
\frac{d{\bf p}}{dt} = -\frac{\partial V({\bf q})}{\partial{\bf q}}
\end{array}
\right\} \Longrightarrow \frac{d{\bf z}}{dt} = L_{_{H}}{\bf z},
\end{equation*}
where ${\bf z} = ({\bf q},{\bf p})$ is a point in the 
phase space and $L_{_{H}}$ is defined 
in the same way as in (\ref{eq:poisson}).
The operator $L_{_{H}}$ can be 
written as $L_{_{H}}=L{_{V}}+L{_{T}}$ 
where the time evolution maps corresponding to $L{_{V}}$ and $L_{_{T}}$ are
\begin{equation}\label{op:var_motiona}
e^{\tau L{_{V}}} = \left\{
\begin{array}{ll}
{\bf q}^\prime = {\bf q} + \tau{\bf p}\\
\\
{\bf p}^\prime = {\bf p}
\end{array}
\right.\qquad
\textnormal{and}\qquad
e^{\tau L{_{T}}} = \left\{
\begin{array}{ll}
{\bf q}^\prime = {\bf q}\\
\\
{\bf p}^\prime = {\bf p} - \tau\frac{\partial V({\bf q})}{\partial{\bf q}}.
\end{array}
\right.
\end{equation}

We use the TM method (Section \ref{sec3:var_eq}) to solve the 
variational equations \eqref{eq:variational}. We consider 
equations (\ref{eq:motionp}) and (\ref{eq:variational}) 
as a unified set of differential equations
\begin{equation*}\label{eq:kg_variational}
    		\left.
                \begin{array}{ll}
                  \frac{d{\bf q}}{dt} = {\bf p}\\
                  \\
                  \frac{d{\bf p}}{dt} = -\frac{\partial V({\bf q})}{\partial{\bf q}}\\
                  \\
			\frac{d{\bf\delta q}}{dt} = {\bf \delta p}\\
			\\
                  \frac{d{\bf\delta p}}{dt} = -D_V^2({\bf q}){\bf \delta p}
                \end{array}
              \right\} \Longrightarrow \frac{d{\bf u}}{dt} = L_{_{H_\mathcal{V}}}{\bf u},
  \end{equation*}
  where ${\bf u} = ({\bf q},{\bf p},{\bf \delta q},{\bf \delta p})$ is a vector formed 
  by the phase space vector $({\bf q},{\bf p})$ and the deviation vector 
  $({\bf \delta q},{\bf \delta p})$. $L_{_{H_\mathcal{V}}}$ is 
  used 
  to solve the 
  whole combined system of equations, in the phase and tangent space.
  The operator $L_{_{H_\mathcal{V}}}$ can be 
  	written as $L_{_{H_\mathcal{V}}}=L{_{V_\mathcal{V}}}+L{_{T_\mathcal{V}}}$ 
  where the maps $L{_{V_\mathcal{V}}}$ and $L{_{T_\mathcal{V}}}$ 
  give the equations
  \begin{equation}\label{op:var}
    		e^{\tau L{_{V_\mathcal{V}}}} = \left\{
                \begin{array}{ll}
                  {\bf q}^\prime = {\bf q} + \tau{\bf p}\\
                  \\
                  {\bf p}^\prime = {\bf p}\\
                  \\
			{\bf\delta q}^\prime = {\bf\delta q} + \tau{\bf\delta p}\\
			\\
                   {\bf\delta p}^\prime = {\bf\delta p}
                \end{array}
              \right.\qquad
              \textnormal{and}\qquad
              e^{\tau L{_{T_\mathcal{V}}}} = \left\{
              \begin{array}{ll}
              {\bf q}^\prime = {\bf q}\\
              \\
              {\bf p}^\prime = {\bf p} - \tau\frac{\partial V({\bf q})}{\partial{\bf q}}\\
              \\
              {\bf\delta q}^\prime = {\bf\delta q}\\
              \\
              {\bf\delta p}^\prime = {\bf\delta p} - \tau D_V^2({\bf q}){\bf \delta p}.
              \end{array}
              \right.
  \end{equation}
The new sets of equations (\ref{op:var_motiona}) and (\ref{op:var}) 
which respectively correspond to the equations of motion and variational 
equations can be 
  used in the SIs and their explicit forms for 
  the DKG Hamiltonians (\ref{eq:kg1dham}) and 
  (\ref{eq:kg2dham}) are given in Appendix \ref{app:A}.
  
\section{Numerical results}
\label{numerical_results}
In this Section we present numerical findings on the performance 
of the 
SIs that have been described in 
Section \ref{sec:numerical_integration} for the integration of the 
equations of motion and 
variational equations of the 1D \eqref{eq:kg1dham} and 2D 
\eqref{eq:kg2dham} DKG models. Firstly, for each of the 
models, we systematically show how we select 
initial deviation vector to use in our investigation. 

We note that all of our 
simulations were performed on an Intel Xeon E$5-2623$ with $3.00$ GHz.
Our programs were written in the FORTRAN $90$ programming language and we 
used the Intel Fortran (ifort) and gfortran  
Compiler Suites with an optimization level $2$ (-O$2$).

\subsection{Integration of the 1D DKG model}
\label{sec3:1d}
\subsubsection{The initial deviation vector}\label{sec3:ini_dvd}
\begin{figure}[H]
	\centering
	\includegraphics[width=0.75\textwidth,keepaspectratio]{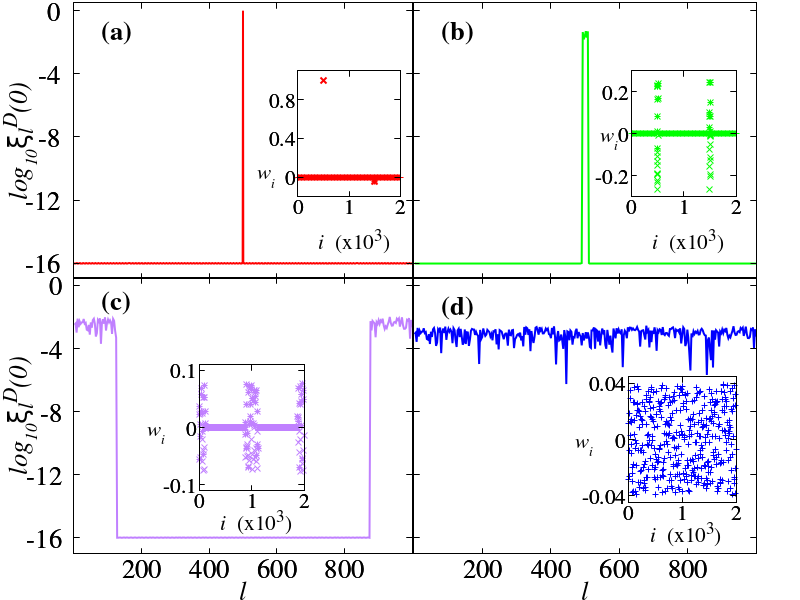}
	\caption{The DVDs $\xi_l^D(0)$ for initial deviation vectors {$\bf 1$} [red curves 
		in {\bf (a)}], 
		{$\bf 2$} [green curves in {\bf (b)}], {$\bf 3$} [purple 
		curves in {\bf (c)}] and 
		{$\bf 4$} [blue curves in {\bf (d)}]. The inset 
		plots show the coordinates $w_i$ of the initial deviation 
		vectors ${\bf w}=(w_1,w_2,\cdots,w_{2N})$ \eqref{eq:dev_vec}. The plots are in linear-log 
		scale and the insets are in log-log scale.}
	\label{ini_dvd_1D}
\end{figure}

Numerical computation of the 
mLCE (\ref{eq:mLCE}), 
involves integration of variational 
equations for 
a very long time interval in order to get conclusive results about the 
nature of the dynamics.
Here we compare various possible general forms 
of initial deviation vectors 
and investigate how they affect the solutions of the variational equations. 
To do this we compute the finite time mLCE and reproduce parts of a similar analysis that was 
presented in \cite{Gkolias2013}. We consider the following different 
initial deviation vectors (\ref{eq:dev_vec}).
%
 \begin{itemize}
  \item[ {$\bf 1$:}] a deviation vector whose central coordinate (corresponding to the 
  middle site of the lattice) of both the position and momentum 
  components is non-zero and all other coordinates of the vector are zero.
  \item[ {$\bf 2$:}] a deviation vector where (a few) more than one centrally 
  positioned coordinates of both the positions and momenta 
  components 
  are non-zero and all other points of the vector are zero.
  \item[ {$\bf 3$:}] a deviation vector where all centre coordinates of both the positions and momenta 
  components are zero 
  except for a (few) number of coordinates at the boundaries of the vector.
  \item[ {$\bf 4$:}] a random deviation vector where all coordinates 
   of both the positions and momenta 
  components are non-zero.
 \end{itemize}
The representative forms of these initial deviation 
vectors {$\bf 1$}, 
{$\bf 2$}, {$\bf 3$} and {$\bf 4$}, whose coordinate component 
numbers have been randomly 
generated, are presented in \autoref{ini_dvd_1D}. The panels {\bf (a)}, 
{\bf (b)}, {\bf (c)} 
and {\bf (d)} in 
\autoref{ini_dvd_1D} show the DVDs ${\bf \xi}_l^D(0)$ (\ref{eq:dvd}) 
plotted against the degrees of freedom (sites) $l$ (where $1\leq l\leq N=1~000$) 
for the initial 
deviation vectors of type {$\bf 1$} (red), 
{$\bf 2$} (green), {$\bf 3$} (purple) and {$\bf 4$} (blue) 
respectively. 
Each inset plot in panels {\bf (a)}, 
{\bf (b)}, {\bf (c)} and {\bf (d)} shows the corresponding actual 
deviation vector ${\bf w}$ (\ref{eq:dev_vec}) 
[shown using vector coordinates 
$w_i$] with the abscissa [labeled 
using $i$] representing 
positions coordinate components ${\bf\delta q}$ on one half to the left
and momenta coordinate components ${\bf\delta p}$ on the other half 
to the right.

\begin{figure}[H]
	\centering
	\includegraphics[width=0.55\textwidth,keepaspectratio]{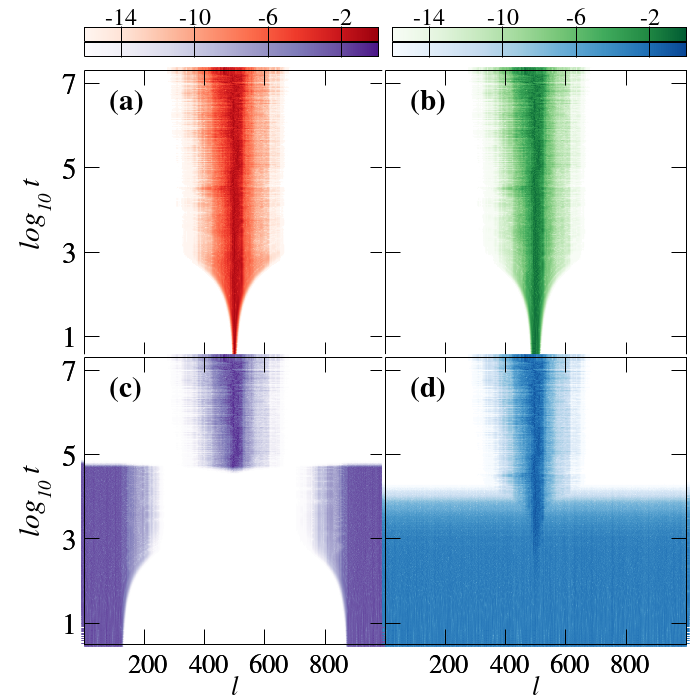}
	\includegraphics[width=0.4\textwidth,keepaspectratio]{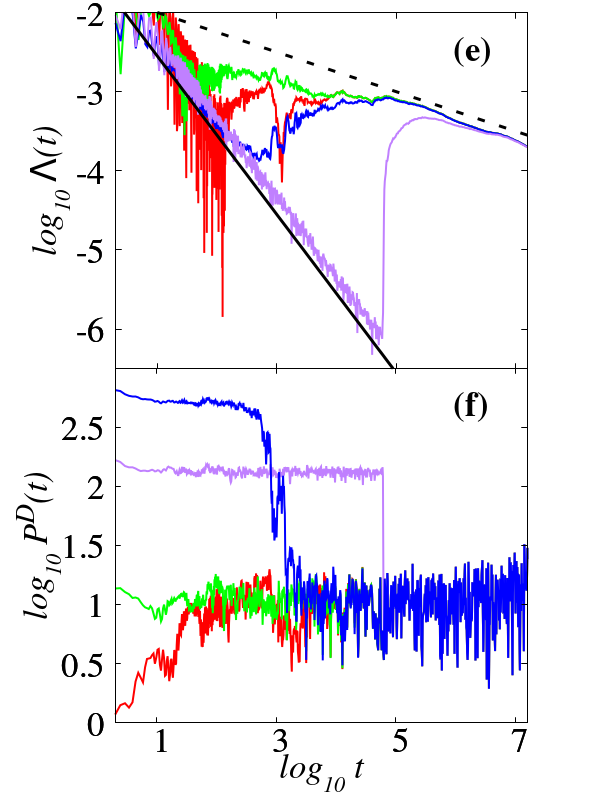}
	\caption{Results for the time evolution of the DVD for initial deviation vectors {$\bf 1$} 
		[in {\bf (a)}], 
		{$\bf 2$} [in {\bf (b)}], {$\bf 3$} [in {\bf (c)}] and 
		{$\bf 4$} [in {\bf (d)}]. Time evolution of the finite time 
		mLCEs  $\Lambda(t)$ \eqref{eq:ftmLCE} [in {\bf (e)}] 
		and DVD participation number $P^D(t)$ \eqref{eq:P_Dvd} 
		[in {\bf (f)}] 
		corresponding to the deviation vectors {$\bf 1$} (red curve), 
		{$\bf 2$} (green curve), {$\bf 3$} (purple curve) and {$\bf 4$} (blue curve). 
		The horizontal colour bars indicate the corresponding 
		values of $\log_{10}\xi^D_l(0)$ (\ref{eq:dvd}). 
		The continuous black and dashed lines in 
		{\bf (e)} respectively guide the eye for slopes $-1$ and 
		$-1/3$. Panels {\bf (a)}-{\bf (d)} are in linear-log scale while {\bf (e)} 
		and {\bf (f)} are in log-log scale.}
	\label{ini_dvd_palette_n_mLCE_1D}
\end{figure}

To study the dependence of $\Lambda(t)$ (\ref{eq:ftmLCE}) on the initial 
deviation vector, we integrate the 1D 
DKG model
(\ref{eq:kg1dham}) with single site excitation (i.e.~participation 
number $P(0)=1$) of the central oscillator using parameters: 
$W=4$, lattice size 
$N=1{,}000$ and total energy $H_{1K}= 0.4$, up 
to a final time $t_f\approx 10^{7.2}$. This set of 
initial conditions have been reported by \cite{Skokos2013} 
to belong to the weak chaos 
regime. We evaluate $\Lambda(t)$ 
and also monitor the evolution of the 
DVDs ${\bf \xi_l}^D(t)$ for each
of the different cases  {$\bf 1$}, 
{$\bf 2$}, {$\bf 3$} and {$\bf 4$}. The initial values 
of the DVD participation number, 
$P^D(0)$, 
for the cases 
{$\bf 1$}, {$\bf 2$}, {$\bf 3$} and {$\bf 4$} are respectively $1$, 
$21$, $251$ and $1~000$. The non-zero coordinates of 
each of these deviation vectors were 
randomly generated using a uniform distribution on the interval 
$(-1,1)$ with random numbers allocated to each of the 
components $\delta q_i$ and $\delta p_i$ of the vector, as seen in the
insets of \autoref{ini_dvd_1D}.
During the integration we monitor the DVD's evolution and compute 
the corresponding finite time mLCEs. 
The time evolution of the DVDs up to $t=t_f\approx 10^{7.2}$ is shown in 
\autoref{ini_dvd_palette_n_mLCE_1D} 
panels 
{\bf (a)} [for {$\bf 1$}], {\bf (b)} [for {$\bf 2$}], {\bf (c)} 
[for {$\bf 3$}] and {\bf (d)} [for {$\bf 4$}]
 where the horizontal colour bars show the 
values of the $\log_{10}\xi^D_l(t)$ (\ref{eq:dvd}). In all cases, 
the deviation vector eventually concentrate in the area of the 
lattice where the initial excitation was performed. The deviation 
vector {$\bf 3$} initially has no (non-zero) components 
concentrated around the 
sites where the excitation takes place.

For this reason, it takes the 
longest time (compared to all the other initial deviation vectors 
considered) before converging to the region where the dynamics takes place 
as shown in \autoref{ini_dvd_palette_n_mLCE_1D}{\bf (c)}.
In panels {\bf (e)}  and {\bf (f)} of \autoref{ini_dvd_palette_n_mLCE_1D} 
we report respectively the time 
evolution of the finite time 
mLCEs $\Lambda(t)$ and DVD participation number $P^D(t)$ 
for the different initial deviation vectors 
 {$\bf 1$} (red curve), {$\bf 2$} 
 (green curve) {$\bf 3$} (purple curve) and {$\bf 4$} (blue curve). 
The straight continuous line in  {\bf (e)} indicates the slope 
of $\Lambda(t)$ corresponding to the law $\Lambda(t) \propto t^{-1}$ 
(representing regular motion)
while the dashed line shows the direction [characteristic slope for 
weak chaos law $\Lambda(t) \propto t^{-1/3}$ [\cite{Skokos2013}]]
to which the finite time mLCEs $\Lambda(t)$ from the different initial deviation vectors 
eventually tend to. After $t=10^6$, 
the evolution of $\Lambda(t)$ and $P^D(t)$ 
is independent of the initial deviation vector used 
in the corresponding 
integration as respectively seen in {\bf (e)}  and {\bf (f)}. However, for 
initial deviation vector {$\bf 2$}, where more than one central coordinates 
of the deviation vector are non-zero and concentrated around 
excited sites, $\Lambda(t)$ 
converges fastest to the direction of 
convergence of all finite time mLCEs (i.e., the dotted line in \autoref{ini_dvd_palette_n_mLCE_1D} {\bf (e)}). The results of 
\autoref{ini_dvd_palette_n_mLCE_1D} also show 
that the finite time mLCE rate of 
convergence to it's asymptotic value 
depends on the number of non-zero 
coordinates of the initial deviation vector covering the 
excited sites and the position of 
these coordinates with respect to the region of excitation.
We can clearly see in \autoref{ini_dvd_palette_n_mLCE_1D}{\bf (e)}
that for $P^D(0)=1$ (initial deviation vector {$\bf 1$}), we require more integration time 
to observe convergence of $\Lambda(t)$ to the direction of the dotted line 
compared to when $P^D(0)=21>P(0)=1$ (initial deviation vector {$\bf 2$}). 
We observe a similar behaviour in 
\autoref{ini_dvd_palette_n_mLCE_1D}{\bf (f)} for the convergence of $P^D(t)$. 
For a vector of form 
{$\bf 3$}, in the 
first stages ($t<10^{4.5}$) of the dynamics 
the deviation vector does not reveal the actual (weak chaos) 
dynamics of the system and so $\Lambda(t)$ shows the behaviour of regular 
motion as seen in \autoref{ini_dvd_palette_n_mLCE_1D}{\bf (e)}. 
However, at a later 
stage (around $t\approx 10^{4.8}$) 
as the energy spreads (due to chaoticity in the system 
[\cite{Skokos2013}]) 
to more lattice sites, the system's true chaotic nature 
is characterised by the deviation vector 
and so $\Lambda(t)$ starts to show the expected behaviour following 
a law $\Lambda(t) \sim t^{-1/3}$. 
For the spatial extent of the DVD, we see in {\bf (f)} that eventually 
for all initial deviation vectors, 
the participation number 
$P^D(t)\approx10$ and case {$\bf 2$} converges 
to this value before any of the other three deviation vectors. 
For case {$\bf 3$}, during the initial integration stages ($t<10^{4.5}$) 
when the system shows regular behaviour, the spatial extent 
of the deviation 
vector as 
characterised by the participation number $P^D$ is practically fixed 
with $P^D(t<10^{4.5})\approx251$. $P^D$ then falls to 
$P^D(t\ge10^{5})\approx10$ when the chaotic behaviour is reflected 
by $\Lambda$. For the fastest 
convergence to the eventual state of chaoticity (quantified 
using $\Lambda$) and DVD (quantified using the participation number 
$P^D$), we use in this work the deviation vector {$\bf 2$}
whose non-zero component of the coordinates fully covers the sites that 
have been initially excited. That is to say, if we give the same 
energy to each initially excited site, then we set the deviation vector 
such that $\frac{P(0)}{P^D(0)}\lesssim1$ for 
the 1D DKG model (\ref{eq:kg1dham}).

\subsubsection{Efficiency of Symplectic integrators}
\label{subsec3:si1D}
Using the various SIs presented in 
Section \ref{sec:numerical_integration} we numerically 
integrate the equations of motion 
of the 1D KG Hamiltonian (\ref{eq:kg1dham}) for different initial 
energy excitations and parameters. 
In this process, for a particular disorder realization 
we compute the quantities $m_2(t)$ (\ref{eq:m2_1dkg}), 
$P(t)$ (\ref{eq:P_1dkg}), the energy distribution 
$\{\xi_{_l}(t)\}$ (\ref{eq:norm_en1d})
and the absolute relative energy error $e_r(t)$ which is given by 
\begin{equation}\label{eq:ree1}
e_r(t)=\frac{|H_{1K}(t)-H_{1K}(0)|}{H_{1K}(0)},
\end{equation} and we evaluate these quantities 
using all SIs. A reproduction of these measures using the 
different SIs ensures that we have the same dynamics or wave packet from 
each of the SIs. We also use the same initial deviation vector and 
we compute 
the time evolution of $\Lambda(t)$ as a way of checking that we 
accurately compute the level of 
chaoticity in the system across all SIs.

For our computations we work with a lattice of size $N=1~000$, 
energy $H_{1K}(0)$, and
disorder strength $W$. We excite a block of 
$L$ adjacent sites at the centre of the lattice by giving each site 
energy $h=H_{1K}(0)/L$.
We consider the following 
different sets of initial parameter conditions:

\begin{itemize}
\item [I:] $L=1$, $W=4$ and $H_{1K}(0)=h=0.4$

\item [II:] $L=1$, $W=4$ and $H_{1K}(0)=h=1.5$

\item [III:] $L=21$, $W=4$, and $H_{1K}(0)=4.2$, $\left(h=0.2\right)$

\item [IV:] $L=37$, $W=3$, and $H_{1K}(0)=0.37$, $\left(h=0.01\right)$

\item [V:] $L=100$, $W=4$, and $H_{1K}(0)=1$, $\left(h=0.01\right)$

\item [VI:] $L=1000$, $W=4$, and $H_{1K}(0)=10$, $\left(h=0.01\right)$.
\end{itemize}

In \cite{Skokos2013} and \cite{Gkolias2013} the cases I and IV 
have been reported to exhibit the weak chaos 
behaviour. In \cite{Gkolias2013}  case II has been reported to belong 
to the strong 
chaos regime and case III to the strong chaos regime where some 
disorder realisations exhibited a crossover to 
weak chaos in a finitely long period of time. For the first four cases 
(I, II, III and IV), we restrict our 
evolutions to a system with fixed boundary conditions 
where we mimic the dynamics of a lattice with an infinite number 
of sites. We include 
cases V and VI as representations 
for general performances of the SIs on situations where 
the energy would eventually spread to all sites of the lattice.
In each of the cases I-VI we initially give the same momentum to each of 
the excited site(s) 
with an allocation of random signs to each of these 
values while setting momentum to 
zero for all other sites. The displacements for all sites are 
initially set to zero. An initial deviation vector of the 
form {$\bf 2$} (\autoref{ini_dvd_1D}{\bf (b)}) is used. We monitor the 
normalized energy distributions $\{\xi_l(t)\}_l$ (\ref{eq:norm_en1d}) for 
$l=1,\cdots,N$ and evaluate their second moment $m_2(t)$ (\ref{eq:m2_1dkg}) 
and participation 
number $P(t)$ (\ref{eq:P_1dkg}) for $t\in[0,10^7]$. 
During the simulations we adjust the integration time step so that the 
absolute relative energy error $e_r$ is at a value smaller than $10^{-4}$, a 
typically accepted 
level [\cite{Flach2009b,Skokos2009,Laptyeva2010,Flach2010,Bodyfelt2011}]. 
That is to say, we set 
$e_r(t)\lesssim 10^{-5}$. 
The computational efficiency of each scheme is evaluated by 
checking its ability to correctly reproduce 
the dynamics of the energy propagation within an acceptable precision range.
 \begin{figure}[H]
 	\centering
 	\includegraphics[width=0.8\textwidth,keepaspectratio]{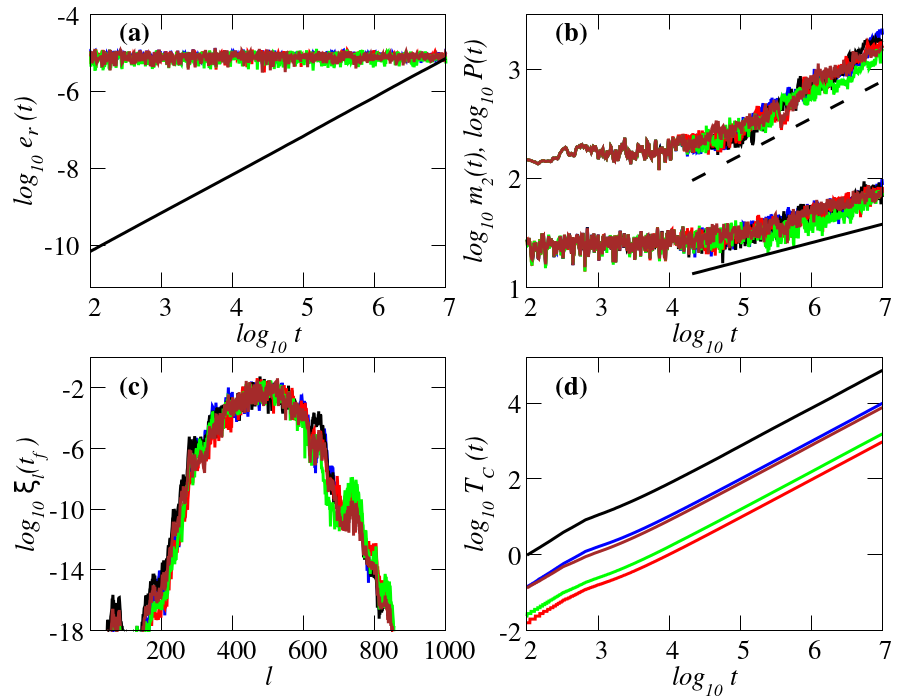}
 	\caption{Results obtained for the integration of equations of motion for 
 		ICs IV (see text) of the 1D DKG Hamiltonian
 		(\ref{eq:kg1dham}) by the integrators $ABA82$ of order two, $ABA864$ of order four,
 		$SABA_2Y6$ of order six, $SABA_2Y8A$ of order eight and 
 		$RK4$ [blue, red,
 		green, brown and black curves respectively]. {\bf (a)} the time 
 		evolution of the absolute
 		relative energy error $e_r(t)$ (\ref{eq:ree1}); {\bf (b)} the time evolution of the second
 		moment $m_2(t)$ (\ref{eq:m2_1dkg}) (upper curves) and participation number $P(t)$ (\ref{eq:P_1dkg}) 
 		(lower curves); {\bf (c)} the normalized energy distribution
 		$\{\xi_l(t_f=10^7)\}_l$ (\ref{eq:norm_en1d}) as a function of lattice site index $l$; and {\bf (d)}
 		the time evolution of the required CPU time $T_C(t)$ in seconds. The
 		straight lines in {\bf (b)} guide the eye for slopes $1/3$ (dashed line)
 		and $1/6$ (solid line). The five different curves
 		practically overlap each other in panels {\bf (a)}, {\bf (b)} and {\bf (c)}.  Panels {\bf (a)}, {\bf (b)} and {\bf (d)} 
 		are in $\log -
 		\log$ scale and {\bf (c)} is in $\textnormal{linear}-\log$ scale.}
 	\label{fig:1D-4best}
 \end{figure}
 We therefore monitor the nature of the energy 
 profiles as quantified by the distribution $\xi_l(t)$, and the time 
 evolution of
 $e_r(t)$, $m_2(t)$, $P(t)$ and $\Lambda(t)$.\newline 
\autoref{fig:1D-4best} shows findings obtained for the evolution of the 
 orbit by four SIs, 
 namely $ABA82$ of order $2$ (blue curves), $ABA864$ of order $4$ (red
 curves), $SABA_2Y6$ of order $6$ (green curves), $SABA_2Y8A$ 
 of order $8$ (brown curves) and the fourth order 
 non symplectic Runge-Kutta $RK4$ 
 [\cite{Runge1895,Weiner1992}] (black curves) integrator for ICs
IV. We include $RK4$ as a representative integrator 
for non symplectic schemes amidst SIs. All integrators studied
produce the same dynamical evolution of the model as the
results for $m_2(t)$ (upper curves of
\autoref{fig:1D-4best}{\bf(b)}), $P(t)$ (lower curves of
\autoref{fig:1D-4best}{\bf(b)}) and the normalized energy
 profiles (\autoref{fig:1D-4best} {\bf(c)}) 
 at the final integration time 
 $t_f=10^7$ practically
 overlap. We note that by the final integration time $t_f=10^7$, the energy 
 had not yet reached the boundaries of the lattice as clearly shown in panel 
 {\bf(c)}. The results of
 \autoref{fig:1D-4best}{\bf(b)} show that the wave packet's $m_2(t)$ and 
 $P(t)$ eventually grow respectively as
 $m_2(t) \propto t^{1/3}$ (direction of black dashed line in 
 \autoref{fig:1D-4best}{\bf(b)}) and $P(t) \propto t^{1/6}$ (direction of 
 black continuous 
 line in 
 \autoref{fig:1D-4best}{\bf(b)}), in
 agreement with previously published works
 [\cite{Flach2009,Flach2009b,Skokos2009,Laptyeva2010,
 	Bodyfelt2011}]. 
The CPU time, $T_C$, time evolution for each integrator 
is shown in \autoref{fig:1D-4best}{\bf(d)}. 
From this result, clearly the $ABA864$ scheme shows
the best performance from all the five integrators shown in the 
figure requiring the least CPU time for 
the integration.
We note that the time step $\tau=0.006$ 
used for the integrator $RK4$ maintains $e_r(t)$ \eqref{eq:ree1} 
below $10^{-5}$ for the whole computation. For shorter integration times 
smaller than $t=10^7$, $RK4$ is more accurate since it has the least change 
in energy $H_{1K}$ (as seen in \autoref{fig:1D-4best}{\bf(a)}) 
compared to the other four schemes presented in \autoref{fig:1D-4best}, 
but 
it requires more 
CPU time (as seen in \autoref{fig:1D-4best}{\bf(d)}).
 Under these 
conditions, integration by $RK4$ beyond $t=10^7$ would lead to the 
system losing it's energy conservation property beyond the acceptable 
accuracy level of $e_r(t)\approx10^{-5}$. This change in energy beyond 
the desired accuracy can 
be avoided for very long integration times by reducing the time 
step, which in turn slows the computations and hence an 
increased CPU time for the $RK4$. 
A more detailed account of the findings for 
this integrator comparison are presented in 
\autoref{tab:1}.
 
More precisely, \autoref{tab:1}, shows information on the 
  performance of the SIs of order $n$
  used for the integration of the equations of motion (\ref{eq:kg1dmotion}) 
  of the 1D KG model (\ref{eq:kg1dham}) up to a final
  time $t_f=10^7$ for the initial excitation of cases I, IV 
  and VI. The number of SI integration steps ($S$) is given and
  for case IV the integration time step $\tau$ which keeps
  $e_r(t) \lesssim 10^{-5}$ for all $t$ is listed. The required CPU time
  in seconds, $T_C$, needed for each integrator is also reported.
   From the results of this table we see that the SIs exhibiting
  the best performance are: the order four
schemes $ABA864$, $ABAH864$ and the order six schemes
  $SABA_2Y6$, $s9SABA_26$, $ABA864Y6$ in descending order of computational 
  efficiency. 
  Only the CPU times for the SIs that are stable 
  (where $e_r(t)$ remains practically constant)
  in each 
  of the cases have been reported in \autoref{tab:1}. 
  In order to get best performance from each of the SIs, we choose a 
  time step 
  $\tau$ which is big enough to keep $e_r(t)$ as close as possible 
  to $10^{-5}$. The SIs of order two require very high CPU times 
  of the order of $10^4$ seconds compared to the many other better performing SIs, 
  so we only include their results for IC $IV$.
  Also, some of the SIs whose theory was discussed in Section \ref{sec3:SIs} 
  have not been included in \autoref{tab:1} as they 
  do not maintain  
  the value of $e_r(t)$ bounded around $10^{-5}$ because they require 
  relatively huge values of $\tau$ which makes them unstable (with values of 
  $e_r(t)$ fluctuating highly and far from the value $10^{-5}$). 
  However, 
  some of these SIs give a fairly 
  and competitively 
  good performance at higher accuracies $e_r(t)<10^{-5}$. 
  Nonetheless, they still do not 
  feature among the top most efficient SIs. For example, the SI $ABA82Y8D$ 
  which is unstable at $e_r(t)\approx10^{-5}$ requires a 
  CPU time of 
  approximately $5\times10^4$ seconds for an accuracy 
  $e_r(t)\approx10^{-6}$ where it is stable. On the other hand, 
  the SI $SABA_2Y8A$, which is stable at $e_r(t)\approx10^{-5}$, requires a 
  larger CPU time of approximately $7\times10^4$ seconds 
  at this accuracy. However, there are 
  still many more SIs that have a final 
  CPU time $T_C$ that is smaller than that of 
  $ABA82Y8D$, and so it does not appear amongst the top SIs. We 
  	note that each of the SIs included in \autoref{tab:1} showed stability for 
  	all the six ICs I - VI considered.
  
\begingroup
\setlength{\tabcolsep}{10pt} 
\renewcommand{\arraystretch}{1.2} 
\begin{table}[H]
	\centering
	\begin{tabular}{lrr|r|lr|r}
		\hline \hline
		&&& I&{\hfill IV} 
		& &VI\\
		\hline \noalign{\smallskip}
		SI&$n$&$S$& $T_C$ & 
		$\tau$  & $T_C$ 
		& $T_C$\\
		\hline
		\noalign{\smallskip}
		$ABA82$&2 &9&&0.04 & 8530& \\
		$SABA_2$& 2&5&&0.02 & 12780&\\
		$SBAB_2$&2 &5&&0.02 & 14430&\\
		$LF$& 2&3&22350&0.01 & 32280&33740\\
		\hline
		$ABA864$&4 &15&820&0.56 & 840&820\\
		$ABAH864$&4 &17&1290& 0.38 & 1350&1300\\
		$ABA82Y4$&4 &25&2310&0.26 & 2630&2700\\
		$SABA_2C$&4 &7&2170&0.19 & 3350&2110\\
		$FR4$&4 &7&3040&0.09 & 3310&3040\\
		$SABA_2Y4$&4 &13&2620& 0.12 & 3560 &3350\\
		$SBAB_2Y4$&4 &13&2900& 0.12 & 3840&3620\\
		$SBAB_2C$&4 &7&2970& 0.14 & 4780&3140\\
		\hline
		$SABA_2Y6$&6 &29&1310&0.55 & 1400&1300\\
		$s9SABA_26$&6 &37&1440&0.67 & 1410&1320\\
		$ABA864Y6$& 6&43&1640&0.65 & 1650&1590\\
		$SBAB_2Y6$& 6&29&1700&0.46 & 1750&1660\\
		$s9ABA82\_6$&6 &73&*4260&0.93 & 1920&*2870\\
		$FR4Y6$&6 &19&2860&0.18 & 3090&3010\\
		$SABA_2CY6$&6 &19&1800&0.37 & 3240&2190\\
		$SABA_2Y4Y6$&6 &37&3380&0.28 & 3370&3380\\
		$SBAB_2Y4Y6$&6 &37&3870&0.28 & 3850&3610\\
		\hline
		$SABA_2Y8A$ &8 &61&6590&0.20 & 7290&6820\\
		$ABA82Y8A$ &8 &121&13600&0.22 & 12470&11700\\
		\hline \hline
	\end{tabular}
	\caption{Results obtained for the integration of the equations of motion 
		\eqref{eq:kg1dmotion} for
		some of the investigated
		ICs I, IV and VI (see text) of the 1D DKG Hamiltonian
		(\ref{eq:kg1dham}). The respective order $n$, 
		number of steps $S$, 
		CPU times $T_C$ (in seconds) required for integrating 
		the equations of motion
		up to a time $t_f=10^7$, the integration time steps $\tau$ used for 
		initial condition IV are reported. (*) shows CPU times 
			obtained at a stable value $e_{\bf r}(t)<10^{-5}$.}
	\label{tab:1}
\end{table}
\endgroup

We clearly see that the order four integrator 
$ABA864$ gives the best performance 
for the integration of the equations of motion of the 1D DKG model 
\eqref{eq:kg1dham} basing on the results 
in \autoref{tab:1}. The performance rankings shown in this table for the 
cases $I$, $IV$ and $VI$ are 
similar to those of cases $II$, $III$ and $V$ with the $ABA864$ giving 
the smallest CPU time. We now extend this study to solving 
variational equations of the 1D DKG model. 

\begin{figure}[H]
	\centering
	\includegraphics[width=0.9\textwidth,keepaspectratio]{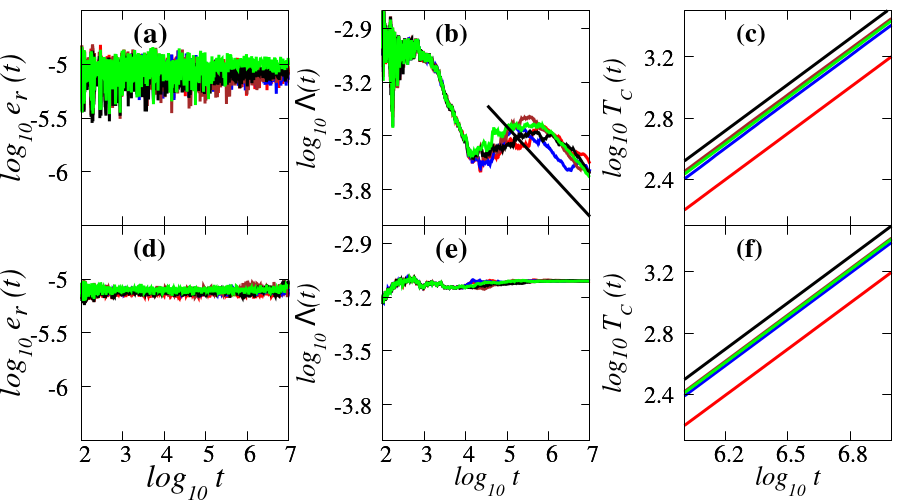}
	\caption{Results obtained for the integration of the variational
		equations \eqref{eq:variational} of the 1D DKG Hamiltonian (\ref{eq:kg1dham}) for cases IV 
		[{\bf (a)}, {\bf (b)} and {\bf (c)}]
		and VI [{\bf (d)}, {\bf (e)} and {\bf (f)}] for 
		the SIs
		$ABA864$, $ABAH864$, $ABA864Y6$, $s9SABA_26$ and $SABA_2Y6$ 
		[red, blue, black, brown and green curves respectively]: the time
		evolution of $e_r(t)$ (\ref{eq:ree1}) [{\bf (a)} and {\bf (d)}],
		$\Lambda(t)$ (\ref{eq:ftmLCE}) [{\bf (b)} and {\bf (e)}] 
		and of the
		required CPU time $T_C(t)$ [ {\bf (c)} and {\bf (f)}]. The straight 
		solid line in 
		{\bf (b)} guides the eye for
		slope $-1/4$. All curves in panels {\bf (a)}, {\bf (d)} 
		and {\bf (e)}, and those 
		for $s9SABA_26$, $SABA_2Y6$ in panels 
		{\bf (c)}, and {\bf (f)} practically overlap. 
		The panels are in $\log -
		\log$ scale.}
	\label{fig:1D-5best_LEs}
\end{figure}

In \autoref{fig:1D-5best_LEs} we
present results based on the numerical integration of the variational
equations \eqref{eq:variational} for the 1D DKG Hamiltonian (\ref{eq:kg1dham}) 
as obtained using the
best five SIs for the weak chaos case IV
[\autoref{fig:1D-5best_LEs}{\bf (a)}, {\bf (b)} and {\bf (c)}], as 
well and the fully chaotic
excitation case VI [\autoref{fig:1D-5best_LEs}{\bf (d)}, {\bf (e)} 
and {\bf (f)}]. 
From panels {\bf (a)} and {\bf (d)} we see that the 
absolute relative energy error $e_r(t)$ is 
practically independent of time at a value $e_r(t)\approx10^{-5}$ 
for all SIs. 
Similarly, for a particular case of initial excitation, 
the time evolution
of $\Lambda(t)$ (\ref{eq:ftmLCE}) is qualitatively the same
for all these SIs as seen in each of the panels 
{\bf (b)} and {\bf (e)}.
In the weak 
chaos case IV, the finite time mLCE $\Lambda(t)$ (panel {\bf (b)}) 
eventually
decreases in a way similar to what was reported 
in \cite{Skokos2013}, following the law $\Lambda(t)\propto
t^{-\alpha_{\Lambda}}$, with $\alpha_{\Lambda}=-1/4$. The observed value of $\alpha_{\Lambda}=-1/4$ is different 
from the value $-1$, a value indicating the behaviour 
for regular orbits. This shows that for this particular disorder realisation, 
the strength of chaoticity which is measured by $\Lambda$ [shown in 
\autoref{fig:1D-5best_LEs}{\bf(b)}] decreases as the wave packet extends 
[spreading shown by $m_2(t)$ in \autoref{fig:1D-4best}{\bf(b)}]
without showing any tendency of the dynamics to exhibit 
regular behaviour for the entire 
duration of the integration
[\cite{Skokos2013}]. In Chapter \ref{chap_1d}, we give a more detailed 
discussion on the 1D DKG model and substantiate on the persistence or 
relaxation of 
chaos using better statistics with more dynamical quantities.

\begin{figure}[H]
	\centering
	\includegraphics[width=0.75\textwidth,keepaspectratio]{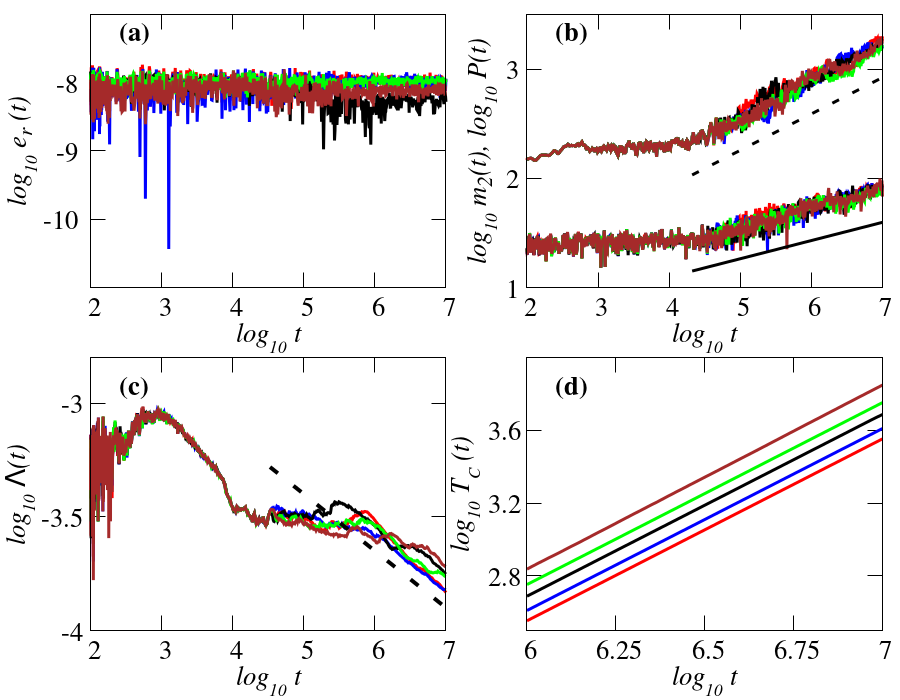}
	\caption{Results obtained for the integration of the variational 
		equations of the 1D DKG Hamiltonian
		(\ref{eq:kg1dham}) for the case of ICs IV by the order six SI $s11ABA82\_6$ for
		$\tau=0.42$, order eight SI $s15ABA82\_8$ for $\tau=0.48$,
		order eight SI $s19ABA82\_8$ for $\tau=0.59$, order six SI 
		$s9ABA82\_6$ for $\tau=0.50$ and order six SI $SABA_2Y6$
		for $\tau=0.18$ [red, blue, black, brown and green curves 
		respectively]. The time
		evolution of {\bf (a)} $e_r(t)$ \eqref{eq:ree1}, {\bf(b)} $m_2(t)$ 
		\eqref{eq:m2_1dkg} and $P(t)$ \eqref{eq:P_1dkg}
		[ upper and lower curves respectively], {\bf (c)} $\Lambda(t)$ 
		\eqref{eq:ftmLCE} and 
		{\bf (d)} CPU time $T_C(t)$. The
		straight lines in {\bf (b)} guide the eye for slopes $1/3$ 
		(black dashed line)
		and $1/6$ (black solid line), while in {\bf (c)} the straight dashed line
		corresponds to slope $-1/4$. All panels are in $\log -
		\log$ scale.}
	\label{fig:1D-5best_low}
\end{figure}

For the fully chaotic case VI, the saturation of 
$\Lambda(t)$ 
[\autoref{fig:1D-5best_LEs}{\bf (e)}] to a constant positive value 
is very fast,  
showing the typical behaviour of
chaos.
The SIs give a similar performance for the different ICs as shown in 
\autoref{fig:1D-5best_LEs}{\bf (c)} and
{\bf (f)}, an indication that their efficiency is independent of the ICs. 
We also need to check if the computational efficiency as presented in 
\autoref{tab:1} and \autoref{fig:1D-5best_LEs} is independent 
of the accuracy level $e_r(t)$.
\newline\newline
For the case when a high accuracy is needed e.g.~for $e_r\approx10^{-8}$, 
the time step $\tau$ is set to smaller values. Through an appropriate 
reduction in the time step $\tau$, all SIs that have been investigated 
are stable 
at the absolute relative energy error $e_r\lesssim10^{-8}$.
We perform an analysis similar to the one performed when creating
\autoref{tab:1} and \autoref{fig:1D-5best_LEs} for the accuracy 
level $e_r\approx10^{-8}$ and we present the results for the 
most efficient SIs on integration of variational equations for 
case IV in \autoref{fig:1D-5best_low}.

From \autoref{fig:1D-5best_low}{\bf (a)} we see that
for each SI considered, the time step $\tau$ was chosen so
that $e_r(t)\approx 10^{-8}$ with all SIs successfully reproducing
the same wave packet characterised by the time evolution laws 
$m_2(t)\propto t^{1/3}$, $P(t)\propto t^{1/6}$
[\autoref{fig:1D-5best_low}{\bf (b)}] and $\Lambda(t)\propto{t^{-1/4}}$
[\autoref{fig:1D-5best_low}{\bf (c)}]. 
The corresponding 
time steps $\tau$ used in ICs case IV to attain this accuracy are also given.
We note that the 
chaotic behaviour observed for $e_r(t)\approx10^{-5}$ 
persists for high accuracy level $e_r(t)\approx10^{-8}$ with 
$\Lambda(t)$ 
following the same power law 
as was observed in \autoref{fig:1D-5best_LEs}.
For each set of the six ICs, all the five SIs 
presented in \autoref{fig:1D-5best_low} require more CPU time than 
that used to obtain $e_r(t) \approx
10^{-5}$ [\autoref{fig:1D-5best_LEs}{\bf (c)}].
At the accuracy $e_r(t)\approx10^{-8}$, the time step is 
very small for the lower order 
SIs especially the SIs of order two and so they are successively 
applied for very high number of times during integration 
leading to a huge growth in the required CPU time.
For the accuracy $e_r(t)\approx10^{-8}$, the five best 
performing SIs 
are $s11ABA82\_6$, $SABA_2Y6$ and $s9ABA82\_6$ of order six, and 
$s15ABA82\_8$ and $s19ABA82\_8$ of
order eight.
%
%
%

We now investigate the generality of our results in integrating 
Hamiltonian lattices of spatial dimension greater than one by considering the 
2D DKG model \eqref{eq:kg2dham} in the next section.
\subsection{Integration of the 2D DKG model}
\label{sec3:2d}
Here we present a numerical study for the 2D DKG model 
\eqref{eq:kg2dham}, similar to the discussion of the 
1D DKG system \eqref{eq:kg1dham} 
reported in Section 
\ref{sec3:1d}.
\subsubsection{The initial deviation vector}
\begin{figure}[H]
	\centering
	\includegraphics[width=0.6\textwidth,keepaspectratio]{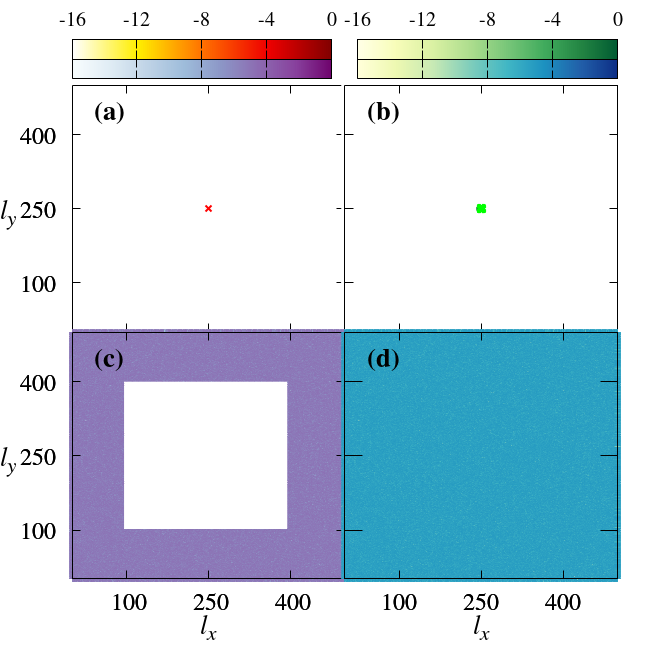}
	\caption{Projection of the initial deviation vectors {$\bf 1$} 
		[{\bf (a)} red], 
		{$\bf 2$} [{\bf (b)} green], {$\bf 3$} [{\bf (c)} purple] and 
		{$\bf 4$} [{\bf (d)} blue]
		on the 2D space. The horizontal colour bars indicate the corresponding 
		values of $\log_{10}\xi^D_{_{\bf r}}(0)$ (\ref{eq:dvd}) where 
		${\bf r}$ is of the form $(l_x,l_y)$ for $l_x,l_y\in\{1, 2,\cdots,501\}$.}
	\label{fig:ini_dvd_2D_palette}
\end{figure}

In order to investigate the dependence of finite time mLCE 
$\Lambda(t)$ \eqref{eq:ftmLCE} and other deviation vector related 
quantity computations on the 
nature/kind of the initial 
deviation vector for 
the 2D DKG model, we performed similar 
tests like the ones of Section 
\ref{sec3:1d}. For this purpose, the following different forms of 
initial deviation vectors are considered:
\begin{itemize} 
	\item[ {$\bf 1$}:] a deviation vector whose central coordinate (corresponding to the 
	middle site of the lattice) of both the position and momentum 
	components is non-zero and all other coordinates of the vector are zero.
	\item[ {$\bf 2$}:] a deviation vector where several centrally 
	positioned coordinates of both the positions and momenta 
	components 
	are non-zero and all other elements of the vector are zero.
	\item[ {$\bf 3$}:] a deviation vector where all centre coordinates of both the positions and momenta 
	components are zero 
	except for a number of coordinates at the boundaries of the lattice.
	\item[ {$\bf 4$}:] a random deviation vector where all coordinates 
	of both the positions and momenta 
	components are non-zero.
\end{itemize} 
In \autoref{fig:ini_dvd_2D_palette} we see the projections of 
these initial deviation vectors {$\bf 1$} 
[red, panel {\bf (a)}], 
{$\bf 2$} [green, panel {\bf (b)}], {$\bf 3$} 
[purple, panel {\bf (c)}] and {$\bf 4$} [blue, panel {\bf (d)}] 
on the 2D space. 
In order to understand the dependence of 
dynamical quantities (for example $\Lambda(t)$ \eqref{eq:ftmLCE} 
and $P^D(t)$ \eqref{eq:P_Dvd}) 
on the initial deviation 
vector, we integrate the variational equations of the 
two dimensional DKG system (\ref{eq:kg2dham}). 
We use a 2D square lattice, $N\times M=501\times501$, of $501$ sites 
in one spatial dimension ($x$) and $501$ sites in a perpendicular 
spatial dimension 
($y$), 
strength of disorder $W=10$ 
and we excite the 
middle site of the lattice with energy $H_{2K}=\xi_{_{(251,251)}}(0) = 0.3$. 
This configuration has been reported in \cite{Laptyeva2012} to 
belong to the weak chaos regime. We compute 
$\Lambda(t)$ (\ref{eq:ftmLCE}) while monitoring 
the evolution of the 
DVDs for each
case:  {$\bf 1$} 
(non-zero values for only the coordinates corresponding to the 
middle site 
of the lattice for both the position and momentum. i.e $P^D(0)=1$), 
{$\bf 2$} ($7\times7$ 
sublattice centre coordinates of both the position and momentum 
components are non-zero. i.e. $P^D(0)=49$), {$\bf 3$} ($301\times301$ 
sublattice centre coordinates of both the position and momentum 
components are zero and all the remaining coordinates are non-zero. i.e. $P^D(0)=1.604\times10^5$) and 
{$\bf 4$} (all coordinates 
of both the position and momentum 
components are non-zero. i.e $P^D(0)=501^2$). 

\begin{figure}[H]
	\centering
	\includegraphics[width=0.2425\textwidth,keepaspectratio]{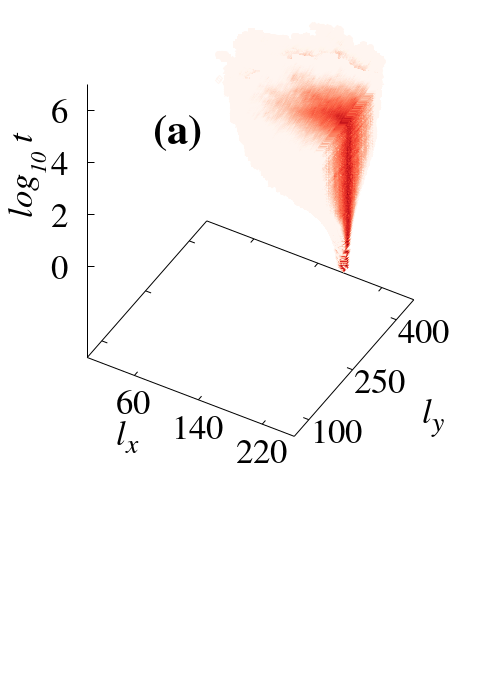}
	\includegraphics[width=0.2425\textwidth,keepaspectratio]{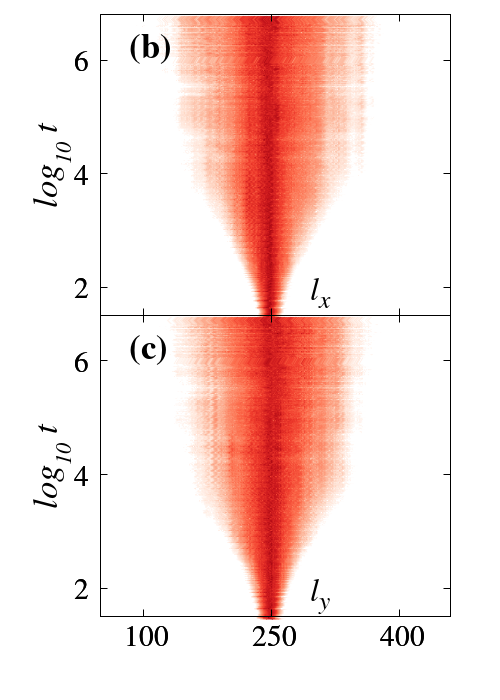}
	\includegraphics[width=0.2425\textwidth,keepaspectratio]{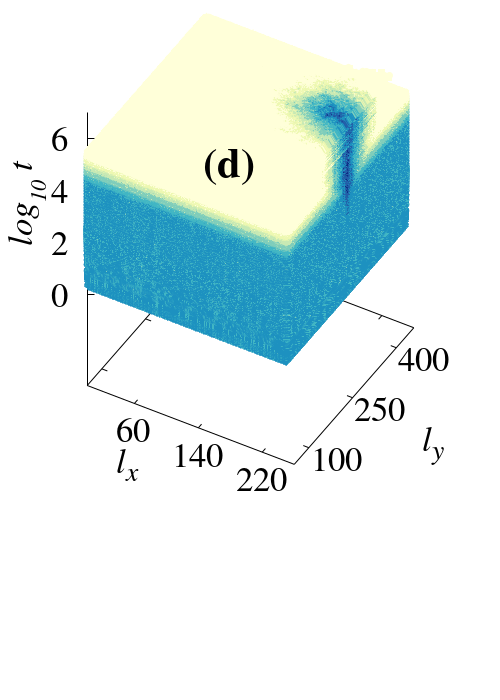}
	\includegraphics[width=0.2425\textwidth,keepaspectratio]{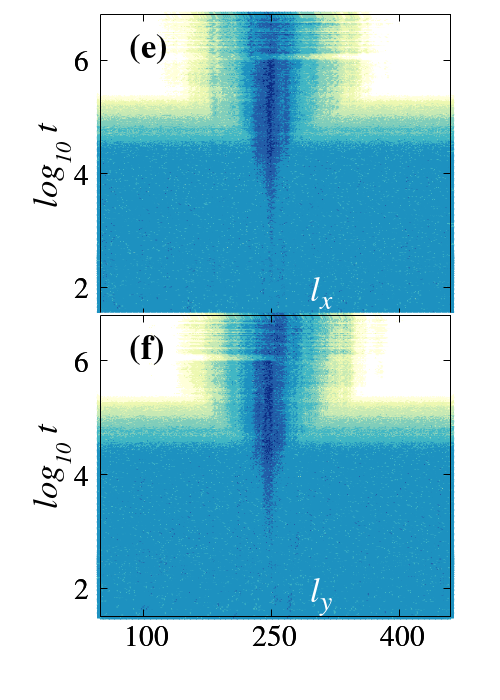}
	\caption{The time evolution of initial deviation vectors 
		{$\bf 1$} (red) and 
		{$\bf 4$} (blue). A 3D time evolution with a cross-section 
		at $l_x=251$ in panels {\bf (a)} and {\bf (d)}. A 1D time evolution 
		at cross-section $l_x=251$ [panels {\bf (c)} and {\bf (f)}] and 
		cross-section $l_y=251$ [panels {\bf (b)} and {\bf (e)}]. 
		The plots are in $\textnormal{linear}-\log$ (for 2D) and 
		$\textnormal{linear}-\textnormal{linear}-\log$ (for 3D) scales.}
	\label{fig:ini_dvd_2D_palette_end}
\end{figure}

\autoref{fig:ini_dvd_2D_palette_end} shows the time evolution of the 
DVDs for deviation 
vector cases {$\bf 1$} and {$\bf 4$} with the colour 
gradients representing the logarithm of the
values of the $\xi^D_{_{\bf r}}(t)$ (\ref{eq:dvd}) as indicated 
in \autoref{fig:ini_dvd_2D_palette}. 
The evolution of {$\bf 1$} is shown in panels {\bf (a)}, {\bf (b)} and 
{\bf (c)} while the evolution of {$\bf 2$} is shown in 
panels {\bf (d)}, {\bf (e)} and 
{\bf (f)}. Panels {\bf (a)} and 
{\bf (d)} show a 3D plot for the time (vertical axis) 
evolution of the DVDs in the 
phase space with the $x-$coordinates considered from $l_x=1$ up 
to $l_x=251$.
The 1D time evolution front faces of panels {\bf (a)} and 
{\bf (d)} which correspond to the 
$x-$coordinate $l_x=251$ are respectively shown in 
panels {\bf (c)} and {\bf (f)}. Similarly the corresponding 1D time 
evolution corresponding to $l_y=251$ are shown in panels {\bf (b)} 
and {\bf (e)}. From \autoref{fig:ini_dvd_2D_palette_end} we 
see that by $t\approx10^6$, the deviation vectors 
are qualitatively the same
for the two cases. For longer times, this is the 
same qualitative behaviour for all cases of initial deviation vectors. 
We also see that just like in the 
1D DKG case, the deviation vector eventually concentrates around the 
degrees of freedom where the excitation takes place.
This shows that the choice of initial deviation vector 
does not affect it's 
asymptotic dynamics. 
However, the kind of deviation vector 
that takes the least integration 
time to give the long term dynamical behaviour is preferred for our 
numerical computations.

\begin{figure}[H]
	\centering
	\includegraphics[width=0.48\textwidth,keepaspectratio]{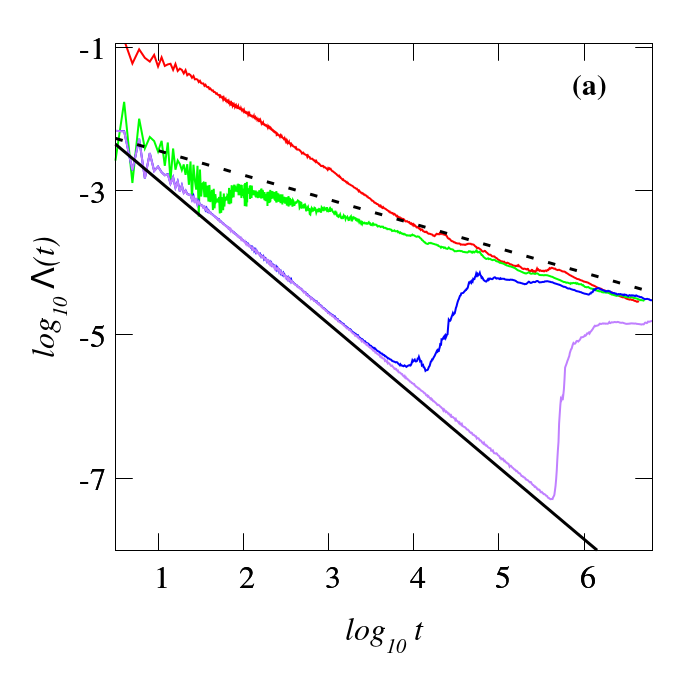}
	\includegraphics[width=0.48\textwidth,keepaspectratio]{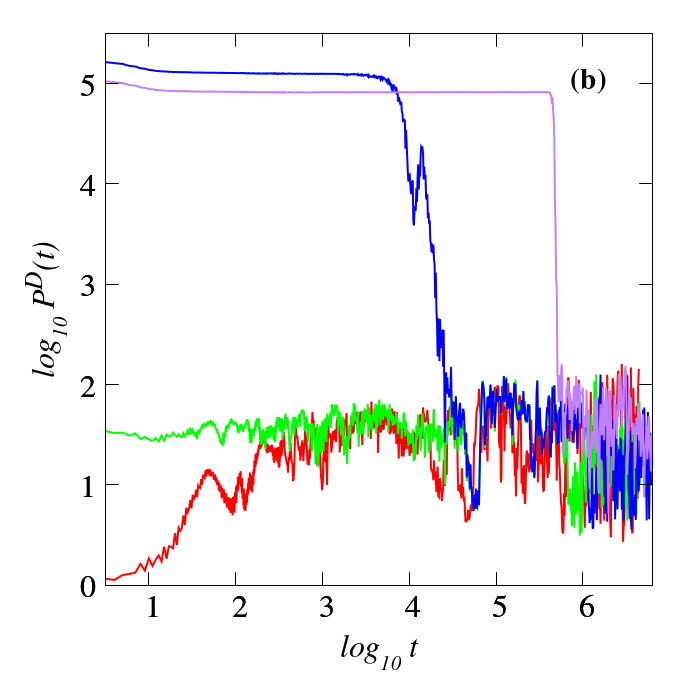}	
	\caption{Results for the time evolution of {\bf (a)} the mLCE 
		$\Lambda(t)$ \eqref{eq:ftmLCE} and {\bf (b)} $P^D(t)$ 
		\eqref{eq:P_Dvd} for the 
		initial deviation vectors {$\bf 1$} (red curve), 
		{$\bf 2$} (green curve), {$\bf 3$} (purple curve) and {$\bf 4$} 
		(blue curve). The black continuous line guides the eye for a slope 
		$-1$ and the black dashed line estimates the direction of convergence of 
		the exponents. The plots are in log-log scale.}
	\label{fig:dvd_mlce}
\end{figure}

In \autoref{fig:dvd_mlce} we report the 
evolution of the quantities $\Lambda(t)$ (\ref{eq:ftmLCE}) and 
$P^D(t)$ \eqref{eq:P_Dvd} whose evolution depends on the deviation vector. 
The black continuous line in \autoref{fig:dvd_mlce}{\bf (a)} guides 
the eye along the slope $-1$, indicating 
$\Lambda(t)$ decay power law for regular dynamics. 
By the integration time $\log_{10}t \approx 6.5$ the 
value of $\Lambda(t)$ is qualitatively independent of the corresponding initial 
deviation vector used in the 
integration for the cases {$\bf 1$}, {$\bf 2$} and 
{$\bf 4$} while case {$\bf 3$} requires additional integration time 
to saturate at the same level as the other cases. 
However, all the exponents for the four cases 
converge to a 
limit direction line (black dashed line) as seen in panel {\bf (a)}. 
For the 
initial deviation vector {$\bf 2$}, where more than one central coordinates 
of the vector are non-zero and concentrated around 
the sites where the initial excitation takes place, $\Lambda(t)$  
converges fastest to the direction of 
the dashed line. Once again, just like in the case of the 1D DKG model, 
the results also show 
that the rate of 
convergence of the finite time 
mLCE depends on the size of the non-zero component 
of the coordinates of the initial deviation vector covering the 
excited site(s) and the concentration of 
these coordinates with respect to the region of energy excitation.
We can clearly see in \autoref{fig:dvd_mlce}{\bf (a)}
that for $P^D(0)=1$ (red curve of case {$\bf 1$}), we require more integration time 
to observe convergence of the finite time 
mLCE to the direction of the dashed line 
compared to when $P^D(0)=49>1$ (green curve of {$\bf 2$}). For the initial 
deviation vector {$\bf 3$}, during the initial stages of the dynamics 
the vector 
behaves like a zero vector around the degrees of freedom where the 
excitation takes place
and so $\Lambda(t)$ shows the behaviour observed in the case 
of regular 
motion as seen in \autoref{fig:dvd_mlce}{\bf (a)} (purple curve). 
At a later stage (around $t\approx 10^{5.8}$) 
of the dynamics 
however, when the energy has spread to more lattice sites, 
the deviation vector behaves as expected (non-zero vector) 
and therefore $\Lambda(t)$ changes and starts showing a chaotic behaviour.
For the fastest 
convergence of the $\Lambda(t)$ to the asymptotic direction [dashed line in 
\autoref{fig:dvd_mlce}{\bf (a)}], we take the initial deviation vector {$\bf 2$}
whose non-zero component of it's coordinates fully covers the sites that 
have been initially excited. That is to say that, if we give the same 
energy to each initially excited site, then we set the deviation vector 
such that it's component with non-zero coordinates just covers the 
sites where the initial 
excitation has taken place and $\frac{P^D(0)}{P(0)}\gtrsim1$.
We note that case {$\bf 4$} is not our preferred choice of vector because 
$\frac{P^D(0)}{P(0)}\gg1$, and this is why we have the kind of 
observed delay in convergence of $\Lambda(t)$.

For the next and proceeding sections of this work, 
we use initial vectors of the form 
{$\bf 2$} whenever our computations involve deviation vectors in 
	two spatial dimensions.

\subsubsection{Efficiency of Symplectic integrators}
\label{subsec3:si2D}
For the 2D DKG model \eqref{eq:kg2dham}, we 
perform an analysis similar to 
the discussion of Section \ref{subsec3:si1D} for the 
1D model \eqref{eq:kg1dham}. 
 We numerically 
integrate the equations of motion (\ref{eq:kg2dmotion})
for different initial 
excitations and parameters. 
In this process, for a particular set of random disorder numbers 
$\epsilon_{\bf r}$, we compute the quantities $m_2(t)$ (\ref{eq:m2_2dkg}), 
$P(t)$ (\ref{eq:P_2dkg}), 
the normalised site energies $\xi_{\bf r}(t)$ (\ref{eq:norm_en2d}) 
and the absolute relative energy error $e_r(t)$ given by 
\begin{equation}\label{eq:kg2d_ree}
e_r(t)=\frac{|H_{2K}(t)-H_{2K}(0)|}{H_{2K}(0)},
\end{equation}
similar to \autoref{eq:ree1}. We compute all these quantities using 
the SIs we presented in Section \ref{sec3:SIs}. 
We also compute 
$\Lambda(t)$ as a way of checking that all SIs 
correctly estimate the same level of 
chaoticity in the system.
For our computations we work with a Hamiltonian lattice of $40{,}000$ 
degrees of freedom 
with dimensions $N=200$ and $M=200$. 
 We excite a rectangular block of $n\cdot m$ sites (with dimensions
$n\leq N$, $m\leq M$) which is positioned at the lattice centre 
for a disorder strength $W$ and 
an energy $H_{2K}(0)$ (or energy per excited site 
$h=\frac{H_{2K}(0)}{nm}$).
We consider the following 
different sets of initial conditions and parameters:
\begin{itemize}
\item [I:] $n=m=1$, $W=10$ and $H_{2K}(0)=h=0.3$,
\item [II:] $n=m=1$, $W=10$ and $H_{2K}(0)=h=2$,
\item [III:] $n=m=200$, $W=10$, and $H_{2K}(0)=10$, $\left(h=2.5\times10^{-4}\right)$.
\end{itemize}
Cases I and II have been reported to belong to the weak 
and selftrapping regimes of chaos respectively [\cite{Laptyeva2012}], while 
case III was included to investigate 
the general performance of SIs for situations 
where the energy of the system reaches the boundaries of the lattice.
The lattice size was chosen such that for the two cases I and II, 
the energy of the 
system never reaches the  
boundary sites.
In each of the cases we give the same momentum to each of the excited site(s)
(with these sites bearing random signs) 
and we set the momentum to 
zero for all other sites. We also start with no displacements for all sites 
of the lattice. As a result of the discussion in the preceeding 
section, we set the initial deviation vector in 
such a way that it 
has non 
zero random numbers at the coordinates which correspond to the 
sites that have been initially
excited. That is to say, the initial deviation vector in the form 
we referred 
to as {$\bf 2$}. 
In particular, we use 
an initial 
deviation vector whose non-zero coordinates just cover a square lattice of 
$16$ sites with dimensions $4\times4$ when 
integrating the variational. 
We consider the normalized energy distributions 
$\{\xi_{\bf r}(t)\}_{\bf r}$ (\ref{eq:norm_en2d}) for 
${\bf r}\in\{1,\cdots,N\}\times\{1,\cdots,M\}$ and evaluate their 
second moment $m_2(t)$ (\ref{eq:m2_2dkg}) and participation 
number $P(t)$ (\ref{eq:P_2dkg}).
For each simulation we set the integration time 
step so that the relative energy error 
$e_r(t)$ is bounded from above by $10^{-5}$. We then evaluate the efficiency 
of the schemes by checking their ability to produce 
the same dynamics of the energy propagation. This is done by considering 
the shape 
of the computed energy 
profiles and the time evolution of 
$e_r(t)$, $m_2(t)$, $P(t)$ and $\Lambda(t)$. 

\begin{figure}[H]
	\centering
	\includegraphics[width=0.9\textwidth,keepaspectratio]{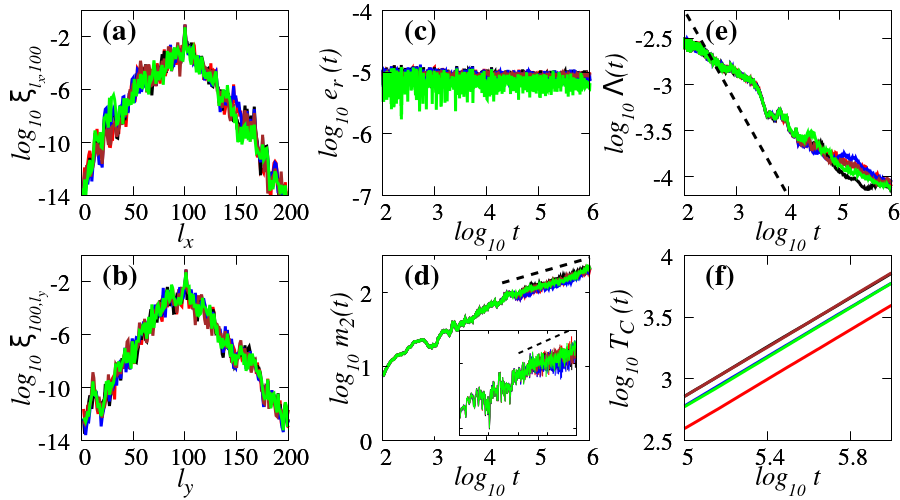}
	\caption{Results obtained for the numerical 
		integration of case I of the 2D DKG model
		(\ref{eq:kg2dham}) by order four SIs $ABA864$ for time step $\tau=0.56$, 
		$ABAH864$ for $\tau=0.40$ and order six schemes $ABA864Y6$ 
		for $\tau=0.68$, $s9SABA_26$ for $\tau=0.60$, 
		$SABA_2Y6$ for $\tau=0.60$ [red, blue, black, brown and green 
		curves]. {\bf (a)} the normalized energy distributions 
		$\xi_{_{l_x,100}}$ and {\bf (b)} $\xi_{_{100,l_y}}$ respectively 
		at the final integration time 
		$t_f=10^6$ as functions of the lattice site indices $l_x$ 
		and $l_y$.
		The time
		evolution of {\bf (c)} $e_r(t)$,  
		{\bf (d)} $m_2(t)$ and $P(t)$ [inset], 
		{\bf (e)} $\Lambda(t)$ and 
		{\bf (f)} $T_C(t)$. The dashed lines in {\bf (d)} 
		and it's inset plot guide the eye for slope $0.2$ while for 
		{\bf (e)} a slope $-1$. The curves in panels 
		{\bf (a)}-{\bf (e)} overlap while for 
		{\bf (f)} the curves for $ABAH864$, $SABA_2Y6$ and 
		$s9SABA_26$, $ABA864Y6$ practically overlap each other. Panels
		{\bf (a)}-{\bf (b)} are in linear-log scale while 
		{\bf (c)}-{\bf (f)} are in log-log scale.}
	\label{fig:2D}
\end{figure}

In \autoref{fig:2D} we present results obtained for case I by implementing the 
five best performing SIs among the studied schemes, for 
$e_r(t)\approx 10^{-5}$ [\autoref{fig:2D}{\bf (c)}]. These are the same five 
SIs which exhibited the best numerical performance in the 1D DKG 
model (\autoref{fig:1D-5best_LEs}), namely, order four SIs $ABA864$, 
$ABAH864$ and order six SIs $SABA_2Y6$, $s9SABA_26$, $ABA864Y6$. 
All these SIs captured the same dynamics of the system as seen by the 
overlapping curves in panels {\bf (a)}, {\bf (b)}, {\bf (c)}, {\bf (d)} and 
{\bf (e)} of \autoref{fig:2D}. 
We present in
\autoref{fig:2D}{\bf (a)} the energy profiles along the $x-$axis for 
sites where the $y-$value is fixed at $l_y=100$ at the final integration time 
$t_f=10^6$. Panel {\bf (b)} shows the 
corresponding energy profile along the $y-$axis for sites where the 
$x-$value is fixed at $l_x=100$ at the final integration time. 
In each of the two cross-sections the curves of the profiles from the different 
SIs overlap. Also, the implemented SIs 
produce the same time evolution of 
$m_2$ and $P$ [\autoref{fig:2D}{\bf (d)}] and $\Lambda$ 
[\autoref{fig:2D}{\bf (e)}].  
The results in \autoref{fig:2D}{\bf (d)} show 
that eventually both the second moment and participation number behave as 
$m_2(t), P(t) \propto t ^{0.2}$, indicating that the 
dynamics belong to the weak chaos regime, a confirmation of the 
findings of 
\cite{Laptyeva2012}.
In panel {\bf (e)} we present the time evolution of 
$\Lambda(t)$, with the 
dashed line guiding the eye
\begingroup
\setlength{\tabcolsep}{10pt} 
\renewcommand{\arraystretch}{1.2} 
\begin{table}[H]
	\centering
	\begin{tabular}{|l|lr|lr|lr|}
		\hline\hline
		&I &   &II & &III &\\
		SI&$\tau$ & $T_C$  & $\tau$  & $T_C$ &$\tau$ & $T_C$ \\
		\hline
		$ABA864$&0.56&3950&0.35&6230&0.73&5550\\
		$ABAH864$&0.40&5980&0.25&9270&0.51&8700\\
		$SABA_2Y6$&0.60&5960&0.38&9230&0.71&9190\\
		$s9SABA_26$&0.60&7230&0.37&10740&0.85&9300\\
		$ABA864Y6$&0.68&7170&0.40&11510&0.84&10960\\
		$SBAB_2Y6$&0.58&11400&0.36&17370&0.73&17130\\
		\noalign{\smallskip}\hline
	\end{tabular}
	\caption{Similar to \autoref{tab:1} but for the 
		2D DKG model (\ref{eq:kg2dham}). Information on the performance 
		of the most efficient SIs 
		used for the integration of the variational equations 
		of the 2D DKG model 
		at $e_r(t)\approx10^{-5}$ up to final
		time $t_f=10^6$ for the initial excitation of cases I-III (see text).
		The integration time step $\tau$ used for obtaining the 
		respective absolute
		relative energy error levels $e_r(t)$ and the required CPU time 
		$T_C(t_f)$
		(seconds) needed for each integrator are reported.}
	\label{tab:2}
\end{table}
\endgroup

 for the slope observed in the case of  
regular dynamics (slope $-1$).
$\Lambda(t)$ therefore shows a power 
law decay different from that of regular 
dynamics with no sign of the dynamics slowing down to exhibit regular behaviour.

In \autoref{tab:2}, we present results for the 2D DKG model, including 
computation of $\Lambda(t)$ for longer times, considering some  
typical representative cases and for various 
system parameters to give a global perspective 
of the system's dynamics. 
From \autoref{fig:2D}{\bf (f)} we see that, for case I and at the accuracy of 
$e_r(t)\approx10^{-5}$ [panel {\bf (c)}], the most efficient SI is 
$ABA864$ followed by 
$ABAH864$, $SABA_2Y6$ and $s9SABA_26$, $ABA864Y6$.
We present, in \autoref{tab:2}, findings for all the three 
considered cases I, II 
and III reporting the CPU times for the top six SIs at 
moderate accuracy $e_r\approx10^{-5}$. The time steps that 
maintain the energy of the systems at the specified accuracy 
are also provided. We also present 
the required 
CPU times 
for solving the variational equations.
From these results, it is clear that the computational 
performance of the SIs for the 2D DKG system
at $e_r\approx10^{-5}$ is similar to what was reported in Section \ref{sec3:1d} 
for the 
1D DKG model.

\section{Summary}
\label{sec3:conclusions}
In this chapter we presented some numerical techniques for 
investigating the dynamics of 1D and 2D 
disordered Hamiltonian systems. We discussed 
theoretical aspects of the dynamics of 
deviation vectors, their distribution and the 
variational equations. We also 
defined the LCEs which are the most widely 
used measures of chaoticity.
In Section \ref{sec:numerical_integration}, we described some existing 
integrators 
and through composition techniques, we constructed some higher 
order SIs.
We then showed the effect of the nature 
of a deviation vector on computation of dynamical 
quantities like the finite time mLCE $\Lambda(t)$ \eqref{eq:ftmLCE} 
and $P^D(t)$ \eqref{eq:P_Dvd}. In particular, 
we saw that an 
initial deviation 
vector whose non-zero coordinates just cover the region of the 
initially excited 
sites gives the fastest convergence of $\Lambda(t)$ to it's asymptotic 
evolution. That is to say, an 
initial deviation vector where $\frac{P(0)}{P^D(0)}\lesssim1$. 

We have also analysed in detail 
the computational performance of a number of symmetric SIs of 
even orders ranging from two up to eight. For each of these SIs we 
integrated the equations of motion 
and the corresponding 
variational equations (\ref{eq:variational}) of 
the 1D (\ref{eq:kg1dham}) and the 2D (\ref{eq:kg2dham}) DKG models.
We performed extensive numerical simulations for $35$ different SIs as we 
used all of them to reproduce the dynamics of the two models for 
several initial excitations. We monitored the evolution 
of the energy and it's distribution focussing our investigation 
on cases with the absolute relative energy error 
$e_r(t)\lesssim10^{-5}$ for the two models and 
$e_r(t)\lesssim10^{-8}$ for the 1D DKG system. 
We followed the evolution of two characteristics 
$m_2(t)$ and $P(t)$ of the energy distribution together with 
$\Lambda(t)$ computed by 
each one of these SIs. From our results, we 
see that the performance of the SIs that were considered is independent of 
the initial excitations as we consistently get the same ranking in terms of
exhibited efficiency by the SIs for 
the different excitations.
Secondly the performance is independent of the spatial 
dimensions of the model being investigated, as the rank in performance for 
both the 1D and 2D DKG models is the same.
For both 1D and 2D DKG models, the order four SIs $ABA864$ and $ABAH864$, 
the order six SIs $SABA_2Y6$, $s9SABA_26$ and $ABA864Y6$, 
exhibited the best performance at an absolute relative energy error 
$e_r\approx 10^{-5}$
with $ABA864$ [\cite{Blanes2013}] requiring the least CPU time for both models.
Therefore our numerical computations involving 
the 1D and 2D 
DKG systems
in the next chapters are performed 
using the $ABA864$ SI over other integrators like the popular Runge-Kutta 
class of integrators or 
the $SABA$ family [\cite{Laskar2001}] which have been used to study 
similar Hamiltonian systems 
[\cite{Flach2009,Skokos2009,Skokos2010a,Laptyeva2010,Bodyfelt2011,Laptyeva2012}].

A number of order six and order eight SIs that we considered 
required a relatively large time step $\tau$ to realise the 
approximate relative 
energy error $e_r\approx10^{-5}$ but, because of the size of $\tau$, 
they could not keep the 
absolute relative energy error $e_r(t)$ constant 
at this 
level of accuracy. However, even at higher accuracy of 
say $e_r\approx10^{-7}$ some 
of these SIs gave a better performance compared to a few others that were stable 
at $e_r\approx10^{-5}$, but they did not feature amongst the top most SIs.
These
higher order SIs have been found to stabilise at smaller values 
of $\tau$, 
and thus give higher accuracy of energy conservation in computations. 
For example, for $e_r\approx 10^{-8}$ the 
best performing schemes have been found to be the order six SIs
$s11ABA82\_6$, $SABA_2Y6$ and $s9ABA82\_6$ and the order eight SIs 
$s15ABA82\_8$ and $s19ABA82\_8$,   
with $s11ABA82\_6$ requiring the least CPU times for the integrations 
of the 1D DKG model.
\chapter{Chaotic behaviour of the 1D DKG model}
\label{chap_1d}

\pagestyle{fancy}
\fancyhf{}
\fancyhead[OC]{\leftmark}
\fancyhead[EC]{\rightmark}
\cfoot{\thepage}

In this chapter we investigate the chaotic dynamics of the 1D DKG 
model \eqref{eq:kg1dham}, extending some previous 
investigations  
on this topic for example by \cite{Gkolias2013} and \cite{Skokos2013}.
We present our work as follows: 
In Section \ref{sec4:num_techs} we highlight the 
numerical techniques used for our computations. In Section 
\ref{sec4:spread} we give a theoretical account of choaticity and 
spreading theories. We then 
present  
our findings 
for the different classifications of chaotic dynamical behaviours, namely the 
weak chaos spreading regime
in Section \ref{sec4:weak}, strong chaos case in Section \ref{sec4:strong} and 
selftrapping behaviour in Section \ref{sec4:self}. Lastly, in Section \ref{sec4:summary} we 
discuss and summarize our findings.
Parts of the results presented in Sections \ref{sec4:weak} and \ref{sec4:strong} 
of this chapter have also been reported 
in \cite{Senyange2018a}.

\section{Numerical techniques}\label{sec4:num_techs}
As part of the numerical techniques, we pay close attention to the 
integration method we use for our simulations and data analysis, the 
specifications of the computational facility to use, the programming software 
including compilers, the lattice size and other statistical analysis techniques.

\subsection{Integration method}
Integrating the equations of motion \eqref{eq:motion} for the 1D DKG system 
\eqref{eq:kg1dham} as well as 
the variational equations \eqref{eq:variational}, involves 
implementing the ABA864  
[\cite{Blanes2013}] integrator, which is a two-part split order four SI 
whose computational efficiency has been shown in 
Chapter \ref{chap:num_tech} to be relatively high compared to other integrators. 
More specifically we apply the TM method 
[Section \ref{sec3:tangent_map}; \cite{Skokos2010,Gerlach2011,Gerlach2012}] 
to integrate the 
variational equations, and thus utilise equations \eqref{op:var}.
In our simulations, the relative energy error  
\begin{equation}\label{eq4:ree1}
	e_r(t)=\frac{|H_{1K}(t)-H_{1K}(0)|}{H_{1K}(0)}
\end{equation}  
of the integration 
is kept at $e_r(t)\approx10^{-5}$ by using time steps $\tau\approx0.2-0.5$. 
Furthermore, in our study we use initial deviation vectors of the form {$\bf 2$} 
[Section \ref{sec3:ini_dvd}, \autoref{ini_dvd_1D}{\bf (b)}] where only the 
position and momentum coordinates of $20-350$ central sites are non-zero.

\subsection{Practical computational consideration}\label{sec4:comp}
Like it was mentioned in Section \ref{numerical_results}, the numerical  
computations were done using a facility with specifications 
Intel Xeon E$5-2623$, $3.00$ GHz 
working in FORTRAN $90$ programming language with compilers 
Intel Fortran (ifort) and GCC gfortran  
at optimization level $2$ (-O$2$). We also used the Open Multi-Processing (OPENMP) parallelization 
application programming interface allocating on average $4-6$ threads 
(CPU-cores) for 
each simulation. This was made possible after eliminating 
data dependencies in the program sections 
where the parallelisation directives were implemented and 
thus the computational efficiency 
(i.e. CPU time)  
was improved by an approximate factor of $2.5$.

\subsection{Lattice size}
Since in our investigation we want to study the asymptotic behaviour 
of initially localized energy excitations in theoretically infinite lattices 
we therefore perform simulations for very large but obviously finite chains.
This setup however is computationally costly in terms of 
CPU time, storage space and 
computer processing memory. 
To counter these challenges, we use lattices 
whose length $N$ increases with time (i.e. $N=N(t)\ge N(0)$) 
depending on the wave packet extent taking special care in avoiding 
boundary effects. 
To do so, we start 
with a relatively 
small lattice of size $N(0)$ and increase it whenever either the wave packet 
or the deviation vector distribution 
gets close to the boundary sites. We check that in the following 
way: for each site $l$ located within the boundary strip of width $N_{W}$, 
we ensure that the site
energy  
\begin{equation}\label{eq4:en_pa_site_1dkg}
	h_l=\frac{p_l^2}{2}+\frac{\tilde{\epsilon_l}q_l^2}{2}+
	\frac{\left(q_{l-1}-q_l\right)^2+\left(q_{l+1}-q_l\right)^2}{4W}+
	\frac{q_l^4}{4},
\end{equation} 
and the square norm component $\delta q_l^2+\delta p_l^2$ of the deviation 
vector at this site are always less 
than $10^{-8}$. That is 
to say, for all $l\in[1,N_W]\cup[N-N_W+1,N]$ 
\begin{equation}\label{eq:bdry_cndtn}
	\max_{_l} \{h_l\}<10^{-8} \qquad\textnormal{and}\qquad\max_{_l}\{\delta q_l^2+\delta p_l^2\}<10^{-8}.
\end{equation} We note that this restriction on the deviation vector 
ensures that its dynamics is not affected by finite lattice size 
effects thereby ensuring that any 
deductions made using the vector are reliable.
In case criteria \eqref{eq:bdry_cndtn} are violated, 
the lattice size 
is then increased by a width dependant on the 
disorder strength $W$ in order to increase 
the lattice size so that the wave packet 
will continue spreading without the boundary sites affecting the dynamics. 
We use $N_W$ with values inversely proportional 
to $W$ since the spatial extent (estimated for example by the localization 
volume) of the NMs 
increases with a decrease in $W$  
as noted in  
Chapter \ref{chap:disor_sys} and \cite{Krimer2010}. 

\subsection{Numerical estimation of slopes}\label{sec4:slope}
In our study, we numerically compute quantities for many different 
disorder realizations over which 
we average the obtained results. For cases where we are interested 
in the power law 
description of the dependence of such a 
quantity (let us generally denote it by $Q$) 
usually on time $t$, we numerically estimate a slope of 
the average curve to find 
this law. This is done as follows: 
From the averaged data we compute a smooth curve using 
the robust locally weighted regression (LOESS) algorithm 
[\cite{Cleveland1979,Cleveland1988,Laptyeva2010,Gkolias2013}]. 
In this process, we smoothen points' trend 
using linear locally weighted regression, 
the weight function 
\begin{equation*}
	W(x)=\begin{cases} 
		(1-|x|^3)^3 & \textnormal{for} \qquad|x| < 1 \\
		0 & \textnormal{for}\qquad |x| \geq 1, 
	\end{cases}
\end{equation*} a single iteration of the robust fitting procedure 
and smoothening proportion $0.1$ [\cite{Cleveland1979}].
We then use the 
central finite difference calculation described in 
\cite{Laptyeva2010}, \cite{Bodyfelt2011} and 
\cite{Gkolias2013} to determine the local derivative $\alpha_Q(t)=dQ/dt$ of 
quantity $Q$ with respect to (usually time) $t$.
For cases where we compute the error in the power law exponent (using say 
error bars), 
we first compute the local derivative from each data realization and then 
average it over all realizations.

\section{Chaoticity and energy spreading theories}\label{sec4:spread}
We now give a theoretical description 
of a comparison between 
chaoticity and the wave packet extent.
As mentioned before in Chapter \ref{chap:num_tech}, 
chaoticity can be quantified using the 
finite time mLCE $\Lambda(t)$ \eqref{eq:ftmLCE}. Therefore, the chaoticity time 
scale $T_{\Lambda}(t)$ (usually referred to as Lyapunov time), 
defined as
\begin{equation}
	\label{eq:TL}
	T_{\Lambda}(t) \sim \frac{1}{\Lambda(t)}
\end{equation} [\cite{Skokos2010}], estimates the time a system 
takes to become chaotic.
Knowing that $\Lambda(t)\propto t^{\alpha_{\Lambda}}$, equation 
\eqref{eq:TL} becomes 
\begin{equation}
	\label{eq:TL2}
	T_{\Lambda}(t) \sim t^{-\alpha_{\Lambda}}.
\end{equation}
Wave packet extent on the other hand can be estimated using 
the second moment $m_2$ [\eqref{eq:m2_1dkg} for 1D and \eqref{eq:m2_2dkg} 
for the 2D models] and participation number 
$P$ [\eqref{eq:P_1dkg} for 1D and \eqref{eq:P_2dkg} 
for the 2D models]. 
We therefore define 
wave packet characteristic spreading time scales 
based on $m_2$ and $P$ as follows: 
The diffusion 
coefficient $D$, which is related to the second moment as 
\begin{equation}\label{eq:D}
	m_2(t) \sim Dt,
\end{equation}
defines a spreading time scale 
$T_{m_2}$
\begin{equation}
	\label{eq:TM}
	T_{m_2}(t) \sim \frac{1}{D}.
\end{equation}
Equations \eqref{eq:D} and \eqref{eq:TM} give 
\begin{equation}
	\label{eq:TM2}
	T_{m_2}(t) \sim t^{1-\alpha_{m_2}}
\end{equation} where $m_2(t) \propto 
t^{\alpha_{m_2}}$.

Secondly, the spreading time scale [we denote by $T_P$] 
can be interpreted as the 
time taken for a marginal 
change in the wave packet participation number $P$. 
Therefore, 
\begin{equation}
	\label{eq:TP}
	T_P(t) \sim \frac{1}{\dot{P}},
\end{equation}
where $\dot{P}$ denotes the derivative of $P$. 
For the selftrapping regime where the number $P$ of highly 
excited sites is practically constant, $T_p$ 
is infinite.
Since $P(t) \propto t^{\alpha_P}$, then 
 \begin{equation}
	\label{eq:TP2}
	T_P(t) \sim t^{1-\alpha_P}.
\end{equation}
Using equations \eqref{eq:TL2}, \eqref{eq:TM2} and \eqref{eq:TP2} we 
get the spreading to chaoticity time scale ratios
\begin{equation}
	\label{eq:T_ratios}
	R_{m_2}:=\frac{T_{m_2}}{T_{\Lambda}}  \sim t^{1+\alpha_{\Lambda}-{\alpha_{m_2}}}, \,\qquad\qquad\, R_P:=\frac{T_P}{T_{\Lambda}}  \sim t^{1+\alpha_{\Lambda}-{\alpha_P}}.
\end{equation}
From the ratios \eqref{eq:T_ratios}, we can compare the rates of 
wave packet's chaotization and extent of subdiffusive spreading.

\section{Procedure for numerical simulations}\label{sec4:procedure}
In our numerical simulations, we give the system of $N$ sites 
a total energy $H_{1K}$ 
\eqref{eq:kg1dham} by initially exciting a block of 
$L$ central sites of the lattice with each site getting the same 
energy $h=h_l=H_{1K}(0)/L$
by setting $p_l = \pm \sqrt{2h}$, where the signs are assigned 
randomly and $p_l=0$ for 
all other sites. The displacements $q_l$ of the lattice are all 
set to $0$. In all cases of the simulations, 
we imposed conditions $p_0=p_{N+1}=q_0=q_{N+1}=0$.

We follow the evolution of several ICs 
for various disorder realizations of the system 
\eqref{eq:kg1dham} where for each setup we obtain 
statistical results for the time evolution of a dynamical quantity, 
say $Q$, 
by averaging its values over $200$ 
different disorder realizations to get $\langle Q \rangle$, where 
$\langle \cdot \rangle$ denotes the average. We then extract the  
rate of change of $Q$ (derivative of $Q$) 
to get 
\begin{equation}
	\alpha_{_Q}(t) = \frac{d\langle Q \rangle}{d t} = \left\langle\frac{d Q }{d t}\right\rangle.
	\label{eq:aQ}
\end{equation}
An almost constant value of $\alpha_{_Q}(t)$ indicates the existence 
of a linear dependence 
$Q(t)\propto\alpha_{_Q}t$ 
of $Q$ on $t$. 
If the fitting is done for the logarithms of $Q$ and $t$ (as repeatedly 
happens in our studies) then implementing the procedure we just discussed 
we obtain a slope 
\begin{equation}
	\alpha'_{_Q}=\frac{d\langle\log Q \rangle}{d(\log t)}.
	\label{eq:bQ}
\end{equation} The practical constancy of $\alpha'_{_Q}$ indicates a power law 
dependence $Q(t)\propto t^{\alpha_{Q}}$ of $Q$ on $t$.

\section{The weak chaos regime}\label{sec4:weak}
\begin{figure}[H]
	\centering
	\includegraphics[width=0.65\textwidth,keepaspectratio]{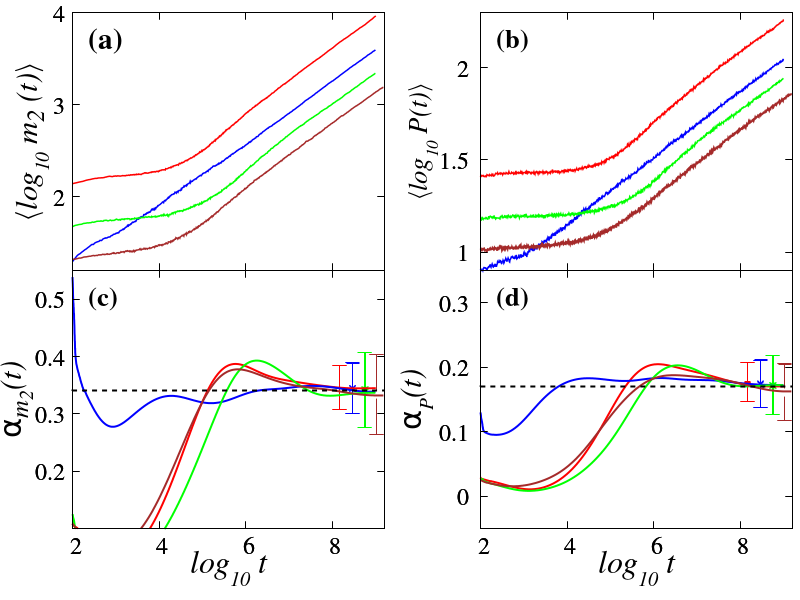}
	\caption{Results for the time evolution of {\bf (a)} $m_2(t)$ \eqref{eq:m2_1dkg}, 
		{\bf (b)} $P(t)$ \eqref{eq:P_1dkg} 
		and the corresponding slopes {\bf (c)} $\alpha_{m_2}(t)$ for $m_2(t)$ 
		{\bf (d)} $\alpha_{P}(t)$ for $P(t)$ for the weak chaos cases $W1_1$ 
		(red curves), $W2_1$ (blue curves), $W3_1$ (green curves) and $W4_1$ (brown curves) of Hamiltonian \eqref{eq:kg1dham}. 
		The dashed lines in panels {\bf (c)} and {\bf(d)} respectively 
		denote values $\alpha_{m_2}=0.34$ and $\alpha_{P}=0.17$.
		The error bars in {\bf(c)} and {\bf(d)} denote 
		1 standard deviation of the computed slopes.
		Plots {\bf(a)} and {\bf (b)} are in log-log scale while {\bf(c)} 
		and {\bf (d)} 
		are in log-linear scale.}
	\label{fig4:weak_m2P}
\end{figure}

We generally perform numerical integration of the system \eqref{eq:kg1dham} 
for time not exceeding 
$t_f \approx 10^9$ time units. The lattice 
size $N(t)$ was increased up to a maximum 
$N(t_f) \approx 700 - 1\,500$ thereby ensuring no energy at the boundary 
sites. 
The energy $H_{1K}$ of the system was conserved at an absolute 
relative energy error \eqref{eq4:ree1} $e_r(t)\approx10^{-5}$ for time 
steps $\tau\approx0.4-0.6$.
We investigate the chaotic behaviour of the system \eqref{eq:kg1dham}
by considering the following four parameter cases 
where each of $L$ central lattice sites is 
excited with energy $h$ \eqref{eq:en_pa_site_1dkg} 
with disorder strength $W$.
\begin{description}
	\item[Case $W1_1$:] $W=3$, $L=37$, $h = 0.01$;
	\item[Case $W2_1$:] $W=4$, $L=1$, $h = 0.4$;
	\item[Case $W3_1$:] $W=4$, $L=21$, $h = 0.01$;
	\item[Case $W4_1$:] $W=5$, $L=13$, $h = 0.02$.
\end{description}
These are representative cases of various $W$ values and initial 
excitation lengths. The cases $W1_1$, $W2_1$ and $W3_1$ were also considered in \cite{Skokos2013} 
and \cite{Gkolias2013} 
and respectively named cases III, I and II while case $W4_1$ 
is a new one. 
In that work, a total of $50$ disorder 
realizations were analysed in each of the cases. In our study, we perform 
an analysis for 
$200$ realizations in each of the four cases $W1_1$, $W2_1$, 
$W3_1$ and $W4_1$, thereby improving the statistical reliability of our 
findings. We also analyse the dynamics in more depth by 
studying observables  
for the deviation vector and it's distribution.\newline
\autoref{fig4:weak_m2P} shows results for the 
time evolution of the second moment 
$m_2(t)$ \eqref{eq:m2_1dkg} and the participation number $P(t)$ 
\eqref{eq:P_1dkg}. 
Panels {\bf (a)} and {\bf (b)} respectively show the evolution of 
$m_2(t)$ and $P(t)$ for the cases $W1_1$ (red curves), $W2_1$ (blue curves), 
$W3_1$ (greens) and 
$W4_1$ (brown curves). 
In panels {\bf (c)} and {\bf (d)} we present 
slopes $\alpha_{m_2}(t)$ of $m_2(t)$ and $\alpha_{P}(t)$ of $P(t)$ 
respectively with the dashed lines indicating the 
values 
$\alpha_{m_2}=0.34$ [panel {\bf (c)}] and $\alpha_{P}=0.17$ 
[panel {\bf (d)}]. The error bars in {\bf(c)} and {\bf(d)} respectively 
denote the numerical error of 
one standard deviation in the computation of slopes 
$\alpha_{m_2}(t)$ and $\alpha_P(t)$.
These results verify that the cases considered
show characteristics of the weak chaos spreading behaviour with 
$m_2(t)$ and $P(t)$ [panels {\bf (a)} and {\bf (b)} respectively] evolving 
as $m_2(t) \propto t^{0.34}$ [panel {\bf (c)}],
$P(t) \propto t^{0.17}$ [panel {\bf (d)}] which is in 
accordance with theoretical estimates discussed in Section 
\ref{sec:theories} and already established results of 
\cite{Flach2009b}, \cite{Skokos2009}, \cite{Laptyeva2010}, 
\cite{Flach2010} and 
\cite{Bodyfelt2011} which emphasize that $m_2(t) \propto t^{1/3}$ and 
$P(t) \propto t^{1/6}$. More specifically, the theoretical predictions for 
the power law exponents lie within an interval, centred around our 
numerically computed values and of width of one standard 
deviation.

\begin{figure}[H]
	\centering
	\includegraphics[scale=0.34]{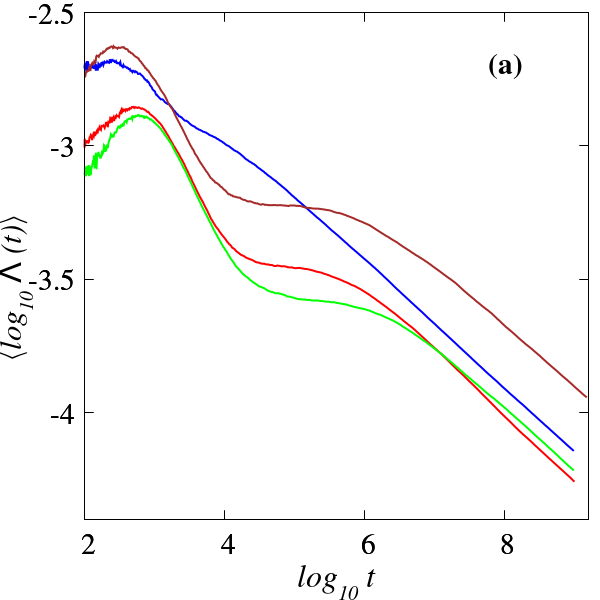}
	\includegraphics[scale=0.34]{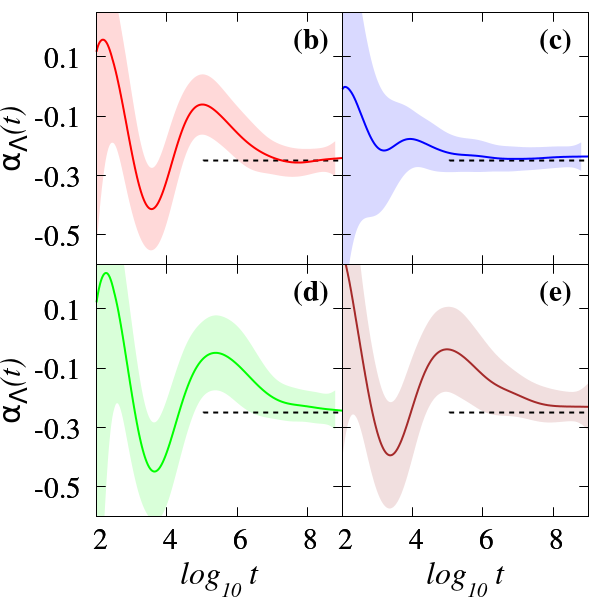}
	\caption{Results for {\bf (a)} the time evolution of the finite time mLCE 
		$\Lambda(t)$ \eqref{eq:ftmLCE} 
		and {\bf (b)}-{\bf (d)} the corresponding numerically computed 
		slopes for the four weak chaos cases of the 
		1D DKG system \eqref{eq:kg1dham} described 
			at the beginning of Section \ref{sec4:weak}. Results are 
		averaged over $200$ disorder realizations. The curve colours correspond to the cases presented in \autoref{fig4:weak_m2P}. The straight 
		dashed lines in {\bf (b)} - {\bf (e)} indicate 
		the value $\alpha_\Lambda =-0.25$. Panel {\bf (a)}
		is in log-log scale while panels {\bf (b)}-{\bf (d)} are in log-linear scale.}
	\label{fig:weak_L}
\end{figure}

We now discuss findings for the chaoticity of the model. 
For each of the four weak chaos cases we investigated, we compute the 
time evolution of 
$\Lambda(t)$ \eqref{eq:ftmLCE} and present the results in 
\autoref{fig:weak_L}. 
The average of $\Lambda(t)$ over 
200 disorder realizations is shown in \autoref{fig:weak_L}{\bf (a)}. 
In the panels {\bf (b)}-{\bf (e)}, we plot the corresponding 
numerically computed 
slopes (\ref{eq:bQ}) of the curves in panel {\bf (a)} together with 
an error (colour shading) defined by one standard 
deviation of the 
computed slopes $\alpha_{\Lambda}$. 
The horizontal dashed lines in {\bf (b)}-{\bf (e)} denote the 
value $\alpha_{\Lambda}=-0.25$.
The results of \autoref{fig:weak_L} {\bf (b)}-{\bf (e)} 
indicate that for each weak chaos case considered, the time 
evolution of $\Lambda(t)$ converges towards the power law 
$\Lambda(t)\propto t^{-0.25}$. 
In this case, the lighter colour shade or clouds correspond 
to one standard deviation of the computed slopes.
This result was also reported by 
\cite{Skokos2013} and \cite{Gkolias2013} where the cases $W1_1$, 
$W2_1$ and $W3_1$ were studied and $\Lambda(t)$ evolved following 
the power law $\Lambda(t)\propto t^{-0.25}$.
The additional case, $W4_1$, we have included in our analysis also complies 
with the 
power law decay of $\Lambda(t)$.
Clearly $\alpha_\Lambda$ is different from 
the value $-1$ [\cite{Benettin1976,Skokos2010c}] 
of regular dynamical behaviour, thereby our findings reveal that 
the dynamics of the 1D DKG model in the weak chaos regime 
shows no tendencies of relaxing into 
regular behaviour. This result was also observed 
and discussed in \cite{Skokos2013}.

Using 
our numerical findings, the spreading to chaoticity time scale
ratios $R_{m_2}$ and $R_P$ of 
\eqref{eq:T_ratios}
give 
\begin{equation*}
	R_{m_2}  \sim t^{0.41}, \,\,\,\, R_P  \sim t^{0.58},
\end{equation*}
where ${\alpha_{m_2}}=0.34$, 
${\alpha_P}=0.17$ and $\alpha_{\Lambda}=-0.25$.
Thus the spreading time scales $T_{m_2}$ and $T_P$ 
remain always larger than the chaoticity time scale $T_{\Lambda}$, 
which implies that wave packet gets more chaotic at a rate faster than 
its spreading.

\subsubsection{Deviation vector distributions}
\label{sec:DVD}
\begin{figure}[H]
	\centering
	\includegraphics[scale=0.35,keepaspectratio]{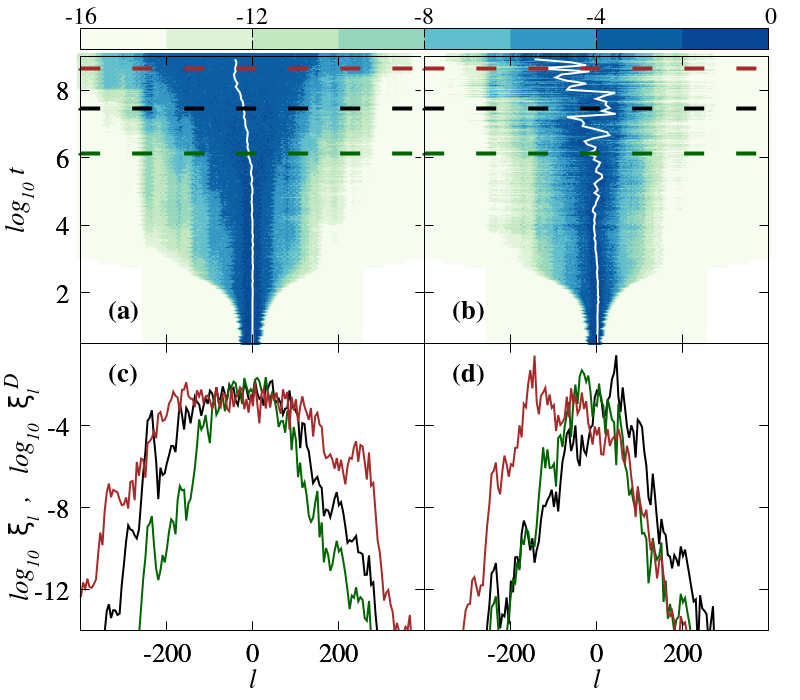}
	\caption{Results for the time evolution of {\bf(a)} the normalized 
		energy distribution $\xi_l(t)$ \eqref{eq4:en_pa_site_1dkg} and 
		{\bf(b)} the corresponding DVD $\xi_l^D(t)$ \eqref{eq:dvd} of case $W1_1$. 
		The colour scales at the top of the figure are used for 
		colouring lattice sites according to their {\bf(a)} $\log_{10} \xi_l(t)$ 
		and {\bf(b)} $\log_{10} \xi_l^D(t)$ values. In both panels a white 
		curve traces the distributions' centre. {\bf(c)} Normalized energy 
		distributions $\xi_l(t)$ and {\bf (d)} DVDs at 
		times $\log_{10}t=6.14$ [green curves], $\log_{10}t=7.47$ 
		[black curves], $\log_{10}t=8.65$ [red curves]. These times are also denoted 
		by similarly coloured horizontal dashed lines in {\bf (a)} and {\bf(b)}. 
	}
	\label{fig:palette_weak_kg}
\end{figure}

In order to further analyse the chaos behaviour of the 1D DKG 
\eqref{eq:kg1dham} model, we investigate the time evolution 
of the deviation vectors \eqref{eq:dev_vec} and their distributions 
\eqref{eq:dvd}. 
As a measure of the separation of trajectories, the deviation vector 
${\bf w}(t)$ \eqref{eq:dev_vec} 
(on which the finite time mLCE depends) eventually aligns 
to the most unstable direction in the system's phase space. 
Therefore large $\xi^D_l$ \eqref{eq:dvd} 
values of the DVD indicate at which lattice 
sites the sensitive dependence on initial conditions is higher. It is for 
this reason that such distributions were used in \cite{Skokos2013} 
to visualize the motion of chaotic seeds (degrees of freedom exhibiting 
more chaos) inside the spreading wave packet. Utilisations of the DVD in 
a similar manner were also reported in \cite{Ngapasare2019}.
In \autoref{fig:palette_weak_kg} we present the evolution 
of the {\bf (a)} energy density $\xi_l(t)$ \eqref{eq4:en_pa_site_1dkg} 
and {\bf (b)} DVD $\xi^D(t)$ \eqref{eq:dvd} 
for a representative realisation belonging to case $W1_1$. 
The centre of the lattice 
has been translated to $0$.
In 
\autoref{fig:palette_weak_kg}{\bf (c)} and {\bf (d)} we present snapshots 
of 
these distributions taken at the instances corresponding 
to the times indicated by the horizontal dashed 
lines of \autoref{fig:palette_weak_kg}{\bf (a)}, {\bf(b)} using matching 
colours.

From the results of \autoref{fig:palette_weak_kg} we see that 
the energy density continuously extends to cover a  
wider region of the chain. This spreading is done more 
or less symmetrically around the position where the initial excitation 
takes place as 
the evolution of the distributions' centre [white curve in 
\autoref{fig:palette_weak_kg}{\bf(a)}] is rather smooth and always 
remains 
close to the middle of the lattice. 
On the other hand, the DVD [panel {\bf(b)}] always stays inside the excited part of 
the lattice as seen in panel 
{\bf (a)} 
where the lattice section traversed by the DVD 
lies within the portion covered by the energy distribution. The DVDs also 
retain a more localised, pointy shape as observed in panel {\bf (d)}.
The DVDs 
are initially 
located in the region of the initial excitation and later 
start moving 
around widely after $\log_{10}t \approx 6$, as clearly 
depicted in the time evolution of each DVD's mean position 
$\bar{l}_w = \sum _l l \xi_l^D$ [white curve in 
\autoref{fig:palette_weak_kg}{\bf(b)}], as $\bar{l}_w$ randomly 
fluctuates with increasing amplitude. 
The authors of \cite{Skokos2013} based on such observations 
and represented the random motion 
of deterministic chaos seeds inside the wave packet using DVDs. These random 
oscillations of the chaotic seeds are essential in homogenizing chaos inside 
the wave packet, leading in this way to the thermalization 
and subdiffusive spreading of the wave packet.

As the wave packet spreads [\autoref{fig:palette_weak_kg} {\bf (c)}], 
the (constant) total energy is 
shared among more degrees of freedom  
as additional lattice sites are excited.
Therefore the energy density of the  excited sites (i.e. the 
effective nonlinearity strength of the system) 
decreases. Also, the system's chaoticity reduces in time 
since the value of $\Lambda(t)$ [\autoref{fig:weak_L} {\bf(a)}] 
follows a power law decay.

\begin{figure}[H]
	\centering
	\includegraphics[scale=0.35,keepaspectratio]{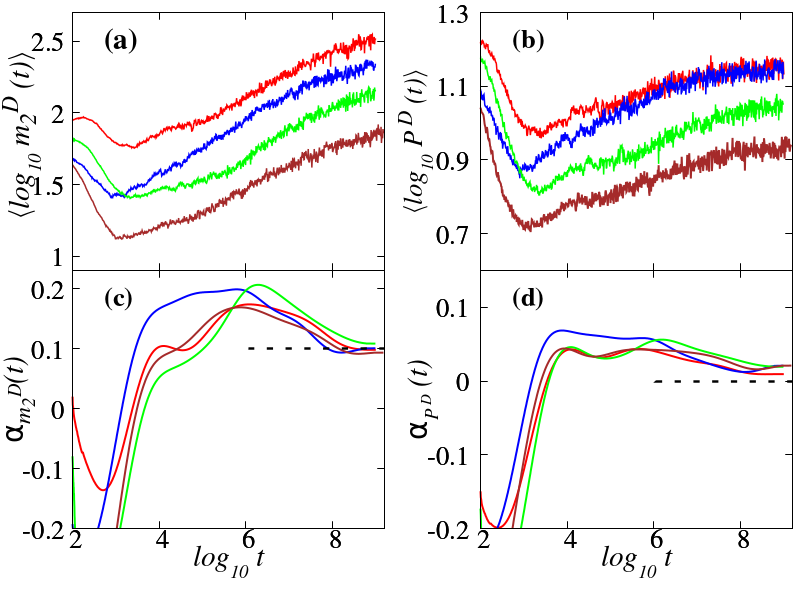}
	\caption{Results averaged over realizations for the time evolution of  
		{\bf (a)} $m_2^D(t)$ \eqref{eq:m2_1dkg}, {\bf (b)} $P^D(t)$ \eqref{eq:P_1dkg}, 
		the corresponding slopes {\bf (c)} $\alpha_{m_2^D}(t)$ for $m_2^D$ 
		and 
		{\bf (d)} $\alpha_{P^D}(t)$ for $P^D$ for the cases $W1_1$ - $W4_1$. 
		The dashed lines in panels {\bf (c)} and {\bf (d)} respectively 
		indicate values $\alpha_{m_2^D}(t)=0.1$ and $\alpha_{P^D}(t)=0$. 
		The curve colours correspond to the cases presented in \autoref{fig4:weak_m2P}. 
		Panels {\bf(a)} and {\bf (b)} are in log-log scale while {\bf(c)}, 
		and {\bf (d)} 
		are in log-linear scale.}
	\label{fig:m2_P_dvd_1D}
\end{figure}

We now generalise the findings of \autoref{fig:palette_weak_kg} on DVDs 
beyond the single realisation reported, by computing some characteristics of 
the distribution. We compute the DVD 
second moment $m_2^D(t)$ and participation number $P^D(t)$.
The obtained average results are presented in \autoref{fig:m2_P_dvd_1D}. 
The DVDs' second moment $m_2^D(t)$ [\autoref{fig:m2_P_dvd_1D}{\bf(a)}] shows an 
asymptotic, slow growth, reaching values 
which are always smaller than the wave packets' $m_2(t)$ 
[\autoref{fig4:weak_m2P}{\bf(a)}] by at least one order of magnitude. This slow 
growth is characterised by the power law $m_2^D(t) \propto t^{0.1}$ 
indicated by the computed derivatives $\alpha_{m_2^D}(t)$ of 
$m_2^D(t)$ in \autoref{fig:m2_P_dvd_1D}{\bf(c)} where the horizontal 
dashed line corresponds to the value $\alpha_{m_2^D}(t)=0.1$.
The fact that the DVDs of \autoref{fig:palette_weak_kg}{\bf (d)} 
remain practically localized 
with a narrow, 
pointy shape is reflected in 
the small and almost constant $P^D$ values 
[\autoref{fig:m2_P_dvd_1D}{\bf(b)} and {\bf (d)}]. 
Their participation 
number $P^D$ attains small values, 
$P^D \gtrsim 8$ 
(the lowest value being $P^D \approx 8$ for case $W4_1$ [brown curves in 
\autoref{fig:m2_P_dvd_1D} {\bf (b)}]), showing a 
tendency to asymptotically saturate to a constant number, since all 
curves of \autoref{fig:m2_P_dvd_1D}{\bf(b)} show signs of an eventual level 
off. This is confirmed by the computation of the slopes in 
\autoref{fig:m2_P_dvd_1D}{\bf(d)} of 
$P^D(t)$ which show a tendency to settle at values close to 
$\alpha_{P^D}=0$ [horizontal dashed line in \autoref{fig:m2_P_dvd_1D} {\bf(d)}].

We note that 
the localized 
DVD [\autoref{fig:palette_weak_kg}{\bf(b)} and {\bf(d)}] meanders over a 
significantly large region of the lattice compared 
to the energy distribution [\autoref{fig:palette_weak_kg}{\bf(a)} and 
{\bf(c)}] whose mean position is more or less fixed at a specific degree of 
freedom.
In order to quantify the width of the lattice region effectively 
covered by the 
DVD, we follow the time evolution of the following
three numerical quantities. Based on the mean position 
$\bar{l}_w(t)$ of 
the DVD \eqref{eq:dvd} at time $t$ we monitor the 
evolution of 

\begin{equation}
	\label{eq:mad}
	M(t)=\max_{t\in\Delta}\left\{||l_{\Delta}-\bar{l}_w(t)||\right\},
\end{equation} where 
\begin{equation*}
	l_\Delta=\frac{1}{|\Delta|}\sum_{t\in\Delta} \bar{l}_w(t)
\end{equation*} with $\Delta$ being a time window of fixed width 
$|\Delta|$. $M(t)$ 
gives us an idea about the mean deviation of the DVD centre 
due to the dynamical evolution within a certain epoch in time. In 
other words, $M$ gives the maximum of the absolute deviation of the elements of 
$\{\bar{l}_w(t)\}$ from their 
average in a time interval $\Delta$. 

A second measure we denote by $R(t)$ is defined as
\begin{equation}
	\label{eq:R}
	R(t)= \max_{0\leq t_1\leq t_2\leq t}\left\{||\bar{l}_w(t_1)-\bar{l}_w(t_2)||\right\},
\end{equation} where $||.||$  denotes the 
Euclidean norm. $R(t)$ gives the largest separation 
between any two mean positions of the distribution at or before time $t$. 
$R(t)$ in the case of the 1D lattice considered here is equivalent to 
\begin{equation*}
	\label{eq:L}
	R(t)= \max_{0\leq t_1 \leq t} \{ \bar{l}_w(t_1) \} - \min_{0\leq t_2 \leq t} \{ \bar{l}_w(t_2) \}.
\end{equation*} 
The quantity
\begin{equation}
	\label{eq:l}
	\iota^D(t)=\sum_t||\bar{l}_w(t)-\bar{l}_w(t-\tau)||,
\end{equation} estimates the magnitude of the total displacement of 
the DVD centre. $\bar{l}_w(t-\tau)$ is the mean 
position of the distribution, a time $\tau$ before the mean 
position 
$\bar{l}_w(t)$. 
From the evolution of the white 
curves showing the mean positions of the energy distribution 
[\autoref{fig:m2_P_dvd_1D} {\bf(a)}]
and DVD [\autoref{fig:m2_P_dvd_1D} {\bf(b)}] we see 
that the DVD centre travels more distance compared to the energy centre and 
so the value of $\iota^D$ is expected to be larger for the DVD.

These three quantities inform us on the dynamics of the deviation vector 
in different ways. $R$ estimates the extent (amplitude of fluctuations 
of the distribution centre) to which the DVD spreads in 
the lattice, $\iota^D$ the total distance covered by the distribution centre 
as it 
meanders in the lattice while $M$ gives the same information as $R$ but for 
short intervals of time during the evolution. We therefore expect that 
at any integration time $t$
\begin{equation}\label{ineq}
	M(t)\leq R(t)\leq \iota^D(t).
\end{equation}

\begin{figure}[H]
	\centering
	\includegraphics[scale=0.45,keepaspectratio]{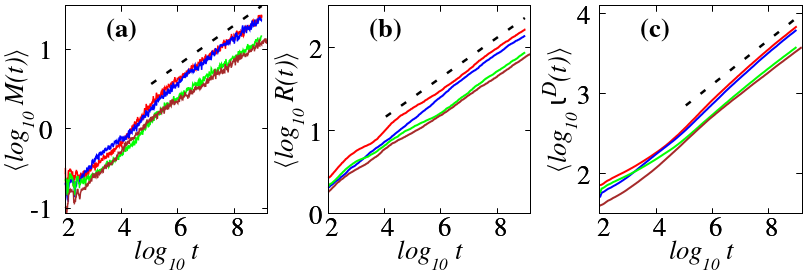}
	\caption{Results for the time evolution of {\bf(a)} $M(t)$ 
		\eqref{eq:mad}, {\bf(b)} $R(t)$ \eqref{eq:R} and {\bf(c)} 
		$\iota^D(t)$ \eqref{eq:l} of the DVD 
		for cases $W1_1$ - $W4_1$ of Hamiltonian \eqref{eq:kg1dham}.
		The curve colours correspond to the 
		cases presented in Figure \ref{fig4:weak_m2P}.
		The straight dashed lines show slopes 
		{\bf (a)} $0.24$, {\bf (b)} $0.25$ and {\bf (c)} $0.28$.
		Panels {\bf(a)}, {\bf (b)} and {\bf(c)} are in log-log 
		scale.}
	\label{fig:mad_diam_weak_1kg}
\end{figure}

\autoref{fig:mad_diam_weak_1kg} shows results obtained for 
the evolution of quantities $M(t)$ 
\eqref{eq:mad} [panel {\bf (a)}], $R(t)$ \eqref{eq:R} 
[panel {\bf(b)}] and $\iota^D(t)$ 
\eqref{eq:l} [panel{\bf(c)}] where the dashed lines guide the eye 
for slopes $0.24$ in {\bf(a)}, $0.25$ in {\bf(b)} and $0.28$ in {\bf(c)}.
Since the wave packet 
continuously spreads, the localized chaotic seeds, which 
are constantly meandering inside it, cover larger lattice 
regions as time increases.
This is evident by the continuously  
increasing values of $M(t)$  
[\autoref{fig:mad_diam_weak_1kg}{\bf(a)}], $R(t)$  [\autoref{fig:mad_diam_weak_1kg}{\bf(b)}] 
and $\iota^D(t)$ [\autoref{fig:mad_diam_weak_1kg}{\bf(c)}]. This increase is 
very well described, for all three measures, by the power laws 
$R(t)\propto t^{0.24}$ [\autoref{fig:mad_diam_weak_1kg}{\bf(a)}], 
$\iota^D(t)\propto t^{0.25}$ [\autoref{fig:mad_diam_weak_1kg}{\bf(b)}]
and  $M(t)\propto t^{0.28}$ [\autoref{fig:mad_diam_weak_1kg}{\bf(c)}].

\section{The strong chaos regime}\label{sec4:strong}
We now turn our attention to the chaotic behaviour of energy 
spreading in the strong chaos regime that was predicted in  
Section \ref{sec:theories}; a study which was 
not considered  in \cite{Skokos2013}. As was 
explained in \cite{Laptyeva2010}, \cite{Flach2010} and \cite{Bodyfelt2011}, 
the strong 
chaos subdiffusive regime is only possible when more than 
one site is initially excited. 
The  wave packet spreading in the initial phase for
this regime is  
faster than in the weak chaos regime with the wave packet 
spreading as $m_2(t) \propto t^{1/2}$ and 
$P(t) \propto t^{1/4}$ compared to the $m_2(t) \propto t^{1/3}$ and 
$P(t) \propto t^{1/6}$ behaviour observed for the weak chaos. 
This initial phase is followed by a 
subsequent slowing down of energy propagation, which asymptotically 
tends to the weak chaos behaviour.

In our study we consider the following four cases that belong to 
the strong chaos regime:
\begin{description}
	\item[Case $S1_1$:] $W=1$, $L=330$, $h = 0.1$;
	\item[Case $S2_1$:] $W=2$, $L=83$, $h = 0.1$;
	\item[Case $S3_1$:] $W=3$, $L=37$, $h = 0.1$;
	\item[Case $S4_1$:] $W=3$, $L=83$, $h = 0.1$,
\end{description}
The cases $S1_1$, $S3_1$ and $S4_1$ were also considered in 
\cite{Gkolias2013} 
where they 
were named respectively as cases III, II and I while case $S2_1$ 
is a new one.
For all these cases, we 
perform investigations based on numerical integrations up to a 
final time 
$t_f \approx 10^{8.2}$ (i.e. $t_f\approx150{,}000{,}000$) time units. 
The lattice size $N(t)$ is gradually increased to prevent the 
boundary sites from interfering with the dynamics of the system. In particular, in our strong chaos simulations we use lattices 
whose size increases up to a maximum 
$N(t_f) \approx 2\,500 - 23\,000$. 
The time steps $\tau \approx 0.25-0.35$ used conserved the energy $H_{1K}$ of 
the system at absolute relative energy 
error $e_r(t)$ \eqref{eq4:ree1} at the level of $10^{-5}$.

\begin{figure}[H]
	\centering
	\includegraphics[width=0.65\textwidth,keepaspectratio]{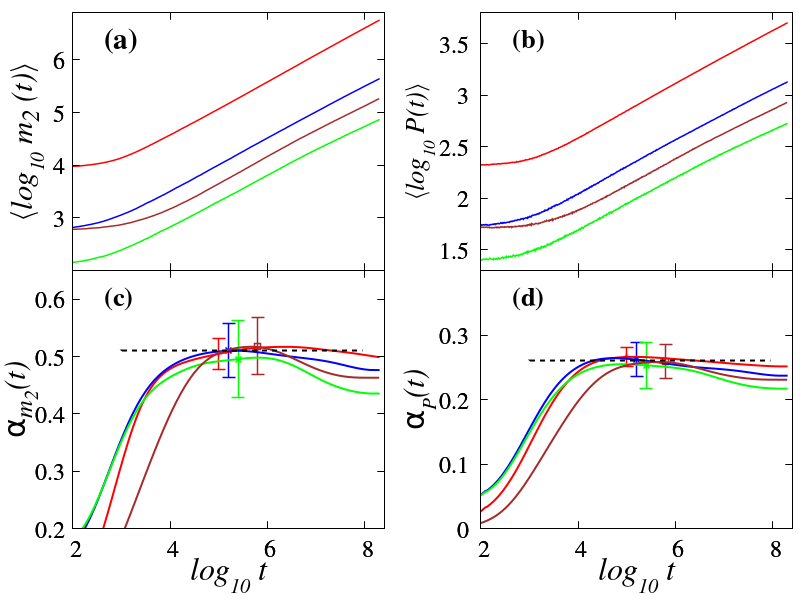}
	\caption{Similar to \autoref{fig:weak_L} but for the strong chaos 
		spreading regime of Hamiltonian \eqref{eq:kg1dham}. Results for the time 
		evolution of {\bf (a)} $m_2(t)$ \eqref{eq:m2_1dkg}, 
		{\bf (b)} $P(t)$ \eqref{eq:P_1dkg}, 
		the corresponding slopes {\bf (c)} $\alpha_{m_2}(t)$ for $m_2$ 
		and 
		{\bf (d)} $\alpha_{P}(t)$ for $P(t)$ for the cases $S1_1$ 
		(red curves), $S2_1$ (blue curves), $S3_1$ (green curves) and 
		$S4_1$ (brown curves). 
		The dashed lines in panels {\bf (c)} and {\bf (d)} respectively 
		indicate values $\alpha_{m_2}=0.5$ and $\alpha_{P}=0.26$.
		The error bars in {\bf (c)} and {\bf (d)} denote the 
		numerical error of 
		one standard deviation in the computed slopes [{\bf (c)} and {\bf (d)}].
		Plots {\bf(a)} and {\bf (b)} are in log-log scale while {\bf(c)} and  
		{\bf (d)} 
		are in log-linear scale.}
	\label{fig:strong_m2P}
\end{figure}

In \autoref{fig:strong_m2P} we present the time evolution of the 
measures of wave packet extent, namely the second moment 
$m_2(t)$ \eqref{eq:m2_1dkg} and participation number $P(t)$ 
\eqref{eq:P_1dkg}. We see clear power law growths 
$m_2(t) \propto t^{\alpha_{m_2}}$ 
and $P(t) \propto t^{\alpha_{P}}$ 
of these two 
quantities. 
The computed values of $\alpha_{m_2}$ and $\alpha_{P}$ are 
respectively shown by straight dashed lines 
shown in \autoref{fig:strong_m2P}{\bf(c)} and {\bf(d)} are 
$\alpha_{m_2}=0.5$ and $\alpha_{P}=0.26$ 
and persist for about $2$ decades of integration time. 
This epoch is generally 
followed by a mild slowing down of the spreading process for 
$\log_{10}t \gtrsim 6$ in accordance with results shown in 
\cite{Laptyeva2010} and \cite{Bodyfelt2011a}.
The error bars in {\bf(c)} and 
{\bf(d)} denote a standard deviation of 
$m_2$ and $P$ slopes.
The findings show that all these cases 
belong to the strong chaos regime since the theoretical estimates 
$\alpha_{m_2}=0.5$, is equal to the numerically computed result, and 
$\alpha_{P}=0.25$ is within a standard deviation of the numerically computed 
slope ($\alpha_{P}=0.26$ )
[panels{\bf (c)} and {\bf (d)}] where the computed values 
$\alpha_{m_2}=0.5$ and $\alpha_{P}=0.26$ for about $2$ decades of integration time. 
We note that for the weak chaos cases 
the wave packet effectively covers approximately 
$800$ sites by $t_f=10^9$ compared to approximately $1\,800$ sites by 
$t_f=10^{8.2}$ in the strong chaos case as estimated using $P$.

\begin{figure}[H]
	\centering
	\includegraphics[scale=0.34]{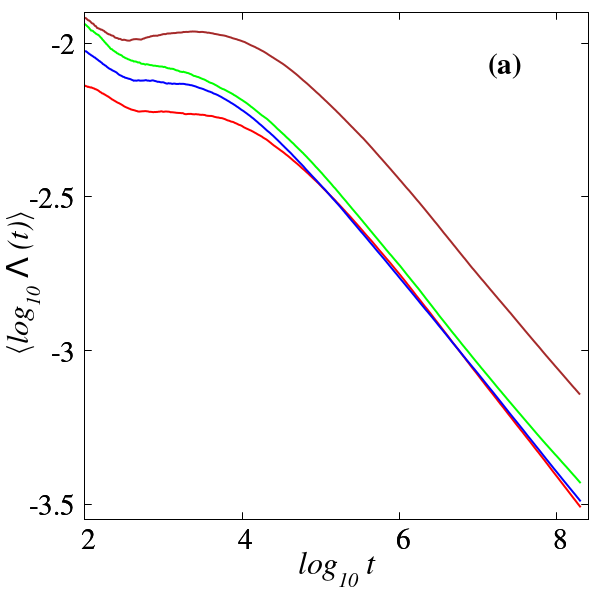}
	\includegraphics[scale=0.34]{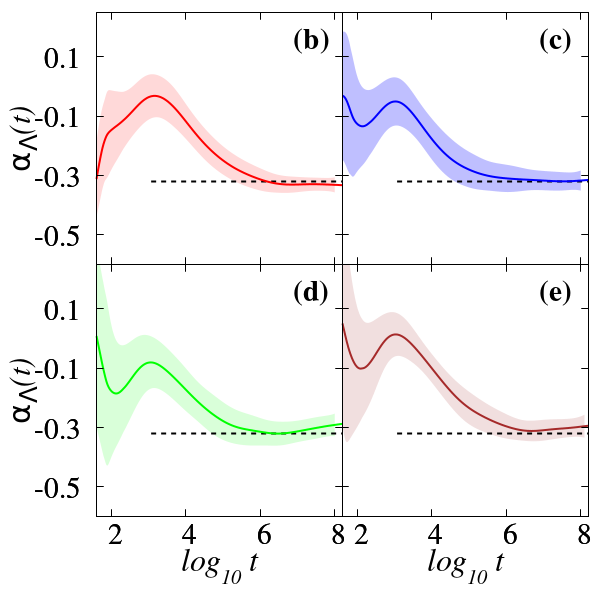}
	\caption{Similar to \autoref{fig:weak_L} but for  the strong chaos 
		spreading regime of Hamiltonian \eqref{eq:kg1dham}. Results for 
		the time evolution of {\bf (a)} the finite time mLCE 
		$\Lambda(t)$ \eqref{eq:ftmLCE}
		and {\bf (b)}-{\bf (e)} the corresponding derivatives for the strong 
		chaos cases. The curve colours correspond to the cases 
		presented in \autoref{fig:strong_m2P}. The straight 
		dashed lines indicate slopes $\alpha_\Lambda =-0.32$. Panel {\bf (a)}
		is in log-log scale while panels {\bf (b)}-{\bf (e)} are in log-linear scale.}
	\label{fig:strong_L}
\end{figure}

We now elaborate on the chaoticity in the strong chaos regime: 
\autoref{fig:strong_L} shows results for the 
evolution of the finite time mLCE $\Lambda(t)$ \eqref{eq:ftmLCE}. 
We observe a power law decay 
$\Lambda(t)\propto t^{\alpha_{\Lambda}}$, with no tendency 
of slowing down to the regular dynamics 
law $\Lambda(t) \propto t^{-1}$, similar to the chaotic 
tendency seen in the weak chaos cases.
The difference is 
that for strong chaos $\alpha_{\Lambda} \approx -0.32$ 
while for weak chaos 
$\alpha_{\Lambda} \approx -0.25$. 
The difference in the values of $\alpha_{\Lambda}$ for the strong 
and weak chaos 
regimes is 
an additional indication of the differences  in the dynamical behaviour for 
these two regimes.
As the strong chaos regime is a transient one, the evolution of $m_2(t)$ and 
$P(t)$ show signs of the crossover to the weak chaos dynamics, as their 
increase becomes slower for 
$\log_{10} t \gtrsim 6$ (\autoref{fig:strong_m2P}). 
This happens because the values of $m_2(t)$ and $P(t)$ are determined by the 
current dynamical state of the wave packet. On the other hand, such changes 
are not well pronounced 
in the evolution of $\Lambda(t)$ (\autoref{fig:strong_L}). 
As the dynamics crosses over from the strong chaos behaviour characterized 
by $\alpha_{\Lambda}=-0.32$ to the asymptotic weak chaos behaviour 
associated with $\alpha_{\Lambda}=-0.25$, one would expect to see some 
change in the values of $\alpha_{\Lambda}$ (panels {\bf (b)}-{\bf(d)} 
of \autoref{fig:strong_L}) indicating this transition. 
Such changes are 
not observed because the value of $\Lambda$ \eqref{eq:ftmLCE} is influenced 
by the whole evolution of the deviation vector 
[i.e.~the ratio $\lvert \lvert\boldsymbol{w}(t)
\rvert \rvert / \lvert \lvert\boldsymbol{w}(0) \rvert \rvert $ 
in \eqref{eq:ftmLCE}] and consequently, the whole history of the dynamics 
(which is predominately influenced by the strong chaos behaviour) 
and not by the current state of the systems. Thus, $\Lambda$ is not 
sensitive to subtle dynamical changes.

Using 
our numerical findings as shown in \autoref{fig:strong_m2P} and 
\autoref{fig:strong_L}, the spreading to chaoticity time scale 
ratios $R_{m_2}$ and $R_P$ of 
\eqref{eq:T_ratios}
give
\begin{equation*}
	R_{m_2}  \sim t^{0.18}, \,\,\,\, R_P  \sim t^{0.42},
\end{equation*}
since  ${\alpha_{m_2}}=0.5$, 
${\alpha_P}=0.26$ and $\alpha_{\Lambda}=-0.32$.
The chaoticity time scale $T_{\Lambda}$ therefore
always remains smaller than the spreading time scales $T_{m_2}$ and $T_P$. 
This means that, just like we observed for the weak chaos case, 
a wave packet gets more chaotic at a rate faster than 
its spreading.

We now investigate the DVD evolution for subtle changes that may not 
have been captured by $m_2$, $P$ or $\Lambda$.

\subsubsection{Deviation vector distributions}
\begin{figure}[H]
	\centering
	\includegraphics[scale=0.35,keepaspectratio]{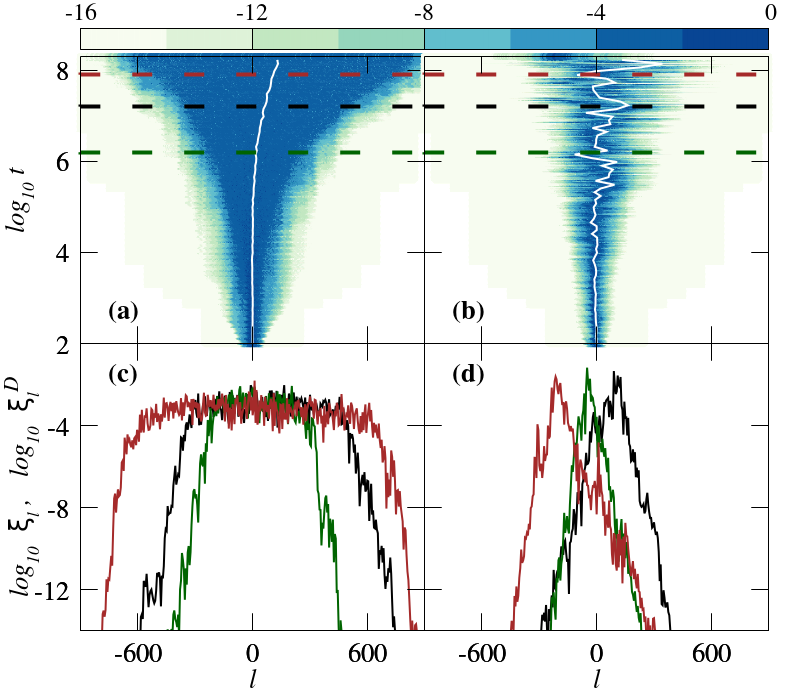}
	\caption{Similar to \autoref{fig:palette_weak_kg} but for  the strong 
		chaos 
		spreading regime case $S3_1$ of Hamiltonian \eqref{eq:kg1dham}. Results for the time evolution of {\bf(a)} the normalized 
		energy distribution $\xi_l(t)$ \eqref{eq4:en_pa_site_1dkg} and 
		{\bf(b)} the corresponding DVD $\xi_l^D(t)$ \eqref{eq:dvd}. 
		The colour scales at the top of the figure are used for 
		colouring lattice sites according to their {\bf(a)} 
		$\log_{10} \xi_l(t)$  
		and {\bf(b)} $\log_{10} \xi_l^D(t)$ values. In each of the 
		panels {\bf (a)} and {\bf(b)} the white 
		curve traces the distributions' centre. {\bf(c)} Normalized energy 
		distributions $\xi_l(t)$ and {\bf (d)} DVDs at 
		times $\log_{10}t=6.2$ [green curve], $\log_{10}t=7.2$ 
		[black curve], $\log_{10}t=7.9$ [red curve]. These times are also denoted 
		by similar coloured horizontal dashed lines in {\bf (a)} and {\bf(b)}. 
	}
	\label{fig:palette_strong_kg}
\end{figure}

We use the deviation vectors to study the behaviour of chaotic seeds  
in this regime of strong chaos. 
In \autoref{fig:palette_strong_kg}\textbf{(a)} and \textbf{(b)} we 
respectively plot the time evolution of the energy density 
\eqref{eq4:en_pa_site_1dkg} and the 
corresponding DVD \eqref{eq:dvd} for an individual set up of case $S3_1$, while snapshots of 
these distributions at some specific times are shown 
in \autoref{fig:palette_strong_kg}\textbf{(c)} and \textbf{(d)}. The centre of the lattice 
has been translated to $0$.

As in the weak chaos cases presented in \autoref{fig:palette_weak_kg},
the energy density spreads smoothly and rather symmetrically around 
the lattice's centre [\autoref{fig:palette_strong_kg}\textbf{(a)}, 
\textbf{(c)}], reaching more distant sites 
from the centre compared to the weak chaos 
cases [\autoref{fig:palette_weak_kg}\textbf{(a)}, \textbf{(c)}]. 

This is because the strong chaos regime is characterized by 
a faster subdiffusive spreading compared to the weak chaos 
regime, which is reflected in the larger exponents in the power law increases 
of $m_2$ and $P$ as presented in \autoref{fig4:weak_m2P} and 
\autoref{fig:strong_m2P}. 
On the other hand, the DVDs again remain localized, exhibiting fluctuations 
in their position, which appear earlier in time and have larger 
amplitudes [\autoref{fig:palette_strong_kg}\textbf{(b)}, \textbf{(d)}] 
with respect to the weak chaos case [\autoref{fig:palette_weak_kg}\textbf{(b)}, 
\textbf{(d)}].

\begin{figure}[H]
	\centering
	\includegraphics[scale=0.35,keepaspectratio]{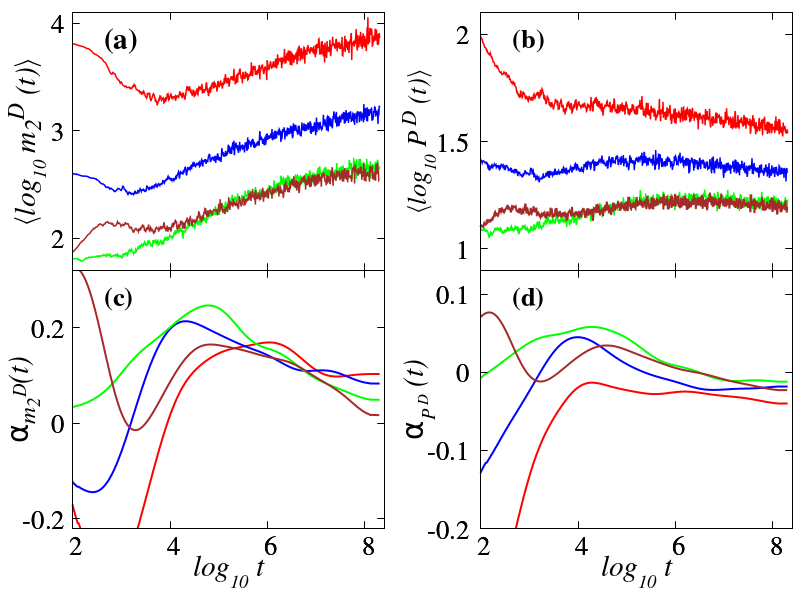}
	\caption{Similar to \autoref{fig:m2_P_dvd_1D} but for  the strong 
		chaos 
		spreading regime cases $S1_1$ - $S4_1$ of Hamiltonian \eqref{eq:kg1dham}.		
		Results averaged over $200$ disorder realizations for the time evolution of  
		{\bf (a)} $m_2^D(t)$ \eqref{eq:m2_1dkg}, {\bf (b)} $P^D(t)$ \eqref{eq:P_1dkg}, 
		the corresponding slopes {\bf (c)} $\alpha_{m_2^D}(t)$ for $m_2^D$ 
		and 
		{\bf (d)} $\alpha_{P^D}(t)$ for $P^D$ for the cases $S1_1$ - $S4_1$.  The curve colours correspond to the cases presented in \autoref{fig:strong_m2P}. 
		Panels {\bf(a)} and {\bf (b)} are in log-log scale while {\bf(c)}, 
		and {\bf (d)} 
		are in log-linear scale.}
	\label{fig:m2_P_dvd_1D_str}
\end{figure}

The DVD second moment $m_2^D$ \eqref{eq:m2_dvd} 
[shown in \autoref{fig:m2_P_dvd_1D_str}\textbf{(a)}] for the strong chaos cases 
$S1_1$ - $S4_1$ 
increases in 
time attaining larger values compared to the weak chaos 
regime [presented in \autoref{fig:m2_P_dvd_1D}\textbf{(a)}], 
although this increase does 
not show signs of a constant power law growth 
rate as observed in the weak chaos 
case where eventually $m_2^D(t) \propto t^{0.1}$. In addition, a slowing down of 
the growth rate is observed at higher times. The DVDs 
remain localized as depicted in \autoref{fig:m2_P_dvd_1D_str}{\bf(b)}, 
with a clear tendency of their participation number $P^D$ \eqref{eq:P_Dvd} to slowly decrease.

The final values of $P^D$ in the cases of strong chaos are 
such that $13\lesssim P^D\lesssim100$, 
thus a bit higher (with signs of a tendency to saturate after 
$t_f$) compared to those observed in the weak chaos case where $0\lesssim P^D\lesssim13$.

\begin{figure}[H]
	\centering
	\includegraphics[scale=0.45,keepaspectratio]{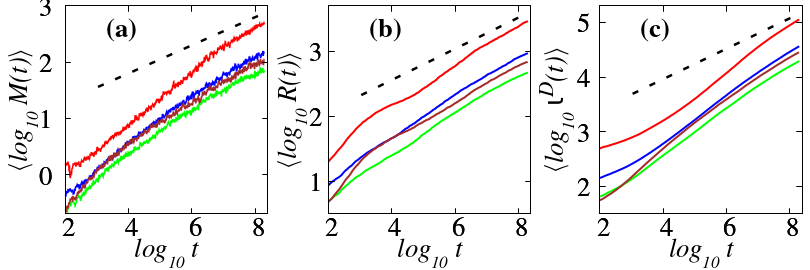}
	\caption{Similar to \autoref{fig:mad_diam_weak_1kg} but for  the strong 
		chaos 
		spreading regime cases $S1_1$ - $S4_1$ of Hamiltonian \eqref{eq:kg1dham}.
		Results for the time evolution of {\bf(a)} $M(t)$ \eqref{eq:mad}, {\bf(b)} $R(t)$ 
		\eqref{eq:R} 
		and {\bf(c)} $\iota^D(t)$ \eqref{eq:l}.
		The curve colours correspond to the 
		cases presented in \autoref{fig:strong_m2P}.
		The straight dashed lines guide the eye for slopes 
		{\bf (a)} $0.24$, {\bf (c)} $0.25$ and {\bf (c)} $0.28$ which 
		correspond to the growth of the quantities as 
		observed and reported in the weak chaos regime 
		[Section \ref{sec4:weak}].
		Panels {\bf(a)}, {\bf (b)} and {\bf(c)} are in log-log 
		scale.}
	\label{fig:R_M_l_cpct_1D_str}
\end{figure}

Since the wave packet spreads faster for the strong chaos regime 
than the 
weak chaos case, while the DVD remains localized, one expects 
faster and wider movements of the chaotic seeds in order to achieve 
chaotization of the 
wave packet. Evolution of the DVD mean position $\bar{l}_w$ 
[white curve in \autoref{fig:palette_strong_kg}\textbf{(b)}], 
$M(t)$ 
(\ref{eq:R}) [\autoref{fig:R_M_l_cpct_1D_str}\textbf{(a)}], 
$R(t)$ (\ref{eq:mad}) [\autoref{fig:R_M_l_cpct_1D_str}\textbf{(b)}] and 
$\iota^D(t)$ (\ref{eq:l}) [\autoref{fig:R_M_l_cpct_1D_str}\textbf{(c)}] 
confirm the chaotic seed dynamics. Each of these quantities grows
reaching values larger by about one 
order of magnitude in comparison with 
the weak chaos regime [\autoref{fig:mad_diam_weak_1kg} \textbf{(a)}, 
\textbf{(b)}, \textbf{(c)}]. The straight dashed lines in panels 
{\bf (a)}, {\bf (b)} and {\bf(c)} of \autoref{fig:R_M_l_cpct_1D_str} respectively show the 
slopes $0.24$, 
$0.25$ and $0.28$ which were observed for the weak chaos cases 
[\autoref{fig:mad_diam_weak_1kg} \textbf{(a)}, 
\textbf{(b)}, \textbf{(c)}]. In each of 
these panels [{\bf(a)}, {\bf(b)} and {\bf(c)}],the quantities 
$M(t)$, $R(t)$ and $\iota^D(t)$ grow faster for strong chaos 
cases compared to weak chaos behaviour represented 
by the dashed lines. 
This analysis also shows that the deviation 
vector meanders at higher speeds in the strong chaos regime 
compared to the weak chaos 
case.

\section{The selftrapping chaos regime}\label{sec4:self}
We consider here the 
dynamics in the selftrapping regime of the 1D DKG model \eqref{eq:kg1dham} 
for which the largest 
portion of the wave packet stays practically localized at the region 
of the initial excitation while some small tails propagate towards the lattice 
edges [\cite{Kopidakis2008,Skokos2009}].
For single site excitations leading to selftrapping behaviour,  it has been 
shown that the wave packet second moment 
$m_2(t)$ grows as 
$m_2(t)\propto t^{1/3}$ while the participation number $P(t)$ remains 
practically constant 
[\cite{Skokos2009}]. 

In our study we initially excite a single site with energy 
$H_{1K}=h$ and integrate the system up to a final time 
$t_f \approx 10^{8.9}$ time units.
We consider the following two parameter set ups which lead to 
selftrapping behaviour:
\begin{description}
	\item[Case $ST1_1$:] $W=4$, $L=1$, $h = 1.5$;
	\item[Case $ST2_1$:] $W=5$, $L=1$, $h = 1$.
\end{description} 
The 
lattice size $N(t)$ is increased up to a maximum value 
$N(t_f) \approx 1\,000$ for $ST1_1$ and $N(t_f) \approx 800$ 
for $ST2_1$ in order to avoid energy spreading to the boundary sites. 
The energy $H_{1K}$ of the system is conserved at an absolute relative energy 
error \eqref{eq4:ree1} $e_r(t)\approx10^{-5}$ using time 
steps $\tau \approx 0.2$ ($ST1_1$) and $\tau \approx 0.25$ ($ST2_1$) for the 
$ABA864$ SI.
Case $ST1_1$ was studied in [\cite{Gkolias2013}] while case $ST2_1$ is new and will 
contribute to the generalization of results for the selftrapping regime in the 
1D DKG model. 
By computing the participation number $P(t)$ \eqref{eq:P_1dkg}, we verify 
that the considered cases belong to 
the selftrapping regime. 

\begin{figure}[H]
	\centering
	\includegraphics[width=0.65\textwidth,keepaspectratio]{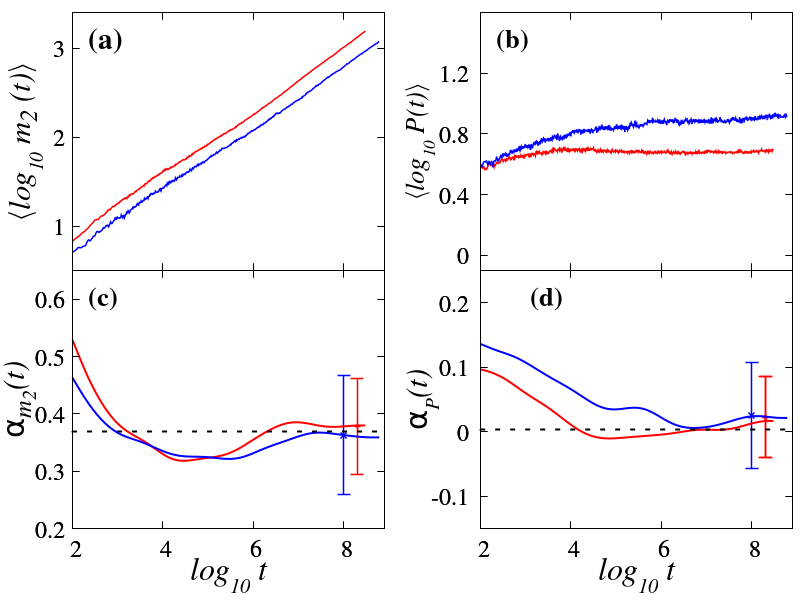}
	\caption{Similar to \autoref{fig4:weak_m2P}. Results for the time evolution of {\bf (a)} $m_2(t)$ \eqref{eq:m2_1dkg}, 
		{\bf (b)}$P(t)$ \eqref{eq:P_1dkg}, 
		and the corresponding slopes {\bf (c)} $\alpha_{m_2}(t)$ for $m_2$ 
		and 
		{\bf (d)} $\alpha_{P}(t)$ for $P$ for the cases $ST1_1$ 
		(red curves) and $ST2_1$ (blue curves). 
		The dashed lines in panels {\bf (c)} and {\bf (d)} respectively guide 
		the eye for values $\alpha_{m_2}(t)=0.37$ and $\alpha_{P}(t)=0$.
		The error bars in {\bf (c)} and {\bf (d)} denote 
		one standard deviation in the computed slopes.
		Plots {\bf(a)} and {\bf (b)} are in log-log scale while {\bf(c)}, 
		and {\bf (d)} 
		are in log-linear scale.}
	\label{fig:slf_m2P}
\end{figure}

In \autoref{fig:slf_m2P} we present the averaged results for the 
time evolution of the second moment 
$m_2(t)$ \eqref{eq:m2_1dkg} and the participation number $P(t)$ 
\eqref{eq:P_1dkg}. 
Panels {\bf (a)} and {\bf (b)} respectively show the time evolution of 
$m_2(t)$ and $P(t)$ for the cases $ST1_1$ (red curves) and $ST2_1$ (blue curves). 
In panels {\bf (c)} and {\bf (d)} we present the 
slopes $\alpha_{m_2}(t)$ of $m_2(t)$ and $\alpha_{P}(t)$ of $P(t)$ 
respectively with the dashed lines indicating the 
values 
$\alpha_{m_2}=0.37$ [panel {\bf (c)}] and $\alpha_{P}=0$ 
[panel {\bf (d)}]. 
The error bars in {\bf(c)} and {\bf(d)} respectively 
denote 
one standard deviation in the computation of slopes 
$\alpha_{m_2}(t)$ and $\alpha_P(t)$. These results verify that the cases considered
belong to the selftrapping chaos regime since 
$P(t)$ [panel {\bf (b)}] gives a practically constant value,  
in 
accordance with established results of \cite{Kopidakis2008} and 
\cite{Skokos2009} which emphasize that the wave packet remains 
localized (i.e.~$P(t) \approx constant$).
The theoretical predictions and 
previously published work [\cite{Skokos2009}] for 
the power law exponents of respectively $P(t)$ and $m_2(t)$ 
lie within the interval centred around our 
numerically computed values and of width equal to one standard 
deviation.
 
\begin{figure}[H]
	\centering
	\includegraphics[scale=0.32,keepaspectratio]{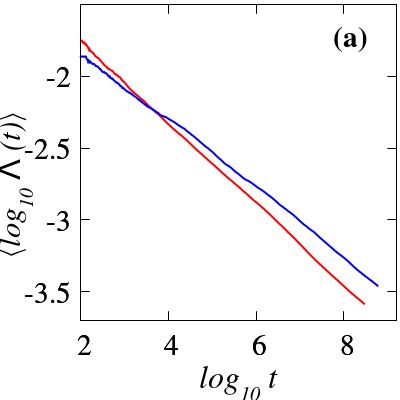}
	\includegraphics[scale=0.32,keepaspectratio]{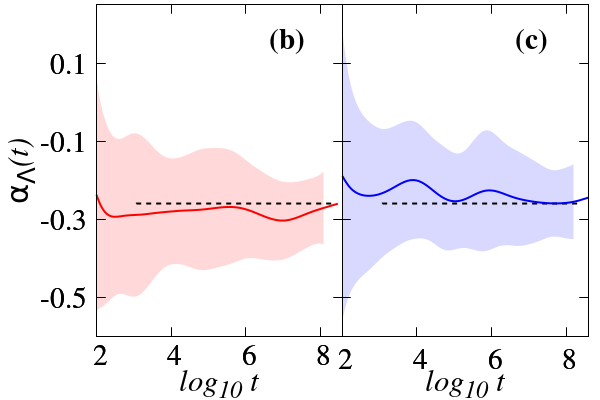}
	\caption{Similar to \autoref{fig:weak_L}. 
		Results for the time evolution of {\bf (a)} the finite time mLCE 
		$\Lambda(t)$ \eqref{eq:ftmLCE} 
		and {\bf (b)}-{\bf (c)} the corresponding numerically computed 
		slopes for the two selftrapping chaos cases of the 
		1D DKG system \eqref{eq:kg1dham} considered in 
		\autoref{fig:slf_m2P}. Results are 
		averaged over $200$ disorder realizations. The curve colours correspond to the cases presented in \autoref{fig:slf_m2P}. The straight 
		dashed lines in {\bf (b)} - {\bf (c)} indicate 
		the value $\alpha_\Lambda =-0.26$. Panel {\bf (a)}
		is in log-log scale while panels {\bf (b)}-{\bf (c)} are in log-linear scale.}
	\label{fig:slf_L}
\end{figure}

 We now investigate the chaoticity of the model. 
 For each of the two selftrapping chaos cases we investigated, we compute the 
 time evolution of 
 $\Lambda(t)$ \eqref{eq:ftmLCE} and present the results in 
 \autoref{fig:slf_L}. 

In panel \textbf{(a)} of \autoref{fig:slf_L}, 
the time evolution of the finite time mLCE $\Lambda(t)$ \eqref{eq:ftmLCE} 
averaged over 200 disorder realisations for the selftrapping 
cases $ST1_1$ and $ST2_1$ is reported. 
The corresponding evolution of derivatives $\alpha_{\Lambda}$ 
is shown in panels 
\textbf{(b)} and \textbf{(c)}.
These results show that the chaoticity of system \eqref{eq:kg1dham}
is characterised by  
the power law $\Lambda(t)\propto t^{-0.26}$, a law similar to what we 
observed for the weak chaos regime,
and does not show any tendencies to crossover to regular dynamics for the entire 
duration of our integration.
From the numerical computations of $m_2$ and $\Lambda$, the  
ratio $R_{m_2}$  
\eqref{eq:T_ratios} gives the value
\begin{equation*}
	R_{m_2}  \sim t^{0.37},
\end{equation*}
since ${\alpha_{m_2}}=0.37$ and $\alpha_{\Lambda}=-0.26$.
The wave packet therefore gets more chaotic much faster 
compared to it's spreading since the chaoticity time scale $T_{\Lambda}$ 
is always smaller than the spreading time scale $T_{m_2}$.

\subsubsection{Deviation vector distributions}
We now investigate the deviation vector dynamics in this regime of chaos. 
Like for the weak and strong chaos cases presented before, we first present 
a representative realisation, after which we give averaged results for 
200 disorder realisations.

\begin{figure}[H]
	\centering
	\includegraphics[scale=0.35,keepaspectratio]{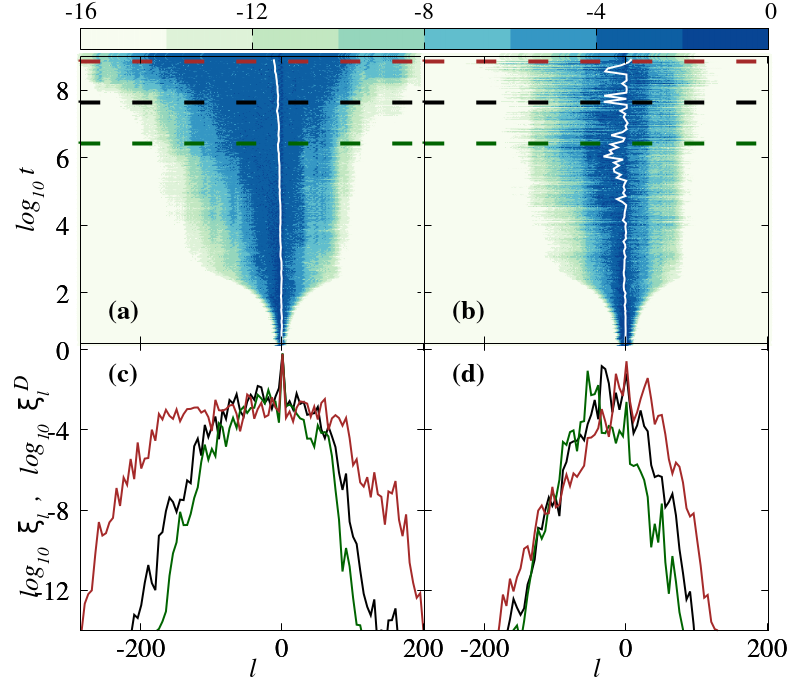}
	\caption{Similar to \autoref{fig:palette_weak_kg}. Results for the time evolution of {\bf(a)} the normalized 
		energy distribution $\xi_l(t)$ \eqref{eq4:en_pa_site_1dkg} and 
		{\bf(b)} the corresponding DVD $\xi_l^D(t)$ \eqref{eq:dvd}. 
		The colour scales at the top of the figure are used for 
		colouring lattice sites according to their {\bf(a)} 
		$\log_{10} \xi_l(t)$  
		and {\bf(b)} $\log_{10} \xi_l^D(t)$ values. In each of the 
		panels {\bf (a)} and {\bf(b)} the white 
		curve traces the distributions' centre. {\bf(c)} Normalized energy 
		distributions and {\bf (d)} DVDs at 
		times $\log_{10}t=6.4$ [green curve], $\log_{10}t=7.6$ 
		[black curve], $\log_{10}t=8.8$ [red curve]. These times are also denoted 
		by similarly coloured horizontal dashed lines in {\bf (a)} and {\bf(b)}.}
	\label{fig:palette_slf}
\end{figure}

In \autoref{fig:palette_slf} we present the evolution 
of the {\bf (a)} energy density $\xi_l(t)$ \eqref{eq4:en_pa_site_1dkg} 
and {\bf (b)} DVD $\xi^D(t)$ \eqref{eq:dvd} 
for a representative realisation belonging to case $ST1_1$. 

In panels 
\textbf{(a)} and \textbf{(b)} the position of the distributions' mean value 
is traced by a thick white curve while in panels \textbf{(c)} and 
\textbf{(d)} we show 
three snapshots of the distributions correspondingly 
taken at times indicated by the 
horizontal lines of panels \textbf{(a)} and \textbf{(b)} with matching colours. 
The centre of the lattice 
has been translated to $0$.
We note that 
the energy distribution has high values 
in the lattice centre, which corresponds to 
the trapped part of the wave packet. This is visible in panel {\bf(c)} 
where the snapshots have very sharp pointy peaks at the centre of the energy 
profiles. The snapshots of \autoref{fig:palette_slf}{\bf(d)} and the 
mean position [traced by white curve in \textbf{(b)}]
show that the DVD distribution centre fluctuates with increasing 
amplitudes as the energy spreads to more sites of the lattice.
The DVD behaviour shown by the single realisation in \autoref{fig:palette_slf} 
is characteristic of the behaviour for the considered cases as revealed 
by the computations of $m_2^D(t)$ \eqref{eq:m2_1dkg} and 
$P^D(t)$ \eqref{eq:P_1dkg} performed for more disorder realisations.

\begin{figure}[H]
	\centering
	\includegraphics[scale=0.35,keepaspectratio]{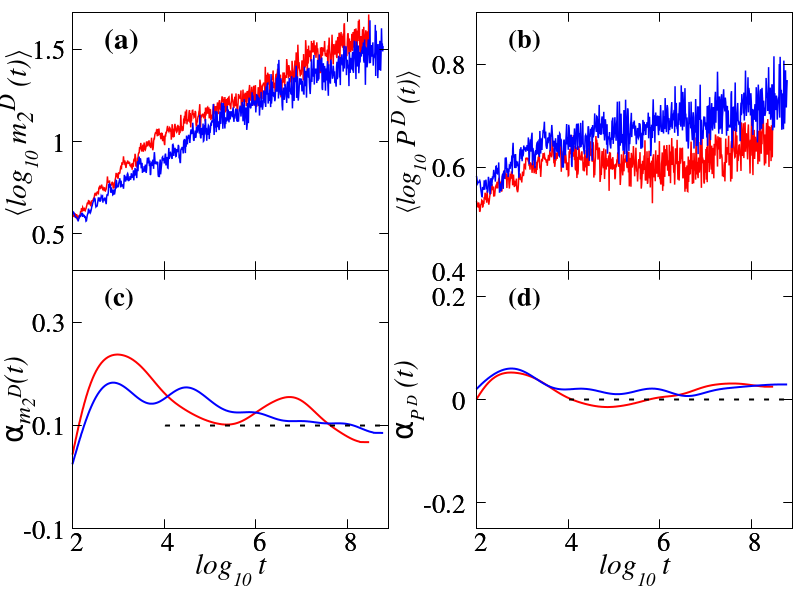}
	\caption{Similar to \autoref{fig:m2_P_dvd_1D}.
		Results averaged over realizations for the time evolution of  
	{\bf (a)} $m_2^D(t)$ \eqref{eq:m2_1dkg}, {\bf (b)} $P^D(t)$ \eqref{eq:P_1dkg}, 
	the corresponding slopes {\bf (c)} $\alpha_{m_2^D}(t)$ for $m_2^D$ 
	and 
	{\bf (d)} $\alpha_{P^D}(t)$ for $P^D$ for the cases $ST1_1$ and $ST2_1$. 
	The dashed lines in panels {\bf (c)} and {\bf (d)} respectively 
	indicate values $\alpha_{m_2^D}(t)=0.1$ and $\alpha_{P^D}(t)=0$. 
	The curve colours correspond to the cases presented in \autoref{fig:slf_m2P}. 
	Panels {\bf(a)} and {\bf (b)} are in log-log scale while {\bf(c)}, 
	and {\bf (d)} 
	are in log-linear scale.}
	\label{fig:m2_P_dvd_1D_slf}
\end{figure}

\autoref{fig:m2_P_dvd_1D_slf} shows results for 
\textbf{(a)} $m_2^D(t)$ \eqref{eq:m2_1dkg}, \textbf{(b)} 
$P^D(t)$ \eqref{eq:P_1dkg} and the corresponding slopes 
\textbf{(c)} $\alpha_{m_2^D}$ of $m_2(t)$ and \textbf{(d)} 
$\alpha_{P^D}$ of $P(t)$.
The straight dashed lines are slope values $\alpha_{m_2^D}(t)=0.1$ 
and $\alpha_{P^D}(t)=0$. We see that, just like for 
the weak chaos case, the second moment has a low 
growth rate, close to $0.1$,  
while the participation number of the DVD is also 
practically constant.

\begin{figure}[H]
	\centering
	\includegraphics[scale=0.45,keepaspectratio]{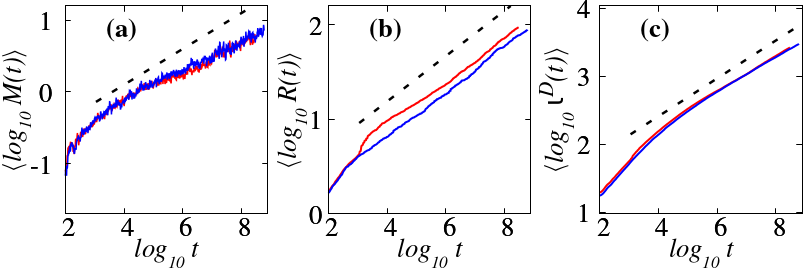}
	\caption{Similar to \autoref{fig:mad_diam_weak_1kg} but for the selftrapping 
		chaos 
		spreading regime cases $ST1_1$ and $ST2_1$ of Hamiltonian \eqref{eq:kg1dham}.
		Results for the time evolution of {\bf(a)} $M(t)$ \eqref{eq:mad}, {\bf(b)} $R(t)$ 
		\eqref{eq:R} 
		and {\bf(c)} $\iota^D(t)$ \eqref{eq:l}.
		The curve colours correspond to the 
		cases presented in \autoref{fig:slf_m2P}.
		The straight dashed lines guide the eye for slopes 
		{\bf (a)} $0.24$, {\bf (c)} $0.25$ and {\bf (c)} $0.28$ which 
		correspond to the growth observed in the weak chaos regime.
		Panels {\bf(a)}, {\bf (b)} and {\bf(c)} are in log-log 
		scale.}
	\label{fig:cpctnss_slf}
\end{figure}

In \autoref{fig:cpctnss_slf} we present results obtained for 
the quantities $M(t)$ 
\eqref{eq:mad} [panel {\bf (a)}], $R(t)$ \eqref{eq:R} 
[panel {\bf(b)}] and $\iota^D(t)$ 
\eqref{eq:l} [panel{\bf(c)}] with the dashed lines guiding the eye 
for the 
slopes $0.24$ in {\bf(a)}, $0.25$ in {\bf(b)} and $0.28$ in {\bf(c)} 
which were obtained for the weak chaos case [\autoref{fig:mad_diam_weak_1kg}].
The values of $M(t)$  
[\autoref{fig:cpctnss_slf}{\bf(a)}], $R(t)$  [\autoref{fig:cpctnss_slf}{\bf(b)}] 
and $\iota^D(t)$ [\autoref{fig:cpctnss_slf}{\bf(c)}] continuously 
increase the entire time of the integration. This is because the 
meandering localized chaotic seeds have to cover wider regions 
of the lattice 
as time increases. 
The quantities 
$M(t)$, $R(t)$ and $\iota^D(t)$ have growth rates which are 
approximately equal to 
those of the weak chaos case (indicated by the straight lines).

\section{Summary}\label{sec4:summary}
We investigated the chaotic behaviour of the 1D DKG model 
\eqref{eq:kg1dham} in the weak, strong and selftrapping chaos regimes. 
We performed 
extensive simulations for chaotic spreading of initially 
localized excitations, for several parameter cases 
and obtained statistical results on ensembles of approximately $200$ 
disorder realizations in each case.
For all three regimes we compute the finite time 
mLCE $\Lambda(t)$ \eqref{eq:ftmLCE}, a widely used chaos indicator, and 
show evidence that although the 
wave packet chaoticity strength decreases in time 
the dynamics shows no tendency to 
cross 
to regular behaviour. $\Lambda(t)$ decreases following a power law  
$\Lambda(t) \propto t^{\alpha_{\Lambda}}$, which is characterized 
by $\alpha_{\Lambda}$ values different from  $\alpha_{\Lambda}=-1$ 
observed for regular motion. In particular, we found that 
$\alpha_{\Lambda} \approx -0.25$ for the  weak chaos regime 
[in agreement with the findings of \cite{Skokos2013} and 
\cite{Gkolias2013}], 
$\alpha_{\Lambda} \approx -0.32$ for the strong chaos regime and 
$\alpha_{\Lambda} \approx -0.26$ for the  selftrapping chaos regime.
From the theoretical explanation of the relation between the 
chaotic regime and the dynamical characteristics we see that 
for all considered regimes of chaos, 
the chaoticity time scale $T_{\Lambda}(t)$ \eqref{eq:TL} 
always remains smaller than the spreading time scales $T_{m_2}(t)$ 
\eqref{eq:TM} and or $T_P(t)$ \eqref{eq:TP}. 
This implies that wave packet gets more chaotic at a rate faster than 
its spreading.
\newline\newline
The computation of the corresponding DVDs created by the deviation vector 
used to compute $\Lambda(t)$ and of quantities related to their dynamics 
($m_2^D$, $P^D$, $R$, $M$, $\iota^D$), allowed us to better capture 
the instantaneous and long term 
features of the underlying chaotic behaviour and to quantify the 
meandering tendencies of chaotic seeds inside the wave packet. In all studied 
cases the DVD retained a localized, pointy shape with its participation 
number $P^D(t)$ remaining asymptotically constant at small values. As 
time increased the DVD exhibited oscillations of larger amplitudes in 
order to visit all regions inside the spreading wave packet. 
As a results, 
the quantities $M$ \eqref{eq:mad}, $R$ (\ref{eq:R}) and 
$\iota^D$ \eqref{eq:l} which estimate the range of the 
lattice region visited by the DVD, increased in time. This increase 
was characterized by power law growths, 
$R(t)\propto t^{0.24}$, $M(t)\propto t^{0.25}$ and $\iota^D(t)\propto t^{0.28}$
in the weak chaos regime. On the 
other hand, in the strong chaos case $R$, $M$ and $\iota^D$ 
grow with higher, but 
non constant, rates since the wave packet spreads faster than in the weak 
chaos case and the DVD visits a wider region. It is worth noting that 
this rate decreases in time, tending to the values observed in 
weak chaos regime. This is a direct consequence of the transient nature 
of the strong chaos regime, as this regime eventually crosses over 
toward the weak chaos dynamics. For the selftrapping case, 
quantity $R(t)$ grows at the same rate 
as observed in the weak chaos regime while $M(t)$ and $\iota^D(t)$ have a 
slightly slower 
growth.
\newline\newline
In conclusion, we numerically 
verified for the weak and strong 
chaos spreading regimes of the 1D DKG model \eqref{eq:kg1dham} 
that (a) the deterministic chaoticity of wave 
packet dynamics persists in time, although its strength decreases;  (b) 
chaotic seeds meander inside the wave packet fast enough to ensure 
its chaotization; and (c) the characteristics of chaos evolution (like 
for example the power law $\Lambda(t) \propto t^{\alpha_{\Lambda}}$) in 
the weak and strong chaos regimes are distinct for each 
case (e.g.~$\alpha_{\Lambda} \approx -0.25$ for weak chaos and 
$\alpha_{\Lambda} \approx -0.32$ for strong chaos).

In Chapter \ref{chap:2D} we present the investigation of the chaotic behaviour 
for disordered lattices in two spatial dimensions as we try to understand 
the behaviour in 
higher spatial dimensions.
\chapter{Chaotic behaviour of the 2D DKG model}
\label{chap:2D}

\pagestyle{fancy}
\fancyhf{}
\fancyhead[OC]{\leftmark}
\fancyhead[EC]{\rightmark}
\cfoot{\thepage}

In this chapter we investigate the chaotic nature of dynamics for the 2D DKG 
model \eqref{eq:kg2dham}, extending the studies of Chapter \ref{chap_1d} 
to two spatial dimensions.
We present our findings as follows: 
In Section \ref{sec5:num_techs} we highlight the 
numerical techniques used for the computation of our results. We then 
present our findings 
for the different classifications of chaotic dynamical behaviours namely, the 
weak chaos spreading regime
in Section \ref{sec5:weak}, the strong chaos case in Section \ref{sec5:strong} and 
the selftrapping behaviour in Section \ref{sec5:self}. Lastly, we 
discuss and summarize our discussion on dynamics of the 2D model 
in Section \ref{sec5:summary}.
Parts of the results presented in Sections \ref{sec5:weak} and \ref{sec5:strong} 
of this chapter have also been reported 
in \cite{ManyManda2020}.

\section{Numerical techniques}\label{sec5:num_techs}
The numerical techniques used for the 2D DKG model \eqref{eq:kg2dham} 
are similar to the ones used for the 1D DKG model \eqref{eq:kg1dham} whose 
dynamics has been presented in 
Chapter \ref{chap_1d}.
In particular, we once again implement the two-part split order four 
SI ABA864  
[Chapter \ref{chap:num_tech}, \cite{Blanes2013}] to solve the variational 
equations 
by applying equations
\eqref{op:var} whose explicit form is given in Appendix (\ref{app:B}).
In our simulations, the absolute relative energy error  
\begin{equation}\label{eq5:ree1}
	e_r(t)=\frac{|H_{2K}(t)-H_{2K}(0)|}{H_{2K}(0)},
\end{equation}  
of the integration 
is kept at $e_r(t)\approx10^{-5}$ by using time steps $\tau\approx0.2-0.9$. 
Furthermore, in our study we use initial deviation vectors of the form {$\bf 2$} 
[Section \ref{sec3:2d}; \autoref{fig:ini_dvd_2D_palette}{\bf (b)}] 
where only the 
position and momentum coordinates of square 
sub-lattices of sizes from $3\times3$ to $25\times25$ at the centre of the 
lattice are non-zero.

	The numerical simulations were performed using the FORTRAN $90$ programming language 
with OPENMP parallelization allocation of an average of 
$4-8$ threads 
(CPU-cores) for 
each simulation.

We use lattices 
whose size, $N\times{M}$, increases with time [i.e.~$N(t)$ 
and $M(t)$ are smallest at $t=0$] 
depending on the wave packet extent while avoiding any 
boundary effects to the dynamics of the system. 
We start 
with a relatively 
small lattice of extent coverage $N(0)\times{M(0)}$ and increase it whenever 
either the wave packet 
or the deviation vector distribution 
gets close to the boundary sites. We ensure this by checking 
that for each site $(l,m)$ in an exterior strip of width $N_{W}$ 
at the lattice boundary, the site
energy
\begin{equation}\label{eq5:en_pa_site_1dkg}
	h_{l,m} =  \frac{p^2_{l,m}}{2} + \frac{\epsilon_{l,m}}{2} q^2_{l,m} +
	\frac{q_{l,m} ^{4}}{4} + \frac{1}{4W}\big[q_{l,m+1} - q_{l,m})^2 +
	((q_{l+1,m} - q_{l,m})^2\big]
\end{equation} 
and the corresponding square norm component $\delta q_{_{l,m}}^2+\delta p_{_{l,m}}^2$ 
of 
the deviation 
vector at this site are always less 
than $10^{-8}$. That is to say, for each site $(l,m)$ in an exterior strip of width $N_{W}$ we ensure  
\begin{equation}\label{eq:bdry_cndtn2d}
	\max_{_{l,m}} \{h_{l,m}\}<10^{-8} \qquad\textnormal{and}\qquad\max_{_l}\{\delta q_{l,m}^2+\delta p_{l,m}^2\}<10^{-8}.
\end{equation}
The lattice size 
is uniformly increased by a width dependant on the particular 
$W$
in case criteria \eqref{eq:bdry_cndtn2d} is not fulfilled. For smaller 
disorder strength
$W$ we increase the lattice by more sites compared to a larger $W$.
For simplicity, we use a square lattice where 
$N(t)=M(t)$ for our study.

We compute the slopes of data-curves using the LOESS procedure and 
the central finite differences described in Section \ref{sec4:slope}.

\section{Procedure for numerical simulations}\label{sec5:procedure}
In all our numerical simulations, we give the system a total energy $H_{2K}$ 
\eqref{eq:kg2dham} by initially exciting a block of 
$L\times{J}$ central sites of the $N(0)\times{M(0)}$ lattice. We 
give each of the sites in the central block 
energy $h=h_{l,m}=H_{2K}(0)/(L\cdot{J})$
by setting $p_{l,m} = \pm \sqrt{2h}$ with random signs and $p_{l,m} = q_{l,m}=0$ 
for all other sites. We use the fixed boundary conditions 
$q_{0,m}=q_{N+1,m}=q_{l,M+1}=q_{l,0}=0$ and $p_{0,m}=p_{N+1,m}=p_{l,M+1}=p_{l,0}=0$ for all $0<l<N+1$ and $0<m<M+1$.
Once again we follow the time evolution of different dynamical 
quantities for several ICs and various disorder realizations of the system
\eqref{eq:kg2dham} where we obtain 
statistical results  
by averaging the values over $50$ 
different disorder realizations.
\section{The weak chaos regime}\label{sec5:weak}
The participation number $P(t)$ 
\eqref{eq:P_2dkg} and second moment $m_2(t)$ 
\eqref{eq:m2_2dkg}, in the 2D DKG model, are known to 
characterize the energy 
spreading extent as 
$P(t)\propto t^{1/5}$ and $m_2(t)\propto t^{1/5}$ [\cite{Laptyeva2012}] 
respectively in the 
weak chaos regime. The system 
is known to behave chaotically and not cross to regular dynamical 
behaviour, 
characterised by $\Lambda(t) \propto t^{-1}$ 
[\cite{Benettin1976,Skokos2010c}]. 

Our numerical integrations are up to a maximum final time 
$t_f \approx 10^8$ time units with 
the lattice $N\times{M}$ increased up to size with dimension lengths in range 
$N(t_f)=M(t_f) \approx 300 - 500$. 
The energy $H_{2K}$ of the system was conserved at an absolute 
relative energy error \eqref{eq5:ree1} $e_r(t)\approx10^{-5}$ for time 
steps of size range $\tau \approx 0.4-0.9$.
For the following four weak chaos parameter cases, 
we investigate chaos in the 
system \eqref{eq:kg2dham}
 where a total of $L\cdot{J}$ central sites 
(which make up sub-lattice $L\times{J}$) of 
the lattice are 
excited with energy density $h$ and 
disorder strength $W$. For simplicity, we take $L=J$ and $N(t)=M(t)$.
\begin{description}
	\item[Case $W1_2$:] $W=10$, $L=1$, $h = 0.05$;
	\item[Case $W2_2$:] $W=10$, $L=1$, $h = 0.3$;
	\item[Case $W3_2$:] $W=10$, $L=3$, $h = 0.0085$;
	\item[Case $W4_2$:] $W=11$, $L=2$, $h = 0.0175$.
\end{description}
For each of these cases, we average 
the computations over 50 disorder realizations.
We first verify that the selected cases $W1_2$, $W2_2$, $W3_2$ and $W4_2$ 
belong to the weak chaos regime. Case $W2_2$ was studied in \cite{Laptyeva2012} 
as belonging to weak chaos with some of its dynamical quantities 
based on the wave packet computed 
and findings reported. In our work, we recalculate these quantities and 
also include some qualitative and statistical results about the 
evolution of the deviation vector for each of the four cases. 
In some sense, case $W2_2$ also acts as a benchmark for our investigation. 

\begin{figure}[H]
	\centering
	\includegraphics[width=0.65\textwidth,keepaspectratio]{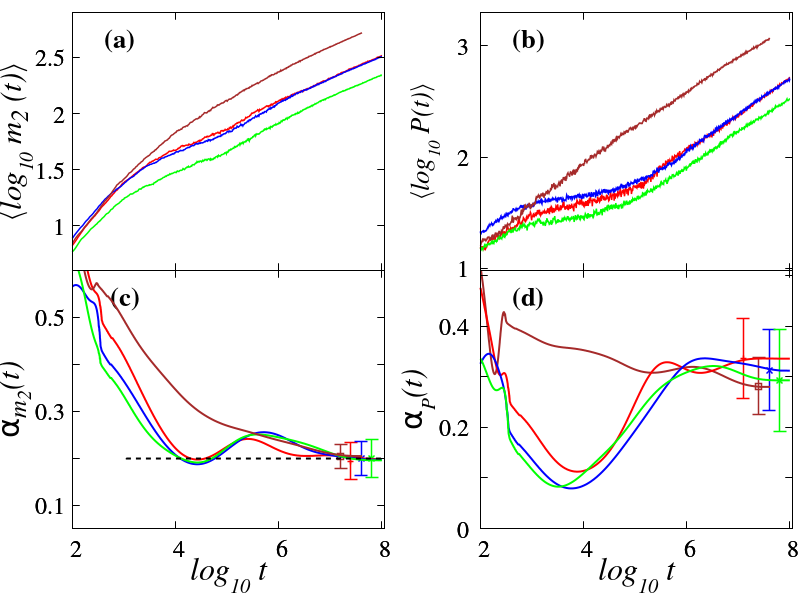}
	\caption{Results for the time evolution of {\bf (a)} $m_2(t)$ \eqref{eq:m2_2dkg}, 
		{\bf (b)} $P(t)$ \eqref{eq:P_2dkg}, 
		the corresponding slopes {\bf (c)} $\alpha_{m_2}(t)$ for $m_2(t)$ 
		and 
		{\bf (d)} $\alpha_{P}(t)$ for $P(t)$ for the weak chaos cases $W1_2$ 
		(red curves), $W2_2$ (brown curves), $W3_2$ (blue curves) and $W4_2$ 
		(green curves) of Hamiltonian \eqref{eq:kg2dham}. 
		The dashed line in {\bf (c)}  
		denotes the value $\alpha_{m_2}=0.2$.
		Each error bar denotes 
		a standard deviation in the corresponding computed slope.
		Plots {\bf(a)} and {\bf (b)} are in log-log scale while {\bf(c)} 
		and {\bf (d)} 
		are in log-linear scale.}
	\label{fig5:weak_m2P}
\end{figure}

In \autoref{fig5:weak_m2P} we present the averaged results for the 
evolution of 
$m_2(t)$ \eqref{eq:m2_2dkg} [panel {\bf (a)}] and $P(t)$ 
\eqref{eq:P_2dkg} [panel {\bf (b)}] 
for the cases $W1_2$ (red curves), $W2_2$ (brown curves), 
$W3_2$ (blue curves), and 
$W4_2$ (green curves). 
The corresponding 
slopes $\alpha_{m_2}(t)$ of $m_2(t)$ and $\alpha_{P}(t)$ of $P(t)$ 
respectively 
are shown in panels {\bf (c)} and {\bf (d)} 
with the dashed line indicating the 
average value $\alpha_{m_2}=0.2$ [panel {\bf (c)}] which is computed 
	for the saturated slopes at the 
end of the integration.
These results verify that the cases considered
show weak chaos behaviour since the evolution of 
$m_2(t)$ [panel {\bf (a)}] follows 
the power laws $m_2 \propto t^{0.2}$ [panel {\bf (c)}], 
in 
accordance with established results of 
\cite{Laptyeva2012}
 which emphasize the same power 
 law growth reported here for $m_2(t)$. 
 $\alpha_P$ on the other
 hand is observed to be approximately $0.31$ by the final time of 
 integration. The theoretical value of $\alpha_P=0.2$ 
 could be possible to attain after 
 longer times $log_{10}t>10^8$ that we could not reach due to 
 computational challenges.
The error bars shown in panels {\bf (c)} and {\bf (d)} 
are computations of 
a standard deviation of the quantities 
$\alpha_{m_2}(t)$ and $\alpha_P(t)$ respectively.

\begin{figure}[H]
	\centering
	\includegraphics[scale=0.34]{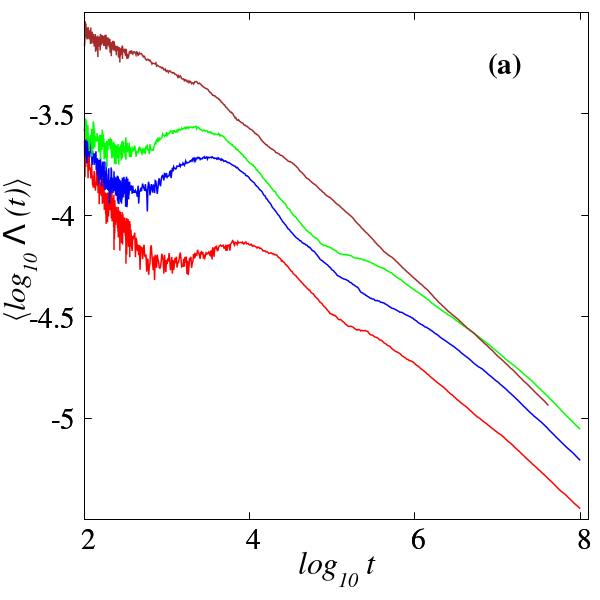}
	\includegraphics[scale=0.34]{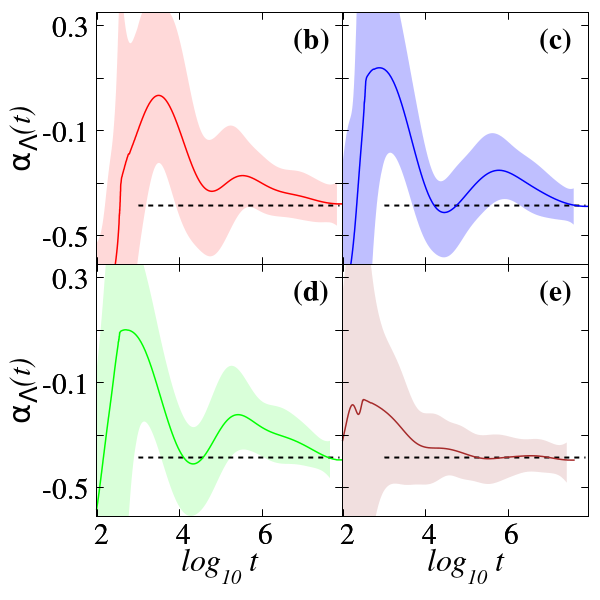}
	\caption{Results for {\bf (a)} the time evolution of the finite time mLCE 
		$\Lambda(t)$ \eqref{eq:ftmLCE} 
		and {\bf (b)}-{\bf (d)} the corresponding numerically computed 
		slopes for the four weak chaos cases of the 
		2D DKG system \eqref{eq:kg2dham} considered in 
		\autoref{fig5:weak_m2P}. Results are 
		averaged over $50$ disorder realizations. The straight 
		dashed lines in {\bf (b)} - {\bf (e)} indicate 
		the value $\alpha_\Lambda =-0.37$. The curve colors correspond to the cases presented in \autoref{fig5:weak_m2P}. Panel {\bf (a)}
		is in log-log scale while panels {\bf (b)}-{\bf (e)} are in log-linear scale.}
	\label{fig5:weak_L}
\end{figure}

We investigate the chaoticity of the system using the 
finite time mLCE $\Lambda(t)$ \eqref{eq:ftmLCE}. For each of the 
cases we numerically compute $\Lambda(t)$ \eqref{eq:ftmLCE} 
and present the 
results, averaged over $50$ disorder realizations, 
in \autoref{fig5:weak_L}{\bf (a)}. 
The corresponding numerically computed 
slopes (\ref{eq:aQ}) of the curves in panel {\bf (a)} together with 
an error (color shading) defined as one standard 
deviation of the 
distribution of computed slopes $\alpha_{\Lambda}$ 
are plotted in panels {\bf (b)}-{\bf (e)}. 
The horizontal dashed lines in {\bf (b)}-{\bf (d)} denote the 
value $\alpha_{\Lambda}=-0.37$.
The results of \autoref{fig5:weak_L} {\bf (b)}-{\bf (d)} 
indicate that the time 
evolution of $\Lambda(t)$ in the weak chaos regime 
eventually follows the power law 
$\Lambda(t) \propto t^{-0.37}$. 
The light color shade correspond 
to an error of one standard deviation in the numerically 
computed slopes.
As we can see from the results in \autoref{fig5:weak_L}, the dynamics 
of the system \eqref{eq:kg2dham} shows no tendency to cross to 
regular behaviour  
as the computed exponent $\alpha_\Lambda$ 
[panels {\bf (b)}-{\bf (d)} of \autoref{fig5:weak_L}] of the finite time 
mLCE \eqref{eq:ftmLCE} saturates around 
$\alpha_\Lambda \approx -0.37 \neq -1$ for all the considered cases.
The 
ratios $R_{m_2}$ and $R_P$ of 
\eqref{eq:T_ratios}
become
\begin{equation*}
	R_{m_2}  \sim t^{0.43} \,\,{\textnormal {and}}\,\, R_P  \sim t^{0.32},
\end{equation*}
since ${\alpha_{m_2}}=0.2$, 
${\alpha_P}=0.31$ and $\alpha_{\Lambda}=-0.37$.
The chaoticity time scale $T_{\Lambda}(t)$ is therefore 
smaller than the spreading time scales $T_{m_2}(t)$ and $T_P(t)$. The growth 
therefore of chaoticity is higher than that of spreading in the weak chaos 
regime of the 2D DKG model \eqref{eq:kg2dham}.

\subsubsection{Deviation vector distributions}
\label{sec5:DVD}
\begin{figure}[H]
	\centering
	\includegraphics[scale=0.39,keepaspectratio]{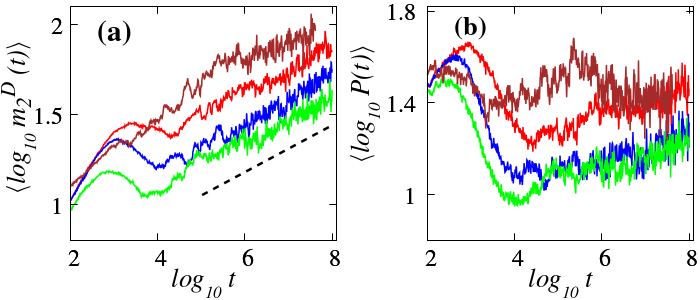}
	\caption{Results averaged over 50 realizations for the time evolution of  
		{\bf (a)} $m_2^D(t)$ \eqref{eq:m2_2dkg} and {\bf (b)} $P^D(t)$ \eqref{eq:P_2dkg} for the cases $W1_2$ - $W4_2$. 
		The dashed straight line in panel {\bf (a)} 
		indicates the slope $0.13$. The 
		curve colors correspond to the cases presented in \autoref{fig5:weak_m2P}.
		Panels {\bf(a)} and {\bf (b)} are in log-log scale.}
	\label{fig:m2_P_dvd_2D}
\end{figure}

We now make an investigation of the DVD and some of it's 
characteristics in relation to the dynamics of the system \eqref{eq:kg2dham}. 
We compute the evolution of $m_2^D(t)$ 
\eqref{eq:m2_dvd} and participation number $P^D(t)$ 
\eqref{eq:P_Dvd} and present their 
average results in \autoref{fig:m2_P_dvd_2D}.
The DVDs' second moment $m_2^D(t)$ [\autoref{fig:m2_P_dvd_2D}{\bf(a)}] shows a 
slow growth following the power law $m_2^D(t)\propto t^{0.13}$ and 
it reaches values 
always smaller than those computed for wave packets' $m_2(t)$ 
[\autoref{fig5:weak_m2P}{\bf(a)}]. 
The participation 
number $P^D(t)$ of the DVD attains relatively smaller values, 
$P^D \lesssim 40$ 
(the lowest value being $P^D \approx 13$ for case $W4_2$ [green curves in 
\autoref{fig:m2_P_dvd_2D} {\bf (b)}]).  
This shows that, compared to the energy distributions, 
the DVDs
are more localized and they keep a 
narrow and pointy shape with a relatively slow growth in 
the regime of weak chaos.

\begin{figure}[H]
	\centering
	\includegraphics[scale=0.4,keepaspectratio]{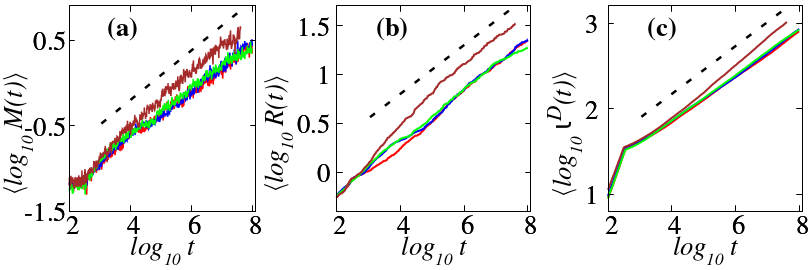}
	\caption{Results for the time evolution of {\bf(a)} $M(t)$ \eqref{eq:mad}, {\bf(b)} 
		$R(t)$ 
		\eqref{eq:R} and {\bf(c)} $\iota^D(t)$ \eqref{eq:l} of the DVD 
		for cases $W1_2$ - $W4_2$ of Hamiltonian \eqref{eq:kg2dham}.
		The curve colours correspond to the 
		cases presented in Figure \ref{fig5:weak_m2P}.
		The straight dashed lines guide the eye for slopes 
		{\bf (a)} $0.29$, {\bf (b)} $0.25$ and {\bf (c)} $0.28$.
		The plots are in log-log scale.}
	\label{fig:mad_diam_weak_2kg}
\end{figure}

We now present findings for the fluctuations of the DVD and it's 
centre by 
following the time evolution of $M(t)$ \eqref{eq:mad}, $R(t)$ \eqref{eq:R} and 
 $\iota^D(t)$ \eqref{eq:l} which were defined in Section 
\ref{sec4:weak}. In \autoref{fig:mad_diam_weak_2kg} we present results for the $M(t)$ 
[panel {\bf (a)}], $R(t)$ [panel {\bf(b)}] and $\iota^D(t)$ 
[panel{\bf(c)}] where the dashed lines guide the eye 
for slopes $0.29$ in {\bf(a)}, $0.25$ in {\bf(b)} and $0.28$ in {\bf(c)}.
The results from the evolution of $M$, $R$ and $\iota^D$ show a 
strict increase in these quantities thereby indicating that 
chaotic seeds constantly meander in the lattice.


\section{The strong chaos regime}\label{sec5:strong}
\begin{figure}[H]
	\centering
	\includegraphics[width=0.65\textwidth,keepaspectratio]{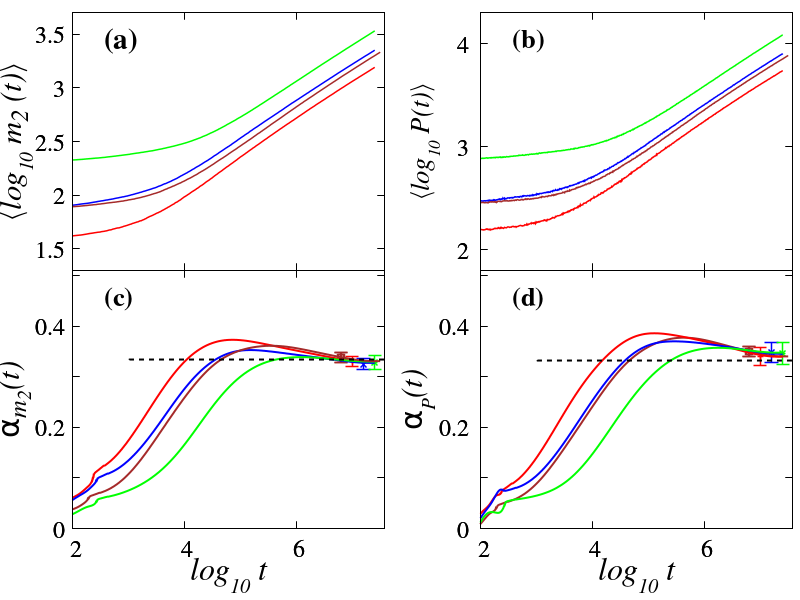}
	\caption{Similar to \autoref{fig5:weak_m2P} but for the strong chaos 
		spreading regime of Hamiltonian \eqref{eq:kg2dham}. Results for the time 
		evolution of {\bf (a)} $m_2(t)$ \eqref{eq:m2_2dkg}, 
		{\bf (b)} $P(t)$ \eqref{eq:P_2dkg}, 
		the corresponding slopes {\bf (c)} $\alpha_{m_2}(t)$ for $m_2(t)$ 
		and 
		{\bf (d)} $\alpha_{P}(t)$ for $P(t)$ for the cases $S1_2$ 
		(green curves), $S2_2$ (blue curves), $S3_2$ (brown curves) 
		and $S4_2$ (red curves). 
		The dashed lines in panels {\bf (c)} and {\bf (d)} 
		indicate values $\alpha_{m_2}=0.33$ and $\alpha_{P}=0.33$ respectively.
		The error bars in {\bf (c)} and {\bf (d)} are 
		a standard deviation of the computed slopes.
		Plots {\bf(a)} and {\bf (b)} are in log-log scale while {\bf(c)}, 
		{\bf (d)} 
		is in log-linear scale.}
	\label{fig5:strong_m2P}
\end{figure}

The time evolution of the second moment $m_2(t)$ 
\eqref{eq:m2_2dkg} and participation number $P(t)$ 
\eqref{eq:P_2dkg} of the 2D DKG model wave packet is known to 
follow as 
$m_2(t)\propto t^{1/3}$ and $P(t)\propto t^{1/3}$ [\cite{Laptyeva2012}] in the 
strong regime of chaos. 
We investigate the 
behaviour of energy 
propagation in this regime. 
We numerically integrate the system \eqref{eq:kg2dham}
up to a final time 
$t_f \approx 10^{7.7}$ (i.e. $t_f\approx50{,}000{,}000$) time units with 
the lattice $N\times{M}$ increased up to a size with dimension lengths in range 
$N(t_f)=M(t_f) \approx 300 - 500$. 
The energy $H_{2K}$ of the system was conserved at an absolute 
relative energy error \eqref{eq5:ree1} $e_r(t)\approx10^{-5}$ for time 
steps of size range $\tau \approx 0.4-0.9$.
A total of $L\cdot{J}$ central sites 
(which make up sub-lattice $L\times{J}$) of 
the lattice are 
excited with energy density $h$ and 
disorder strength $W$. Taking $L=J$ and $N(t)=M(t)$, we
consider the following four parameter cases:
\begin{description}
	\item[Case $S1_2$:] $W=9$, $L=35$, $h = 0.006$;
	\item[Case $S2_2$:] $W=10$, $L=21$, $h = 0.0135$;
	\item[Case $S3_2$:] $W=12$, $L=21$, $h = 0.01$;
	\item[Case $S4_2$:] $W=12.5$, $L=15$, $h = 0.035$.
\end{description}
The results of \autoref{fig5:strong_m2P} show that each of the cases 
$S1_2$-$S4_2$
exhibits the characteristics of strong chaos, where $m_2(t)$ 
\eqref{eq:m2_2dkg} and $P(t)$ \eqref{eq:P_2dkg} respectively follow the 
power laws 
$m_2(t) \propto t^{\alpha_{m_2}}$ 
[panels {\bf (a)} and {\bf (c)}] and $P(t) \propto t^{\alpha_P}$ 
[panels {\bf (b)} and {\bf (d)}], for values $\alpha_P=\alpha_{m_2}=0.33$ 
[shown by dashed lines in {\bf (c)} and {\bf (d)}] and 
in agreement with the theoretical estimates [\cite{Laptyeva2012}].
The error bars in {\bf (c)} and {\bf (d)} are each 
one standard deviation of the 
slopes $\alpha_{m_2}$ and 
$\alpha_{P}$ respectively.

\begin{figure}[H]
	\centering
	\includegraphics[scale=0.34]{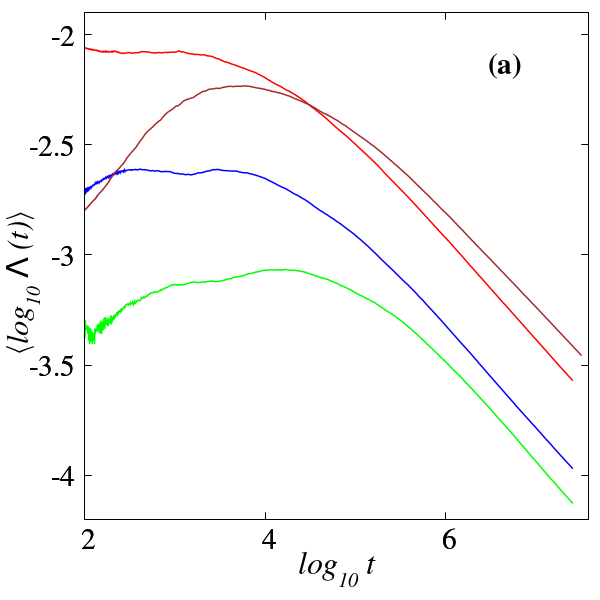}
	\includegraphics[scale=0.34]{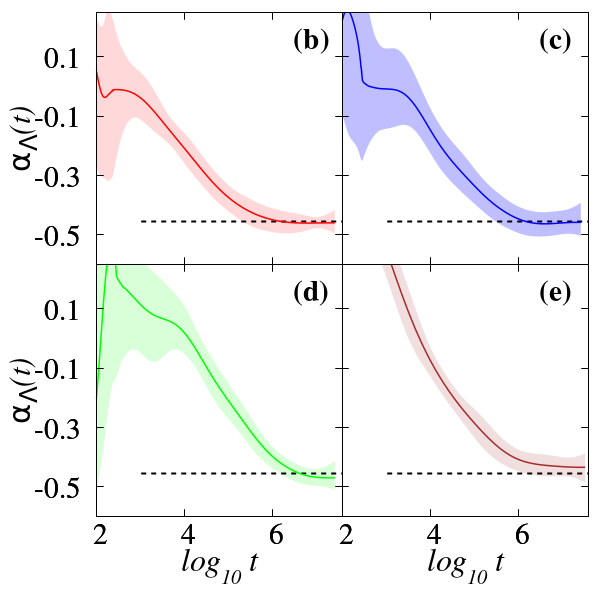}
	\caption{Similar to \autoref{fig5:weak_L} but for  the strong chaos 
		spreading regime of Hamiltonian \eqref{eq:kg2dham}. Results for 
		the time evolution of {\bf (a)} the finite time mLCE 
		$\Lambda(t)$ \eqref{eq:ftmLCE}
		and {\bf (b)}-{\bf (e)} the corresponding derivatives for the cases $S1_2$ 
		(green curves), $S2_2$ (blue curves), $S3_2$ (brown curves) and $S4_2$ (red curves). The straight 
		dashed lines indicate slopes $\alpha_\Lambda =-0.46$. Panel {\bf (a)}
		is in log-log scale while panels {\bf (b)}-{\bf (e)} are in log-linear scale.}
	\label{fig5:strong_L}
\end{figure}

In \autoref{fig5:strong_L}, we present results for the 
evolution of $\Lambda(t)$ \eqref{eq:ftmLCE} 
where we
observe an 
eventual power law decay 
$\Lambda(t)\propto t^{\alpha_{\Lambda}}$ with no tendency 
to cross to the regular dynamics 
law $\Lambda(t) \propto t^{-1}$.
The exponent $\alpha_{\Lambda} \approx -0.46$ 
is different from the one of the weak chaos case where 
$\alpha_{\Lambda} \approx -0.37$. This shows the difference in the 
chaoticity of the system \eqref{eq:kg2dham} for the two regimes.

The spreading to chaoticity time scale
ratios $R_{m_2}$ and $R_P$ of 
\eqref{eq:T_ratios}
are therefore
\begin{equation*}
	R_{m_2},~R_P  \sim t^{0.21},
\end{equation*}
since ${\alpha_{m_2}}={\alpha_P}=0.33$ and $\alpha_{\Lambda}=-0.46$.
$T_{\Lambda}(t)$ is therefore 
smaller than either of $T_{m_2}(t)$ or $T_P(t)$.

\subsubsection{Deviation vector distributions}
\begin{figure}[H]
	\centering
	\includegraphics[scale=0.39,keepaspectratio]{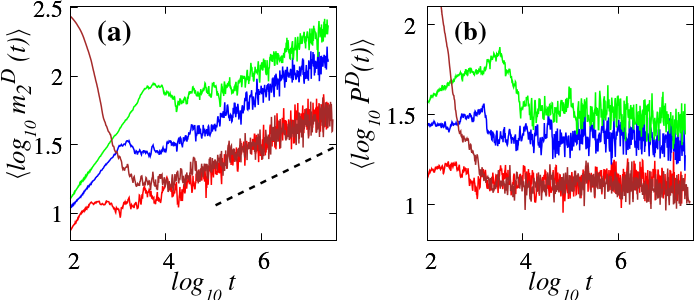}
	\caption{Similar to \autoref{fig:m2_P_dvd_2D} but for  the strong 
		chaos 
		spreading regime cases $S1_2$ - $S4_2$ of Hamiltonian \eqref{eq:kg2dham}.		
		Results averaged over $50$ disorder realizations for the time 
		evolution of  
		{\bf (a)} $m_2^D(t)$ \eqref{eq:m2_2dkg} and {\bf (b)} $P^D(t)$ \eqref{eq:P_2dkg}. The curve colors correspond to the cases presented in \autoref{fig5:strong_m2P}. The straight dashed line in 
		{\bf (a)} guides the eye 
		for slope $0.17$. 
		Panels {\bf(a)} and {\bf (b)} are in log-log scale while {\bf(c)}, 
		and {\bf (d)} 
		are in log-linear scale.}
	\label{fig5:m2_P_dvd_2D_str}
\end{figure}

In \autoref{fig5:m2_P_dvd_2D_str} we present averaged results for the 
DVDs' second moment $m_2^D(t)$ [\autoref{fig5:m2_P_dvd_2D_str}\textbf{(a)}] 
and the participation number $P^D(t)$ [\autoref{fig5:m2_P_dvd_2D_str}\textbf{(b)}] for the strong chaos cases 
$S1_2$ - $S4_2$. 

The $m_2^D(t)$ follows a power law 
$m_2^D(t) \propto t^{0.17}$ which shows a faster growth in the spatial 
extent of the DVD compared to 
revelations from the weak chaos cases whose power law was 
$m_2^D(t) \propto t^{0.13}$. 
The DVD participation number $P^D(t)$ on the other hand is practically 
constant as seen in \autoref{fig5:m2_P_dvd_2D_str}\textbf{(b)}. 
These findings show that in the strong chaos regime, the DVD keeps spreading 
just like in the weak chaos case, but gets a more pointy shape.

\begin{figure}[H]
	\centering
	\includegraphics[scale=0.4,keepaspectratio]{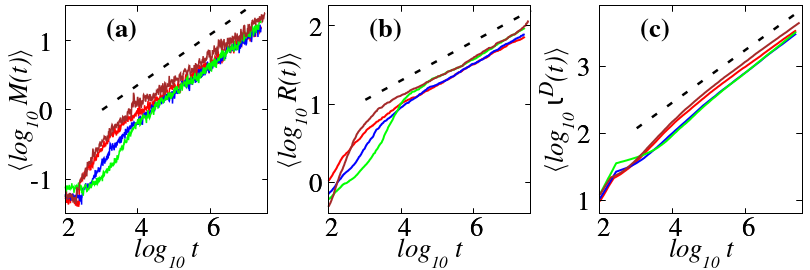}
	\caption{Similar to \autoref{fig:mad_diam_weak_2kg} but for  the strong 
		chaos 
		spreading regime cases $S1_2$ - $S4_2$ of Hamiltonian \eqref{eq:kg2dham}.
		Results for the time evolution of {\bf(a)} $M(t)$ 
		\eqref{eq:mad}, {\bf(b)} $R(t)$ \eqref{eq:R} and 
		{\bf (c)} 
		$\iota^D(t)$ \eqref{eq:l}.
		The curve colours correspond to the 
		cases presented in \autoref{fig5:strong_m2P}.
		The straight dashed lines guide the eye for slopes 
		{\bf (a)} $0.37$, {\bf (b)} $0.25$ and {\bf (c)} $0.39$.
		The panels are in log-log 
		scale.}
	\label{fig5:R_M_l_cpct_2D_str}
\end{figure}
In \autoref{fig5:R_M_l_cpct_2D_str} we show the results for the evolution 
of DVD 
quantities 
$M(t)$ \eqref{eq:mad}, $R(t)$ 
\eqref{eq:R} and $\iota^D(t)$ \eqref{eq:l}. The straight dashed lines 
show that these quantities grow following 
power laws $M(t)\propto{t}^{0.37}$ [\autoref{fig5:R_M_l_cpct_2D_str}{\bf (a)}], 
$R(t)\propto{t}^{0.25}$ [\autoref{fig5:R_M_l_cpct_2D_str}{\bf (b)}] and 
$\iota^D(t)\propto{t}^{0.39}$ [\autoref{fig5:R_M_l_cpct_2D_str}{\bf (c)}].
This shows that the quantities 
$R(t)$, $M(t)$ and $\iota^D(t)$ in the strong chaos 
regime grow with the same or higher 
power law exponents compared to the weak chaos case where results showed 
that  
$M(t)\propto{t}^{0.29}$ [\autoref{fig:mad_diam_weak_2kg}{\bf (a)}], 
$R(t)\propto{t}^{0.25}$ [\autoref{fig:mad_diam_weak_2kg}{\bf (b)}] and 
$\iota^D(t)\propto{t}^{0.28}$ [\autoref{fig:mad_diam_weak_2kg}{\bf (c)}].
Since $M(t)\leq R(t)$ \eqref{ineq} and our results show 
$M(t)$ is growing faster than $R(t)$, it means that at some point $M(t)$ 
will have to slow down so as to satisfy the bound relation \eqref{ineq}. 

\section{The selftrapping chaos regime}\label{sec5:self}

In the selftrapping chaos regime, the 
participation number $P(t)$ \eqref{eq:P_2dkg} of the wave packet  
remains practically constant 
as the system evolves. 
For our investigation in this regime, 
we numerically integrate the system \eqref{eq:kg2dham}
up to a final time 
$t_f \approx 10^{7.7}$ (i.e. $t_f\approx50{,}000{,}000$) time units with 
the lattice $N\times{M}$ increased up to size with dimension lengths in range 
$N(t_f)=M(t_f) \approx 350$. 
The energy $H_{2K}$ of the system was conserved at an absolute energy 
relative error \eqref{eq5:ree1} $e_r(t)\approx10^{-5}$ for time 
step $\tau \approx 0.7$.
A single site located at the centre of the lattice is 
excited with energy density $h=2.0$ and 
disorder strength $W=10$ is used. We refer to this configuration 
	as case $ST_2$.
$ST_2$ was studied in \cite{Laptyeva2012} and classified 
as leading to chaotic behaviour in the selftrapping regime. In our work we 
investigate this 
parameter set and we compare the resultant dynamics with that of 
the weak and strong 
chaos regimes. 

\begin{figure}[H]
	\centering
	\includegraphics[width=0.65\textwidth,keepaspectratio]{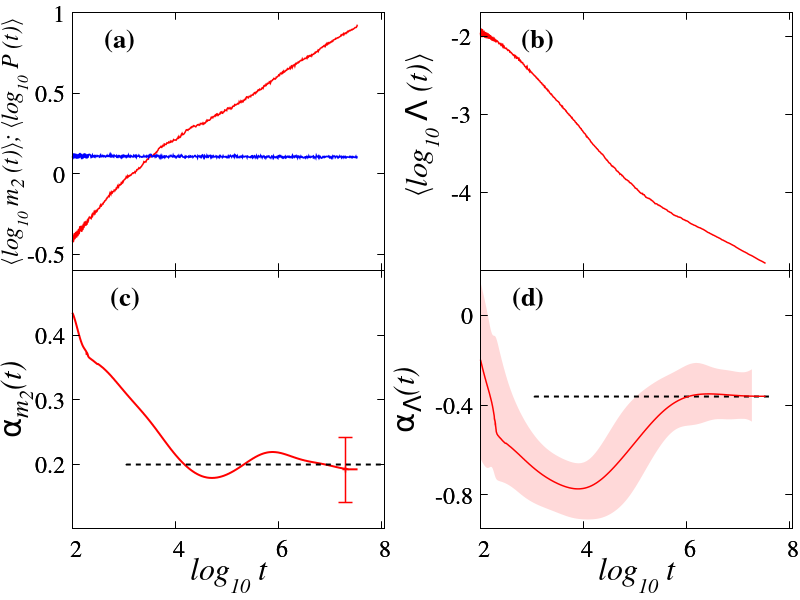}
	\caption{Results for the time 
		evolution of {\bf (a)} $m_2(t)$ \eqref{eq:m2_2dkg} [red curve] and $P(t)$ \eqref{eq:P_2dkg} [blue curve], 
		{\bf (b)} the finite time mLCE 
		$\Lambda(t)$ \eqref{eq:ftmLCE},  
		the corresponding slopes {\bf (c)} $\alpha_{m_2}(t)$ for $m_2$ 
		and 
		{\bf (d)} $\alpha_{\Lambda}(t)$ for $\Lambda(t)$ for the selftrapping 
		regime case $ST_2$ of Hamiltonian \eqref{eq:kg2dham}. 		
		Results averaged over $50$ disorder realizations.
		The dashed lines in panels {\bf (c)} and {\bf (d)} 
		indicate values $\alpha_{m_2}=0.2$ and $\alpha_\Lambda =-0.36$ respectively.
		The error bar in {\bf (c)} and the 
			light color shade in {\bf(d)} denote the 
		numerical error of 
		one standard deviation in the computed slopes $\alpha_{m_2}(t)$ 
		and $\alpha_{\Lambda}(t)$ respectively.
		Plots {\bf(a)} and {\bf (b)} are in log-log scale while {\bf(c)} 
		and 
		{\bf (d)} 
		are in log-linear scale.}
	\label{fig5:slf_m2P}
\end{figure}

In \autoref{fig5:slf_m2P} we present the time evolution of 
$m_2(t)$ \eqref{eq:m2_2dkg} and $P(t)$ 
\eqref{eq:P_2dkg} [panel {\bf (a)}] and 
$\Lambda(t)$ \eqref{eq:ftmLCE}
[panel {\bf (b)}]. 

$ST_2$ 
exhibits characteristics of selftrapping chaos with 
the $m_2(t)$ evolving following a 
power law $m_2(t) \propto t^{\alpha_{m_2}}$ where 
$\alpha_{m_2}\approx0.2$
[\autoref{fig5:slf_m2P}{\bf (c)}], a finding consistent with what was 
reported in \cite{Laptyeva2012} and similar to the behaviour of the 
$m_2(t)$ for the weak chaos regime. Also $P(t)$, as expected, remains 
constant at very small values.

For the finite time mLCE $\Lambda(t)$ \eqref{eq:ftmLCE}, which 
was not computed in \cite{Laptyeva2012}, we
observe an eventual power law decay 
$\Lambda(t)\propto t^{\alpha_{\Lambda}}$ which is different from 
$\Lambda(t) \propto t^{-1}$ of regular dynamics.
The computed exponent $\alpha_{\Lambda} \approx -0.36$ is close to 
the average $\alpha_{\Lambda} \approx -0.37$ observed in the weak chaos case. 
This shows that the chaoticity for the selftrapping case is similar to 
that of the weak chaos case but different from the one of strong chaos case (where $\alpha_{\Lambda} \approx -0.46$).
The spreading to chaoticity time scale
ratio $R_{m_2}$ of 
\eqref{eq:T_ratios}
is therefore
\begin{equation*}
	R_{m_2} \sim t^{0.44},
\end{equation*}
since ${\alpha_{m_2}}=0.2$ 
and $\alpha_{\Lambda}=-0.36$, indicating that 
$T_{\Lambda}(t) < T_{m_2}(t)$.

\subsubsection{Deviation vector distributions}
We present our findings for a representative realisation of case $ST_2$ 
in this regime, after 
which we give results similar to the outcomes 
of the DVD discussion given in the cases of weak 
and strong chaos.

\begin{figure}[H]
	\centering
	\includegraphics[scale=0.28]{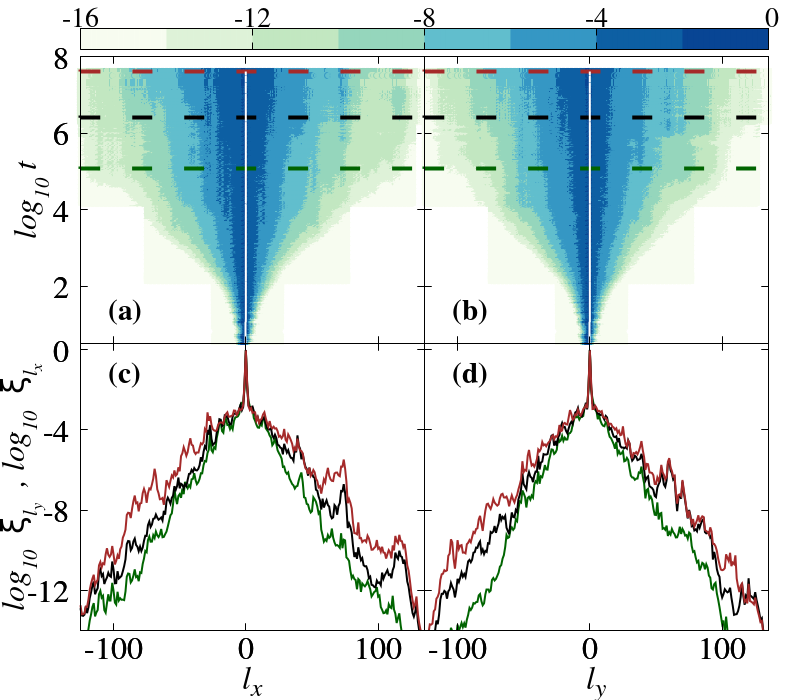}
	\includegraphics[scale=0.28]{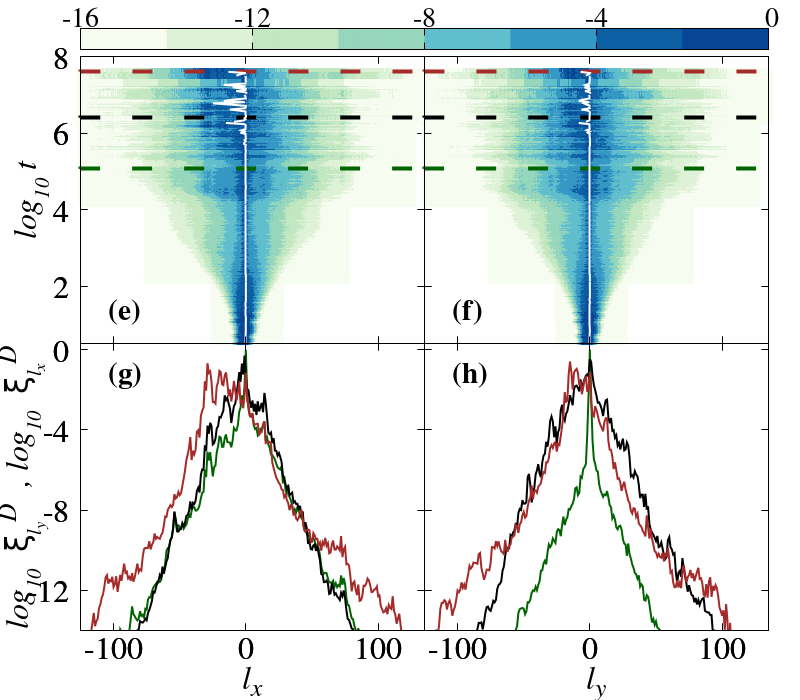}
	\caption{Results for the time evolution of the normalized 
		energy distribution {\bf(a)} $\xi_{l_x}(t)$ and 
		{\bf(b)} $\xi_{l_y}(t)$ \eqref{eq:marg_en} and 
		the DVD {\bf(e)} $\xi_{l_x}^D(t)$ and 
		{\bf(f)} $\xi_{l_y}^D(t)$ \eqref{eq:marg_en} for a single 
		realisation of case $ST_2$. 
		The colour scales at the top of the figure are used for 
		colouring lattice sites according to their {\bf(a)} 
		$\log_{10} \xi_{l_x}(t)$, {\bf(b)} $\log_{10} \xi_{l_y}(t)$,  
		{\bf(e)} $\log_{10} \xi_{l_x}^D(t)$ and  
		{\bf(f)} $\log_{10} \xi_{l_y}^D(t)$ values. In each of the 
		panels {\bf (a)}, {\bf(b)}, {\bf(c)} and {\bf(d)} the white 
		curve traces the distributions' centre. {\bf(c)} 
		Normalized energy 
		distributions $\xi_{l_x}(t)$, {\bf(d)} 
		Normalized energy 
		distributions $\xi_{l_y}(t)$, {\bf(g)} 
		DVD $\xi^D_{l_x}(t)$ and {\bf(h)} 
		DVD $\xi^D_{l_y}(t)$ at 
		times $\log_{10}t=5$ [green curve], $\log_{10}t=6.4$ 
		[black curve], $\log_{10}t=7.7$ [red curve]. These times are also denoted 
		by similarly coloured horizontal dashed lines in {\bf (a)}, 
		{\bf(b)}, {\bf(e)} and {\bf(f)}.}
	\label{fig5:slf_palette}
\end{figure}

In \autoref{fig5:slf_palette} we show the time evolution of the 
normalised energy density \eqref{eq:norm_en2d} and 
DVDs \eqref{eq:dvd}. 
For this analysis we 
use the marginal 
energy densities 
\begin{equation}\label{eq:marg_en}
	\xi_{l_x}=\sum_{m}\xi_{l,m} \qquad {\textnormal{and}} \qquad \xi_{l_y}=\sum_{l}\xi_{l,m}
\end{equation} and 
\begin{equation}\label{eq:marg_dv}
	\xi^D_{l_x}=\sum_{m}\xi^D_{l,m} \qquad {\textnormal{and}}\qquad  \xi^D_{l_y}=\sum_{l}\xi^D_{l,m}.
\end{equation} 
  The centre of the lattice 
 has been translated to $0$ in each of the two dimensions.
In \autoref{fig5:slf_palette}\textbf{(a)} and \textbf{(b)} we 
respectively plot the time evolution of $\xi_{l_x}(t)$ \eqref{eq:marg_en} 
and $\xi_{l_y}(t)$ \eqref{eq:marg_en}, while in \textbf{(e)} and 
\textbf{(f)} we show 
the evolution of $\xi^D_{l_x}(t)$ \eqref{eq:marg_dv} 
and $\xi^D_{l_y}(t)$ \eqref{eq:marg_dv}.
Once again the energy density exhibits a symmetric 
spreading behaviour around 
the mean position of the wave packet 
[\autoref{fig5:slf_palette}\textbf{(a)}, 
\textbf{(b)}, \textbf{(c)} and \textbf{(d)}] 
while the DVD centre fluctuates as shown in 
\autoref{fig5:slf_palette}\textbf{(e)}, 
\textbf{(f)}, \textbf{(g)} and \textbf{(h)}, within the region covered by the 
wave packet. Snapshots of 
the energy distributions at specific integration 
times $\log_{10}t=5$, $\log_{10}t=6.4$ and $\log_{10}t=7.7$ are shown 
in \textbf{(c)} and \textbf{(d)} and those for the DVDs are presented in 
\textbf{(g)} and \textbf{(h)}.
As expected of the selftrapping behaviour, there are some sites which are highly 
excited and this manifests in form of sharp peaks of the snapshots in 
panels \textbf{(c)} and \textbf{(d)}. 

\begin{figure}[H]
	\centering
	\includegraphics[scale=0.39,keepaspectratio]{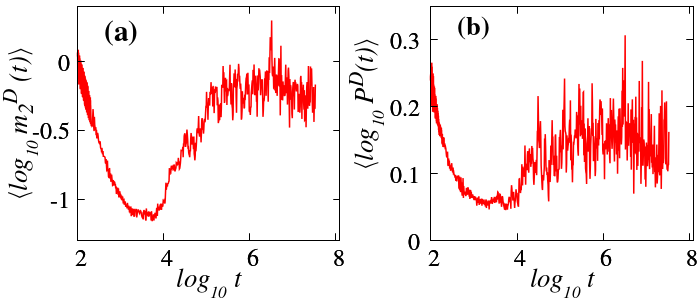}
	\caption{Similar to \autoref{fig:m2_P_dvd_2D} but for  the 
		selftrapping chaos 
		spreading regime case $ST_2$ of Hamiltonian \eqref{eq:kg2dham}. 		
		Results averaged over $50$ disorder realizations for the time 
		evolution of  
		{\bf (a)} $m_2^D(t)$ \eqref{eq:m2_2dkg} and {\bf (b)} $P^D(t)$ \eqref{eq:P_2dkg}.
		The plots are in log-log scale.}
	\label{fig5:m2_P_dvd_2D_slf}
\end{figure}

In \autoref{fig5:m2_P_dvd_2D_slf} we present averaged results for the 
second moment $m_2^D(t)$ \eqref{eq:m2_dvd} 
and the participation number $P^D(t)$ \eqref{eq:P_Dvd} of the 
DVD \eqref{eq:dvd} for the selftrapping 
case 
$ST_2$. 

The $m^D_2(t)$ [\autoref{fig5:m2_P_dvd_2D_slf}\textbf{(a)}] 
and $P^D(t)$ [\autoref{fig5:m2_P_dvd_2D_slf}\textbf{(b)}] 
show a tendency to eventually become constant at very small values, 
indicating that the DVD eventually stops spreading and gets a 
pointy shape.

\begin{figure}[H]
	\centering
	\includegraphics[scale=0.4,keepaspectratio]{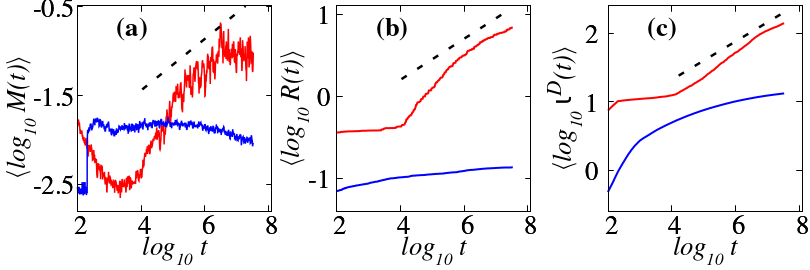}
	\caption{Similar to \autoref{fig:mad_diam_weak_2kg} but for the 
		selftrapping chaos 
		regime case $ST_2$ of Hamiltonian \eqref{eq:kg2dham}.
		Results for the time evolution of {\bf(a)} $M(t)$ 
		\eqref{eq:mad}, {\bf(b)} $R(t)$ \eqref{eq:R} and 
		{\bf (c)} 
		$\iota^D(t)$ \eqref{eq:l} for the 
		DVD (red curves) and energy distribution (blue curves).
		The straight dashed lines guide the eye for slopes 
		{\bf (a)} $0.29$, {\bf (b)} $0.25$ and {\bf (c)} $0.28$.
		The panels are in log-log 
		scale.}
	\label{fig5:R_M_l_cpct_2D_slf}
\end{figure}

\autoref{fig5:R_M_l_cpct_2D_slf} shows the results for the quantities 
{\bf (a)} $M(t)$ \eqref{eq:mad}, {\bf (b)} $R(t)$ 
\eqref{eq:R} and {\bf (c)} $\iota^D(t)$ \eqref{eq:l} computed for the 
DVD (red curves) and energy distribution (blue curves).  
The results show that mean position fluctuations for the energy 
distribution are of smaller 
magnitude compared to those of the DVD mean position. 
This result 
for the comparison of DVD and energy distribution mean 
position fluctuations, 
which is not different from the behaviour in both the weak 
and strong chaos cases, confirms the 
behaviour we see 
from the representative distribution plots in 
\autoref{fig5:slf_palette} about meandering of mean positions.
The straight dashed lines 
correspond to slopes
$0.29$ [\autoref{fig5:R_M_l_cpct_2D_slf}{\bf (a)}], 
$0.25$ [\autoref{fig5:R_M_l_cpct_2D_slf}{\bf (b)}] and 
$0.28$ [\autoref{fig5:R_M_l_cpct_2D_slf}{\bf (c)}] observed in the case 
of weak chaos in Section \ref{sec5:weak}.
The DVD quantities 
$R(t)$, $M(t)$ and $\iota^D(t)$ in this case continue to increase 
although they eventually 
grow away from and smaller than what was observed 
for the weak chaos regime, showing that the DVD mean position fluctuations 
in the 
selftrapping regime eventually grow slower than those in the weak chaos case.

\section{Summary}\label{sec5:summary}
We studied the dynamical behaviour of the 2D DKG model 
\eqref{eq:kg2dham} in the weak, strong and selftrapping chaotic regimes. 
For a number of chaotic parameter cases, we 
obtained statistical results from approximately $50$ 
disorder realizations in each case. 
Like for the case of the 1D DKG model \eqref{eq:kg1dham} 
studied in Chapter \ref{chap_1d}, we computed the finite time 
mLCE $\Lambda(t)$ \eqref{eq:ftmLCE} and 
verified that 
the level of 
chaoticity of the propagating wave packets decreases in time 
with the system maintaining its chaotic nature without any tendency 
of the dynamics to crossover 
to regular behaviour. $\Lambda(t)$ follows a power law  
$\Lambda(t) \propto t^{\alpha_{\Lambda}}$ characterized 
by $\alpha_{\Lambda}\ne-1$. We found that 
in the  weak, strong and selftrapping chaos regimes
$\alpha_{\Lambda} \approx -0.37$, 
$\alpha_{\Lambda} \approx -0.46$ and 
$\alpha_{\Lambda} \approx -0.36$ respectively.
The wave packet gets more chaotic at a rate higher than 
its spreading since 
for all considered regimes of chaos, 
the chaoticity time scale $T_{\Lambda}(t)$ \eqref{eq:TL} 
always remains smaller than the spreading time scales $T_{m_2}(t)$ 
\eqref{eq:TM} and or $T_P(t)$ \eqref{eq:TP}.
\newline\newline
The DVD \eqref{eq:dvd} kept a localized, pointy shape with small values 
of the participation 
number $P^D(t)$ \eqref{eq:P_Dvd} compared to $P(t)$ \eqref{eq:P_2dkg} 
of the normalised energy distribution \eqref{eq:norm_en2d}. 
However, the  
DVD centre fluctuations amplitude for all weak, strong and 
selftrapping chaos cases 
increased with time as it (DVD centre) moved 
to regions inside the spreading wave packet. 
Therefore the quantities $R(t)$ (\ref{eq:R}), $M(t)$ \eqref{eq:mad} and 
$\iota^D(t)$ \eqref{eq:l} which estimate the DVD centre fluctuations 
increased with time following power laws 
$R(t)\propto t^{0.25}$, $M(t)\propto t^{0.29}$ and $\iota^D(t)\propto t^{0.28}$
in the weak chaos regime. 
For the strong chaos case on the other hand, $M(t)$ and $\iota^D(t)$ 
grow with higher 
non constant exponents as $M(t)\propto t^{0.37}$ and 
$\iota^D(t)\propto t^{0.39}$ showing that the chaotic seeds 
spread faster with the DVD covering wider areas compared to the weak 
chaos case. Just like in the weak chaos case, 
$R(t)\propto t^{0.25}$ for the strong chaos regime, 
showing that the growth rate of  
the fluctuation amplitude of the DVD centre is similar for both regimes but with 
higher amplitudes observed for the strong chaos case.
\newline\newline
In summary, we showed numerically 
for the 2D DKG model \eqref{eq:kg2dham} 
that (a) the strength of chaotic dynamics of a wave packet decreases in time 
but it does not crossover to regular dynamics, (b) chaotization 
is due to the chaotic seeds meandering in the wave packet
and (c) the characteristics of chaos evolution show a clear 
distinction between weak and strong chaos behaviour.

\chapter{Summary and Conclusions}\label{chap:summary}
In this thesis, the study of energy propagation in 
disordered nonlinear Hamiltonian 
systems of many degrees of freedom was discussed. 
Firstly, normal 
mode (NM) properties for a 1D modified Klein-Gordon model 
\eqref{eq:LDKG} with parameters 
$D$, which 
allows the system to be ordered for $D=0$ and increase disorder in the system 
up to when $D=1/2$, and $\mathcal{W}$ which 
regulates the contribution of nearest neighbour interactions, were studied. 
For a fixed $\mathcal{W}=4$, the NMs become less 
spatially extended [with spatial extent measured using the localization volume $V$
\eqref{eq:Vmod} and participation number 
$P$ \eqref{eq:Pmod}] in the lattice 
as values of $D$ increase towards $1/2$ following the laws 
$P \propto  D^{-2}$ and $V \propto  D^{-2}$ 
and a participation number 
fluctuation correction scaling 
given by 
$V \approx 2.6 P$.
Therefore, the NMs of linear systems with 
strong enough disorder only extend to a finite number of sites and 
so, an initially localized wave packet 
will not spread to infinitely many sites. 
Nonlinearity, however, introduces chaos to the system and destroys AL as the 
NMs interact with each other. 
Spreading energy distribution characteristics 
for different dynamical behaviours in nonlinear 
disordered lattices 
including the weak, strong and selftrapping 
regimes of chaos for $D=1/2$ were theoretically explained 
using the DDNLS system 
\eqref{eq:dnls1ham}. 
\newline\newline
In chapter \ref{chap:num_tech}, numerical techniques for 
investigating the dynamical behaviour of one- and two-dimensional Hamiltonian 
chains were discussed. These included 
aspects of deviation vector dynamics, 
the corresponding DVDs, Lyapunov characteristic exponents, the 
variational equations and integration methods. 
For both the $1D$ and $2D$ DKG models, an initial deviation vector 
where all coordinates of both the positions and momenta components 
are zero except for the coordinates which just cover the degrees of freedom 
where the initial excitation takes place, gives the fastest convergence of the 
DVD behaviour [monitored using finite time mLCE $\Lambda(t)$ and DVD 
participation number $P^D(t)$] to it's asymptotic 
state compared to other forms of initial deviation vector. 
A number of existing 
integrators were described and higher order symplectic schemes 
constructed 
through composition techniques.
The computational efficiency of these schemes in integrating 
the equations of motion [\eqref{eq:kg1dmotion} and 
\eqref{eq:kg2dmotion}] and the corresponding 
variational equations \eqref{eq:variational} of 
the one \eqref{eq:kg1dham} and two \eqref{eq:kg2dham} spatial 
dimension DKG systems was analysed.
The performance of the integrators was found to be independent of the spatial 
dimensions of the model being investigated, as the rank in performance over 
both the 1D and 2D DKG models was the same.
The order four SIs $ABA864$ and $ABAH864$, 
the order six SIs $s9SABA_26$, $SABA_2Y6$ and $ABA864Y6$, 
gave the best performance at an absolute relative energy error 
$e_r\approx 10^{-5}$
with $ABA864$ [\cite{Blanes2013}] requiring the least CPU time in both models.
At a high accuracy level $e_r\approx 10^{-8}$, the order six SIs
$s11ABA82\_6$, $SABA_2Y6$ and $s9ABA82\_6$ and the order eight SIs 
$s15ABA82\_8$ and $s19ABA82\_8$ were found to be the most computationally 
efficient schemes.
\newline\newline
In chapters \ref{chap_1d} and \ref{chap:2D}, systematic numerical calculations
of finite time maximum Lyapunov characteristic exponents $\Lambda(t)$ 
\eqref{eq:ftmLCE} were presented for respectively 
the 1D and 2D disordered nonlinear Hamiltonian lattices with the DKG 
models \eqref{eq:kg1dham} and \eqref{eq:kg2dham} as prototype 
representative systems. The dynamical behaviour in the spreading wave packet 
was found to remain 
chaotic for the entire duration of the simulation. In one spatial dimension, 
$\Lambda(t)$ evolved following the laws 
$\Lambda(t)\propto t^{-0.25}$, 
$\Lambda(t)\propto t^{-0.32}$ and 
$\Lambda(t)\propto t^{-0.26}$ for respectively
the weak, strong and selftrapping regimes of chaos. 
There was however a slowing down in the chaotic behaviour for 
the two spatial dimensions system with 
$\Lambda(t)$ evolving as 
$\Lambda(t)\propto t^{-0.37}$, 
$\Lambda(t)\propto t^{-0.46}$ and 
$\Lambda(t)\propto t^{-0.36}$ for respectively
the weak, strong and selftrapping regimes of chaos.
For all initial system configurations presented in the $1D$ 
	and $2D$ models, there was no signs of a tendency by the dynamics to slow 
down and cross to 
ordered motion which is characterised by  
$\Lambda(t)\propto t^{-1}$. 
The chaoticity timescale $T_{\Lambda}(t)$ \eqref{eq:TL} was found to be 
shorter than 
the spreading timescales $T_{m_2}(t)$ 
\eqref{eq:TM} and/or $T_P(t)$ \eqref{eq:TP} with the corresponding 
chaoticity-spreading timescale 
ratios $R_{m_2}(t)$ and/or $R_P(t)$ \eqref{eq:T_ratios} diverging as a power law. 
Thus a 
confirmation of the subdiffusive spreading theories assumption necessitating 
chaos to be persistent and fast enough 
in disordered nonlinear lattices.
Additionally, computations of the $m_2(t)$, 
\eqref{eq:m2_1dkg} and \eqref{eq:m2_2dkg}, 
showed that the wave packet continues spreading for the different 
chaos regimes in the one and two 
spatial dimensions models, contrary to the reports of \cite{Johansson2010} 
and \cite{Aubry2011}.
\newline\newline
For all initial conditions considered in the $1D$ and $2D$ 
models, 
the energy density was observed to spread 
evenly with time about the distribution centre while  
the DVD remained localized with it's peak position meandering to 
cover distances approximately equal to the 
energy distribution width.
As estimators for fluctuation amplitudes of the DVD mean position, quantities 
$M(t)$ \eqref{eq:mad} [estimates the mean deviation of the DVD centre 
in a fixed epoch progressively in time], 
$R(t)$ \eqref{eq:R} [estimates the maximum fluctuation amplitudes 
of the DVD mean position] and $\iota^D(t)$ \eqref{eq:l} 
[estimates the magnitude of the total displacement of 
the DVD centre], were introduced. The amplitude of the fluctuations 
increased in time for all regimes of chaos 
with characteristic power law growths 
$R(t)\propto t^{0.24}$, $M(t)\propto t^{0.25}$ and $\iota^D(t)\propto t^{0.28}$
in the weak and selftrapping chaos regimes of the 1D DKG model. 
The strong chaos regime however, 
exhibited faster growth of these quantities with an 
eventual tendency to settle at the same power laws as the 
weak chaos regime, a good indication of the transient nature 
of the strong to weak chaos 
behaviour. 
In the regime of weak chaos for the 2D DKG model, $R(t)$ and 
$\iota^D(t)$ grow at rates similar to those observed in the same chaotic regime 
for the 1D DKG system.
$M(t)$ on the other hand grows faster for the 2D DKG model \eqref{eq:kg2dham} 
following the laws 
$M(t)\propto t^{0.29}$ for the weak chaos case and $M(t)\propto t^{0.37}$ 
for the strong chaos case, an indication that fluctuations of the DVD 
mean position in small time intervals of the dynamics
are stronger in two dimensions compared to one dimension. 
The total distance covered by the fluctuations of the DVD mean position, 
which is estimated using $\iota^D(t)$, was found to have a faster growth 
in the strong chaos case of two dimensions 
compared to one dimension.  
For both models, the quantities $M(t)$, $R(t)$, and $\iota^D(t)$ 
keep growing up to the 
largest simulation times, showing no sign of becoming constant as would be for 
regular motion.
This verifies another subdiffusive spreading theories assumption 
which requires the seeds of chaos to meander fast enough through the 
wave packet of a disordered Hamiltonian lattice thereby ensuring 
its chaotization.
\section{Outlook}
In this work, the chaotic behaviour of disordered nonlinear 
lattices has been 
discussed using a typical example of the DKG models with quartic nonlinearity. 
Our results agree with the 
findings for the quartic discrete nonlinear Schr\"{o}dinger equations 
(which have two conserved quantities) presented 
in \cite{Senyange2018a} and \cite{ManyManda2020}, thereby enabling a 
generalization of this behaviour to nonlinear disordered lattices. 
However, there are some open questions that have emerged as a follow up 
of this work. We mention here some of them. 
Our first example is the question on a thorough analytical and numerical 
investigation of linear NM properties in models 
of 2D spatial dimensions starting with the purely ordered case and gradually 
increasing the systems' heterogeneity to get a fully disordered model.
Further more, how the dynamical behaviour of the models (1D and 2D) 
is affected by the 
width of the interval containing on-site potentials. That is to say, 
how the dynamics of nonlinear disordered models 
changes for 
DKG model disorder selected from an interval $[1-D,1+D]$ where $D$ 
takes on different values, i.e.~$D\in(0,1/2]$. 

The second open question is how the findings of this work are affected 
when the 1D and 2D 
models have a generalised (including fractional) nonlinearity. A good 
starting point is the investigation of 
\cite{Skokos2010a} and \cite{Laptyeva2012} 
where a theoretical and 
numerical dynamical 
study of respectively the generalised 1D and 
2D Klein-Gordon models is reported. In both papers, it was shown that 
the wave packet continues to grow with the second moment following 
a power law. However there is need for further inquiry 
on the time evolution of chaos 
for these models compared to the presentation of Chapters \ref{chap_1d} 
and \ref{chap:2D}.


\begin{appendix}
\chapter{Symplectic integration of the DKG models}\label{app:A}

We present here the explicit form of the variational equations 
\eqref{eq:variational}, 
operators $e^{\tau L_{T_{\mathcal{V}}}}$ and 
$e^{\tau L_{V_{\mathcal{V}}}}$
used for the time propagation of a
deviation vector with initial conditions $({\bf q}, {\bf p},
{\delta {\bf q}}, {\delta {\bf p}})$ at time $t$ to their final values
$({\bf q}', {\bf p}', {\delta {\bf q}}', 
{\delta {\bf p}}')$ at time $t+
\tau$ for Hamiltonians \eqref{eq:kg1dham} and \eqref{eq:kg2dham} 
TDH \eqref{eq:tdh}.


\section{The 1D DKG model}\label{app:A1}

The  variational equations \eqref{eq:kg_variational} of the 1D DKG model (\ref{eq:kg1dham}) are
\begin{equation}
\displaystyle
\begin{array}{lll}
\displaystyle \frac{dq_i}{dt}&=& \displaystyle p_{i},
\,\,\,\mbox{for}\,\,\, 1\leq i\leq N\\
& & \\ 
\displaystyle
\frac{dp_1}{dt}&=& \displaystyle - \left[\epsilon_{1}q_{1} +
q_{1}^3 +
\frac{1}{W}\left(2q_{1}-q_{2}\right)\right]\\
& & \\ 
\displaystyle
\frac{dp_i}{dt}&=& \displaystyle - \left[\epsilon_{i}q_{i} +
q_{i}^3 + \frac{1}{W}\left(2q_{i}-q_{i-1}
-q_{i+1}\right)\right], \,\,\,\mbox{for}\,\,\, 2\leq i\leq
N-1\\
& & \\ 
\displaystyle \frac{dp_N}{dt}&=& \displaystyle -
\left[\epsilon_{N}q_{N} + q_{N}^3 +
\frac{1}{W}\left(2q_{N}-q_{N-1}\right)\right]\\
\displaystyle \frac{d\delta q_i}{dt}&=& \displaystyle \delta p_{i}, \,\,\,\mbox{for}\,\,\,
1\leq i\leq N \\
& & \\
\displaystyle \frac{d\delta p_1}{dt}&=& \displaystyle - \left[
\delta q_{1}\left(\epsilon_{1}+3q_{1}^2\right)
+ \frac{1}{W}\left(2\delta q_{1}-\delta q_{2}\right)\right]\\
& & \\
\displaystyle \frac{d\delta p_i}{dt}&=& \displaystyle - \left[
\delta q_{i}\left(\epsilon_{i}+3q_{i}^2\right)
+ \frac{1}{W}\left(2\delta q_{i}-\delta q_{i-1}
-\delta q_{i+1}\right)\right], \,\,\,\mbox{for}\,\,\,  2\leq i\leq N-1\\
& & \\
\displaystyle \frac{d\delta p_N}{dt}&=& \displaystyle - \left[
\delta q_{N}\left(\epsilon_{N}+3q_{N}^2\right)
+ \frac{1}{W}\left(2\delta q_{N}-\delta q_{N-1}\right)\right].
\end{array}
\label{eq:1eqmot}
\end{equation}

In order to implement the SIs discussed in 
Section \ref{sec:numerical_integration} for the
integration of equations (\ref{eq:1eqmot}) 
we split Hamiltonian (\ref{eq:kg1dham}) in two
integrable parts
\begin{equation}
\label{eq:ham1_split}
T({\bf p}) = \sum_{i=1}^N \frac{p_i^2}{2}\, , \,\,\,\,\,\,
V({\bf q}) =\sum_{i=1}^N \left[
\frac{\epsilon_i}{2}q_i^2 + \frac{q_i^4 }{4}+
\frac{1}{2W}\left(q_{i+1} - q_i\right)^2 \right],
\end{equation}
i.e. respectively the kinetic and potential energies respectively of the system. The
solution of the variational equations
for integrable Hamiltonians $T$ and $V$ are obtained through the action of
the operators
\begin{equation}
e^{\tau L_{V_{\mathcal{V}}}}: \left\{
\begin{array}{lll}
\displaystyle q'_{i}&=& \displaystyle q_{i} + \tau p_{i} \\
\displaystyle p'_{i}&=& \displaystyle p_{i}\\
\displaystyle \delta q'_{i}&=& \displaystyle \delta q_{i} + \tau \delta p_{i} \\
\displaystyle \delta p'_{i}&=& \displaystyle \delta p_{i}
\end{array}
\right. , \,\,\,\mbox{for}\,\,\,  1\leq i\leq N,
\label{eq:AV1}
\end{equation}
and
\begin{equation}
e^{\tau L_{T_{\mathcal{V}}}}: \left\{
\begin{array}{lll}
\displaystyle q'_{i}&=& \displaystyle q_{i}, \,\,\,\mbox{for}\,\,\,  1\leq i\leq N\\
& & \\
\displaystyle p'_{1}&=& \displaystyle p_{1} - \tau\left[\epsilon_{1}q_{1} +
q_{1}^3 + \frac{1}{W}\left(2q_{1}-q_{2}\right)\right]\\
& & \\
\displaystyle p'_{i}&=& \displaystyle p_{i} - \tau\left[\epsilon_{i}q_{i} +
q_{i}^3 + \frac{1}{W}\left(2q_{i}-q_{i-1}
-q_{i+1}\right)\right], \,\,\,\mbox{for}\,\,\,  2\leq i\leq N-1\\
& & \\
\displaystyle p'_{N}&=& \displaystyle p_{N} - \tau\left[\epsilon_{N}q_{N} +
q_{N}^3 + \frac{1}{W}\left(2q_{N}-q_{N-1}\right)\right]\\
& & \\
\displaystyle \delta q'_{i}&=& \displaystyle \delta q_{i}, \,\,\,\mbox{for}\,\,\,  1\leq i\leq N\\
& & \\
\displaystyle \delta p'_{1}&=& \displaystyle \delta p_{1} - \tau\left[
\delta q_{1}\left(\epsilon_{1}+3q_{1}^2\right)
+ \frac{1}{W}\left(2\delta q_{1}-\delta q_{2}\right)\right]\\
& & \\
\displaystyle \delta p'_{i}&=& \displaystyle \delta p_{i} - \tau\left[
\delta q_{i}\left(\epsilon_{i}+3q_{i}^2\right)
+ \frac{1}{W}\left(2\delta q_{i}-\delta q_{i-1}
-\delta q_{i+1}\right)\right], \,\,\,\mbox{for}\,\,\,  2\leq i\leq N-1\\
& & \\
\displaystyle \delta p'_{N}&=& \displaystyle \delta p_{N} - \tau\left[
\delta q_{N}\left(\epsilon_{N}+3q_{N}^2\right)
+ \frac{1}{W}\left(2\delta q_{N}-\delta q_{N-1}\right)\right].
\end{array}
\right.
\label{eq:BV1}
\end{equation}
\section{The 2D DKG model}\label{app:B}
The 2D DKG Hamiltonian (\ref{eq:kg2dham}) can also be written as the sum
of the kinetic energy $T({\bf p})$ and the
potential energy $V({\bf q})$. The propagation
operators for the solution of the variational equations 
are given by the expressions
\begin{equation}
e^{\tau L_{V_{\mathcal{V}}}}: \left\{
\begin{array}{lll}
\displaystyle q'_{i,j}&=& \displaystyle q_{i,j} + \tau p_{i,j} \\
\displaystyle p'_{i,j}&=& \displaystyle p_{i,j}\\
\displaystyle \delta q'_{i,j}&=& \displaystyle \delta q_{i,j} + \tau \delta p_{i,j} \\
\delta p'_{i,j}&=& \displaystyle \delta p_{i,j}
\end{array}
\right. , \,\,\,\mbox{for}\,\,\,  1\leq i\leq N, \,\, 1\leq j\leq M,
\label{eq:AV2}
\end{equation}
and
\begin{equation}
e^{\tau L_{T_{\mathcal{V}}}}: \left\{
\begin{array}{lll}
\displaystyle q'_{i,j}&=& \displaystyle q_{i,j}, \,\,\,\mbox{for}\,\,\,  1\leq i\leq N, \,\, 1\leq j\leq M\\
\displaystyle p'_{1,1}&=& \displaystyle p_{1,1} - \tau\left[\epsilon_{1,1}q_{1,1} + q_{1,1}^3 + \frac{1}{W}\left(4q_{1,1}-q_{2,1}-q_{1,2}\right)\right]\\
\displaystyle p'_{1,M}&=& \displaystyle p_{1,M} - \tau\left[\epsilon_{1,M}q_{1,M} +  q_{1,M}^3 + \frac{1}{W}\left(4q_{1,M}-q_{1,M-1}-q_{2,M}\right)\right]\\
\displaystyle p'_{N,1}&=& \displaystyle p_{N,1} - \tau\left[\epsilon_{N,1}q_{N,1} +  q_{N,1}^3 + \frac{1}{W}\left(4q_{N,1}-q_{N-1,1}
-q_{N,2}\right)\right]\\
\displaystyle p'_{N,M}&=& \displaystyle p_{N,M} - \tau\left[\epsilon_{N,M}q_{N,M} +  q_{N,M}^3 + \frac{1}{W}\left(4q_{N,M}-q_{N-1,M}
-q_{N,M-1}\right)\right]\\
\displaystyle p'_{i,j}&=& \displaystyle p_{i,j} - \tau\left[\epsilon_{i,j}q_{i,j} +  q_{i,j}^3 +\frac{1}{W}\left(4q_{i,j}- q_{i-1,j}  -q_{i,j-1}\right. \right.\displaystyle  \left. \left. -q_{i+1,j}-q_{i,j+1}\right) \bigg] \right. ,\\
& &  \,\,\,\mbox{for}\,\,\,  2\leq i\leq N-1,\,\,2\leq j\leq M-1\\
\displaystyle p'_{i,1}&=& \displaystyle p_{i,1} - \tau\left[\epsilon_{i,1}q_{i,1} +  q_{i,1}^3 + \frac{1}{W}\left(4q_{i,1}-q_{i-1,1} -q_{i+1,1}\right. \right.\displaystyle \left. \left.-q_{i,2}\right)\bigg] \right., \\
& & \,\,\,\mbox{for}\,\,\,   2\leq i\leq N-1\\
\displaystyle p'_{i,M}&=& \displaystyle p_{i,M} - \tau\left[\epsilon_{i,M}q_{i,M} +  q_{i,M}^3 + \frac{1}{W}\left(4q_{i,M}-q_{i-1,M}  -q_{i,M-1}\right. \right.\displaystyle \left. \left.-q_{i+1,M}\right)\bigg] \right.,\\
& &  \,\,\,\mbox{for}\,\,\,   2\leq i\leq N-1\\
\displaystyle p'_{1,j}&=& \displaystyle p_{1,j} - \tau\left[\epsilon_{1,j}q_{1,j} +  q_{1,j}^3 +\frac{1}{W}\left(4q_{1,j}-q_{1,j-1}-  q_{2,j}\right. \right.\displaystyle \left. \left.-q_{1,j+1}\right)\bigg] \right.,\\
& &  \,\,\,\mbox{for}\,\,\,  2\leq j\leq M-1\\
\displaystyle p'_{N,j}&=& \displaystyle p_{N,j} - \tau\left[\epsilon_{N,j}q_{N,j} +  q_{N,j}^3 +\frac{1}{W}\left(4q_{N,j}-q_{N-1,j}  -q_{N,j-1}\right. \right.\displaystyle \left. \left.-q_{N,j+1}\right)\bigg] \right.,\\
& &  \,\,\,\mbox{for}\,\,\,    2\leq j\leq M-1\\

\displaystyle \delta q'_{i,j}&=& \displaystyle \delta q_{i,j}, \,\,\,\mbox{for}\,\,\,    1\leq i\leq N,\,\, 1\leq j\leq M\\
\displaystyle \delta p'_{1,1}&=& \displaystyle \delta p_{1,1} - \tau\left[      \delta q_{1,1}\left(\epsilon_{1,1}+3q_{1,1}^2\right)  +  \frac{1}{W}\left(4\delta q_{1,1}-\delta q_{2,1}- \delta q_{1,2}\right)\right]\\
\displaystyle \delta p'_{1,M}&=& \displaystyle \delta p_{1,M} - \tau\left[  \delta q_{1,M}\left(\epsilon_{1,M}+3q_{1,M}^2\right)   +  \frac{1}{W}\left(4\delta q_{1,M}-\delta q_{1,M-1}-  \delta q_{2,M}\right)\right]\\
\displaystyle \delta p'_{N,1}&=& \displaystyle \delta p_{N,1} - \tau\left[  \delta q_{N,1}\left(\epsilon_{N,1}+3q_{N,1}^2\right)    + \frac{1}{W}\left(4\delta q_{N,1}-\delta q_{N-1,1}  -\delta q_{N,2}\right)\right]\\
\displaystyle \delta p'_{N,M}&=& \displaystyle \delta p_{N,M} - \tau\left[   \delta q_{N,M}\left(\epsilon_{N,M}+3q_{N,M}^2\right)    + \frac{1}{W}\left(4\delta q_{N,M}-\delta q_{N-1,M}   -\delta q_{N,M-1}\right)\right]\\
\displaystyle \delta p'_{i,j}&=& \displaystyle \delta p_{i,j} - \tau\left[  \delta q_{i,j}\left(\epsilon_{i,j}+3q_{i,j}^2\right)   +\frac{1}{W}\left(4\delta q_{i,j}-\delta q_{i-1,j}  -\delta q_{i,j-1}\right. \right.\displaystyle \left. \left.-\delta q_{i+1,j}-\delta q_{i,j+1}\right) \bigg] \right.,\\
& &  \,\,\,\mbox{for}\,\,\,   2\leq i\leq N-1, \,\, 2\leq j\leq M-1 \\
\displaystyle \delta p'_{i,1}&=& \displaystyle \delta p_{i,1} - \tau\left[   \delta q_{i,1}\left(\epsilon_{i,1}+3q_{i,1}^2\right)    + \frac{1}{W}\left(4\delta q_{i,1}-\delta q_{i-1,1}  -\delta q_{i+1,1}\right. \right.\displaystyle \left. \left.-\delta q_{i,2}\right)\bigg] \right.,\\
& &  \,\,\,\mbox{for}\,\,\,  2\leq i\leq N-1\\
\displaystyle \delta p'_{i,M}&=& \displaystyle \delta p_{i,M} - \tau\left[   \delta q_{i,M}\left(\epsilon_{i,M}+3q_{i,M}^2\right)   +\frac{1}{W}\left(4\delta q_{i,M}-\delta q_{i-1,M} -\delta q_{i,M-1}\right. \right.\displaystyle \left. \left.-\delta q_{i+1,M}\right)\bigg] \right.,\\
& &  \,\,\,\mbox{for}\,\,\,    2\leq i\leq N-1\\
\displaystyle \delta p'_{1,j}&=& \displaystyle \delta p_{1,j} - \tau\left[  \delta q_{1,j}\left(\epsilon_{1,j}+3q_{1,j}^2\right)     +\frac{1}{W}\left(4\delta q_{1,j}   -\delta q_{1,j-1}-\delta q_{2,j}\right. \right.\displaystyle \left. \left.-\delta q_{1,j+1}\right)\bigg] \right.,\\
& &  \,\,\,\mbox{for}\,\,\,    2\leq j\leq M-1\\
\displaystyle \delta p'_{N,j}&=& \displaystyle \delta p_{N,j} - \tau\left[   \delta q_{N,j}\left(\epsilon_{N,j}+3q_{N,j}^2\right)    +\frac{1}{W}\left(4\delta q_{N,j}-\delta q_{N-1,j}  -\delta q_{N,j-1}\right. \right.\displaystyle \left. \left.-\delta q_{N,j+1}\right)\bigg] \right.,\\
& &  \,\,\,\mbox{for}\,\,\,    2\leq j\leq M-1.\\
\end{array}
\right.
\label{eq:BV2}
\end{equation}

\end{appendix}


\bibliographystyle{abbrvnat}
\bibliography{library}

\printindex


\end{document}